# A Taxonomy and Archetypes of Business Analytics in Smart Manufacturing

**Wanner, Jonas**

University of Würzburg

**Wissuchek, Christopher**

Friedrich-Alexander-Universität Erlangen-Nürnberg (FAU)

**Welsch, Giacomo**

University of Würzburg

**Janiesch, Christian**

TU Dortmund University

## Abstract

*Fueled by increasing data availability and the rise of technological advances for data processing and communication, business analytics is a key driver for smart manufacturing. However, due to the multitude of different local advances as well as its multidisciplinary complexity, both researchers and practitioners struggle to keep track of the progress and acquire new knowledge within the field, as there is a lack of a holistic conceptualization. To address this issue, we performed an extensive structured literature review, yielding 904 relevant hits, to develop a quadripartite taxonomy as well as to derive archetypes of business analytics in smart manufacturing. The taxonomy comprises the following meta-characteristics: application domain, orientation as the objective of the analysis, data origins, and analysis techniques. Collectively, they comprise eight dimensions with a total of 52 distinct characteristics. Using a cluster analysis, we found six archetypes that represent a synthesis of existing knowledge on planning, maintenance (reactive, offline, and online predictive), monitoring, and quality management. A temporal analysis highlights the push beyond predictive approaches and confirms that deep learning already dominates novel applications. Our results constitute an entry point to the field but can also serve as a reference work and a guide with which to assess the adequacy of one's own instruments*

**Keywords:** Business Analytics, Smart Manufacturing, Literature Review, Industry 4.0, Taxonomy, Archetypes.

## Introduction

Driven by technological innovations such as cyber-physical systems (CPS) and the (Industrial) Internet of Things (IoT), the age of the fourth industrial revolution is bringing disruptive and radical changes to value networks, business models, and business operations in manufacturing. Thereby, the physical world has become linked to digital entities. Smart objects communicate with each other and humans in real time (Hermann, Pentek, & Otto, 2016), leaving massive amounts of digital traces, also termed "big data", which are the vehicle for intelligent automation (Grover, 2019) and novel business analytics (BA) applications to enable so-called smart manufacturing. The successful application of BA in smart manufacturing can lead to cost advantages, increased customer satisfaction, and improvements in production effectiveness and quality (Fay & Kazantsev, 2018).

Studies confirm that a majority of organizations are aware of the importance of BA in smart manufacturing (Haas, 2018). However, only a small proportion apply advanced analytics or fully exploit its potential (Bughin, Manyika, & Woetzel, 2016; Derwisch & Iffert, 2017). This can be attributed to high technological and organizational barriers as well as substantial implementation costs (LaValle, Lesser, Shockley, Hopkins, & Krschwitz, 2011; Z.-H. Zhou, Chawla, & Jin, 2014). While there has been a delay in the adoption of advanced analytics in practice, researchers have already embraced applications of BA in smart manufacturing with a variety of different research efforts (e.g., Arık & Toksarı, 2018; Aydın, Karaköse, & Akın, 2015; O. Aydin & Guldamlasioglu, 2017) that have not yet been systematized.

With our research, we tackle this issue and identify two underlying problems: i) lack of a holistic conceptualization of the area of research and ii) formidable complexity-related barriers for practitioners.

As a way of structuring areas of interest, taxonomies have been proven useful in both the natural and social sciences. They are increasingly being used in information systems research as they enable the conceptualization and classification of objects (Nickerson, Varshney, & Muntermann, 2013) and can serve as theories (Gregor, 2006) or for theory building (Bapna, Goes, Gupta, & Jin, 2004; Doty & Glick, 1994). Hence, we deem them a suitable artifact for our research objectives. Furthermore, the identification of recurring patterns and the establishment of reusable knowledge can support both researchers and practitioners (A. Brodsky et al., 2017; Zschech, 2018). Recurring patterns represent guidelines or templates and enable the synthesis of existing knowledge in a cumulative form to extract specific archetypes (Russo, 2016).

As of today, there is an apparent knowledge gap with respect to common ground for central concepts to integrate perspectives and capabilities, as well as to assess the transformative potential of BA for smart manufacturing. To address this gap, we extend the conceptualization of BA in this context by discerning distinct archetypes of analytics applications based on their common characteristics and exploring their importance for smart manufacturing. Taxonomies are not static and set in stone, but rather evolve over time as a research domain shifts its priorities (Nickerson et al., 2013). How a research domain has changed in the past, how interest has shifted over the years, and the direction of current research trends are not readily apparent from the taxonomy itself. To address this, we perform a temporal analysis to both demonstrate and visualize how the taxonomy's inherent structure and the derived archetypes or recurrent patterns have evolved. While enabling a coherent understanding of the past and current priorities, this also reveals research gaps and possible research trends that may gain traction, empowering both researchers and practitioners to position their work as well as anticipate future considerations.

We summarize these objectives with the following three research questions (RQ):

**RQ1:** What is a taxonomy that enables to conceptualize and structure the application of business analytics in smart manufacturing?

**RQ2:** What archetypes of business analytics in smart manufacturing can we derive from the taxonomy and its underlying data?

**RQ3:** Which temporal variations or trends can be distinguished for business analytics in smart manufacturing?

We analyzed 904 articles on smart manufacturing to build and validate our taxonomy, derive archetypes, and understand temporal variations in research. Through our research, we address two user groups: researchers and practitioners who create and assess BA artifacts and theories. Novices especially can benefit from the comprehensiveness of our survey when first entering the field. By answering our research question, we provide the following concrete contributions to the fields of BA and smart manufacturing.

1. We create a taxonomy that acts as a theory-building artifact for BA in smart manufacturing research to support structuring the scientific discussion and to enable scientists new to the field to access a comprehensive overview of relevant dimensions and characteristics. For example, these can be used as a codebook for labeling the data and for automated explorative analysis.
2. We use cluster analysis to derive and analyze archetypes of BA in smart manufacturing present in contemporary research. These help to bring structure to the field and enable us to understand which facets of BA exist in smart manufacturing, how they are composed, and how distinct areas (e.g., production monitoring and security surveillance) may inspire each other through their use of similar techniques such as real-time analysis.
3. Finally, our survey enables us to see how research has evolved over time and to discover that certain methods, such as deep learning, have experienced an unprecedented uptick of research innovation. Discussion of the results allows for a better understanding of past and current priorities in the field. Likewise, it becomes possible to evaluate research trends, which helps researchers and practitioners alike to classify, structure, and evaluate their work. In addition, our results can serve as a guide for both customers and vendors of BA in smart manufacturing to classify, compare, and evaluate existing applications, as well as design new applications with explicit consideration of relevant features for the intended type of application.

For this purpose, we provide the theoretical background necessary to establish the context of our work in Section 2. Section 3 comprises our research methodology. We answer RQ1 in Section 4 by describing the data collection, building the taxonomy, and introducing the taxonomy artifact. We answer RQ2 in Section 5 based on a cluster analysis and discussion of the identified archetypes. Subsequently, we answer RQ3 in Section 6 by using time analysis on both the taxonomies' characteristics and the respective archetypes. In Section 7, we discuss our contributions before closing with a conclusion and an outlook.

## Context and Theoretical Background

While we consider all kinds of BA methods for our study, we only examine those relevant to the context of our research (Johns, 2006), that is, smart manufacturing. This contextualization enables us to link our observations to the point of view of smart manufacturing and its related facts (Rousseau & Fried, 2001). In order to do so, in the following we highlight those facts that define this field and determine the significance and appearance of BA applications. Context, in this sense, comprises organizational and technical characteristics such as application areas, data origin and handling, as well as the intentions underlying the use of BA. This context provides the constant that surrounds our research (Johns, 2006). Consequently, as context we introduce the age of the fourth industrial revolution and the cornerstones of smart manufacturing. Furthermore, we introduce the topic of business analytics and its approaches, as well as an overview of related scientific surveys.

### Fourth Industrial Revolution

Technical innovations have continuously advanced industrialization. Figure 1 classifies these advances into four evolutionary steps: i) the introduction of mechanical production facilities using hydro- and steam power at the end of the 18th century, ii) the introduction of the division of labor and mass production using electric energy at the end of the 19th century, and iii) the use of electronics and IT to further automate production in the mid-20th century.

Today, the iv) fourth revolutionary industrial change is already in progress and has enabled the ubiquitous connectivity of CPS through the Industrial IoT (Bauer, Schlund, Marrenbach, & Ganschar, 2014).

CPS link the physical world with the digital world. They comprise objects, devices, robots, buildings, and products, as well as production facilities that feature communication capabilities through system and user interfaces, capture their environment through sensors, process data, and interact with the physical world through actuators (Bauernhansl, 2014; Broy, 2010). CPS are interconnected, enabling communication and data exchange (Broy, 2010; Vaidya, Ambad, & Bhosle, 2018) to control and monitor the physical processes (Broy, 2010; E. A. Lee, 2008). The Industrial IoT is the underlying communication and interaction network for physical objects on shop floors (Gubbi, Buyya, Marusic, & Palaniswami, 2013).

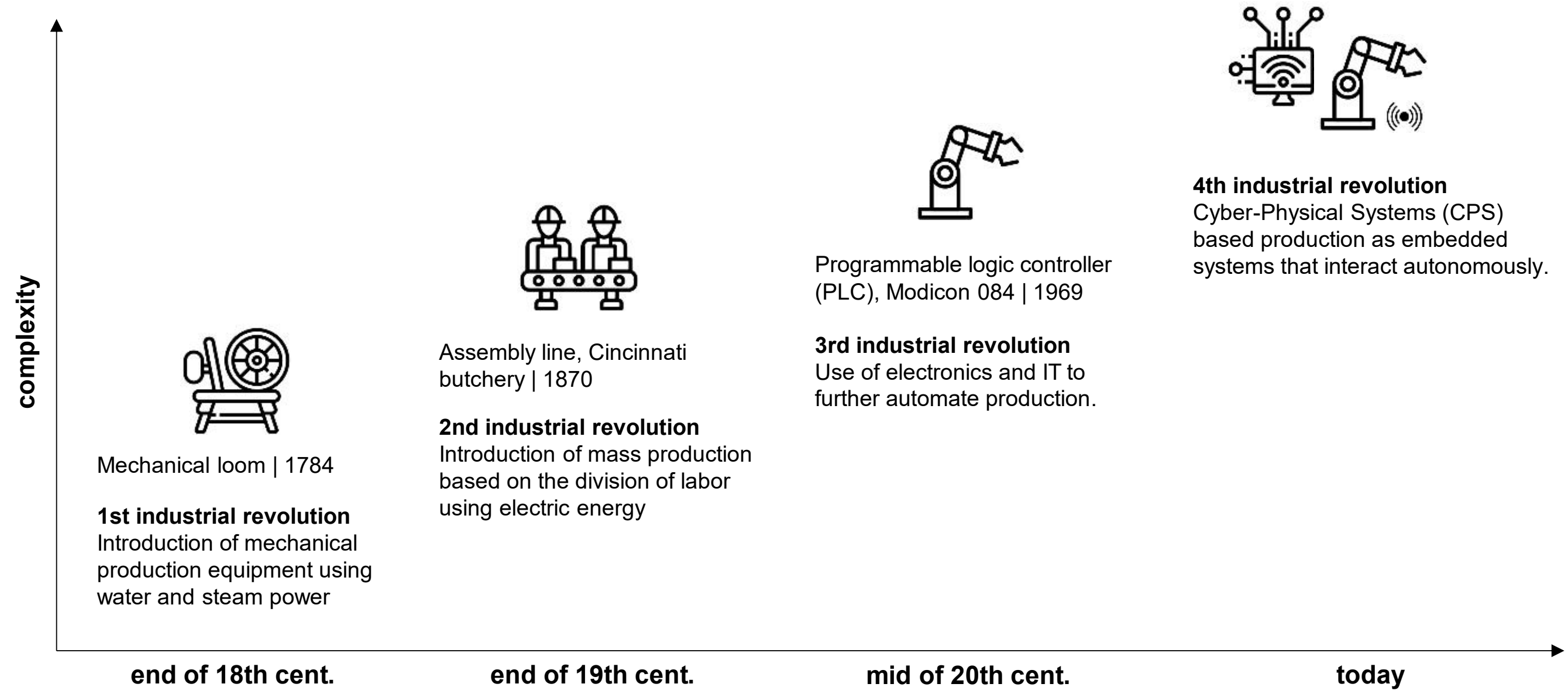

**Figure 1. Evolution of Industrial Revolutions (cf. Bauer et al., 2014)**

Although several definitions for this fourth industrial revolution exist, there is no consensus on its constituent properties (Bauer et al., 2014; Bauernhansl, 2016; Hermann et al., 2016). In addition to the German (*Industry 4.0*) and U.S. (*Advanced Manufacturing*) initiatives, comparable efforts to establish high-tech manufacturing are taking place in France (Conseil national de l'industrie, 2013), the UK (Foresight, 2013), South Korea (Kang et al., 2016), the EU (Europäische Kommission, 2016), Singapore (National Research Foundation, 2016), and China (K. Li, 2015).

### Smart Manufacturing

While Industry 4.0 and Advanced Manufacturing designate initiatives undertaken as part of the fourth industrial revolution, the term *smart manufacturing* predominates as a description of production in so-called smart factories (e.g., Chiang, Jiang, Zhu, Huang, & Braatz, 2015; C.-F. Chien, Chang, & Wang, 2014; C.-F. Chien, Diaz, & Lan, 2014; C.-F. Chien, C.-Y. Hsu, & P.-N. Chen, 2013a). Other terms denoting related concepts are the *Industrial Internet* (Evans & Annuziata, 2012; J.-Q. Li et al., 2017), *integrated industry* (Bauernhansl, 2016), *intelligent manufacturing* (R. Y. Zhong, Xu, Klotz, & Newman, 2017), *cloud manufacturing* (Y. Liu & Xu, 2017), and *smart industry* (Haverkort & Zimmermann, 2017; N. Kaur & Sood, 2015). Utilizing Hermann et al.'s quantitative and qualitative literature analysis to structure the field (Hermann, Pentek, & Otto, 2015; Hermann et al., 2016), we define smart manufacturing as "a collective term for technologies and concepts of value chain organization. Within the modular structured smart factories of Industry 4.0, CPS monitor physical processes, create a virtual copy of the physical world, and make decentralized decisions. Over the IoT, CPS communicate and cooperate with each other and humans in real-time" (Hermann et al., 2015).

Thus, smart manufacturing takes place in an interconnected industrial environment based on CPS and the Industrial IoT. It exhibits several additional characteristics of interest for researchers and practitioners. First and foremost, smart manufacturing takes place in a value network, that is a system of individual value creation processes, which is realized by autonomous, legally independent actors. Its integration provides flexibility, new business opportunities, chances for (partial) automation, as well as intelligence and interchangeability to address greater manufacturing complexity, dynamics-based economics, and radically different performance objectives. Its networked, real-time, information-based setup transforms reactive responses into predictive or even prescriptive approaches, shifts the focus from incident response to prevention, and replaces vertical decision-making with distributed intelligence in order to enable local decision-making with global impact (Davis, Edgar, Porter, Bernaden, & Sarli, 2012).

Smart manufacturing relies on several key technologies. Apart from CPS and the IoT, most authors include cloud computing (Mell & Grance, 2011) and (big) data analytics (Müller, Junglas, Brocke, & Debortoli, 2016) as central enablers for smart manufacturing. Furthermore, technologies such as cybersecurity (Wells, Camelio, Williams, & White, 2014), additive manufacturing (Kang et al., 2016; Thiesse et al., 2015), the Internet of Services (Terzidis, Oberle, Friesen, Janiesch, & Barros, 2012), visualization technologies (Paelke, 2014; Posada et al., 2015), and simulation (Smith, 2003) are crucial, as well.

### Business Analytics

Analytics is a collection of methods, technologies, and tools for creating knowledge and insight from data to solve complex problems and make better and faster decisions (Delen & Zolbanin, 2018). A definition of analytics can be abstracted into three dimensions (Holsapple, Lee-Post, & Pakath, 2014 ): i) domain, ii) methods, and iii) orientation. The domain is the field of application – i.e., the context – of analytics: for example the supply chain (Trkman, McCormack, de Oliveira, & Ladeira, 2010) or manufacturing (R. Y. Zhong et al., 2017). Methods comprise different techniques for analyzing data. Orientation describes the direction of thinking or the objective of the analysis. It is not idiosyncratic towards a domain and is considered the core perspective of analytics (Delen & Zolbanin, 2018; Holsapple et al., 2014).

Prior to the 1970s, domain experts collected and evaluated data using traditional mathematical and statistical methods. These systems were referred to as operations research systems. Due to the spread of integrated enterprise information systems, so-called enterprise resource planning (ERP) systems, analytics has attracted more and more attention and has been consistently further developed (Delen, 2014). Later, the ability to derive insights by descriptively analyzing historical data became valuable under the umbrella term of business intelligence.

Today, analytics is a multifaceted and interdisciplinary field of research. Holsapple et al. (2014) generated a taxonomic and definitional framework consisting of six classes to describe the term: i) analytics is a corporate cultural movement in which, among other things, fact-based decisions are made. In addition, ii) analytics includes a collection of technologies and approaches, iii) a transformative process, and iv) an assortment of (organizational) skills, as well as v) specific activities, which subsequently lead to vi) optimized decisions and insights. Mortenson, Doherty, and Robinson (2015) complement these classes' interdisciplinarity with technological approaches rooted in computer science and engineering; quantitative methods from mathematics, statistics, and econometrics; and decision-supporting aspects from psychology and behavioral science. Analytics is further based on the integration of the interdisciplinary research domains of artificial intelligence and machine learning (ML), information systems, and operations research (Delen & Zolbanin, 2018; Mortenson et al., 2015).

### Business Analytics Approaches

Categorization into maturity levels is a widely used classification system to illustrate the orientations and objectives of BA approaches. Their characteristics are related to their increasing complexity and their business potential (see Figure 2). Based on Davenport and Harris (2007) and Lustig, Dietrich, Johnson, and Dziekan (2010), analytics can be categorized into three types: i) descriptive analytics, ii) predictive analytics, and iii) prescriptive analytics. New works extend these categories by iv) diagnostic analytics positioned between descriptive analytics and predictive analytics (Banerjee, Bandyopadhyay, & Acharya, 2013; Delen & Zolbanin, 2018).

**Descriptive Analytics** seeks to create transparency, often through data visualization. It aims at providing answers to the questions "What is happening?" or "What is happening right now?" Tools for analysis include periodic/ad hoc and dynamic/interactive online analytical processing (OLAP) and reporting, as well as exploratory ML algorithms such as clustering. Descriptive analytics typically uses historical data to uncover business potentials and problems that can support decision-making. Business intelligence is frequently used as a synonym for descriptive analytics (Delen & Demirkan, 2013; Delen & Zolbanin, 2018).

**Diagnostic Analytics** processes data to answer the question "Why did it happen?" It leverages techniques from descriptive analytics and methods from data discovery and statistics, including explanatory statistical modeling (Shmueli, 2010) to determine the cause of problems or incidents (Banerjee et al., 2013; Delen & Zolbanin, 2018).

Due to its emphasis on the past, insights from diagnostic analytics may not be appropriate for predicting future events (Shmueli, 2010).

**Predictive Analytics** determine phenomena that are likely to happen (e.g., trends and associations). Consequently, it aims at answering the question "What will happen?" To this end, it typically uses methods from ML and forecasting, such as decision trees, random forests, and artificial neural networks (Breiman, 2001; Shmueli, 2010). In ML, the intention is to find generalizable associations between the predictors (i.e., the independent variables) and the target (i.e., the dependent variable), an approach usually referred to as supervised learning (Fayyad, Piatetsky-Shapiro, & Smyth, 1996; Marsland, 2015).

**Prescriptive Analytics** processes data to estimate the best possible alternative under certain constraints by using methods of mathematical optimization that calculate the optimal measure or decision. The question to answer is "What must be done?" Prescriptive analytics can help to take advantage of future events or mitigate potential risks by presenting the implications of the various options for action (Delen, 2014; Delen & Zolbanin, 2018; Lustig et al., 2010). Both predictive analytics and prescriptive analytics are referred to as advanced analytics (Delen, 2014).

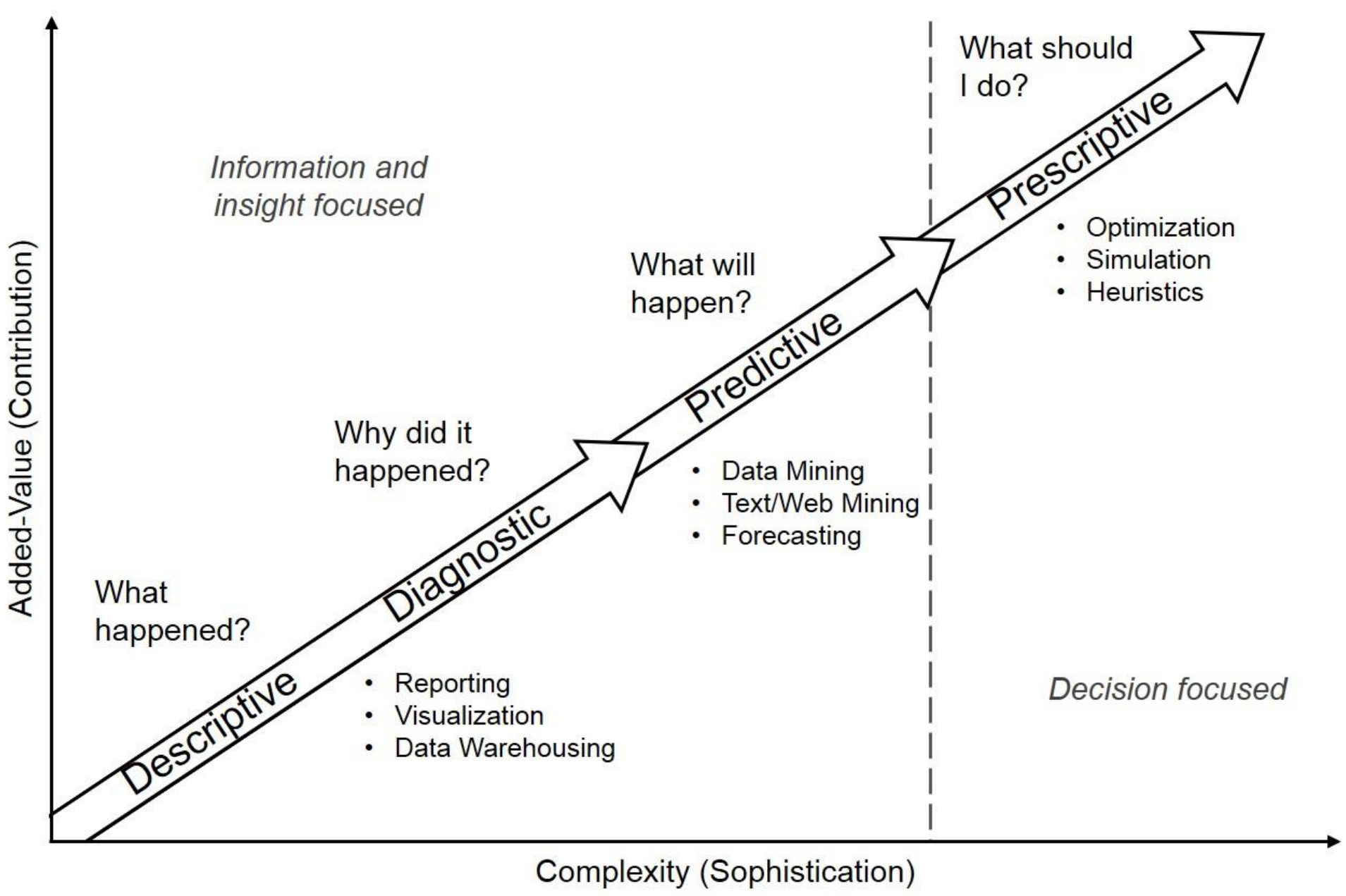

**Figure 2. Maturity Level of Analytics (cf. Delen & Zolbanin, 2018)**

### Business Analytics in Smart Manufacturing

Interdisciplinary research on data-driven smart manufacturing has existed for quite some time. However, there is still an inadequate structuring of this knowledge and thus a barrier to implementation in practice. Using a systematic search of survey articles on BA in smart manufacturing as a pre-test (see our search process in Section 4.1), we identified 39 publications that survey BA in smart manufacturing. In these surveys, we identified (1) a lack of a holistic synthesis of the research area and determined that there are (2) complexity-related barriers to the practical application. Table 1 uses Harvey balls to illustrate the degree to which these research gaps have been addressed, differentiating between the categories *fully addressed*, *partially addressed*, and *not addressed*.[1]

None of the publications fully addresses the first research gap of a (1) holistic synthesis. Most authors examine only certain topics of BA and position their work in the broader context or sub-areas of smart manufacturing. This means that they focus on specific issues or topics such as the business value of applying BA (Bordeleau, Mosconi, & Santa-Eulalia, 2018; Fay & Kazantsev, 2018; T. Wuest, Weimer, Irgens, & Thoben, 2016), technological or algorithmic aspects concerning the use of ML (Diez-Olivan, Del Ser, Galar, & Sierra, 2019; Sharp, Ak, & Hedberg, 2018; T. Wuest et al., 2016), or selective applications in smart manufacturing (O'Donovan, Leahy, Bruton, & O'Sullivan, 2015b; X. Y. Xu & Hua, 2017). Some authors review the use of BA in specific domains such as maintenance (X. Y. Xu & Hua, 2017; Zschech, 2018), process monitoring (Sutharssan, Stoyanov, Bailey, & Yin, 2015; Y. Zhou & Xue, 2018), or production planning (Cadavid, Lamouri, Grabot, Pellerin, & Fortin, 2020; M. S. Reis & Gins, 2017; Sutharssan et al., 2015). Some offer a categorization schema when doing so.

[1] In describing the comprehensiveness of each survey, (1) *fully addressed* reflects that the publication takes a broad view on smart manufacturing, *partially addressed* means that it focuses on a specific area or key technology, and *not addressed* suggests that it does not try to synthesize the research area at all. Regarding the reduction of complexity, (2) we classified each survey as *fully addressed* if the goal is to transfer the synthesized research results to industry and if it is structured accordingly to make these insights applicable in practice (e.g., to present recurring patterns or applications of analytics in industrial practice). *Partially addressed* means results are prepared for practical use, but not in a structured and reusable form. Lastly, *not addressed* suggests that the contribution does not provide for the utilization of analytics in industrial practice.

**Table 1. Verification of Problem Relevance and Research Gaps**

| **Research Gap** | Bang, Ak, Narayanan, Lee, and Cho (2019) | Baum, Laroque, Oeser, Skoogh, and Subramaniyan (2018) | Bordeleau et al. (2018) | Bousdekis, Magoutas, and Mentzas (2015) | Cadavid et al. (2020) | Çaliş and Bulkan (2015) | Cardin et al. (2017) | Cerrada et al. (2018) | Y. Cheng, Chen, Sun, Zhang, and Tao (2018) | Dalzochio et al. (2020) | Diez-Olivan et al. (2019) | H. Ding et al. (2020) | Dowdeswell, Sinha, and MacDonell (2020) | Fay and Kazantsev (2018) | Gölzer, Cato, and Amberg (2015) | Gölzer and Fritzsche (2017) | Khan and Yairi (2018) | D.-H. Kim et al. (2018) | S. L. Kumar (2017) | Y.-H. Kuo and Kusiak (2018b) | J. Lee, Wu, et al. (2014) | G. Y. Lee et al. (2018) | Yaguo Lei et al. (2018) | Nath, Udmale, and Singh (2020) | O'Donovan, Leahy, Bruton, and O'Sullivan (2015a) | Precup, Angelov, Costa, and Sayed-Mouchaweh (2015) | Priore, Gómez, Pino, and Rosillo (2014) | M. S. Reis and Gins (2017) | Sharp et al. (2018) | Sutharssan et al. (2015) | T. Wuest et al. (2016) | Y. Xu, Sun, Wan, Liu, and Song (2017) | X. Y. Xu and Hua (2017) | Zarandi, Asl, Sotudian, and Castillo (2018a) | G. Zhao, Zhang, Ge, and Liu (2016) | Y. Zhou and Xue (2018) | W. Zhang, Yang, and Wang (2019) | Zonta et al. (2020) | Zschech (2018) |
|---|---|---|---|---|---|---|---|---|---|---|---|---|---|---|---|---|---|---|---|---|---|---|---|---|---|---|---|---|---|---|---|---|---|---|---|---|---|---|---|
| **1** | ◐ | ◐ | ◐ | ◐ | ◐ | ◐ | ◐ | ◐ | ◐ | ◐ | ◐ | ◐ | ◐ | ○ | ◐ | ◐ | ◐ | ◐ | ○ | ○ | ◐ | ◐ | ○ | ◐ | ◐ | ○ | ◐ | ◐ | ◐ | ○ | ○ | ◐ | ◐ | ◐ | ◐ | ◐ | ○ | ◐ | ◐ |
| **2** | ○ | ○ | ○ | ○ | ○ | ○ | ○ | ○ | ○ | ● | ○ | ○ | ○ | ○ | ● | ● | ○ | ○ | ○ | ○ | ● | ○ | ○ | ○ | ○ | ○ | ○ | ○ | ○ | ○ | ○ | ○ | ○ | ○ | ○ | ○ | ○ | ○ | ● |

● *Research gap fully addressed* | ◐ *Research gap partially addressed* | ○ *Research gap not addressed*

We identified five surveys that adequately address (2) complexity-based barriers. Zschech (2018) and Dalzochio et al. (2020) each develop a taxonomy for maintenance analytics to provide researchers and practitioners with a reusable knowledge base of analytical techniques, data characteristics, and objectives. Gölzer et al. (2015) and Gölzer and Fritzsche (2017) identify organizational and technical requirements for data and analytical processing to prepare practitioners for smart manufacturing. J. Lee, Wu, et al. (2014) develop a taxonomical framework for selecting optimal algorithms for prognosis and health management of rotary machinery. These contributions reduce complexity for researchers and practitioners. Nevertheless, a comprehensive examination of BA in smart manufacturing – not limited to specific scenarios, domains, or techniques – is still missing.

## Methodology

To address both research gaps, we develop a taxonomy as well as archetypes and examine them for temporal variations and trends. To do so, we apply a multi-step approach. Our methodology can be divided into three main parts: i) taxonomy development, ii) archetypes derivation, and iii) temporal analysis.

**Taxonomy Development.** First, we carry out the necessary preliminary work to identify suitable foundations for our taxonomy by employing a structured literature search process (Section 4.1) following the method proposed by vom Brocke et al. (2015). We used these publications as input for taxonomy building (Section 4.2). Taxonomies are a structured result of classifying things or concepts, including the principles underlying such classification. Nickerson et al. (2013) define a taxonomy $T$ as a set of $n$ dimensions $D_i$ (i=1, …, n), each consisting of $k_i$ ($k_i \geq 2$) mutually exclusive and complete characteristics $C_{ij}$ (j=1, …, $k_i$), so that each object has only one $C_{ij}$ for each $D_i$:

$$T = \{D_i, i = 1, \dots n \mid D_i = \{C_{ij}, j = 1, \dots, k_i; k_i \geq 2\}\}$$

Here, 'mutually exclusive' indicates that no object can have two different characteristics in one dimension, while 'complete' states that objects must have at least one characteristic for each dimension. A few authors, who have developed taxonomies according to Nickerson et al. (2013), criticize this restriction because some objects clearly have hierarchical and combinatorial relationships between characteristics. This could lead to confusing taxonomies by introducing additional dimensions or characteristics. Accordingly, these authors recommend performing the taxonomy development without the restriction (e.g., Jöhnk, Röglinger, Thimmel, & Urbach, 2017; L. Püschel, Schlott, & Röglinger, 2016a; Zschech, 2018). We followed the restrictions of Nickerson et al. (2013) in the first iteration, but renounced it in the subsequent iterations (for more details on the taxonomy building process, see Appendix B).

After completing the development of a taxonomy and describing it (Section 4.3), it is necessary to evaluate its applicability in the domain under consideration (Nickerson et al., 2013). Szopinski, Schoormann, and Kundisch (2019) suggest the use of illustrative scenarios as a way of demonstrating the usefulness of a taxonomy (Section 4.4). We strengthen this by additionally performing a cluster analysis to derive archetypes so that researchers can position their approaches in, or between, research streams – or so that practitioners can select a more specific purpose domain to guide them in real-world implementations initially.

**Archetypes Derivation.** Cluster analysis is an appropriate means by which to gain understanding from data (Jahirabadkar & Kulkarni, 2013). It aims at dividing a dataset into subsets (i.e., clusters), each containing observations that are similar to each other and dissimilar to observations in different clusters (Jahirabadkar & Kulkarni, 2013; Kaufman & Rousseeuw, 2009). Agglomerative hierarchical clustering is a popular clustering approach that performs a hierarchical separation of observations by their degree of similarity. It is frequently used in information systems research for deriving archetypes from taxonomies and corresponding data (e.g., Oses, Legarretaetxebarria, Quartulli, Garcia, & Serrano, 2016; Pandiyan, Caesarendra, Tjahjowidodo, & Tan, 2018; Jiten Patel & Choi, 2014). As "clustering is a subjective process in nature, which precludes an absolute judgment as to the relative efficacy of all clustering techniques" (R. Xu & Wunsch, 2005), it is necessary to explore the suitability of different clustering algorithms and distance measures.

Therefore, we applied different reasonable combinations of popular agglomerative clustering algorithms (i.e., single linkage, complete linkage, group average linkage, centroid linkage, and Ward's method) (R. Xu & Wunsch, 2005) and distance measures suitable for binary input streams (i.e., Euclidean distance, squared Euclidean distance, Jaccard distance, Hamming distance, Dice distance, and Yule distance) (S.-S. Choi, Cha, & Tappert, 2010). Subsequently, we evaluated the different outcomes (i.e., dendrograms and crosstab analyses) by intensively discussing them within a group of four proficient researchers. Ultimately, we agreed that the combination of Ward's algorithm and the Euclidean distance led to the most reasonable clusters (Section 5.1). This result is in line with several contributions in the information systems literature that were able to derive meaningful archetypes from taxonomies and corresponding data by applying Ward's algorithm and a Euclidean distance metric (Ragab, Ouali, Yacout, & Osman, 2016; Ragab, Yacout, Ouali, & Osman, 2017; Ranjit et al., 2015). Subsequently, we introduce and explore the characteristics of the archetypes (Section 5.2).

**Temporal Analysis.** Temporal statistical analysis can be used to examine and model changes in variables of a dataset over a period of time. This allows researchers to draw conclusions about changes of concentrations. In temporal analysis, the behavior of the variable under consideration is modeled as a function of its previous data points in the same series using common tabular representations. A further visualization of this data presents a superficial view of the variety of data by providing quick ways of grasping important areas. As such, it supports trend analysis through the identification of rare anomalies that are increasing in frequency (Vogel, 2020).

In our analysis, we take a descriptive analytics approach. We do not include consideration and explanations for influencing factors (Vogel, 2020). We use time series plots of the raw data. Here, the time is plotted on the *x*-axis and the observation(s) of the data series on the *y*-axis. For our temporal trends of BA applications in smart manufacturing, we employ a two-staged approach. First, we consider changes within the dimensions and characteristics of the created taxonomy. In this way, we derive thematic trends that point towards future developments within the research field. Second, we consider changes within the derived archetypes. In this way, we enable the reader – be it as a researcher or as a practitioner – to develop awareness of future trends in specific coherent clusters of application.

## Taxonomy Development

During our taxonomy development, we pass through the following steps: i) data collection; ii) taxonomy building; iii) taxonomy representation; and, finally, iv) showcasing the taxonomy using three illustrative examples.

**Literature Search Process and Data Collection**

Our systematic literature search process follows the methodology recommendations of vom Brocke et al. (2015). Before conducting the search process, we first define the scope of the search and corresponding keywords.

**Search Scope.** To define and present the scope of our structure literature review, we made use of the taxonomy of Cooper (1988): Our focus is on research and applications. The aim of the research is the integration of the BA research topics in the context smart manufacturing, as well as the exploration of related topics and key technologies. Furthermore, we chose a neutral representation with a representative coverage. The results are to be methodically organized according to Nickerson et al. (2013). We define technology and management-oriented audiences as our target groups. To conduct the search process, we chose interdisciplinary databases – i.e., *IEEE Xplore*, *AISeL*, and *ACM Digital Library* – to cover IT-related research areas. For business research, we included *Business Source Premier* (EBSCO). Lastly, we added *Web of Science* and *ScienceDirect*.

**Table 2. Keywords for Literature Review**

| Domain | Group | Keywords |
|---|---|---|
| **Analytics** | General | Analytics<br>Data Science<br>Business Intelligence<br>Big Data |
| | Methods | Data Mining<br>Machine Learning<br>Cognitive Computing<br>Statisti*<br>Artificial Intelligence |
| | **AND** | |
| **Smart Manufacturing** | General and Related Terms | Industrie 4.0<br>Industry 4.0<br>Smart Manufacturing<br>Integrated Industry<br>Industrial Internet<br>Smart Industry<br>Smart Factory<br>Advanced Manufacturing<br>Intelligent Manufacturing |
| | Key Technologies | Industrial Internet of Things<br>*Cyber-physical Systems |

**Keywords.** To create a comprehensive and uniform search string, we defined the search terms shown in Table 2. This is in line with the explanation of terms and connections given in Section 2. For our search of surveys (Section 2.5), we combined the keywords for surveys and reviews with keywords from Table 2.

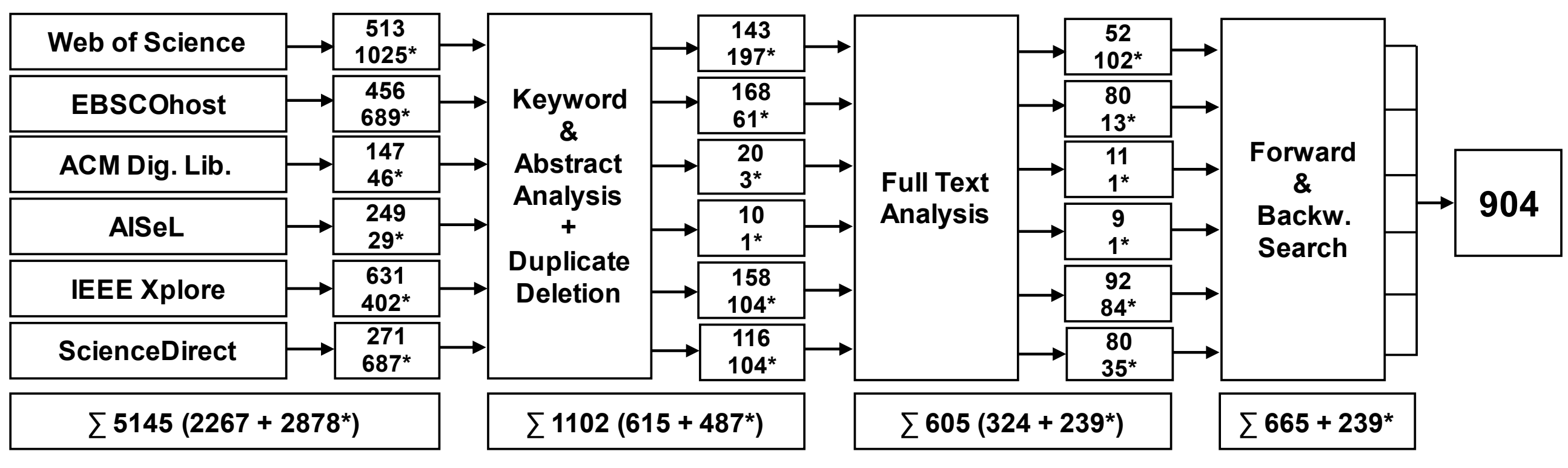

**Figure 3. Literature Search Results by Database**

As Nickerson et al. (2013) conclude, taxonomies are not static but evolve over time as new objects are developed or identified. After an initial search, we revisited the taxonomy at a later point. In total, we screened 5145 publications in two searches and finally considered 904 papers relevant for taxonomy building and evaluation. In the following, we summarize the search processes (see also Figure 3).

**Initial search (date: 21 Nov. 2018).** To ensure timeliness and relevance, we limited the earliest year of publication to 2013. This is the year in which the article of Kagermann, Wahlster, and Helbig (2013), now regarded as the seminal publication for the development of the research field of smart manufacturing, first appeared. In our initial search process, we identified 2267 results. After keyword and abstract analysis, as well as deleting duplicates, 615 publications remained. Following the full-text screening, we considered 324 publications to be relevant. Finally, we conducted a forward and backward search, which yielded another 341 relevant search results for a total of 665 publications. We performed the forward search by analyzing citation data on Google Scholar. We carried out the backward search manually via the bibliography of the contributions.

**Second search (date: 01 Nov. 2020).** We used the date of the initial search as a starting point with unaltered keywords, filters, and databases. This resulted in 4179 hits, reflecting the rising amount of research interest in the domain under consideration. Due to the large number of results, we decided to filter the results for the sake of manageability and excluded all publications except peer-reviewed journals, which were represented in 2878 hits. The abstract and keyword analysis resulted 487 relevant articles. For further manageability, we eliminated all articles with a journal impact factor of less than 2.000 in 2020[2] and decided to forgo the backward and forward search, since we already carried out a broad coverage of the relevant time frame in the initial search. After a full-text screening, 239 publications remained, resulting in a total of 904 publications. The full bibliography is available in Appendix A.

**Table 3. Meta-characteristics**

| # | Meta-characteristic | Related Practitioner's Questions of Interest |
|---|---|---|
| MC.1 | Domain | Which smart manufacturing domains are affected by BA? |
| MC.2 | Orientation | What is the orientation of the use of BA in smart manufacturing? |
| MC.3 | Data | What are the characteristics of the processed data? |
| MC.4 | Technique | Which analytical techniques are used? |

### Taxonomy Building

**Meta-Characteristics.** Meta-characteristics form the basis for the assignment of further characteristics. This prevents naive empiricism in which the researcher assigns a large number of related and unrelated characteristics for the purpose of identifying undiscovered patterns (Nickerson et al., 2013). The objective of our contribution is to structure the research area of BA in smart manufacturing holistically and to reduce complexity barriers for practitioners by creating reusable knowledge. The latter objective is achieved if the taxonomy allows for the derivation of analytics patterns according to Russo (2016). We decided on quadripartite meta-characteristics to classify BA patterns as objects (see Table 3).

Holsapple et al. (2014) define three dimensions of relevance that we use: i) *domain* is the area of application of BA in smart manufacturing (e.g., production or quality control); ii) *orientation* refers to the direction of thought (e.g., predict vs. prescribe) and is concerned with the objective of applying BA (e.g., cost reduction); iv) *technique* describes the way a BA task is performed (e.g., using ML). Furthermore, we added a meta-characteristic based on C.-W. Tsai, Lai, Chiang, and Yang (2013): iii) *data* describes the underlying properties of the data available for BA in smart manufacturing.

[2] https://clarivate.com/webofsciencegroup/solutions/
journal-citation-reports/

**Table 4. Performed Iterations in Taxonomy Development**

| It. | Approach | Summary | Quantities | Input |
|---|---|---|---|---|
| I | **Conceptual-to-Empirical** | Due to limited availability of real-world data, we analyzed 10 articles that structure a part of the research field and employ categorization schemes to derive an initial set of dimensions and characteristics (cf. Appendix A for details on the selected articles). | D=13; C=53; P=10 | Initial Literature Survey |
| II | **Conceptual-to-Empirical** | We performed an analysis of related work that synthesizes a specific domain of smart manufacturing (e.g., maintenance), but does not offer a categorization scheme. | D=16; C=69; P=16 | |
| III | **Empirical-to-Conceptual** | Following the example of Schoormann, Behrens, and Knackstedt (2017), we considered a research article as an object. To maintain relevance for our taxonomy's objectives, we only selected contributions that address at least three dimensions of the meta-characteristic. A total of 633 publications met this condition, from which we randomly selected approximately 30%. | D=13; C=53; P=189 | |
| IV | **Empirical-to-Conceptual** | We selected another random 30%. The fourth iteration enabled us to confirm the third iteration's structure, and we conclusively met the specified ending conditions. | D=13; C=52; P=189 | |
| V | **Conceptual-to-Empirical** | We repeated the analysis of related work from our second literature survey. We found seven additional papers that synthesize a specific domain of smart manufacturing (cf. again Appendix A). | D=13; C=52; P=7 | Second Literature Survey |
| VI | **Empirical-to-Conceptual** | Finally, we again followed the example of Schoormann et al. (2017) and considered a research article as an object. We analyzed a total of 232 articles from our second literature survey. This confirmed the structure and characteristics of the taxonomy. | D=13; C=52; P=232 | |

*It.: Iteration | D: (Sub-)Dimensions | C: Characteristics | P: Publications analyzed*

**Ending Conditions.** Step two of the method of taxonomy development is the determination of the ending conditions (EC). Nickerson et al. (2013) propose both objective and subjective termination conditions, some of which have been adopted and adapted from Sowa and Zachman (1992).

We define the following objective EC: (1.1) all objects or a representative sample of objects have been examined; (1.2) no object was merged with a similar object or split into multiple objects in the last iteration; (1.3) at least one object must have been identified per characteristic and dimension; (1.4) no new dimensions or characteristics were added in the last iteration; (1.5) no dimensions or characteristics were merged or split in the last iteration; (1.6) every dimension is unique and not repeated; and (1.7) every characteristic is unique within its dimension.

Furthermore, we define the following subjective EC: (2.1) concise; (2.2) robust; (2.3) comprehensive; and (2.4) extendible. Nickerson et al. (2013) suggest another restriction that requires characteristics to be mutually exclusive in dimensions. Following other recently developed taxonomies, we include non-exclusive characteristics (Jöhnk et al., 2017; L. Püschel, Schlott, & Röglinger, 2016b; Zschech, 2018) to avoid a verbose and non-comprehensive taxonomy with a bloated number of combined characteristics.

**Dimensions and Characteristics.** In the third step, we carried out an iterative selection of the dimensions and the assignment of the characteristics. A taxonomy is never set in stone, as research progresses or new objects are identified or discovered (Nickerson et al., 2013). In line with this notion, we revisited our taxonomy at a later point in time. As summarized in Table 4, we performed four iterations with our initial data set and added two iterations using the text corpus from our second literature search. The details on all iterations can be reviewed in Appendix B.

### Final Taxonomy

In the following, we present our final taxonomy on BA in smart manufacturing (see Table 5). The taxonomy is structured based on the quadripartite meta-characteristics of domain, orientation, data, and technique. For conciseness and comprehensiveness, we added auxiliary dimensions to these meta-characteristics (e.g., production or maintenance) that are purely descriptive and do not contradict the definition of the taxonomy (Witte & Zarnekow, 2018). They are not considered actual dimensions, but only a means by which to enhance clarity and enable thematic grouping within large dimensions. Table shows the final taxonomy and includes the number of observations of the characteristics.

#### *Meta-Characteristic Domain*

The meta-characteristic *domain* refers to the functions to which BA is applied (Holsapple et al., 2014). We identified a variety of functional areas in smart manufacturing. We distinguish a total of 14 mutually exclusive characteristics in one dimension with five auxiliary dimensions.

**Product Development & Management.** Product development covers all aspects that lead to a marketable product, from product design to preparation for production (Brown & Eisenhardt, 1995). Product management deals with subsequent activities along the product life cycle (Murphy & Gorchels, 1996). Combined, these represent only 2.9% of applications for BA in smart manufacturing.

Product Development & Management comprises *design analysis* to improve product development based on customer requirements and design data. In contrast, product development is associated with uncertainties since customer preferences regarding product configurations are traditionally non-transparent (Afshari & Peng, 2015; H. Ma, Chu, Xue, & Chen, 2017). In contrast, *product life cycle optimization* includes further aspects of the product life cycle, going beyond the design phase by using BA, for example, to optimize product repair processes, customer support, remanufacturing, and spare parts management (Y. Cheng, K. Chen, et al., 2018; Yingfeng Zhang, Ren, Liu, & Si, 2017).

**Production.** Production comprises all aspects directly related to manufacturing and represents about 29.9 % of all applications. The production cycle starts with planning, covers production and its monitoring, and ends with the finished product (H. L. Lee & Rosenblatt, 1987).

*Production planning* covers the efficient allocation of production resources such as machines or personnel. The improvement brought about by BA is not only based on historical data but partly on real-time data, which enables a dynamic adaptation to current circumstances of production (Nouiri, Bekrar, Jemai, Niar, & Ammari, 2018; Oses et al., 2016; Shiue, Lee, & Su, 2018; Shijin Wang & Liu, 2015). BA is also used for *monitoring* of operations. The primary objectives are transparency, fault detection, and the identification of anomalies in processes (Caggiano, 2018; S.-G. He, He, & Wang, 2013; Dominik Kozjek, Kralj, & Butala, 2017; Peres, Dionisio Rocha, Leitao, & Barata, 2018; Sanchez, Conde, Arriandiaga, Wang, & Plaza, 2018; Susto, Terzi, & Beghi, 2017). Production planning and monitoring represent the major applications with a share of 9.2% and 13.6%, respectively. In addition to monitoring, *performance analysis* aims to measure production performance (e.g., throughput time) to support decision-making (Kumru & Kumru, 2014a; Lingitz et al., 2018; Subramaniyan, Skoogh, Gopalakrishnan, & Hanna, 2016; Wedel, von Hacht, Hieber, Metternich, & Abele, 2015). *Performance optimization* utilizes this analysis in order to identify performance weaknesses for subsequent improvement (Chao-Chun et al., 2016; Khakifirooz, Chien, & Chen, 2018; L. Zheng et al., 2014).

**Maintenance.** Maintenance is about the necessary measures of a unit, as a specific part of a machinery, to keep it in – or restore it to – a state in which it can perform the intended functions (Pawellek, 2016). Thus, maintenance addresses all aspects of servicing machines and tools and represents the most significant auxiliary dimension, with roughly 42.1% of applications.

*Condition analysis* analyzes the condition of machines or tools. It is the most extensive characteristic and comprises a quarter of all applications. The focus of the analysis can be on the overall condition (Villalonga, Beruvides, Castaño, Haber, & Novo, 2018; Yunusa-Kaltungo & Sinha, 2017), the condition of specific components (Soualhi, Medjaher, & Zerhouni, 2015), or the degree of wear (Shaban, Yacout, Balazinski, & Jemielniak, 2017; B. Zhang, Katinas, & Shin, 2018). Condition analysis can lead to the identification of potential machine faults. However, there are objects in which BA is used explicitly for fault analysis. Therefore, these analytics patterns are classified separately under the function *defect analysis*. The objective is not the monitoring itself, but explicitly identifying faults, anomalies, or defects (Chakravorti et al., 2018; Dou & Zhou, 2016; C. Lu, Wang, & Zhou, 2017; Shijin Wang & Liu, 2015; G. Zhao et al., 2016; X. Zhu, Xiong, & Liang, 2018). In both condition and defect analysis, the ultimate judgment concerning the required maintenance strategy continues to be based on human decision-making (J. Lee, Kao, & Yang, 2014). *Maintenance planning* aims to enhance this by recommending the optimal maintenance intervals and actions (Luangpaiboon, 2015; Mbuli, Trentesaux, Clarhaut, & Branger, 2017). It is a comparably small function with only 3.2% of all applications.

**Table 5. Final Taxonomy**

<table>
<tr><th>MC</th><th colspan="2">Dimension(s)</th><th colspan="4">Characteristics</th><th>EX</th></tr>
<tr><td rowspan="5">Domain</td><td rowspan="5">Function</td><td><i>Product Development & Management</i></td><td colspan="2">Design Analysis<br>20 | 2.3%</td><td colspan="2">Product Life Cycle Optimization<br>5 | 0.6%</td><td rowspan="5">ME</td></tr>
<tr><td><i>Production</i></td><td>Production Planning<br>80 | 9.2%</td><td>Monitoring<br>118 | 13.6%</td><td>Performance Analysis<br>21 | 2.4%</td><td>Performance Optimization<br>41 | 4.7%</td></tr>
<tr><td><i>Maintenance</i></td><td>Condition Analysis<br>215 | 24.9%</td><td colspan="2">Defect Analysis<br>121 | 14.0%</td><td>Maintenance Planning<br>28 | 3.2%</td></tr>
<tr><td><i>Quality Management</i></td><td colspan="2">Quality Control<br>115 | 13.3%</td><td colspan="2">Quality Optimization<br>18 | 2.1%</td></tr>
<tr><td><i>Sustainability/ Security</i></td><td>Energy Consumption Analysis<br>15 | 1.7%</td><td colspan="2">Energy Consumption Optimization<br>17 | 2.0%</td><td>Security/Risk Analysis<br>40 | 4.6%</td></tr>
<tr><td rowspan="3">Orientation</td><td colspan="2">Maturity</td><td>Descriptive Analytics<br>227 | 26.2%</td><td>Diagnostic Analytics<br>144 | 16.6%</td><td>Predictive Analytics<br>348 | 40.2%</td><td>Prescriptive Analytics<br>146 | 16.9%</td><td>ME</td></tr>
<tr><td colspan="2" rowspan="2">Objective</td><td>Time<br>687 | 79.4%</td><td>Cost<br>722 | 83.2%</td><td>Conformance<br>170 | 19.7%</td><td>Flexibility<br>60 | 6.9%</td><td rowspan="2">NE</td></tr>
<tr><td>Security<br>58 | 6.7%</td><td colspan="2">Sustainability<br>47 | 5.4%</td><td>Customer Satisfaction<br>83 | 9.6%</td></tr>
<tr><td rowspan="4">Data</td><td colspan="2" rowspan="2">Source</td><td>Machine/Tool<br>547 | 63.2%</td><td>Process<br>253 | 29.2%</td><td>Product<br>129 | 14.9%</td><td>Customer<br>26 | 3.0%</td><td rowspan="2">NE</td></tr>
<tr><td>Reference<br>48 | 5.5%</td><td>ERP<br>55 | 6.4%</td><td>Environment<br>34 | 3.9%</td><td>Human<br>29 | 3.4%</td></tr>
<tr><td colspan="2">Integration</td><td>No Integration<br>570 | 65.9%</td><td>Vertical<br>186 | 21.5%</td><td>Horizontal<br>107 | 12.4%</td><td>End-to-End<br>35 | 3.9%</td><td>NE</td></tr>
<tr><td colspan="2">Frequency</td><td colspan="2">Real-time/Stream<br>342 | 39.5%</td><td colspan="2">Historical/Batch<br>523 | 60.5%</td><td>ME</td></tr>
<tr><td rowspan="4">Technique</td><td rowspan="4">Method</td><td rowspan="2"><i>Machine Learning</i></td><td>Classification<br>290 | 33.5%</td><td>Regression<br>104 | 12.0%</td><td>Probabilistic Method<br>36 | 4.2%</td><td>Clustering<br>42 | 4.9%</td><td rowspan="4">NE</td></tr>
<tr><td>Dimensionality Reduction<br>50 | 5.8%</td><td colspan="2">Deep Learning<br>154 | 17.8%</td><td>Reinforcement Learning<br>37 | 4.3%</td></tr>
<tr><td><i>Optimization</i></td><td>Mathematical Optimization<br>57 | 6.6%</td><td colspan="2">Evolutionary Algorithm<br>27 | 3.1%</td><td>Swarm Intelligence<br>20 | 2.3%</td></tr>
<tr><td><i>Others</i></td><td>Multi-Agent Systems<br>13 | 1.5%</td><td colspan="2">Fuzzy Logic<br>28 | 3.2%</td><td>Custom Development<br>222 | 25.7%</td></tr>
</table>

*MC: Meta-Characteristic | EX: Exclusivity | ME: Mutually Exclusive | NE: Non-Exclusive*

**Quality Management.** Manual inspection of product quality is time-consuming, yet even minor irregularities can lead to lower customer satisfaction (Saucedo-Espinosa, Escalante, & Berrones, 2017; Shatnawi & Al-Khassaweneh, 2014). It represents 15.4% of applications.

Within this auxiliary dimension, q*uality control* focuses on monitoring quality and identifying defects, typically in finished products or materials. It is the larger area of application with a share of 13.3%. *Quality optimization* goes beyond this and tries to improve the quality proactively (Luangpaiboon, 2015).

**Sustainability and Security.** Sustainability and security summarize BA approaches that focus on the conscious and responsible use of production facilities. It has been a side-topic so far, with a combined share of 8.3% (more than half of which is security/risk analysis).

*Energy consumption analysis* monitors the energy consumption of industrial systems to create transparency and detect anomalies (Ak & Bhinge, 2015; C. Li, Tao, Ao, Yang, & Bai, 2018; Oses et al., 2016; Ouyang, Sun, Chen, Yue, & Zhang, 2018; Tristo, Bissacco, Lebar, & Valentinčič, 2015). *Energy consumption optimization* can be used to optimize the energy use to achieve higher energy efficiency (Liang, Lu, Li, & Wang, 2018; G. Shao, Brodsky, Shin, & Kim, 2017; S.-J. Shin, Kim, Shao, Brodsky, & Lechevalier, 2017). *Security/risk analysis* describes analytic patterns that focus on safety-relevant aspects, risk, and compliance. An example is cybersecurity (Anton, Kanoor, Fraunholz, & Schotten, 2018; Gawand, Bhattacharjee, & Roy, 2017; Xun, Zhu, Zhang, Cui, & Xiong, 2018; Jun Yang, Zhou, Yang, Xu, & Hu, 2018).

***Meta-Characteristic Orientation***

For BA, orientation refers to the direction of thought (Holsapple et al., 2014). We identified two mutually exclusive dimensions with a total of 11 characteristics. While those of maturity are mutually exclusive, as the more mature characteristic subsumes the prior characteristics, the characteristics of objectives are non-exclusive.

**Maturity.** Maturity describes the complexity and expected business value of BA. It is concerned with what BA offers at different levels of sophistication (see also Figure 2). We identified four characteristics (Baum et al., 2018; Diez-Olivan et al., 2019; O'Donovan et al., 2015b; Zschech, 2018). Most applications in the surveyed time frame are predictive (40.2%), followed by descriptive analytics (26.2%), with diagnostic and prescriptive approaches being almost on par for the remainder.

As outlined above, *descriptive analytics* is purely delineative and answers questions through data integration, navigation, and visualization (Delen & Demirkan, 2013; Delen & Zolbanin, 2018). *Diagnostic analytics* goes beyond the mere creation of transparency and attempts to identify the root cause of incidents (Y. He, Zhu, He, Gu, & Cui, 2017; Yinhua Liu & Jin, 2013; J. Tian, Azarian, Pecht, Niu, & Li, 2017). While descriptive and diagnostic analytics focus on the past, *predictive analytics* aims to predict future events (Bousdekis, Papageorgiou, Magoutas, Apostolou, & Mentzas, 2017; Chao-Yung Hsu, Kang, & Weng, 2016; Kanawaday & Sane, 2017; Wanner, Wissuchek, & Janiesch, 2019). *Prescriptive analytics* builds on the previous characteristics to recommend concrete measures or alternative courses of action (Delen & Demirkan, 2013; Delen & Zolbanin, 2018).

**Objective.** Objective as the second dimension describes a positive impact, benefit, measure of performance, or value, which the use of BA can achieve for businesses. From an economic perspective, there are four areas of performance in manufacturing environments: time, cost, quality (i.e., conformance to specifications), and flexibility (Neely, Gregory, & Platts, 1995). These can be attributed directly to the application of BA in smart manufacturing (Kagermann et al., 2013) and are confirmed as characteristics in taxonomy development (Bordeleau et al., 2018; Fay & Kazantsev, 2018; T. Wuest et al., 2016). Besides these four fundamental performance measures, we identified the objectives of security, sustainability, and customer satisfaction. All objectives are non-exclusive.

*Time* is a source of competitive advantage and can be considered a fundamental measure of performance in manufacturing (Neely et al., 1995). It is used in 79.4% of approaches. The objective of *cost* defines the reduction of monetary expenses achieved by using BA. It is the most important objective at 83.2%. Furthermore, it yields the best possible allocation or combination of manufacturing resources. BA in smart manufacturing also addresses the improvement of *conformance* to predefined specifications (Neely et al., 1995), i.e., the quality of operations (19.7%). All other objectives account for less than 10% each. Thereby, *flexibility* describes the ability to adapt efficiently to new circumstances and requirements in daily manufacturing operations (Neely et al., 1995). Work *security* and safety can be an objective at the production site (Domova & Dagnino, 2017; Lavrova, Poltavtseva, & Shtyrkina, 2018; Xiaoya Xu, Zhong, Wan, Yi, & Gao, 2016). In the age of climate change, the objective of *sustainability* has gained importance and describes the application BA for the purpose of increasing sustainability in manufacturing operations (Dutta, Mueller, & Liang, 2018; S.-J. Shin et al., 2017). Finally, the primary goal of *customer satisfaction* is to improve the customer experience (Shatnawi & Al-Khassaweneh, 2014; Y. Zhang, S. Ren, Y. Liu, & S. Si, 2017).

### *Meta-Characteristic Data*

The meta-characteristic data describes the properties of the processed data, and all are non-exclusive without any limitations for object classification. We identified three mutually exclusive dimensions with a total of 12 non-exclusive characteristics and two exclusive characteristics concerning data frequency.

**Source.** Source describes the origin of data used for analysis (Kwon, Lee, & Shin, 2014; Sahay & Ranjan, 2008). We distinguished eight prevalent data sources. The most frequently used data source in smart manufacturing is machines/tool with 63.2%. Processes (29.2%) and product (14.9%) represent additional important data sources, with the remainder not exceeding 6.4%

*Machine/tool* data comprises properties, conditions, parameters, and states. It is collected primarily by CPS-connected sensors (Bousdekis, Magoutas, & Mentzas, 2015; Bousdekis et al., 2017; Gölzer et al., 2015; Gölzer & Fritzsche, 2017; Zschech, 2018). *Processes* generate data while connected tasks of production or maintenance are executed (D.-H. Kim et al., 2018; M. S. Reis & Gins, 2017). The *product*'s data comprises product specifications and information regarding the usage behavior of customers (Adly, Yoo, Muhaidat, & Al-Hammadi, 2014; Kai Ding & Jiang, 2016). Product data is closely related to *customer* data, such as their requirements or preferences (Lou, Feng, Zheng, Gao, & Tan, 2018b; Saldivar, Goh, Chen, & Li, 2016). *Reference* data, such as test results and specifications, is another data source (Librantz et al., 2017; Lei Ren, Sun, Cui, & Zhang, 2018). In addition to product quality, the data is relevant for quality control of manufacturing processes (S. Hu, Zhao, Yao, & Dou, 2016; L. Zhao, Yan, Wang, & Yao, 2018). Furthermore, BA uses *ERP* system data for evaluation. Examples are orders and production resources for scheduling and planning (Gölzer & Fritzsche, 2017; X. C. Zhu, Qiao, & Cao, 2017). BA in smart manufacturing also takes *human* data (e.g., working hours, health monitoring, activities, and location) into account (X. C. Zheng, Wang, & Ordieres-Mere, 2018). Finally, *environment* data is collected in the vicinity and does not necessarily have to be related directly to manufacturing operations (Filonenko & Jo, 2018; Molka-Danielsen, Engelseth, & Wang, 2018).

**Integration.** Three forms of integration for BA in smart manufacturing are commonly discussed: vertical, horizontal, and end-to-end integration (Kagermann et al., 2013; K. Zhou, Liu, & Liang, 2016). The integration of IT resources and systems should naturally include the exchange of data, which enables new applications of data analysis.

*Vertical* integration (21.5%) comprises the integration of data among various hierarchical levels for smart manufacturing, from the actuator and sensor level to the corporate planning level (Gölzer & Fritzsche, 2017; Kagermann et al., 2013). *Horizontal* (12.4%) integration encompasses not only the integration of data along the value chain but also across company boundaries (Dou & Zhou, 2016; Kagermann et al., 2013). In contrast, *end-to-end* integration describes data integration along the entire product life cycle, from product development to recycling (Dou & Zhou, 2016). Applications of this type are comparatively rare, with 3.9%. In addition, we include the characteristic *no integration*, which has been identified as a research gap by Sharp et al. (2018). For instance, in maintenance, most analytics patterns are limited to individual machines, tools, or components. The effects on the overall system, the production process, or other machines are not considered (H. Shao et al., 2018; Yuwono et al., 2016). It is the de facto situation for most applications, with 65.9%.

**Frequency.** This dimension comprises the temporal perspective of data availability and collection as well as its subsequent processing (Sahay & Ranjan, 2008).

*Real-time/stream* analytics can be considered a fundamental requirement of smart manufacturing. Saldivar, Goh, Chen, et al. (2016) go as far as to describe the ability to analyze data in real time as the 'key' to smart manufacturing. The data is continuously collected and evaluated (Verma, Kawamoto, Fadlullah, Nishiyama, & Kato, 2017). Continuous evaluation of the data enables a dynamic reaction to new requirements and accruing problems, addressing the objectives of time, costs, and flexibility. In contrast, within *historical/batch* analysis, the data is not processed continuously but periodically or irregularly (Carbone et al., 2015). The distribution is roughly 40% to 60%.

### *Meta-Characteristic Technique*

In this context, the term 'technique' refers to the way a BA task is performed (Holsapple et al., 2014). We identified a wide array of techniques applied in smart manufacturing, which are summarized in the single-dimension method with a total of 13 non-exclusive characteristics that can be grouped into the auxiliary dimensions of ML, optimization, and others (with ML being the most significant).

**Machine Learning.** ML is a paradigm that comprises algorithms, which learn from experience automatically without being explicitly programmed to perform a task such as making decisions or predictions (Marsland, 2015; Samuel, 1959). It can be divided into three types: supervised learning using labeled data, unsupervised learning using unlabeled data, and reinforcement learning using reward functions.

*Classification* is the predominant method in supervised learning and used by roughly a third of all applications. It comprises the assignment of data to predefined classes (S.-G. He et al., 2013; Ragab, Yacout, Ouali, & Osman, 2016; Ray & Mishra, 2016). In *regression* analysis, a function establishes relationships between variables to make predictions (12%). The aim of *clustering* in unsupervised learning is to group similar data in order to recognize undiscovered patterns or correlations. Another unsupervised learning method is *dimensionality reduction*. Its task is to reduce complexity by mapping data from a higher to a lower dimensional level (Marsland, 2015). This is also referred to as feature learning or representation learning. Task-specific techniques, such as regression and classification, often require data that is easy to process mathematically (Argyriou, Evgeniou, & Pontil, 2008). *Probabilistic methods* are applied in supervised and unsupervised approaches in a wide array of applications such as anomaly detection (C. Y. Park, Laskey, Salim, & Lee, 2017), monitoring (S. Windmann, Jungbluth, & Niggemann, 2015), and remaining useful lifetime estimation (Shuai Zhang, Zhang, & Zhu, 2018). Lastly, we identified *reinforcement learning* with an explorative character to create new knowledge (Ishii, Yoshida, & Yoshimoto, 2002). None of the above goes beyond 5.8% in use. In contrast, *deep learning* is a comprehensive class of ML algorithms that combines feature learning with task-specific approaches (L. Deng & Yu, 2014). In smart manufacturing, deep learning is used when information from the real world (e.g., images, videos, or sensor data) is transferred into the digital world (Sonntag, Zillner, van der Smagt, & Lörincz, 2017; Srivastava & Salakhutdinov, 2014). Its share accounts for 17.8% of applications.

**Optimization.** Optimization is concerned with the selection of the best element (according to specified criteria) from a set of available options. We identified three common groups of optimization methods in smart manufacturing.

*Mathematical optimization*, known as nonlinear programming or numerical optimization, can be described as the science of determining the optimal solution to mathematically definable problems. The problems are often models from production and management systems (Shaw, Park, & Raman, 1992). Biologically inspired optimization methods are unique optimization methods inspired by biological processes and phenomena. On the one hand, general *evolutionary algorithms* access collective phenomena such as reproduction, mutation, and selection. On the other hand, there is *swarm intelligence*, based on the collective social behavior of organisms. It entails the implementation of collective intelligence based on many individual agents as inspired by the behavior of insect swarms (Binitha & Sathya, 2012).

**Others.** In addition to optimization and ML methods, other approaches are employed in smart manufacturing which, due to their specificity, cannot be assigned to the other auxiliary dimensions or justify a dimension of their own.

*Multi-agent systems* (MAS), also known as distributed artificial intelligence, are systems in which several interacting intelligent software agents pursue specific goals or solve collective problems (Ferber & Weiss, 1999). MAS can be related to agent-based paradigms such as reinforcement learning or swarm intelligence (Oses et al., 2016; Xiao Wang, Wang, & Qi, 2016). The application scenarios of MAS are manifold: agent mining, for example, is an approach in which MAS is used for decision-making problems (L. Cao, Weiss, & Philip, 2012). *Fuzzy logic* is a mathematical system to model manifestations of human decision-making (Bothe, 1995). Fuzzy logic is a many-valued logic. It can process partial truths, whereby the truth-value can lie between entirely true or false. Although fuzzy logic can be deployed with ML or optimization methods (Aydın et al., 2015; H. Ma et al., 2017), we consider it a separate characteristic (Andonovski, Mušič, & Škrjanc, 2018; Arık & Toksarı, 2019; Aydın et al., 2015; Baban, Baban, & Suteu, 2016; Y. Lv & Lin, 2017; Niu & Li, 2017; Dawei Sun, Lee, & Lu, 2016; J. Wu et al., 2018; Zurita, Delgado, Carino, Ortega, & Clerc, 2016). Lastly, the characteristic *custom development* summarizes applications which use an undisclosed or custom-developed method such as, for example, manual data selection, but also expert surveys or architecture concepts (Bekar, Skoogh, Cetin, & Siray, 2018; C.-Y. Tsai, Chen, & Lo, 2014). About a quarter of all applications use custom development, which hints at the diversity and lack of consolidation of the field.

## Illustrative Application

Nickerson et al. (2013) suggest evaluating a taxonomy's applicability after the development. One option is the illustrative application of the taxonomy on objects (Szopinski et al., 2019). The application of a taxonomy to research real-world objects (in our case articles studying specific approaches to BA in smart manufacturing) enables a reflection on the current state of research on a certain type of object (Khalilijafarabad, Helfert, & Ge, 2016). Furthermore, it helps to uncover commonalities and discrepancies between studies on this type of object (Szopinski et al., 2019). Hence, the application of our taxonomy enables researchers to position their contribution in a larger context and identify potential research gaps (Hummel, Schacht, & Maedche, 2016).

**Table 6. Application of Taxonomy, Examples I to III**

| MC | DIMENSIONS | CHARACTERISTICS | | | |
|---|---|---|---|---|---|
| Domain | Function | Design Analysis | Product Life Cycle Optimization | | |
| | | Production Planning | Monitoring | Performance Analysis | Performance Optimization |
| | | Condition Analysis | Defect Analysis | Maintenance Planning | |
| | | Quality Control | Quality Optimization | | |
| | | Energy Consumption Analysis | Energy Consumption Optimization | Security/Risk Analysis | |
| Orientation | Maturity | Descriptive Analytics | Diagnostic Analytics | Predictive Analytics | Prescriptive Analytics |
| | Objective | Time | Cost | Conformance | Flexibility |
| | | Security | Sustainability | Customer Satisfaction | |
| Data | Source | Machine/Tool | Process | Product | Customer |
| | | Reference | ERP | Environment | Human |
| | Integration | No Integration | Vertical | Horizontal | End-to-End |
| | Frequency | Real-time/Stream | Historical/Batch | | |
| Technique | Method | Classification | Regression | Probabilistic Method | Clustering |
| | | Dimensionality Reduction | Deep Learning | Reinforcement Learning | |
| | | Mathematical Optimization | Evolutionary Algorithm | Swarm Intelligence | |
| | | Multi-Agent Systems | Fuzzy Logic | Custom Development | |
| *Coloring Scheme* | | Example I | Example II | Example III | |

To illustrate usefulness of our taxonomy, we applied it to three objects, following other taxonomy research in the information systems discipline (e.g., Lis & Otto, 2021; L. C. Püschel, Röglinger, & Brandt, 2020). To provide and ensure a sufficient level of variance, we randomly chose three applications. We show a classification of the respective examples with the taxonomy in Table 6.

**Example I.** Ståhl, Mathiason, Falkman, and Karlsson (2019) employ BA to detect problematic slab shapes in steel rolling. Their application's function is to model the dependencies between the measured product shape and control the product's conformance (*function*). As they predict the ratio of flawed products even before a critical manufacturing operation, their approach is of predictive maturity (*maturity*). The application's goals are to increase customer satisfaction through higher product quality and reduce manufacturing costs (*objective*), as defective products can be detected before further process steps are initialized. There is no data integration, as the approach collects the data at a specific point in the process (*integration*). The data, slab width, and deviation from the targeted position are collected through multiple sensors directly from the product (*data source*). The authors do not employ real-time analytics, but rather use a data set collected before the analysis in their experimental set-up (*frequency*). They apply a classification approach to decide between defective and non-defective steel slabs. Specifically, the application uses recurrent neural networks with long-term memory cells, that is deep learning (*method*).

**Example II.** H. Hu, Jia, He, Fu, and Liu (2020) use BA for production planning, specifically the scheduling of automated guided vehicles (AGVs) for material handling (*function*). The goal of their analysis is to find the optimal mixed rule policy to identify the best course of action, which reflects a prescriptive analytics problem (*maturity*). The goal of the analysis is to improve time efficiency, specifically the delay ratio of AGVs, while simultaneously increasing their flexibility by enabling them to participate in various jobs, which also leads to makespan optimization. The data used is multifaceted, including machines (AGVs) and the production process (RFID and Industrial IoT sensors) (*data source*). The data is integrated vertically, as various sources are tapped into and passed on to the analysis (*integration*) continuously in a real-time fashion (*frequency*). Finally, the application employs a deep reinforcement learning-based approach to schedule the AGVs and find the optimal mixed rule policy (*method*).

**Example III.** In contrast to the other examples, Arpaia, Moccaldi, Prevete, Sannino, and Tedesco (2020) place the factory worker's safety at the center of their application (*function*) by predicting worker stress levels (*maturity*). The proposed solution increases the worker's safety and generally fosters a more secure manufacturing environment. Positive side effects are an increase in product quality (less stress translates to more focused workers) as well as a reduction of the cost of the production process (*objective*). Data is collected through a brain-computer interface, specifically a wearable electroencephalography instrument monitoring the brainwaves of the workers (*data source*). The proposed setup is not integrated (*integration*) and the data is transferred in real time (*frequency*) to the analytical system. Classification algorithms are used to predict the workers' stress level (*method*).

## Derivation of Archetypes

In this section, we perform a cluster analysis to derive archetypes for smart manufacturing that comprise similar applications as those illustrated above. After detailing the derivation process, we explore the resulting archetypes' characteristics.

### Cluster Analysis

To gain more insights into different research streams and archetypes, we performed a cluster analysis. Before applying the clustering algorithm, we transformed the collected literature into binary vector representations of the individual articles, each consisting of 52 binary values (i.e., the number of characteristics in the taxonomy). In total, we clustered 854 vectors. This excludes all survey papers ($n$=39) and 11 vectors, which we removed as we could not assign characteristics in the dimension function.

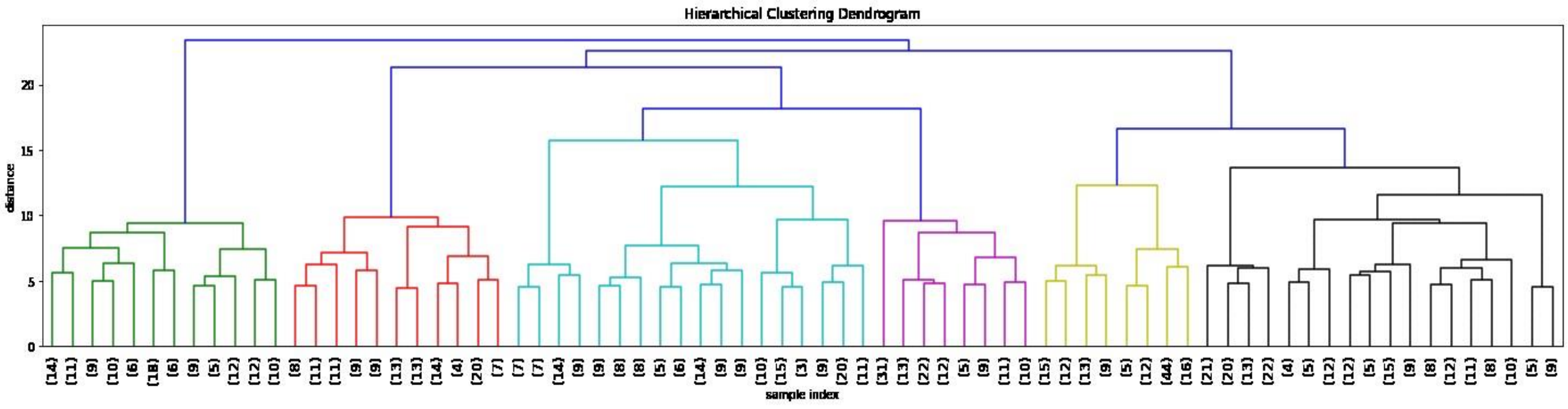

**Figure 4. Dendrogram of Clusters**

We applied different admissible clustering methods and distance measures as outlined above. Figure 4 illustrates the dendrogram of the clustering. For the sake of clarity, the figure only shows the last 75 merged clusters. On the horizontal axis, we illustrate the numbers of elements within the clusters in parentheses. Observing the dendrogram, we uncovered six meaningful clusters, visualized in different colors.

### Derived Archetypes

The six clusters identified by the clustering represent the major archetypes of BA applications in smart manufacturing. These correspond to six branches of research, which in turn correspond to main intentions for practitioners with similar approaches and/or infrastructural requirements. Due to distinct functional differences, the number of observations contained in each cluster varies. Clusters C3 and C6 are the largest, with 173 and 201 articles, respectively. Clusters C1, C2, C4, and C5 comprise 122, 119, 113, and 126 articles. Following an iterative exploratory approach, changes to the defined allowed Euclidean distance between related objects, and thus a smaller or larger number of clusters, did not significantly change our results of six major archetypes. They highlighted or masked minor yet distinct characteristics, which we discuss in the respective final clusters.

We summarize these results in Table 7 to provide a multi-perspective analysis of the identified archetypes. We extracted and analyzed the impact of each characteristic to scrutinize the strength and direction of its influence on the clusters. In the following, we elaborate on the archetypes and their contained subthemes.

**MRO Planning (C2).** This cluster, with 13.9% (*n*=119) of all research, focuses on Maintenance, Repair and Operations (MRO) planning activities. MRO planning ensures overall manufacturing effectiveness, including addressing scheduling problems for production or maintenance as well manufacturing performance optimization. Most applications are prescriptive, as the tasks require various optimization techniques to find optimal policies. As with the maintenance-based archetypes, time and cost dominate the objectives. The archetype has a noticeable share of the flexibility objective. Here, optimized planning seems to lead to a more flexible production environment. Production planning and maintenance planning are the main consumers for ERP data. Furthermore, applications use machine and process data. All frequencies are present, but historical analyses outweigh real-time applications. Methods focus on typical prescriptive techniques like mathematical optimization but also include reinforcement learning, swarm intelligence, and multi-agent systems.

**Reactive Maintenance (C5).** This cluster is also medium-sized, with 14.8% (*n*=126) of all research. It includes reactive maintenance approaches of the functions condition analysis and defect analysis, with the latter dominating this archetype. It constitutes a counterpart to both predictive maintenance archetypes (C4, C6). Here, the focus is not on proactive but reactive industrial maintenance measures (Pawellek, 2016). The distribution of characteristics is mostly comparable with the other two archetypes. Two distinct exceptions exist: First, the analytical maturity is only diagnostic. That is, future behavior is not considered, and the applications are rather reactive in their detection of existing defects or conditions in machine operations. Second, analysis is performed mainly on historical data. That is, the maintenance task takes place after a malfunction has occurred. This is in contrast to the processing of real-time data streams and predictive measures to avoid such malfunctions proactively (Pawellek, 2016).

**Offline Predictive Maintenance (C6).** This is the largest cluster, with 23.5% (n=*201*), and it reflects maintenance-based approaches that are of a proactive nature. This archetype exhibits great similarities to C4. The distribution of the characteristics is close to identical. That is, analytics is primarily of predictive maturity in combination with functional applications of condition analysis followed by defect analysis, both of which are typical maintenance operations. The main objectives are related to cost and time. The machine is the central data source in this archetype, and the data is mostly not integrated, which suggests that the overall production system is not considered. However, this archetype relies on historical data collected up to a certain point in time. That is, in contrast to C4, approaches in this category generally do not employ real-time analytics, but rather fall back on historical or batch-based data frequencies. Furthermore, we see regression analysis as the primary method. From a functional perspective, energy consumption analysis, energy consumption optimization, and design analysis can be distinguished as distinct domains. In line with this, the sustainability objective is of a higher frequency. Consequently, we see similarities between both use cases. Energy-related applications also predominantly utilize predictive analytics, tap into the machine directly as a data source, and feature no integration, indicating a possible novel research stream or gap in smart manufacturing. Finally, the objective of customer satisfaction, aligned with the function design analysis, is present and refers to another (smaller) comparable research stream.

**Online Predictive Maintenance (C4).** The fourth archetype represents the smallest cluster with 13.2% (*n*=113) of all research and is the online rather than offline sibling of C6. In contrast to C6, all approaches employ BA techniques in real-time, and thus online, having an intact and operational connection to (smart) manufacturing objects. We mostly see the application of classification and deep learning methods, two characteristics that are closely related, followed by regression analysis. In summary, this archetype has a clear focus on the machines and equipment by predicting machine conditions and defects in an online fashion.

## Table 2. Archetypes of Business Analytics in Smart Manufacturing

| D | Characteristics | *n* | MRO Planning (C2) | | Reactive Maintenance (C5) | | Offline Predictive Maintenance (C6) | | Online Predictive Maintenance (C4) | | MRO Monitoring (C3) | | Quality Management (C1) | |
|---|---|---|---|---|---|---|---|---|---|---|---|---|---|---|
| | | | 119 | | 126 | | 201 | | 113 | | 173 | | 122 | |
| Function | Design Analysis | *20* | 0 | 0.00 | 0 | 0.00 | 19 | 0.09 | 0 | 0.00 | 0 | 0.00 | 1 | 0.01 |
| | Product Life Cycle Opt. | *5* | 2 | 0.02 | 0 | 0.00 | 0 | 0.00 | 0 | 0.00 | 2 | 0.01 | 1 | 0.01 |
| | Production Planning | *80* | 66 | 0.55 | 0 | 0.00 | 10 | 0.05 | 0 | 0.00 | 4 | 0.02 | 0 | 0.00 |
| | Monitoring | *118* | 1 | 0.01 | 18 | 0.14 | 16 | 0.08 | 3 | 0.03 | 77 | 0.45 | 3 | 0.02 |
| | Performance Analysis | *21* | 0 | 0.00 | 1 | 0.01 | 10 | 0.05 | 0 | 0.00 | 10 | 0.06 | 0 | 0.00 |
| | Performance Opt. | *41* | 17 | 0.14 | 0 | 0.00 | 9 | 0.04 | 3 | 0.03 | 9 | 0.05 | 3 | 0.02 |
| | Condition Analysis | *215* | 2 | 0.02 | 32 | 0.25 | 89 | 0.44 | 76 | 0.67 | 14 | 0.08 | 2 | 0.02 |
| | Defect Analysis | *121* | 1 | 0.01 | 74 | 0.59 | 21 | 0.10 | 20 | 0.18 | 5 | 0.03 | 0 | 0.00 |
| | Maintenance Planning | *28* | 21 | 0.18 | 0 | 0.00 | 4 | 0.02 | 0 | 0.00 | 3 | 0.02 | 0 | 0.00 |
| | Quality Control | *115* | 2 | 0.02 | 0 | 0.00 | 1 | 0.00 | 9 | 0.08 | 3 | 0.02 | 100 | 0.82 |
| | Quality Opt. | *18* | 4 | 0.03 | 0 | 0.00 | 0 | 0.00 | 2 | 0.02 | 1 | 0.01 | 11 | 0.09 |
| | Energy Cons. Analysis | *15* | 0 | 0.00 | 0 | 0.00 | 11 | 0.05 | 0 | 0.00 | 4 | 0.02 | 0 | 0.00 |
| | Energy Cons. Opt. | *17* | 3 | 0.03 | 1 | 0.01 | 11 | 0.05 | 0 | 0.00 | 1 | 0.01 | 1 | 0.01 |
| | Security/Risk Analysis | *40* | 0 | 0.00 | 0 | 0.00 | 0 | 0.00 | 0 | 0.00 | 40 | 0.23 | 0 | 0.00 |
| Maturity | Descriptive | *226* | 1 | 0.01 | 48 | 0.38 | 3 | 0.01 | 34 | 0.30 | 97 | 0.56 | 43 | 0.35 |
| | Diagnostic | *142* | 2 | 0.02 | 77 | 0.61 | 20 | 0.10 | 10 | 0.09 | 19 | 0.11 | 14 | 0.11 |
| | Predictive | *347* | 4 | 0.03 | 1 | 0.01 | 169 | 0.84 | 68 | 0.60 | 47 | 0.27 | 58 | 0.48 |
| | Prescriptive | *139* | 112 | 0.94 | 0 | 0.00 | 10 | 0.05 | 1 | 0.01 | 10 | 0.06 | 6 | 0.05 |
| Objective | Time | *680* | 113 | 0.95 | 126 | 1.00 | 155 | 0.77 | 107 | 0.95 | 119 | 0.69 | 60 | 0.49 |
| | Cost | *710* | 109 | 0.92 | 124 | 0.98 | 178 | 0.89 | 105 | 0.93 | 125 | 0.72 | 69 | 0.57 |
| | Conformance | *169* | 10 | 0.08 | 1 | 0.01 | 6 | 0.03 | 17 | 0.15 | 14 | 0.08 | 121 | 0.99 |
| | Flexibility | *58* | 33 | 0.28 | 1 | 0.01 | 8 | 0.04 | 1 | 0.01 | 12 | 0.07 | 3 | 0.02 |
| | Security | *57* | 1 | 0.01 | 3 | 0.02 | 5 | 0.02 | 8 | 0.07 | 40 | 0.23 | 0 | 0.00 |
| | Sustainability | *44* | 5 | 0.04 | 1 | 0.01 | 24 | 0.12 | 2 | 0.02 | 10 | 0.06 | 2 | 0.02 |
| | Customer Satisfaction | *83* | 4 | 0.03 | 0 | 0.00 | 19 | 0.09 | 0 | 0.00 | 5 | 0.03 | 55 | 0.45 |
| Data | Machine/Tool | *544* | 56 | 0.47 | 123 | 0.98 | 140 | 0.70 | 112 | 0.99 | 77 | 0.45 | 36 | 0.30 |
| | Process | *250* | 78 | 0.66 | 6 | 0.05 | 27 | 0.13 | 3 | 0.03 | 100 | 0.58 | 36 | 0.30 |
| | Product | *129* | 19 | 0.16 | 2 | 0.02 | 29 | 0.14 | 2 | 0.02 | 9 | 0.05 | 68 | 0.56 |
| | Customer | *25* | 4 | 0.03 | 0 | 0.00 | 14 | 0.07 | 0 | 0.00 | 4 | 0.02 | 3 | 0.02 |
| | Reference | *46* | 5 | 0.04 | 3 | 0.02 | 6 | 0.03 | 4 | 0.04 | 4 | 0.02 | 24 | 0.20 |
| | ERP | *50* | 33 | 0.28 | 0 | 0.00 | 5 | 0.02 | 1 | 0.01 | 11 | 0.06 | 0 | 0.00 |
| | Environment | *34* | 4 | 0.03 | 0 | 0.00 | 13 | 0.06 | 7 | 0.06 | 9 | 0.05 | 1 | 0.01 |
| | Human | *29* | 6 | 0.05 | 0 | 0.00 | 4 | 0.02 | 2 | 0.02 | 13 | 0.08 | 4 | 0.03 |
| Integration | No Integration | *566* | 50 | 0.42 | 110 | 0.87 | 137 | 0.68 | 109 | 0.96 | 65 | 0.38 | 95 | 0.78 |
| | Vertical | *183* | 47 | 0.39 | 11 | 0.09 | 35 | 0.17 | 2 | 0.02 | 74 | 0.43 | 14 | 0.11 |
| | Horizontal | *101* | 28 | 0.24 | 10 | 0.08 | 11 | 0.05 | 0 | 0.00 | 40 | 0.23 | 12 | 0.10 |
| | End-to-End | *35* | 3 | 0.03 | 0 | 0.00 | 21 | 0.10 | 1 | 0.01 | 7 | 0.04 | 3 | 0.02 |
| Freq. | Real-time/Stream | *339* | 44 | 0.37 | 9 | 0.07 | 18 | 0.09 | 113 | 1.00 | 122 | 0.71 | 33 | 0.27 |
| | Historical/Batch | *515* | 75 | 0.63 | 117 | 0.93 | 183 | 0.91 | 0 | 0.00 | 51 | 0.29 | 89 | 0.73 |
| Method | Classification | *289* | 16 | 0.13 | 66 | 0.52 | 45 | 0.22 | 55 | 0.49 | 51 | 0.29 | 56 | 0.46 |
| | Regression | *102* | 6 | 0.05 | 4 | 0.03 | 50 | 0.25 | 16 | 0.14 | 9 | 0.05 | 17 | 0.14 |
| | Probabilistic Methods | *35* | 6 | 0.05 | 3 | 0.02 | 8 | 0.04 | 6 | 0.05 | 8 | 0.05 | 4 | 0.03 |
| | Clustering | *41* | 1 | 0.01 | 6 | 0.05 | 12 | 0.06 | 8 | 0.07 | 7 | 0.04 | 7 | 0.06 |
| | Dim. Reduction | *49* | 3 | 0.03 | 8 | 0.06 | 9 | 0.04 | 13 | 0.12 | 9 | 0.05 | 7 | 0.06 |
| | Deep Learning | *154* | 12 | 0.10 | 28 | 0.22 | 53 | 0.26 | 22 | 0.19 | 18 | 0.10 | 21 | 0.17 |
| | Reinforcement Learning | *37* | 30 | 0.25 | 0 | 0.00 | 3 | 0.01 | 2 | 0.02 | 2 | 0.01 | 0 | 0.00 |
| | Mathematical Opt. | *57* | 37 | 0.31 | 1 | 0.01 | 8 | 0.04 | 5 | 0.04 | 2 | 0.01 | 4 | 0.03 |
| | Evolutional Algorithm | *27* | 9 | 0.08 | 1 | 0.01 | 5 | 0.02 | 2 | 0.02 | 2 | 0.01 | 8 | 0.07 |
| | Swarm Intelligence | *18* | 10 | 0.08 | 0 | 0.00 | 2 | 0.01 | 2 | 0.02 | 3 | 0.02 | 1 | 0.01 |
| | Multi-Agent Systems | *12* | 5 | 0.04 | 1 | 0.01 | 0 | 0.00 | 2 | 0.02 | 2 | 0.01 | 2 | 0.02 |
| | Fuzzy Logic | *26* | 6 | 0.05 | 6 | 0.05 | 5 | 0.02 | 4 | 0.04 | 3 | 0.02 | 2 | 0.02 |
| | Custom Development | *220* | 25 | 0.21 | 19 | 0.15 | 41 | 0.20 | 26 | 0.23 | 86 | 0.50 | 23 | 0.19 |

*total count | relative count of objects in the respective cluster*

**MRO Monitoring (C3).** This is the second largest cluster, with 20.3 % (*n*=173), and focuses mainly on the functions of monitoring and security/risk analysis, followed by a small number of performance-related functions and condition analysis. The maturity of the approaches is mostly descriptive, followed by predictive, with some diagnostic objects. The frequency exhibits a distinct trend towards real-time analysis, as a defining trait of monitoring applications. Compared to other archetypes, the process is a dominant data source. In addition, the forms of data integration are much more diverse, that is, it has vertically integrated solutions, includes data from various machines or steps in a production process, and monitors manufacturing operations more holistically. From a methodological perspective, we see many custom-developed solutions as well as classifications to discern between normal and anomalous states in manufacturing. Lastly, we observe a connection between more generic monitoring operations (e.g., process monitoring) and the more specific security/risk analysis (which, e.g., includes applications such IoT network security monitoring). Both show a similar distribution of characteristics (except the security objective).

**Quality Management (C1).** The final archetype represents one of the medium-sized clusters with 14.3% (*n*=122) of all research. It covers typical quality management tasks such as checking whether the products and processes conform to specification, which is decidedly different from planning, MRO, and monitoring. Analyses are mostly descriptive or predictive with the objective of improving conformance (quality) as well as time, cost, and customer satisfaction (albeit to a lesser degree in the latter case). Much of the data is product and reference data with the obvious reliance on machine and process data. Most applications are not integrated, although vertical and horizontal integration does sometimes occur. Quality management methods use more historical than real-time data. Furthermore, methods center on classification – especially image recognition – with some applications of regression and deep learning in addition to custom development.

## Analysis of Temporal Variations and Trends

In this section, we examine trends and highlight temporal variations in our data. First, we apply a temporal trend analysis to the taxonomy's dimensions and characteristics. Second, we apply this analysis to the derived archetypes. To create a consistent data set for the temporal analysis that does not complicates the interpretation, we have only included those research articles from the first iteration that match the filters we used for the second iteration from 2019 onwards (i.e., journals articles; Impact Factor >= 2.000).

### Variation and Trends by Dimension and Characteristics

In the following, we present temporal trends and variations for the three dimensions of function, maturity, and method. These dimensions contained the most interesting findings. A more detailed visualization of all characteristics can be found in Appendix C.

**Function.** The upper part of the section *Function* of Figure 5 illustrates the publications related to this dimension and its five auxiliary dimensions. Here, data shows that *maintenance* and *production* are being most actively researched. While the number of publications in both areas were similar up to and including 2015, the research focus seems to have shifted to *maintenance* thereafter, which dominated until 2018. Subsequently, in 2019, production took over and continues to lead as the most researched function in smart manufacturing. It is also noticeable that in contrast to all other functions, *sustainability/security* has been increasing steadily since 2017. The area of *product development and management* seems to be a marginal topic.

**Maintenance and Production.** A more detailed analysis of *production* and *maintenance* as the two key auxiliary dimensions reveals that research has focused predominantly on *condition analysis* and *defect analysis* (cf. the lower part of the section *Function* of Figure 5). From 2017 on, a propensity towards *condition analysis* has emerged. The two topics of *performance optimization* and *monitoring* show a strong trend of growing interest. While both have been researched most heavily from 2015 onwards, an increasing tendency towards research into optimization is evident. In contrast, *maintenance planning* and *production planning* seem to be specialized and relatively stable topics after 2015. From 2018 onwards, we also observe a more diversified research interest as compared to the years before, when *maintenance* dominated.

**Maturity.** The middle part of Figure 5, *Maturity*, demonstrates the temporal development of the respective dimension of the taxonomy. *Descriptive analytics* approaches were especially popular at the beginning of Industry 4.0 initiatives. Despite reaching a new high in 2018, the number has decreased significantly in the last two years. *Diagnostic analytics* approaches, on the other hand, have grown slightly over time, but without reaching a comparable number of publications. *Predictive analytics* approaches were already showing strong growth in the early years and, as of 2017, were the dominant BA approach in smart manufacturing. *Prescriptive analytics* research was less pronounced, but it grew steadily to become the second most applied approach. This suggests that as BA in smart manufacturing research matures, we may see it moving from predominately predictive to prescriptive approaches in the next couple of years.

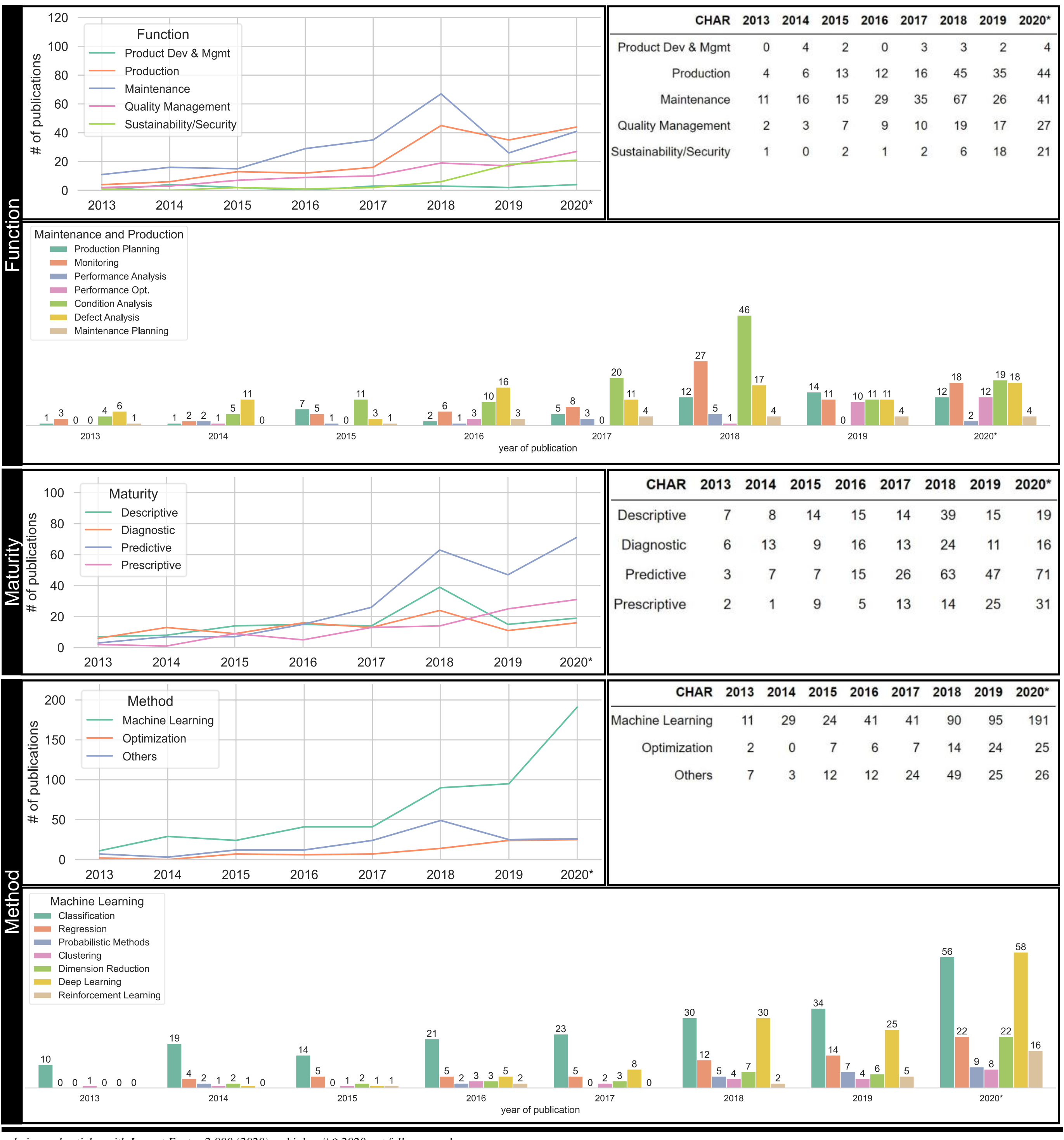

| CHAR | 2013 | 2014 | 2015 | 2016 | 2017 | 2018 | 2019 | 2020* |
|---|---|---|---|---|---|---|---|---|
| Product Dev & Mgmt | 0 | 4 | 2 | 0 | 3 | 3 | 2 | 4 |
| Production | 4 | 6 | 13 | 12 | 16 | 45 | 35 | 44 |
| Maintenance | 11 | 16 | 15 | 29 | 35 | 67 | 26 | 41 |
| Quality Management | 2 | 3 | 7 | 9 | 10 | 19 | 17 | 27 |
| Sustainability/Security | 1 | 0 | 2 | 1 | 2 | 6 | 18 | 21 |

| CHAR | 2013 | 2014 | 2015 | 2016 | 2017 | 2018 | 2019 | 2020* |
|---|---|---|---|---|---|---|---|---|
| Descriptive | 7 | 8 | 14 | 15 | 14 | 39 | 15 | 19 |
| Diagnostic | 6 | 13 | 9 | 16 | 13 | 24 | 11 | 16 |
| Predictive | 3 | 7 | 7 | 15 | 26 | 63 | 47 | 71 |
| Prescriptive | 2 | 1 | 9 | 5 | 13 | 14 | 25 | 31 |

| CHAR | 2013 | 2014 | 2015 | 2016 | 2017 | 2018 | 2019 | 2020* |
|---|---|---|---|---|---|---|---|---|
| Machine Learning | 11 | 29 | 24 | 41 | 41 | 90 | 95 | 191 |
| Optimization | 2 | 0 | 7 | 6 | 7 | 14 | 24 | 25 |
| Others | 7 | 3 | 12 | 12 | 24 | 49 | 25 | 26 |

*only journal articles with Impact Factor 2.000 (2020) or higher || * 2020 not fully covered.*

**Figure 1. Temporal Development of BA in Smart Manufacturing**

**Method.** The upper part of the section *Method* of Figure 5 illustrates the publications related to this dimension. We have subdivided it into the three auxiliary dimensions of *machine learning*, *optimization*, and *others*. From 2013, it shows that there has been a strong trend towards the application of *ML,* peaking and dominating in 2020. We recorded a smaller number of publications for the auxiliary dimension *others*, but we see a small drop in interest after 2018, possibly pointing towards a standardization of the ML methods being used. *Optimization* shows a lower uptick, which points to the fact that the rise of performance optimization is fueled by methods from ML.

**Machine Learning.** The focus in the auxiliary dimension is on supervised ML techniques (cf. lower part of the section *Method* of Figure 5). Here, the use of *classification* techniques dominates in all years. Several waves are discernible, with the first from 2013 to 2014, the second from 2015 to 2017, and the third from 2018 onwards. *Regression* has not experienced a comparable uptick. In contrast, *deep learning* has rapidly gained in popularity since 2017 and now represents the most popular method. This trend seems likely to continue. In addition, *reinforcement learning* has been more widely explored in the last two years, which has a rather prescriptive BA maturity. This observation is in line with the trends identified in the section *Maturity*.

### Variation and Trends by Archetypes

In the following, we present variations and trends on archetype level. Additionally, we further examine two archetypes – namely MRO Monitoring (C3) and Offline Predictive Maintenance (C6) – due to their diversity in certain dimensions (see Table 7). Figure 6 provides a visualization for the discussion. The visualizations of all characteristics for each archetype can be found in Appendix C.

**Archetypes.** The top section of Figure 6 illustrates the temporal development of the six archetypes. Throughout the years, we see a steady increase for all archetypes, while both *reactive maintenance* and *offline predictive maintenance* dominate 2013 and 2014. From 2014 to 2017, it is noticeable that *reactive maintenance* and *online predictive maintenance* gain traction. After 2017, *MRO monitoring* and *offline predictive maintenance* grow particularly rapidly. In contrast, the interest in *reactive maintenance* significantly decreases after 2017. *MRO planning*, *quality management*, and *online predictive maintenance* are rather steady in their growth. In 2019, we can observe a drop of research interest in *MRO monitoring* and *offline predictive maintenance*. After 2019, we see continuous growth over nearly all archetypes, with *online predictive maintenance* growing most rapidly. Again, both *MRO monitoring* and *offline predictive maintenance* outperform the other clusters, but their growth rate has dropped considerably. *Reactive maintenance* is the least considered in research.

**MRO Monitoring.** This archetype is relatively diverse regarding the *function* dimension. Our temporal analysis confirms the novel research stream or gap in *security/risk analysis*. The characteristic *monitoring* is dominant in our first search iteration with data until 2018. *Security/risk analysis*, on the other hand, exhibits rapid growth through 2020. Thus, s*ecurity/risk analysis* constitutes a novel research trend in smart manufacturing and may even constitute a distinct archetype in the future. Regarding integration, this archetype shows a significant percentage of vertical integration, pointing to multi-level monitoring. In general, this archetype implemented more integration than others did.

**Offline Predictive Maintenance**. As the name suggests, this archetype relies not on *integration*, but rather data that is copied for offline analysis. While we observe a slow increase in integration options, it does not seem relevant for implementation in the near future. Furthermore, the archetype displays dissimilar characteristics in the dimension *method*. The distribution and growth of *classification*, *regression*, and *custom development* are steady with no clear frontrunner until 2017, when a new contender appears. *Deep learning* exhibits rapid growth and surpasses all other methods by 2018. It is by far the most popular method today.

## Discussion and Conclusion

Our study contributes to the descriptive knowledge of BA in smart manufacturing, as it explores a diverse, evolving, and not-yet-well-conceptualized domain. We conceptualized a taxonomy and derived archetypes for BA in smart manufacturing and discussed their appearance as well as temporal trends in this field. Our taxonomy, which comprises 52 characteristics, is based on the four meta-characteristics of domain, orientation, data, and technique. Our archetypes summarize the field's foci as MRO planning, maintenance (reactive, offline predictive, online predictive), MRO monitoring, and quality management. Our results reduce complexity for scientists and practitioners, in particular for those who are new to the field, by organizing research into archetypes, which can serve as orientation and context for new artifacts. Our work has revealed several theoretical and practical implications.

Our main contributions are a theoretically well-founded and empirically validated multi-dimensional taxonomy, which focuses on the functional and non-functional characteristics of BA applications in smart manufacturing, and archetypes of applications, derived exploratorily through cluster analysis of our comprehensive, coded bibliography.

As a first theoretical contribution, our taxonomy can be used to systematize BA in smart manufacturing for later analysis and provides a new level of understanding of the novel innovations and technologies within this emerging field. Scientists can use the taxonomy to study and hypothesize about relationships among techniques and applications as well as their characteristics. Our taxonomy not only supports the discussion about the area of research, but also provides a more profound knowledge of the emerging patterns of BA in smart manufacturing by providing a scheme to classify BA applications into archetypes.

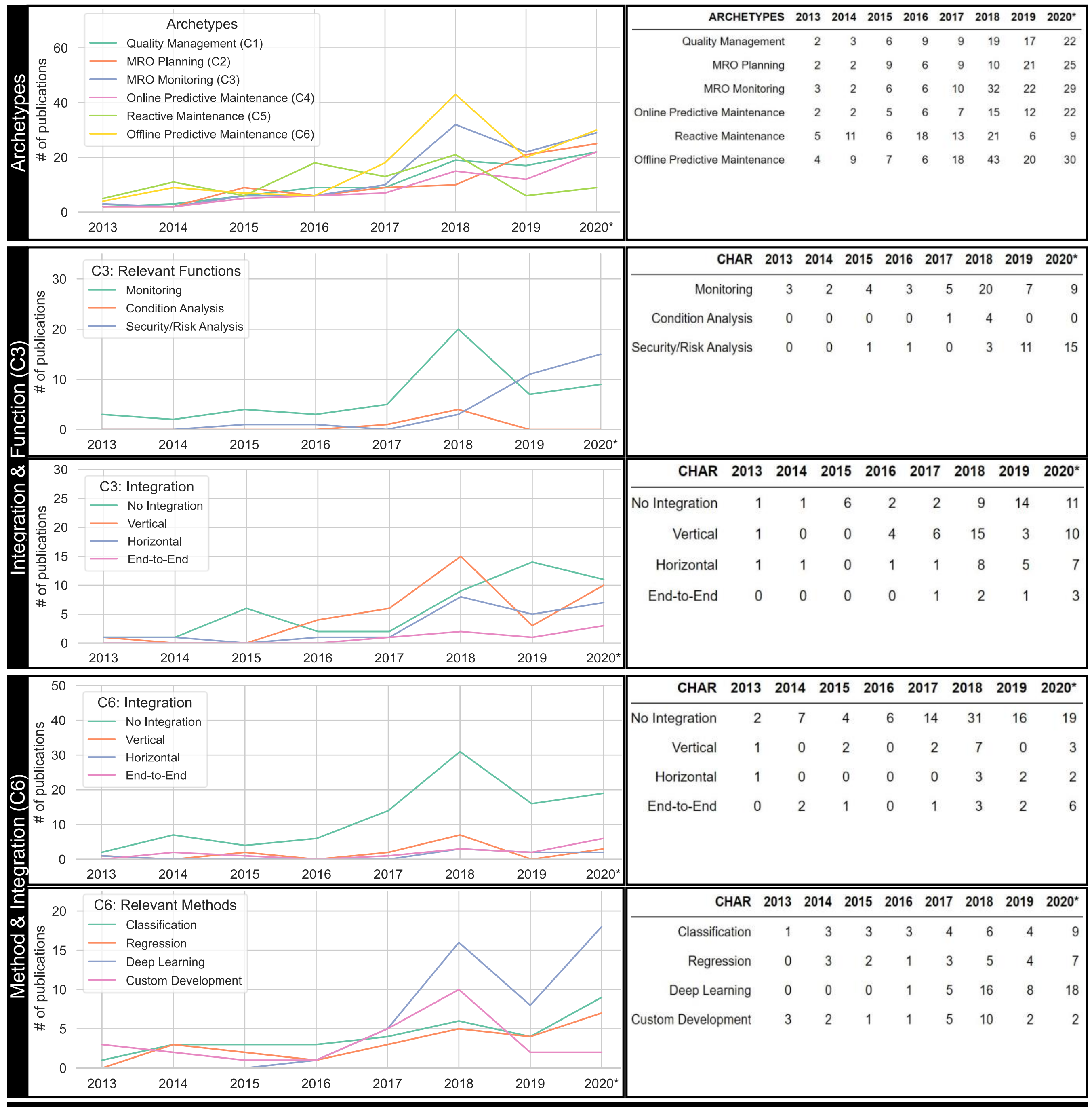

| ARCHETYPES | 2013 | 2014 | 2015 | 2016 | 2017 | 2018 | 2019 | 2020* |
|---|---|---|---|---|---|---|---|---|
| Quality Management | 2 | 3 | 6 | 9 | 9 | 19 | 17 | 22 |
| MRO Planning | 2 | 2 | 9 | 6 | 9 | 10 | 21 | 25 |
| MRO Monitoring | 3 | 2 | 6 | 6 | 10 | 32 | 22 | 29 |
| Online Predictive Maintenance | 2 | 2 | 5 | 6 | 7 | 15 | 12 | 22 |
| Reactive Maintenance | 5 | 11 | 6 | 18 | 13 | 21 | 6 | 9 |
| Offline Predictive Maintenance | 4 | 9 | 7 | 6 | 18 | 43 | 20 | 30 |

| CHAR | 2013 | 2014 | 2015 | 2016 | 2017 | 2018 | 2019 | 2020* |
|---|---|---|---|---|---|---|---|---|
| Monitoring | 3 | 2 | 4 | 3 | 5 | 20 | 7 | 9 |
| Condition Analysis | 0 | 0 | 0 | 0 | 1 | 4 | 0 | 0 |
| Security/Risk Analysis | 0 | 0 | 1 | 1 | 0 | 3 | 11 | 15 |

| CHAR | 2013 | 2014 | 2015 | 2016 | 2017 | 2018 | 2019 | 2020* |
|---|---|---|---|---|---|---|---|---|
| No Integration | 1 | 1 | 6 | 2 | 2 | 9 | 14 | 11 |
| Vertical | 1 | 0 | 0 | 4 | 6 | 15 | 3 | 10 |
| Horizontal | 1 | 1 | 0 | 1 | 1 | 8 | 5 | 7 |
| End-to-End | 0 | 0 | 0 | 0 | 1 | 2 | 1 | 3 |

| CHAR | 2013 | 2014 | 2015 | 2016 | 2017 | 2018 | 2019 | 2020* |
|---|---|---|---|---|---|---|---|---|
| No Integration | 2 | 7 | 4 | 6 | 14 | 31 | 16 | 19 |
| Vertical | 1 | 0 | 2 | 0 | 2 | 7 | 0 | 3 |
| Horizontal | 1 | 0 | 0 | 0 | 0 | 3 | 2 | 2 |
| End-to-End | 0 | 2 | 1 | 0 | 1 | 3 | 2 | 6 |

| CHAR | 2013 | 2014 | 2015 | 2016 | 2017 | 2018 | 2019 | 2020* |
|---|---|---|---|---|---|---|---|---|
| Classification | 1 | 3 | 3 | 3 | 4 | 6 | 4 | 9 |
| Regression | 0 | 3 | 2 | 1 | 3 | 5 | 4 | 7 |
| Deep Learning | 0 | 0 | 0 | 1 | 5 | 16 | 8 | 18 |
| Custom Development | 3 | 2 | 1 | 1 | 5 | 10 | 2 | 2 |

**Figure 6. Temporal Development of BA Archetypes in Smart Manufacturing**

While we said that Industry 4.0 and smart manufacturing represent the context as a constant for our research, our results enable us to turn the tables. Our taxonomy is a concise conceptual representation of the rich body of knowledge of BA in smart manufacturing literature. As Gregor (2006) points out, taxonomies can serve as a theoretical basis for analysis beyond their descriptive and classificational purpose. In their most basic form, they serve as taxonomic theory (Nickerson, Varshney, & Muntermann, 2017), which can be the basis for more advanced theories that aim to explain and predict. Thus, our taxonomy can itself be used as context for further investigations as it enables one to understand the domain and to describe elements that take part in the phenomena of examination. Summarizing, Bapna et al. (2004) assert that “a robust taxonomy can then be used to perform ex post theory building”. Given the comprehensiveness of our data collection and analysis, we consider our taxonomy robust enough to serve such purposes.

Second, our cluster analysis revealed six archetypes, which explain different foci and applications of BA, as well as their respective roles in smart manufacturing. This distinction can be used to differentiate among the scientific results of BA and their uses. It supports clarifying terminology and thus can make the academic discourse more concise. Furthermore, it helps to better interpret findings by being able to judge them in the context of comparable applications. In sum, the taxonomical systematization of BA and the uncovering of archetypes in smart manufacturing provide new tools for the scientific community to understand better application opportunities. For example, it can help to uncover applications within archetypes that exhibit comparable infrastructure and skill requirements despite being functionally distinct at first glance.

Moreover, our archetypes enable a more detailed analysis of the temporal variations of trends in research topics of BA for smart manufacturing. Beyond the general observation of the importance of offline maintenance and monitoring topics, as well as the demise of reactive maintenance research, we found a generally stable relationship between the BA domains. Our archetype analysis further revealed that security topics have gained more traction in recent years. In addition, the maturity of BA research has moved from descriptive to predictive approaches and may eventually transition to a predominant focus on prescriptive approaches. While the data dimension did not reveal insights per se, a look at the archetypes revealed that the distribution of characteristics is rather specific to them. Lastly, machine learning, and in particular deep learning, has come to dominate contemporary method use in recent years. From a scientific – but even more from a practical – point of view, this enables people to question and adjust their priorities to address contemporary challenges with their research or in their work environment.

From a practical perspective, our taxonomy provides dimensions and characteristics that can serve as a blueprint for BA selection and application in concrete real-world scenarios. Our taxonomy provides measures to evaluate the different options for BA, to address a specific problem, area, or requirement. We provide a distinction among archetypes that enables practitioners to find or review suitable configurations of BA to address their requirements. Assuming that research is several years ahead of productive practical applications, identifying these trends can create the awareness necessary to shape organizational and technical developments towards these future opportunities, for example by acquiring the requisite workforce skills.

As with any research, our study is not without its limitations. Taxonomies are never complete and should be considered as a starting point for further contextualization. Ours is the result of a design search process, which we have further documented in the appendices. While our data collection was rigorous, there may be further relevant applications of BA in smart manufacturing that have appeared in other domains of research and which we did not uncover in our data collection. The interpretation of data was not without issues, either. Clustering algorithms and segmentation metrics rarely point to the consistent segmentation of binary data. We consistently scrutinized our results qualitatively to ensure coherence and applicability.

In conclusion, in moving to establish smart manufacturing, businesses must adapt to disruptive and radical changes in value networks, business models, and operations due to the ever-emerging technological manufacturing innovations of the fourth industrial revolution. A massive increase in data, computing power, and connectivity is fueling novel applications of BA that can lead to cost advantages, increased customer satisfaction, and improvements in production effectiveness and quality. Future research could examine the archetypes in more detail and, for example, focus on security/risk analysis, both energy consumption characteristics, and design analysis, as they represent relatively uniform subclusters that we have not yet explored in greater depth. In addition, research should revisit trending areas and, for example, explore the use of machine learning in smart manufacturing in more detail as this will shed light on important future topics such as transfer learning and explainable artificial intelligence.

## References

Abbasi, T., Lim, K. H., Rosli, N., Ismail, I., & Ibrahim, R. (2018). *Development of Predictive Maintenance Interface Using Multiple Linear Regression.* Paper presented at the 2018 International Conference on Intelligent and Advanced System.

Abidi, M. H., Alkhalefah, H., Mohammed, M. K., Umer, U., & Qudeiri, J. E. A. (2020). Optimal Scheduling of Flexible Manufacturing System Using Improved Lion-Based Hybrid Machine Learning Approach. *IEEE Access, 8*, 96088-96114.

Adly, F., Yoo, P. D., Muhaidat, S., & Al-Hammadi, Y. (2014). *Machine-learning-based identification of defect patterns in semiconductor wafer maps: An overview and proposal.* Paper presented at the 2014 IEEE International Parallel & Distributed Processing Symposium Workshops.

Afshari, H., & Peng, Q. (2015). *Using Big Data to minimize uncertainty effects in adaptable product design.* Paper presented at the 2015 International Design Engineering Technical Conferences and Computers and Information in Engineering Conference.

Agarwal, K., & Shivpuri, R. (2014). Knowledge discovery in steel bar rolling mills using scheduling data and automated inspection. *Journal of Intelligent Manufacturing, 25*(6), 1289-1299. doi:10.1007/s10845-013-0730-5

Agarwal, K., & Shivpuri, R. (2015). On line prediction of surface defects in hot bar rolling based on Bayesian hierarchical modeling. *Journal of Intelligent Manufacturing, 26*(4), 785-800.

Ahmad, W., Khan, S. A., Islam, M. M., & Kim, J.-M. (2018). A reliable technique for remaining useful life estimation of rolling element bearings using dynamic regression models. *Reliability Engineering & System Safety, 184*, 67-76.

Ai-ming, X., Jian-min, G., & Kun, C. (2016). Excavation of critical resource node for quality control of multi-variety mixed production shopfloor based on complex network property. *Proceedings of the Institution of Mechanical Engineers, Part B: Journal of Engineering Manufacture, 230*(1), 169-177.

Ak, R., & Bhinge, R. (2015). *Data analytics and uncertainty quantification for energy prediction in manufacturing.* Paper presented at the International Conference on Big Data.

Åkerman, M., Lundgren, C., Bärring, M., Folkesson, M., Berggren, V., Stahre, J., . . . Friis, M. (2018). Challenges Building a Data Value Chain to Enable Data-Driven Decisions: A Predictive Maintenance Case in 5G-Enabled Manufacturing. *Procedia Manufacturing, 17*, 411-418. doi:https://doi.org/10.1016/j.promfg.2018.10.064

Akhavei, F., & Bleicher, F. (2016). *Predictive Modeling to Increase the Reliability of Production Planning in Single-item Production.* Paper presented at the Proceedings of the World Congress on Engineering and Computer Science.

Al Sunny, S. N., Liu, X., & Shahriar, M. R. (2018). *Remote Monitoring and Online Testing of Machine Tools for Fault Diagnosis and Maintenance Using MTComm in a Cyber-Physical Manufacturing Cloud.* Paper presented at the 11th International Conference on Cloud Computing.

Alavian, P., Eun, Y., Meerkov, S. M., & Zhang, L. (2020). Smart production systems: automating decision-making in manufacturing environment. *International Journal of Production Research, 58*(3), 828-845.

Alexopoulos, K., Makris, S., Xanthakis, V., Sipsas, K., & Chryssolouris, G. (2016). A concept for context-aware computing in manufacturing: the white goods case. *International Journal of Computer Integrated Manufacturing, 29*(8), 839-849. doi:10.1080/0951192X.2015.1130257

Alexopoulos, K., Nikolakis, N., & Chryssolouris, G. (2020). Digital twin-driven supervised machine learning for the development of artificial intelligence applications in manufacturing. *International Journal of Computer Integrated Manufacturing, 33*(5), 429-439.

Alfeo, A. L., Cimino, M. G., Manco, G., Ritacco, E., & Vaglini, G. (2020). Using an autoencoder in the design of an anomaly detector for smart manufacturing. *Pattern Recognition Letters, 136*, 272-278.

AlThobiani, F., & Ball, A. (2014). An approach to fault diagnosis of reciprocating compressor valves using Teager–Kaiser energy operator and deep belief networks. *Expert Systems with Applications, 41*(9), 4113-4122.

Amarnath, M., Sugumaran, V., & Kumar, H. (2013). Exploiting sound signals for fault diagnosis of bearings using decision tree. *Measurement, 46*(3), 1250-1256.

Andonovski, G., Mušič, G., & Škrjanc, I. (2018). Fault detection through evolving fuzzy cloud-based model. *IFAC-PapersOnLine, 51*(2), 795-800.

Ansari, F., Glawar, R., & Nemeth, T. (2019). PriMa: a prescriptive maintenance model for cyber-physical production systems. *International Journal of Computer Integrated Manufacturing, 32*(4-5), 482-503.

Anton, S. D., Kanoor, S., Fraunholz, D., & Schotten, H. D. (2018). *Evaluation of Machine Learning-based Anomaly Detection Algorithms on an Industrial Modbus/TCP Data Set.* Paper presented at the Proceedings of the 13th International Conference on Availability, Reliability and Security, Hamburg, Germany.

Aqlan, F., Ramakrishnan, S., & Shamsan, A. (2017). *Integrating data analytics and simulation for defect management in manufacturing environments.* Paper presented at the 2017 Winter Simulation Conference.

Aqlan, F., Saha, C., & Ramakrishnan, S. (2015). *Defect analytics in a high-end server manufacturing environment.* Paper presented at the IIE Annual Conference. Proceedings.

Arabzad, S. M., Ghorbani, M., Razmi, J., & Shirouyehzad, H. (2015). Employing fuzzy TOPSIS and SWOT for supplier selection and order allocation problem. *The International Journal of Advanced Manufacturing Technology, 76*(5-8), 803-818.

Arachchige, P. C. M., Bertok, P., Khalil, I., Liu, D., Camtepe, S., & Atiquzzaman, M. (2020). A trustworthy privacy preserving framework for machine learning in industrial iot systems. *IEEE Transactions on Industrial Informatics, 16*(9), 6092-6102.

Arellano-Espitia, F., Delgado-Prieto, M., Martinez-Viol, V., Saucedo-Dorantes, J. J., & Osornio-Rios, R. A. (2020). Deep-Learning-Based Methodology for Fault Diagnosis in Electromechanical Systems. *Sensors, 20*(14), 3949.

Argyriou, A., Evgeniou, T., & Pontil, M. (2008). Convex multi-task feature learning. *Machine Learning, 73*(3), 243-272.

Arık, O. A., & Toksarı, M. D. (2018). Multi-objective fuzzy parallel machine scheduling problems under fuzzy job deterioration and learning effects. *International Journal of Production Research, 56*(7), 2488-2505.

Arık, O. A., & Toksarı, M. D. (2019). Fuzzy Parallel Machine Scheduling Problem Under Fuzzy Job Deterioration and Learning Effects With Fuzzy Processing Times. In M. Ram (Ed.), *Advanced Fuzzy Logic Approaches in Engineering Science* (pp. 49-67). Hershey, PA: IGI Global.

Arpaia, P., Moccaldi, N., Prevete, R., Sannino, I., & Tedesco, A. (2020). A wearable EEG instrument for real-time frontal asymmetry monitoring in worker stress analysis. *IEEE Transactions on Instrumentation and Measurement, 69*(10), 8335-8343.

Ayad, S., Terrissa, L. S., & Zerhouni, N. (2018, 22-25 March 2018). *An IoT approach for a smart maintenance.* Paper presented at the 2018 International Conference on Advanced Systems and Electric Technologies (IC_ASET).

Aydemir, G., & Paynabar, K. (2019). Image-Based Prognostics Using Deep Learning Approach. *IEEE Transactions on Industrial Informatics, 16*(9), 5956-5964.

Aydin, I., Karakose, M., & Akin, E. (2014). An approach for automated fault diagnosis based on a fuzzy decision tree and boundary analysis of a reconstructed phase space. *ISA transactions, 53*(2), 220-229.

Aydın, İ., Karaköse, M., & Akın, E. (2015). Combined intelligent methods based on wireless sensor networks for condition monitoring and fault diagnosis. *Journal of Intelligent Manufacturing, 26*(4), 717-729.

Aydin, O., & Guldamlasioglu, S. (2017). *Using LSTM networks to predict engine condition on large scale data processing framework.* Paper presented at the 2017 4th International Conference on Electrical and Electronic Engineering.

Azadeh, A., Seif, J., Sheikhalishahi, M., & Yazdani, M. (2016). An integrated support vector regression–imperialist competitive algorithm for reliability estimation of a shearing machine. *International Journal of Computer Integrated Manufacturing, 29*(1), 16-24.

Baban, C. F., Baban, M., & Suteu, M. D. (2016). Using a fuzzy logic approach for the predictive maintenance of textile machines. *Journal of Intelligent & Fuzzy Systems, 30*(2), 999-1006.

Badurdeen, F., Shuaib, M., Wijekoon, K., Brown, A., Faulkner, W., Amundson, J., . . . Boden, B. (2014). Quantitative modeling and analysis of supply chain risks using Bayesian theory. *Journal of Manufacturing Technology Management, 25*(5), 631-654.

Bagheri, B., Yang, S., Kao, H.-A., & Lee, J. (2015). Cyber-physical Systems Architecture for Self-Aware Machines in Industry 4.0 Environment. *IFAC-PapersOnLine, 48*(3), 1622-1627. doi:https://doi.org/10.1016/j.ifacol.2015.06.318

Bai, Y., Sun, Z., Zeng, B., Long, J., Li, L., de Oliveira, J. V., & Li, C. (2018). A comparison of dimension reduction techniques for support vector machine modeling of multi-parameter manufacturing quality prediction. *Journal of Intelligent Manufacturing*, https://doi.org/10.1007/s10845-10017-11388-10841

Bakdi, A., Kouadri, A., & Bensmail, A. (2017). Fault detection and diagnosis in a cement rotary kiln using PCA with EWMA-based adaptive threshold monitoring scheme. *Control Engineering Practice, 66*, 64-75.

Balogun, V. A., & Mativenga, P. T. (2013). Modelling of direct energy requirements in mechanical machining processes. *Journal of Cleaner Production, 41*, 179-186.

Balsamo, V., Caggiano, A., Jemielniak, K., Kossakowska, J., Nejman, M., & Teti, R. (2016). Multi sensor signal processing for catastrophic tool failure detection in turning. *Procedia CIRP, 41*, 939-944.

Banerjee, A., Bandyopadhyay, T., & Acharya, P. (2013). Data Analytics: Hyped Up Aspirations or True Potential? *Vikalpa, 38*(4), 1-12.

Bang, S. H., Ak, R., Narayanan, A., Lee, Y. T., & Cho, H. (2019). A survey on knowledge transfer for manufacturing data analytics. *Computers in Industry, 104*, 116-130.

Bapna, R., Goes, P., Gupta, A., & Jin, Y. (2004). User heterogeneity and its impact on electronic auction market design: An empirical exploration. *MIS quarterly*, 21-43.

Baraldi, P., Cannarile, F., Di Maio, F., & Zio, E. (2016). Hierarchical k-nearest neighbours classification and binary differential evolution for fault diagnostics of automotive bearings operating under variable conditions. *Engineering Applications of Artificial Intelligence, 56*, 1-13.

Barde, S. R., Yacout, S., & Shin, H. (2019). Optimal preventive maintenance policy based on reinforcement learning of a fleet of military trucks. *Journal of Intelligent Manufacturing, 30*(1), 147-161.

Bastani, K., Barazandeh, B., & Kong, Z. J. (2018). Fault Diagnosis in Multistation Assembly Systems Using Spatially Correlated Bayesian Learning Algorithm. *Journal of Manufacturing Science and Engineering, 140*(3), 031003.

Bastania, K., Rao, P. K., & Zhenyu, K. (2016). An online sparse estimation-based classification approach for real-time monitoring in advanced manufacturing processes from heterogeneous sensor data. *IIE Transactions, 48*(7), 579-598. doi:10.1080/0740817X.2015.1122254

Batista, L., Badri, B., Sabourin, R., & Thomas, M. (2013). A classifier fusion system for bearing fault diagnosis. *Expert Systems with Applications, 40*(17), 6788-6797.

Bauer, W., Schlund, S., Marrenbach, D., & Ganschar, O. (2014). *Industrie 4.0 – Volkswirtschaftliches Potenzial für Deutschland*. Retrieved from

Bauernhansl, T. (2014). Die Vierte Industrielle Revolution – Der Weg in ein wertschaffendes Produktionsparadigma. In T. Bauernhansl, M. T. Hompel, & B. Vogel-Heuser (Eds.), *Industrie 4.0 in Produktion, Automatisierung und Logistik* (pp. 5-36). Wiesbaden: Springer Vieweg.

Bauernhansl, T. (2016). Wake-up Call for Enterprises - Why we Need a Common Understanding of Industrie 4.0. *ZWF Zeitschrift für wirtschaftlichen Fabrikbetrieb, 111*(7-11), 453-457

Baum, J., Laroque, C., Oeser, B., Skoogh, A., & Subramaniyan, M. (2018). Applications of Big Data analytics and Related Technologies in Maintenance—Literature-Based Research. *Machines, 6*(4), 54.

Bauza, M. B., Tenboer, J., Li, M., Lisovich, A., Zhou, J., Pratt, D., . . . Knebel, R. (2018). Realization of Industry 4.0 with high speed CT in high volume production. *CIRP Journal of Manufacturing Science and Technology, 22*, 121-125. doi:https://doi.org/10.1016/j.cirpj.2018.04.001

Beghi, A., Brignoli, R., Cecchinato, L., Menegazzo, G., Rampazzo, M., & Simmini, F. (2016). Data-driven fault detection and diagnosis for HVAC water chillers. *Control Engineering Practice, 53*, 79-91.

Bekar, E. T., Skoogh, A., Cetin, N., & Siray, O. (2018). *Prediction of Industry 4.0's Impact on Total Productive Maintenance Using a Real Manufacturing Case.* Paper presented at the The International Symposium for Production Research.

Benkedjouh, T., Medjaher, K., Zerhouni, N., & Rechak, S. (2015). Health assessment and life prediction of cutting tools based on support vector regression. *Journal of Intelligent Manufacturing, 26*(2), 213-223.

Benmoussa, S., & Djeziri, M. A. (2017). Remaining Useful Life estimation without needing for prior knowledge of the degradation features. *IET Science, Measurement & Technology, 11*(8), 1071-1078.

Berger, C., Berlak, J., & Reinhart, G. (2016). *Service-based Production Planning and Control of Cyber-Physical Production Systems*. Paper presented at the BLED 2016 Proceedings

Besenhard, M. O., Scheibelhofer, O., François, K., Joksch, M., & Kavsek, B. (2018). A multivariate process monitoring strategy and control concept for a small-scale fermenter in a PAT environment. *Journal of Intelligent Manufacturing, 29*(7), 1501-1514. doi:10.1007/s10845-015-1192-8

Bevilacqua, M., Ciarapica, F. E., Diamantini, C., & Potena, D. (2017). Big data analytics methodologies applied at energy management in industrial sector: A case study. *International Journal of Rf Technologies-Research and Applications, 8*(3), 105-122. doi:10.3233/rft-171671

Bhinge, R., Biswas, N., Dornfeld, D., Park, J., Law, K. H., Helu, M., & Rachuri, S. (2014). *An intelligent machine monitoring system for energy prediction using a Gaussian Process regression.* Paper presented at the International Conference on Big Data.

Bi, Q., Wang, X., Wu, Q., Zhu, L., & Ding, H. (2019). Fv-SVM-based wall-thickness error decomposition for adaptive machining of large skin parts. *IEEE Transactions on Industrial Informatics, 15*(4), 2426-2434.

Bi, Z. M., Liu, Y. F., Krider, J., Buckland, J., Whiteman, A., Beachy, D., & Smith, J. (2018). Real-time force monitoring of smart grippers for Internet of Things (IoT) applications. *Journal of Industrial Information Integration, 11*, 19-28. doi:10.1016/j.jii.2018.02.004

Binitha, S., & Sathya, S. S. (2012). A survey of bio inspired optimization algorithms. *International Journal of Soft Computing and Engineering, 2*(2), 137-151.

Bink, R., & Zschech, P. (2018). Predictive Maintenance in der industriellen Praxis. *HMD Praxis der Wirtschaftsinformatik, 55*(3), 552-565.

Bordeleau, F.-E., Mosconi, E., & Santa-Eulalia, L. A. (2018). *Business Intelligence in Industry 4.0: State of the art and research opportunities.* Paper presented at the Proceedings of the 51st Hawaii International Conference on System Sciences, Waikoloa, HI.

Borgi, T., Hidri, A., Neef, B., & Naceur, M. S. (2017, 14-17 Jan. 2017). *Data analytics for predictive maintenance of industrial robots.* Paper presented at the 2017 International Conference on Advanced Systems and Electric Technologies.

Bothe, H.-H. (1995). *Fuzzy Logic: Einführung in Theorie und Anwendungen* (Vol. Aufl. 2). Berlin Heidelberg: Springer.

Bouazza, W., Sallez, Y., & Beldjilali, B. (2017). A distributed approach solving partially flexible job-shop scheduling problem with a Q-learning effect. *IFAC-PapersOnLine, 50*(1), 15890-15895.

Bousdekis, A., Magoutas, B., Apostolou, D., & Mentzas, G. (2015). A proactive decision making framework for condition-based maintenance. *Industrial Management & Data Systems, 115*(7), 1225-1250.

Bousdekis, A., Magoutas, B., & Mentzas, G. (2015). Review, analysis and synthesis of prognostic-based decision support methods for condition based maintenance. *Journal of Intelligent Manufacturing, 29*(6), 1303-1316.

Bousdekis, A., Magoutas, B., Mentzas, G., & (2018), L. e. a. (2015). Review, analysis and synthesis of prognostic-based decision support methods for condition based maintenance. *Journal of Intelligent Manufacturing, 29*(6), 1303-1316. doi:10.1007/s10845-015-1179-5

Bousdekis, A., & Mentzas, G. (2017). *Condition-based predictive maintenance in the frame of industry 4.0.* Paper presented at the IFIP International Conference on Advances in Production Management Systems.

Bousdekis, A., Papageorgiou, N., Magoutas, B., Apostolou, D., & Mentzas, G. (2017). A proactive event-driven decision model for joint equipment predictive maintenance and spare parts inventory optimization. *Procedia CIRP, 59*, 184-189.

Bousdekis, A., Papageorgiou, N., Magoutas, B., Apostolou, D., & Mentzas, G. (2018). Enabling condition-based maintenance decisions with proactive event-driven computing. *Computers in Industry, 100*, 173-183.

Bowler, A. L., Bakalis, S., & Watson, N. J. (2020). Monitoring Mixing Processes Using Ultrasonic Sensors and Machine Learning. *Sensors, 20*(7), 1813.

Brandenburger, J., Colla, V., Nastasi, G., Ferro, F., Schirm, C., & Melcher, J. (2016). Big Data Solution for Quality Monitoring and Improvement on Flat Steel Production. *IFAC-PapersOnLine, 49*(20), 55-60. doi:https://doi.org/10.1016/j.ifacol.2016.10.096

Breiman, L. (2001). Statistical modeling: The two cultures (with comments and a rejoinder by the author). *Statistical science, 16*(3), 199-231.

Brodsky, A., Krishnamoorthy, M., Menascé, D. A., Shao, G., & Rachuri, S. (2014, 27-30 Oct. 2014). *Toward smart manufacturing using decision analytics.* Paper presented at the International Conference on Big Data

Brodsky, A., Shao, G., & Riddick, F. (2016). Process analytics formalism for decision guidance in sustainable manufacturing. *Journal of Intelligent Manufacturing, 27*(3), 561-580. doi:10.1007/s10845-014-0892-9

Brodsky, A., Shao, G. D., Krishnamoorthy, M., Narayanan, A., Menasce, D., & Ak, R. (2017). Analysis and optimization based on reusable knowledge base of process performance models. *International Journal of Advanced Manufacturing Technology, 88*(1-4), 337-357.

Brown, S. L., & Eisenhardt, K. M. (1995). Product development: Past research, present findings, and future directions. *The Academy of Management Review, 20*(2), 343-378.

Broy, M. (2010). Cyber-physical Systems - wissenschaftliche Herausforderungen bei der Entwicklung. In M. Broy (Ed.), *Cyber-physical Systems - Innovation durch softwareintensive eingebettete Systeme* (pp. 17-33). Berlin Heidelberg: Springer-Verlag.

Bughin, J., Manyika, J., & Woetzel, J. (2016). *The Age of Analytics: Competing in a data-driven world*. Retrieved from https://www.mckinsey.com/~/media/McKinsey/Industries/Public%20and%20Social%20Sector/Our%20Insights/The%20age%20of%20analytics%20Competing%20in%20a%20data%20driven%20world/MGI-The-Age-of-Analytics-Full-report.pdf

Bulnes, F. G., Usamentiaga, R., Garcia, D. F., & Molleda, J. (2016). An efficient method for defect detection during the manufacturing of web materials. *Journal of Intelligent Manufacturing, 27*(2), 431-445.

Bumblauskas, D., Gemmill, D., Igou, A., & Anzengruber, J. (2017). Smart Maintenance Decision Support Systems (SMDSS) based on corporate big data analytics. *Expert Systems with Applications, 90*, 303-317.

Bustillo, A., Urbikain, G., Perez, J. M., Pereira, O. M., & Lopez de Lacalle, L. N. (2018). Smart optimization of a friction-drilling process based on boosting ensembles. *Journal of Manufacturing Systems, 48*, 108-121. doi:https://doi.org/10.1016/j.jmsy.2018.06.004

Cachada, A., Barbosa, J., Leitño, P., Gcraldcs, C. A. S., Deusdado, L., Costa, J., . . . Romero, L. (2018, 4-7 Sept. 2018). *Maintenance 4.0: Intelligent and Predictive Maintenance System Architecture.* Paper presented at the 2018 IEEE 23rd International Conference on Emerging Technologies and Factory Automation

Cadavid, J. P. U., Lamouri, S., Grabot, B., Pellerin, R., & Fortin, A. (2020). Machine learning applied in production planning and control: a state-of-the-art in the era of industry 4.0. *Journal of Intelligent Manufacturing*, 1-28.

Caggiano, A. (2018). Cloud-based manufacturing process monitoring for smart diagnosis services. *International Journal of Computer Integrated Manufacturing, 31*(7), 612-623.

Cai, B., Liu, H., & Xie, M. (2016). A real-time fault diagnosis methodology of complex systems using object-oriented Bayesian networks. *Mechanical Systems and Signal Processing, 80*, 31-44.

Cai, H., Guo, Y., & Lu, K. (2017). A location prediction method for work-in-process based on frequent trajectory patterns. *Proceedings of the Institution of Mechanical Engineers, Part B: Journal of Engineering Manufacture, 233* (1), 306-320.

Cakir, M., Guvenc, M. A., & Mistikoglu, S. (2020). The experimental application of popular machine learning algorithms on predictive maintenance and the design of IIoT based condition monitoring system. *Computers & Industrial Engineering*, 106948.

Çaliş, B., & Bulkan, S. (2015). A research survey: review of AI solution strategies of job shop scheduling problem. *Journal of Intelligent Manufacturing, 26*(5), 961-973.

Candanedo, I. S., Nieves, E. H., González, S. R., Martín, M. T. S., & Briones, A. G. (2018). *Machine learning predictive model for industry 4.0.* Paper presented at the International Conference on Knowledge Management in Organizations.

Canito, A., Fernandes, M., Conceição, L., Praça, I., Santos, M., Rato, R., . . . Marreiros, G. (2017). *An Architecture for proactive maintenance in the machinery industry.* Paper presented at the International Symposium on Ambient Intelligence.

Canizo, M., Conde, A., Charramendieta, S., Minon, R., Cid-Fuentes, R. G., & Onieva, E. (2019). Implementation of a large-scale platform for cyber-physical system real-time monitoring. *IEEE Access, 7*, 52455-52466.

Canizo, M., Onieva, E., Conde, A., Charramendieta, S., & Trujillo, S. (2017, 19-21 June 2017). *Real-time predictive maintenance for wind turbines using Big Data frameworks.* Paper presented at the International Conference on Prognostics and Health Management.

Canizo, M., Triguero, I., Conde, A., & Onieva, E. (2019). Multi-head CNN–RNN for multi-time series anomaly detection: An industrial case study. *Neurocomputing, 363*, 246-260.

Cao, L., Weiss, G., & Philip, S. Y. (2012). A brief introduction to agent mining. *Autonomous Agents and Multi-Agent Systems, 25*(3), 419-424.

Cao, Z., Zhou, L., Hu, B., & Lin, C. (2019). An adaptive scheduling algorithm for dynamic jobs for dealing with the flexible job shop scheduling problem. *Business & Information Systems Engineering, 61*(3), 299-309.

Cao, Z., Zhou, P., Li, R., Huang, S., & Wu, D. (2020). Multiagent deep reinforcement learning for joint multichannel access and task offloading of mobile-edge computing in industry 4.0. *IEEE Internet of Things Journal, 7*(7), 6201-6213.

Carbery, C. M., Woods, R., & Marshall, A. H. (2018). *A Bayesian network based learning system for modelling faults in large-scale manufacturing.* Paper presented at the 2018 IEEE International Conference on Industrial Technology (ICIT).

Carbone, P., Katsifodimos, A., Ewen, S., Markl, V., Haridi, S., & Tzoumas, K. (2015). Apache flink: Stream and batch processing in a single engine. *Bulletin of the IEEE Computer Society Technical Committee on Data Engineering, 36*(4), 28-38.

Cardin, O., Trentesaux, D., Thomas, A., Castagna, P., Berger, T., & Bril El-Haouzi, H. (2017). Coupling predictive scheduling and reactive control in manufacturing hybrid control architectures: state of the art and future challenges. *Journal of Intelligent Manufacturing, 28*(7), 1503-1517.

Caricato, P., & Grieco, A. (2017). An Application of Industry 4.0 to the Production of Packaging Films. *Procedia Manufacturing, 11*, 949-956. doi:https://doi.org/10.1016/j.promfg.2017.07.199

Carino, J. A., Delgado-Prieto, M., Iglesias, J. A., Sanchis, A., Zurita, D., Millan, M., . . . Romero-Troncoso, R. (2018). Fault Detection and Identification Methodology Under an Incremental Learning Framework Applied to Industrial Machinery. *IEEE Access, 6*, 49755-49766.

Carino, J. A., Delgado-Prieto, M., Zurita, D., Millan, M., Redondo, J. A. O., & Romero-Troncoso, R. (2016). Enhanced Industrial Machinery Condition Monitoring Methodology Based on Novelty Detection and Multi-Modal Analysis. *IEEE Access, 4*, 7594-7604. doi:10.1109/ACCESS.2016.2619382

Carstensen, J., Carstensen, T., Pabst, M., Schulz, F., Friederichs, J., Aden, S., . . . Ortmaier, T. (2016). Condition monitoring and cloud-based energy analysis for autonomous mobile manipulation-smart factory concept with LUHbots. *Procedia Technology, 26*, 560-569.

Carvajal Soto, J., Tavakolizadeh, F., & Gyulai, D. (2019). An online machine learning framework for early detection of product failures in an Industry 4.0 context. *International Journal of Computer Integrated Manufacturing, 32*(4-5), 452-465.

Cerrada, M., Sánchez, R.-V., Li, C., Pacheco, F., Cabrera, D., de Oliveira, J. V., & Vásquez, R. E. (2018). A review on data-driven fault severity assessment in rolling bearings. *Mechanical Systems and Signal Processing, 99*, 169-196.

Chaki, S., Bathe, R. N., Ghosal, S., & Padmanabham, G. (2018). Multi-objective optimisation of pulsed Nd:YAG laser cutting process using integrated ANN-NSGAII model. *Journal of Intelligent Manufacturing, 29*(1), 175-190. doi:10.1007/s10845-015-1100-2

Chakravorti, N., Rahman, M. M., Sidoumou, M. R., Weinert, N., Gosewehr, F., & Wermann, J. (2018). Validation of PERFoRM reference architecture demonstrating an application of data mining for predicting machine failure. *Procedia CIRP, 72*, 1339-1344.

Chamnanlor, C., Sethanan, K., Gen, M., & Chien, C.-F. (2017). Embedding ant system in genetic algorithm for re-entrant hybrid flow shop scheduling problems with time window constraints. *Journal of Intelligent Manufacturing, 28*(8), 1915-1931. doi:10.1007/s10845-015-1078-9

Chao-Chun, C., Min-Hsiung, H., Po-Yi, L., Jia-Xuan, L., Yu-Chuan, L., & Chih-Jen, L. (2016, 21-25 Aug. 2016). *Development of a cyber-physical-style continuous yield improvement system for manufacturing industry.* Paper presented at the International Conference on Automation Science and Engineering.

Che, P., Liu, Y., Che, L., & Lang, J. (2020). Co-Optimization of Generation Self-Scheduling and Coal Supply for Coal-Fired Power Plants. *IEEE Access, 8*, 110633-110642.

Chen, B., & Chang, J.-Y. J. (2017). Dynamic Analysis of Intelligent Coil Leveling Machine for Cyber-physical Systems Implementation. *Procedia CIRP, 63*, 390-395. doi:https://doi.org/10.1016/j.procir.2017.03.115

Chen, B., Wan, J., Lan, Y., Imran, M., Li, D., & Guizani, N. (2019). Improving cognitive ability of edge intelligent IIoT through machine learning. *IEEE Network, 33*(5), 61-67.

Chen, C., Liu, Y., Kumar, M., & Qin, J. (2018). Energy Consumption Modelling Using Deep Learning Technique — A Case Study of EAF. *Procedia CIRP, 72*, 1063-1068. doi:https://doi.org/10.1016/j.procir.2018.03.095

Chen, C., Liu, Y., Kumar, M., Qin, J., & Ren, Y. (2019). Energy consumption modelling using deep learning embedded semi-supervised learning. *Computers & Industrial Engineering, 135*, 757-765.

Chen, C., Xia, B., Zhou, B.-h., & Xi, L. (2015). A reinforcement learning based approach for a multiple-load carrier scheduling problem. *Journal of Intelligent Manufacturing, 26*(6), 1233-1245. doi:10.1007/s10845-013-0852-9

Chen, J., Zhang, Z., & Wu, F. (2020). A data-driven method for enhancing the image-based automatic inspection of IC wire bonding defects. *International Journal of Production Research*, 1-15.

Chen, M. (2019). The influence of big data analysis of intelligent manufacturing under machine learning on start-ups enterprise. *Enterprise Information Systems*, 1-16.

Chen, X., Shen, Z., He, Z., Sun, C., & Liu, Z. (2013). Remaining life prognostics of rolling bearing based on relative features and multivariable support vector machine. *Proceedings of the Institution of Mechanical Engineers, Part C: Journal of Mechanical Engineering Science, 227*(12), 2849-2860.

Chen, Y.-J., Fan, C.-Y., & Chang, K.-H. (2016). Manufacturing intelligence for reducing false alarm of defect classification by integrating similarity matching approach in CMOS image sensor manufacturing. *Computers & Industrial Engineering, 99*, 465-473.

Chen, Y.-J., Lee, Y.-H., & Chiu, M.-C. (2018). *Construct an Intelligent Yield Alert and Diagnostic Analysis System via Data Analysis: Empirical Study of a Semiconductor Foundry.* Paper presented at the IFIP International Conference on Advances in Production Management Systems.

Chen, Y., Jin, Y., & Jiri, G. (2018). Predicting tool wear with multi-sensor data using deep belief networks. *The International Journal of Advanced Manufacturing Technology, 99*(5-8), 1917-1926.

Cheng, Y., Chen, K., Sun, H. M., Zhang, Y. P., & Tao, F. (2018). Data and knowledge mining with big data towards smart production. *Journal of Industrial Information Integration, 9*, 1-13.

Cheng, Y., Chen, M., Cheng, F., Cheng, Y., Lin, Y., & Yang, C. (2018, 13-17 April 2018). *Developing a decision support system (DSS) for a dental manufacturing production line based on data mining.* Paper presented at the International Conference on Applied System Invention.

Cheng, Y., Zhu, H., Wu, J., & Shao, X. (2018). Machine Health Monitoring Using Adaptive Kernel Spectral Clustering and Deep Long Short-Term Memory Recurrent Neural Networks. *IEEE Transactions on Industrial Informatics, 15*(2), 987 - 997.

Cheng, Z., & Cai, B. (2018). Predicting the remaining useful life of rolling element bearings using locally linear fusion regression. *Journal of Intelligent & Fuzzy Systems, 34*(6), 3735-3746.

Chiang, L. H., Jiang, B., Zhu, X., Huang, D., & Braatz, R. D. (2015). Diagnosis of multiple and unknown faults using the causal map and multivariate statistics. *Journal of Process Control, 28*, 27-39.

Chien, C.-F., Chang, K.-H., & Wang, W.-C. (2014). An empirical study of design-of-experiment data mining for yield-loss diagnosis for semiconductor manufacturing. *Journal of Intelligent Manufacturing, 25*(5), 961-972. doi:10.1007/s10845-013-0791-5

Chien, C.-F., Diaz, A. C., & Lan, Y.-B. (2014). A data mining approach for analyzing semiconductor MES and FDC data to enhance overall usage effectiveness. *International Journal of Computational Intelligence Systems, 7*(sup2), 52-65.

Chien, C.-F., Hsu, C.-Y., & Chen, P.-N. (2013a). Semiconductor fault detection and classification for yield enhancement and manufacturing intelligence. *Flexible Services and Manufacturing Journal, 25*(3), 367-388.

Chien, C.-F., Hsu, S.-C., & Chen, Y.-J. (2013b). A system for online detection and classification of wafer bin map defect patterns for manufacturing intelligence. *International Journal of Production Research, 51*(8), 2324-2338.

Chien, C., Chen, Y., Han, Y., Hsieh, M., Lee, C., Shih, T., . . . Yang, W. (2018, 7-7 Sept. 2018). *AI and Big Data Analytics for Wafer Fab Energy Saving and Chiller Optimization to Empower Intelligent Manufacturing.* Paper presented at the 2018 e-Manufacturing & Design Collaboration Symposium.

Cho, H.-W. (2015). Enhanced real-time quality prediction model based on feature selected nonlinear calibration techniques. *The International Journal of Advanced Manufacturing Technology, 78*(1-4), 633-640.

Cho, S., May, G., Tourkogiorgis, I., Perez, R., Lazaro, O., de la Maza, B., & Kiritsis, D. (2018). *A Hybrid Machine Learning Approach for Predictive Maintenance in Smart Factories of the Future.* Paper presented at the IFIP International Conference on Advances in Production Management Systems.

Choi, J.-H., Lee, K.-W., Jung, H., & Cho, E.-S. (2017). *Runtime Anomaly Detection Method in Smart Factories using Machine Learning on RDF Event Streams: Grand Challenge*. Paper presented at the Proceedings of the 11th ACM International Conference on Distributed and Event-based Systems, Barcelona, Spain.

Choi, S.-S., Cha, S.-H., & Tappert, C. C. (2010). A survey of binary similarity and distance measures. *Journal of Systemics, Cybernetics and Informatics, 8*(1), 43-48.

Choi, S., Youm, S., & Kang, Y.-S. (2019). Development of scalable on-line anomaly detection system for autonomous and adaptive manufacturing processes. *Applied Sciences, 9*(21), 4502.

Choo, B. Y., Adams, S., & Beling, P. (2017, 19-21 June 2017). *Health-aware hierarchical control for smart manufacturing using reinforcement learning.* Paper presented at the International Conference on Prognostics and Health Management.

Choo, B. Y., Adams, S. C., Weiss, B. A., Marvel, J. A., & Beling, P. A. (2016). Adaptive multi-scale prognostics and health management for smart manufacturing systems. *International journal of prognostics and health management, 7*, PMID: 28736651.

Chou, C., & Su, Y. (2017). A Block Recognition System Constructed by Using a Novel Projection Algorithm and Convolution Neural Networks. *IEEE Access, 5*, 23891-23900. doi:10.1109/ACCESS.2017.2762526

Chouhal, O., Mouss, H. L., Benaggoune, K., & Mahdaoui, R. (2016). A Multi-Agent Solution to Distributed Fault Diagnosis of Preheater Cement Cyclone. *Journal of Advanced Manufacturing Systems, 15*(4), 209-221. doi:10.1142/S0219686716500153

Chouliaras, S., & Sotiriadis, S. (2019). Real-Time Anomaly Detection of NoSQL Systems Based on Resource Usage Monitoring. *IEEE Transactions on Industrial Informatics, 16*(9), 6042-6049.

Chu, W.-L., Xie, M.-J., Wu, L.-W., Guo, Y.-S., & Yau, H.-T. (2020). The Optimization of Lathe Cutting Parameters Using a Hybrid Taguchi-Genetic Algorithm. *IEEE Access, 8*, 169576-169584.

Codjo, L., Jaafar, M., Makich, H., Knittel, D., & Nouari, M. (2018, 4-7 Sept. 2018). *Honeycomb Core Milling Diagnosis using Machine Learning in the Industry 4.0 Framework.* Paper presented at the 2018 IEEE 23rd International Conference on Emerging Technologies and Factory Automation.

Conde, A., Arriandiaga, A., Sanchez, J. A., Portillo, E., Plaza, S., & Cabanes, I. (2018). High-accuracy wire electrical discharge machining using artificial neural networks and optimization techniques. *Robotics and Computer-Integrated Manufacturing, 49*, 24-38. doi:https://doi.org/10.1016/j.rcim.2017.05.010

Cong, T., & Baranowski, J. (2018). Binary Classifier for Fault Detection Based on Gaussian Model and PCA. *IFAC-PapersOnLine, 51*(24), 1317-1323.

Conseil national de l'industrie. (2013). *The New Face of Industry In France*. Retrieved from Paris:

Cooper, H. M. (1988). Organizing knowledge syntheses. A taxonomy of literature reviews. *Knowledge in Society, 1*(1), 104-126.

Cui, J., Ren, L., Wang, X., & Zhang, L. (2019). Pairwise comparison learning based bearing health quantitative modeling and its application in service life prediction. *Future Generation Computer Systems, 97*, 578-586.

D'Addona, D. M., Ullah, A. S., & Matarazzo, D. (2017). Tool-wear prediction and pattern-recognition using artificial neural network and DNA-based computing. *Journal of Intelligent Manufacturing, 28*(6), 1285-1301.

da Silva, P. R. N., Gabbar, H. A., Junior, P. V., & da Costa Junior, C. T. (2018). A new methodology for multiple incipient fault diagnosis in transmission lines using QTA and Naïve Bayes classifier. *International Journal of Electrical Power & Energy Systems, 103*, 326-346.

Dai, J., Wang, J., Huang, W., Shi, J., & Zhu, Z. (2020). Machinery health monitoring based on unsupervised feature learning via generative adversarial networks. *IEEE/ASME Transactions on Mechatronics, 25*(5), 2252-2263.

Dalzochio, J., Kunst, R., Pignaton, E., Binotto, A., Sanyal, S., Favilla, J., & Barbosa, J. (2020). Machine learning and reasoning for predictive maintenance in Industry 4.0: Current status and challenges. *Computers in Industry, 123*, 103298.

Dan, Y., Dong, R., Cao, Z., Li, X., Niu, C., Li, S., & Hu, J. (2020). Computational Prediction of Critical Temperatures of Superconductors Based on Convolutional Gradient Boosting Decision Trees. *IEEE Access, 8*, 57868-57878.

Davenport, T. H., & Harris, J. G. (2007). *Competing on Analytics: The New Science of Winning*. Brighton, Massachusetts: Harvard Business Review Press.

Davis, J., Edgar, T., Porter, J., Bernaden, J., & Sarli, M. (2012). Smart manufacturing, manufacturing intelligence and demand-dynamic performance. *Computers & Chemical Engineering, 47*, 145-156.

de Farias, A., de Almeida, S. L. R., Delijaicov, S., Seriacopi, V., & Bordinassi, E. C. (2020). Simple machine learning allied with data-driven methods for monitoring tool wear in machining processes. *The International Journal of Advanced Manufacturing Technology, 109*(9), 2491-2501.

de Sa, A. O., Carmo, L. F. d. C., & Machado, R. C. (2017). Bio-inspired active system identification: a cyber-physical intelligence attack in networked control systems. *Mobile Networks and Applications*, 1-14.

Delen, D. (2014). *Real-Word Data Mining - Applied Business Analytics and Decision Making*. Upper Saddle River, NJ: Pearson Education, Inc.

Delen, D., & Demirkan, H. (2013). Data, information and analytics as services. *Decision Support Systems, 55*(1), 359-363.

Delen, D., & Zolbanin, H. M. (2018). The analytics paradigm in business research. *Journal of Business Research, 90*, 186-195.

Demertzis, K., Iliadis, L., Tziritas, N., & Kikiras, P. (2020). Anomaly detection via blockchained deep learning smart contracts in industry 4.0. *Neural Computing and Applications, 32*(23), 17361-17378.

Demetgul, M., Yildiz, K., Taskin, S., Tansel, I., & Yazicioglu, O. (2014). Fault diagnosis on material handling system using feature selection and data mining techniques. *Measurement, 55*, 15-24.

Deng, F., Huang, Y., Lu, S., Chen, Y., Chen, J., Feng, H., . . . Lam, T. L. (2020). A Multi-Sensor Data Fusion System for Laser Welding Process Monitoring. *IEEE Access, 8*, 147349-147357.

Deng, L., & Yu, D. (2014). Deep learning: methods and applications. *Foundations and Trends in Signal Processing, 7*(3–4), 197-387.

Denkena, B., Bergmann, B., & Witt, M. (2019). Material identification based on machine-learning algorithms for hybrid workpieces during cylindrical operations. *Journal of Intelligent Manufacturing, 30*(6), 2449-2456.

Denkena, B., Schmidt, J., & Krüger, M. (2014). Data Mining Approach for Knowledge-based Process Planning. *Procedia Technology, 15*, 406-415. doi:https://doi.org/10.1016/j.protcy.2014.09.095

Denno, P., Dickerson, C., & Harding, J. A. (2018). Dynamic production system identification for smart manufacturing systems. *Journal of Manufacturing Systems, 48*, 192-203. doi:https://doi.org/10.1016/j.jmsy.2018.04.006

Derwisch, S., & Iffert, L. (2017). Advanced & Predictive Analytics - Data Science im Fachbereich - Anwenderstudie. In: CXP Group.

Deuse, J., Lenze, D., Klenner, F., & Friedrich, T. (2016). Manufacturing Data Analytics zur Identifikation dynamischer Engpässe in Produktionssystemen mit hoher wertschöpfender Variabilität. In C. M. Schlick (Ed.), *Megatrend Digitalisierung-Potenziale der Arbeits-und Betriebsorganisation* (pp. 11-26). Berlin.

Deutsch, J., He, M., & He, D. (2017). Remaining useful life prediction of hybrid ceramic bearings using an integrated deep learning and particle filter approach. *Applied Sciences, 7*(7), 649.

Di, Y., Song, W., Liu, L., & Wang, H. (2017, 25-26 March 2017). *A data mining approach for intelligent equipment fault diagnosis.* Paper presented at the 2017 IEEE 2nd Advanced Information Technology, Electronic and Automation Control Conference (IAEAC).

Diao, G., Zhao, L., & Yao, Y. (2015). A dynamic quality control approach by improving dominant factors based on improved principal component analysis. *International Journal of Production Research, 53*(14), 4287-4303.

Diaz-Rozo, J., Bielza, C., & Larrañaga, P. (2017). Machine Learning-based CPS for Clustering High throughput Machining Cycle Conditions. *Procedia Manufacturing, 10*, 997-1008. doi:https://doi.org/10.1016/j.promfg.2017.07.091

Diez-Olivan, A., Del Ser, J., Galar, D., & Sierra, B. (2019). Data fusion and machine learning for industrial prognosis: Trends and perspectives towards Industry 4.0. *Information Fusion, 50*, 92-111.

Diez-Olivan, A., Pagan, J. A., Khoa, N. L. D., Sanz, R., & Sierra, B. (2018). Kernel-based support vector machines for automated health status assessment in monitoring sensor data. *The International Journal of Advanced Manufacturing Technology, 95*(1-4), 327-340.

Dimitriou, N., Leontaris, L., Vafeiadis, T., Ioannidis, D., Wotherspoon, T., Tinker, G., & Tzovaras, D. (2019). Fault diagnosis in microelectronics attachment via deep learning analysis of 3-D laser scans. *IEEE Transactions on Industrial Electronics, 67*(7), 5748-5757.

Ding, H., Gao, R. X., Isaksson, A. J., Landers, R. G., Parisini, T., & Yuan, Y. (2020). State of AI-based monitoring in smart manufacturing and introduction to focused section. *IEEE/ASME Transactions on Mechatronics, 25*(5), 2143-2154.

Ding, K., & Jiang, P. (2016). Incorporating Social Sensors and CPS Nodes for Personalized Production under Social Manufacturing Environment. *Procedia CIRP, 56*, 366-371.

Ding, K., & Jiang, P. (2018). RFID-based production data analysis in an IoT-enabled smart job-shop. *IEEE/CAA Journal of Automatica Sinica, 5*(1), 128-138. doi:10.1109/JAS.2017.7510418

Ding, K., Zhang, X., Chan, F. T., Chan, C.-Y., & Wang, C. (2019). Training a hidden Markov model-based knowledge model for autonomous manufacturing resources allocation in smart shop floors. *IEEE Access, 7*, 47366-47378.

Dolata, P., Mrzygłód, M., & Reiner, J. (2017). Double-stream Convolutional Neural Networks for Machine Vision Inspection of Natural Products. *Applied Artificial Intelligence, 31*(7/8), 643-659. doi:10.1080/08839514.2018.1428491

Doltsinis, S., Krestenitis, M., & Doulgeri, Z. (2019). A machine learning framework for real-time identification of successful snap-fit assemblies. *IEEE Transactions on Automation Science and Engineering, 17*(1), 513-523.

Domova, V., & Dagnino, A. (2017, 6-9 June 2017). *Towards intelligent alarm management in the Age of IIoT.* Paper presented at the 2017 Global Internet of Things Summit.

Dong, C.-L., Zhang, Q., & Geng, S.-C. (2014). A modeling and probabilistic reasoning method of dynamic uncertain causality graph for industrial fault diagnosis. *International Journal of Automation and Computing, 11*(3), 288-298.

Doty, D. H., & Glick, W. H. (1994). Typologies as a unique form of theory building: Toward improved understanding and modeling. *Academy of management review, 19*(2), 230-251.

Dou, D., & Zhou, S. (2016). Comparison of four direct classification methods for intelligent fault diagnosis of rotating machinery. *Applied Soft Computing, 46*, 459-468.

Dowdeswell, B., Sinha, R., & MacDonell, S. G. (2020). Finding faults: A scoping study of fault diagnostics for Industrial Cyber–Physical Systems. *Journal of Systems and Software, 168*, 110638.

Du, S., Liu, C., & Xi, L. (2015). A selective multiclass support vector machine ensemble classifier for engineering surface classification using high definition metrology. *Journal of Manufacturing Science and Engineering, 137*(1), 011003.

Du, W., Kang, M., & Pecht, M. (2019). Fault diagnosis using adaptive multifractal detrended fluctuation analysis. *IEEE Transactions on Industrial Electronics, 67*(3), 2272-2282.

Du, Y., & Du, D. (2018). Fault detection and diagnosis using empirical mode decomposition based principal component analysis. *Computers & Chemical Engineering, 115*, 1-21.

Duan, C., Deng, C., Gong, Q., & Wang, Y. (2018). Optimal failure mode-based preventive maintenance scheduling for a complex mechanical device. *The International Journal of Advanced Manufacturing Technology, 95*(5-8), 2717-2728.

Duong, B., Khan, S., Shon, D., Im, K., Park, J., Lim, D.-S., . . . Kim, J.-M. (2018). A Reliable Health Indicator for Fault Prognosis of Bearings. *Sensors, 18*(11), 3740.

Dutta, R., Mueller, H., & Liang, D. (2018, 23-26 April 2018). *An interactive architecture for industrial scale prediction: Industry 4.0 adaptation of machine learning.* Paper presented at the 2018 Annual IEEE International Systems Conference (SysCon).

Eiskop, T., Snatkin, A., Kõrgesaar, K., & Søren, J. (2014). *Development and application of a holistic production monitoring system.* Paper presented at the Proc. 9th International Conference of DAAAM Baltic Industrial Engineering.

Elgendi, I., Hossain, M. F., Jamalipour, A., & Munasinghe, K. S. (2019). Protecting cyber physical systems using a learned MAPE-K model. *IEEE Access, 7*, 90954-90963.

Ellefsen, A. L., Bjørlykhaug, E., Æsøy, V., Ushakov, S., & Zhang, H. (2018). Remaining Useful Life Predictions for Turbofan Engine Degradation Using Semi-Supervised Deep Architecture. *Reliability Engineering & System Safety, 183*, 240-251.

Elsheikh, A., Yacout, S., Ouali, M.-S., & Shaban, Y. (2020). Failure time prediction using adaptive logical analysis of survival curves and multiple machining signals. *Journal of Intelligent Manufacturing, 31*(2), 403-415.

Emec, S., Krüger, J., & Seliger, G. (2016). Online fault-monitoring in machine tools based on energy consumption analysis and non-invasive data acquisition for improved resource-efficiency. *Procedia CIRP, 40*, 236-243.

Engeler, M., Treyer, D., Zogg, D., Wegener, K., & Kunz, A. (2016). Condition-based Maintenance: Model vs. Statistics a Performance Comparison. *Procedia CIRP, 57*, 253-258.

Epureanu, B. I., Li, X., Nassehi, A., & Koren, Y. (2020). Self-repair of smart manufacturing systems by deep reinforcement learning. *CIRP Annals, 69*(1), 421-424.

Essien, A., & Giannetti, C. (2020). A deep learning model for smart manufacturing using convolutional LSTM neural network autoencoders. *IEEE Transactions on Industrial Informatics, 16*(9), 6069-6078.

Europäische Kommission. (2016). *Factories of the Future PPP: towards competitive EU manufacturing*. Retrieved from Brüssel:

Evans, P. C., & Annuziata, M. (2012). *Industrial Internet: Pushing the Boundaries of Minds and Machines*. Retrieved from Boston, MA:

Ezeme, O. M., Mahmoud, Q. H., & Azim, A. (2019). Dream: deep recursive attentive model for anomaly detection in kernel events. *IEEE Access, 7*, 18860-18870.

Fan, X., Zhu, X., Kuo, K. C., Lu, C., & Wu, J. (2017). *Big data analytics to improve photomask manufacturing productivity.* Paper presented at the 2017 IEEE International Conference on Industrial Engineering and Engineering Management.

Fang, P., Yang, J., Zheng, L., Zhong, R. Y., & Jiang, Y. (2020). Data analytics-enable production visibility for Cyber-Physical Production Systems. *Journal of Manufacturing Systems, 57*, 242-253.

Fang, X., Luo, J., Luo, G., Wu, W., Cai, Z., & Pan, Y. (2019). Big data transmission in industrial IoT systems with small capacitor supplying energy. *IEEE Transactions on Industrial Informatics, 15*(4), 2360-2371.

Faraci, G., Raciti, A., Rizzo, S. A., & Schembra, G. (2020). Green wireless power transfer system for a drone fleet managed by reinforcement learning in smart industry. *Applied Energy, 259*, 114204.

Farivar, F., Haghighi, M. S., Jolfaei, A., & Alazab, M. (2019). Artificial intelligence for detection, estimation, and compensation of malicious attacks in nonlinear cyber-physical systems and industrial IoT. *IEEE Transactions on Industrial Informatics, 16*(4), 2716-2725.

Fay, M., & Kazantsev, N. (2018). *When Smart Gets Smarter: How Big Data Analytics Creates Business Value in Smart Manufacturing.* Paper presented at the Proceedings of the 39th International Conference on Information Systems (ICIS), San Francisco, CA.

Fayyad, U., Piatetsky-Shapiro, G., & Smyth, P. (1996). From data mining to knowledge discovery in databases. *AI magazine, 17*(3), 37.

Feng, J., Li, F., Xu, C., & Zhong, R. Y. (2018). Data-driven analysis for RFID-enabled smart factory: A case study. *IEEE Transactions on Systems, Man, and Cybernetics: Systems, 50*(1), 81-88.

Feng, Y., & Huang, B. (2018). Cloud manufacturing service QoS prediction based on neighbourhood enhanced matrix factorization. *Journal of Intelligent Manufacturing*, https://doi.org/10.1007/s10845-10018-11409-10848.

Ferber, J., & Weiss, G. (1999). *Multi-agent systems: an introduction to distributed artificial intelligence* (Vol. Aufl. 1). Boston, MA: Addison-Wesley Reading.

Fernandes, M., Canito, A., Bolón-Canedo, V., Conceição, L., Praça, I., & Marreiros, G. (2018). Data analysis and feature selection for predictive maintenance: A case-study in the metallurgic industry. *International Journal of Information Management, In Press, Corrected Proof*, https://doi.org/10.1016/j.ijinfomgt.2018.1010.1006.

Filonenko, A., & Jo, K. (2018, 15-18 May 2018). *Fast fire flame detection on videos using AdaBoost and parallel processing.* Paper presented at the Industrial Cyber-Physical Systems.

Fink, O., Zio, E., & Weidmann, U. (2014). Predicting component reliability and level of degradation with complex-valued neural networks. *Reliability Engineering & System Safety, 121*, 198-206.

Flath, C. M., & Stein, N. (2018). Towards a data science toolbox for industrial analytics applications. *Computers in Industry, 94*, 16-25.

Fleischmann, H., Kohl, J., & Franke, J. (2016, 12-16 June 2016). *A reference architecture for the development of socio-cyber-physical condition monitoring systems.* Paper presented at the 2016 11th System of Systems Engineering Conference (SoSE).

Fleischmann, H., Spreng, S., Kohl, J., Kißkalt, D., & Franke, J. (2016, 30 Nov.-1 Dec. 2016). *Distributed condition monitoring systems in electric drives manufacturing.* Paper presented at the 6th International Electric Drives Production Conference.

Foresight. (2013). *The Future of Manufacturing: A New Era of Oppurtunity and Challenge for the UK - Summary Report*. Retrieved from London:

Foresti, R., Rossi, S., Magnani, M., Bianco, C. G. L., & Delmonte, N. (2020). Smart society and artificial intelligence: big data scheduling and the global standard method applied to smart maintenance. *Engineering, 6*(7), 835-846.

Frieß, U., Kolouch, M., Friedrich, A., & Zander, A. (2018). Fuzzy-clustering of machine states for condition monitoring. *CIRP Journal of Manufacturing Science and Technology, 23*, 64-77. doi:https://doi.org/10.1016/j.cirpj.2018.09.001

Frumosu, F. D., Khan, A. R., Schiøler, H., Kulahci, M., Zaki, M., & Westermann-Rasmussen, P. (2020). Cost-sensitive learning classification strategy for predicting product failures. *Expert Systems with Applications, 161*, 113653.

Fu, L., Wei, Y., Fang, S., Zhou, X., & Lou, J. (2017). Condition Monitoring for Roller Bearings of Wind Turbines Based on Health Evaluation under Variable Operating States. *Energies, 10*(10), 1-21.

Fu, Y., Zhou, M., Guo, X., & Qi, L. (2019). Artificial-molecule-based chemical reaction optimization for flow shop scheduling problem with deteriorating and learning effects. *IEEE Access, 7*, 53429-53440.

Fumagalli, L., Macchi, M., Colace, C., Rondi, M., & Alfieri, A. (2016). A Smart Maintenance tool for a safe Electric Arc Furnace. *IFAC-PapersOnLine, 49*(31), 19-24. doi:https://doi.org/10.1016/j.ifacol.2016.12.155

Gajjar, S., Kulahci, M., & Palazoglu, A. (2018). Real-time fault detection and diagnosis using sparse principal component analysis. *Journal of Process Control, 67*, 112-128. doi:10.1016/j.jprocont.2017.03.005

Gan, M., & Wang, C. (2016). Construction of hierarchical diagnosis network based on deep learning and its application in the fault pattern recognition of rolling element bearings. *Mechanical Systems and Signal Processing, 72*, 92-104.

Gan, M., Wang, C., & Zhu, C. a. (2018). Fault feature enhancement for rotating machinery based on quality factor analysis and manifold learning. *Journal of Intelligent Manufacturing, 29*(2), 463-480.

Gao, X.-C., Zhang, J.-K., Chen, H., Dong, Z., & Vucetic, B. (2018). Energy-efficient and low-latency massive SIMO using noncoherent ML detection for industrial IoT communications. *IEEE Internet of Things Journal, 6*(4), 6247-6261.

García, V., Sánchez, J. S., Rodríguez-Picón, L. A., Méndez-González, L. C., & de Jesús Ochoa-Domínguez, H. (2018). Using regression models for predicting the product quality in a tubing extrusion process. *Journal of Intelligent Manufacturing*, DOI: 10.1007/s10845-10018-11418-10847.

Gawand, H. L., Bhattacharjee, A. K., & Roy, K. (2017). Securing a Cyber Physical System in Nuclear Power Plants Using Least Square Approximation and Computational Geometric Approach. *Nuclear Engineering and Technology, 49*(3), 484-494.

Genge, B., Haller, P., & Enăchescu, C. (2019). Anomaly detection in aging industrial internet of things. *IEEE Access, 7*, 74217-74230.

Germen, E., Başaran, M., & Fidan, M. (2014). Sound based induction motor fault diagnosis using Kohonen self-organizing map. *Mechanical Systems and Signal Processing, 46*(1), 45-58.

Ghadimi, P., Toosi, F. G., & Heavey, C. (2018). A multi-agent systems approach for sustainable supplier selection and order allocation in a partnership supply chain. *European Journal of Operational Research, 269*(1), 286-301.

Ghahramani, M., Qiao, Y., Zhou, M., Hagan, A. O., & Sweeney, J. (2020). AI-based modeling and data-driven evaluation for smart manufacturing processes. *IEEE/CAA Journal of Automatica Sinica, 7*(4), 1026-1037.

Giannetti, C., & Ransing, R. S. (2016). Risk based uncertainty quantification to improve robustness of manufacturing operations. *Computers & Industrial Engineering, 101*, 70-80. doi:10.1016/j.cie.2016.08.002

Glawar, R., Kemeny, Z., Nemeth, T., Matyas, K., Monostori, L., & Sihn, W. (2016). A holistic approach for quality oriented maintenance planning supported by data mining methods. *Procedia CIRP, 57*(1), 259-264.

Gölzer, P., Cato, P., & Amberg, M. (2015). *Data Processing Requirements of Industry 4.0 - Use Cases for Big Data Applications*. Paper presented at the ECIS 2015 Research-in-Progress Papers.

Gölzer, P., & Fritzsche, A. (2017). Data-driven operations management: organisational implications of the digital transformation in industrial practice. *Production Planning & Control, 28*(16), 1332-1343.

Goryachev, A., Kozhevnikov, S., Kolbova, E., Kuznetsov, O., Simonova, E., Skobelev, P., . . . Shepilov, Y. (2013). “Smart Factory”: Intelligent System for Workshop Resource Allocation, Scheduling, Optimization and Controlling in Real Time. *Advanced Materials Research, 630*, 508-513.

Gou, L., Zeng, X., Wang, Z., Han, G., Lin, C., & Cheng, X. (2019). A linearization model of turbofan engine for intelligent analysis towards industrial Internet of Things. *IEEE Access, 7*, 145313-145323.

Gouarir, A., Martínez-Arellano, G., Terrazas, G., Benardos, P., & Ratchev, S. (2018). In-Process Tool Wear Prediction System Based on Machine Learning Techniques and Force Analysis. *Procedia CIRP, 77*, 501-504.

Gowid, S., Dixon, R., & Ghani, S. (2015). A novel robust automated FFT-based segmentation and features selection algorithm for acoustic emission condition based monitoring systems. *Applied Acoustics, 88*, 66-74.
Granados, G. E., Lacroix, L., & Medjaher, K. (2018). Condition monitoring and prediction of solution quality during a copper electroplating process. *Journal of Intelligent Manufacturing*, https://doi.org/10.1007/s10845-10018-11445-10844.
Gregor, S. (2006). The nature of theory in information systems. *MIS quarterly*, 611-642.
Gröger, C., Kassner, L., Hoos, E., Königsberger, J., Kiefer, C., Silcher, S., & Mitschang, B. (2016). *The Data-driven Factory - Leveraging Big Industrial Data for Agile, Learning and Human-centric Manufacturing.* Paper presented at the 18th International Conference on Enterprise Information Systems.
Gröger, C., Stach, C., Mitschang, B., & Westkämper, E. (2016). A mobile dashboard for analytics-based information provisioning on the shop floor. *International Journal of Computer Integrated Manufacturing, 29*(12), 1335-1354. doi:10.1080/0951192X.2016.1187292
Grover, V. (2019). Surviving and thriving in the evolving digital age: A peek into the future of IS research and practice. *ACM SIGMIS Database: the DATABASE for Advances in Information Systems, 50*(1), 25-34.
Grzenda, M., & Bustillo, A. (2019). Semi-supervised roughness prediction with partly unlabeled vibration data streams. *Journal of Intelligent Manufacturing, 30*(2), 933-945.
Gubbi, J., Buyya, R., Marusic, S., & Palaniswami, M. (2013). Internet of Things (IoT): A vision, architectural elements, and future directions. *Future generation computer systems, 29*(7), 1645-1660.
Gugulothu, N., TV, V., Malhotra, P., Vig, L., Agarwal, P., & Shroff, G. (2017). Predicting Remaining Useful Life using Time Series Embeddings based on Recurrent Neural Networks. *arXiv preprint arXiv:1709.01073*.
Guo, L., Lei, Y., Xing, S., Yan, T., & Li, N. (2018). Deep Convolutional Transfer Learning Network: A New Method for Intelligent Fault Diagnosis of Machines with Unlabeled Data. *IEEE Transactions on Industrial Electronics, Early Access*, DOI: 10.1109/TIE.2018.2877090.
Guo, L., Li, N., Jia, F., Lei, Y., & Lin, J. (2017). A recurrent neural network based health indicator for remaining useful life prediction of bearings. *Neurocomputing, 240*, 98-109.
Guo, Z., Ngai, E., Yang, C., & Liang, X. (2015). An RFID-based intelligent decision support system architecture for production monitoring and scheduling in a distributed manufacturing environment. *International Journal of Production Economics, 159*, 16-28.
Gururajapathy, S. S., Mokhlis, H., Illias, H. A. B., & Awalin, L. J. (2017). Support vector classification and regression for fault location in distribution system using voltage sag profile. *IEEJ Transactions on Electrical and Electronic Engineering, 12*(4), 519-526.
Haas, M. (2018). *Germany Industry 4.0 Index 2018*. Retrieved from Köngen/Stuttgart:
Haasbroek, A., Strydom, J. J., McCoy, J. T., & Auret, L. (2018). Fault Diagnosis for an Industrial High Pressure Leaching Process with a Monitoring Dashboard. *IFAC-PapersOnLine, 51*(21), 117-122.
Halawa, F., Dauod, H., Lee, I. G., Li, Y., Yoon, S. W., & Chung, S. H. (2020). Introduction of a real time location system to enhance the warehouse safety and operational efficiency. *International Journal of Production Economics, 224*, 107541.
Hammer, M., Somers, K., Karre, H., & Ramsauer, C. (2017). Profit Per Hour as a Target Process Control Parameter for Manufacturing Systems Enabled by Big Data Analytics and Industry 4.0 Infrastructure. *Procedia CIRP, 63*, 715-720. doi:https://doi.org/10.1016/j.procir.2017.03.094
Han, B.-A., & Yang, J.-J. (2020). Research on Adaptive Job Shop Scheduling Problems Based on Dueling Double DQN. *IEEE Access, 8*, 186474-186495.
Han, D., Zhao, N., & Shi, P. (2017). A new fault diagnosis method based on deep belief network and support vector machine with Teager–Kaiser energy operator for bearings. *Advances in Mechanical Engineering, 9*(12), https://doi.org/10.1177%1172F1687814017743113.
Han, Q., Li, H., Dong, W., Luo, Y., & Xia, Y. (2017, 26-28 July 2017). *On fault prediction based on industrial big data.* Paper presented at the 2017 36th Chinese Control Conference (CCC).
Han, W., Tian, Z., Shi, W., Huang, Z., & Li, S. (2019). Low-power distributed data flow anomaly-monitoring technology for industrial internet of things. *Sensors, 19*(12), 2804.
Hang, L. (2016, 23-26 Oct. 2016). *An approach to improve flexible manufacturing systems with machine learning algorithms.* Paper presented at the 42nd Annual Conference of the IEEE Industrial Electronics Society.
Hao, L., Bian, L., Gebraeel, N., & Shi, J. (2017). Residual life prediction of multistage manufacturing processes with interaction between tool wear and product quality degradation. *IEEE Transactions on Automation Science and Engineering, 14*(2), 1211-1224.
Hao, R., Lu, B., Cheng, Y., Li, X., & Huang, B. (2020). A steel surface defect inspection approach towards smart industrial monitoring. *Journal of Intelligent Manufacturing*, 1-11.

Harris, K., Triantafyllopoulos, K., Stillman, E., & McLeay, T. (2016). A Multivariate Control Chart for Autocorrelated Tool Wear Processes. *Quality and Reliability Engineering International, 32*(6), 2093-2106. doi:10.1002/qre.2032

Hassan, M. M., Gumaei, A., Huda, S., & Almogren, A. (2020). Increasing the trustworthiness in the industrial iot networks through a reliable cyberattack detection model. *IEEE Transactions on Industrial Informatics, 16*(9), 6154-6162.

Haverkort, B. R., & Zimmermann, A. (2017). Smart Industry: How ICT Will Change the Game! *IEEE Internet Computing, 21*(1), 8-10.

He, K., Zhang, M., Zuo, L., Alhwiti, T., & Megahed, F. (2017). Enhancing the monitoring of 3D scanned manufactured parts through projections and spatiotemporal control charts. *Journal of Intelligent Manufacturing, 28*(4), 899-911. doi:10.1007/s10845-014-1025-1

He, K., Zhao, Z., Jia, M., & Liu, C. (2018). Dynamic Bayesian Network-based Approach by Integrating Sensor Deployment for Machining Process Monitoring. *IEEE Access, 6*, 33362-33375.

He, Q. P., & Wang, J. (2017, 24-26 May 2017). *Statistical process monitoring in the era of smart manufacturing.* Paper presented at the 2017 American Control Conference.

He, Q. P., & Wang, J. (2018). Statistical process monitoring as a big data analytics tool for smart manufacturing. *Journal of Process Control, 67*, 35-43. doi:https://doi.org/10.1016/j.jprocont.2017.06.012

He, Q. P., & Wang, J. (2018). Statistics Pattern Analysis: A Statistical Process Monitoring Tool for Smart Manufacturing. *Computer Aided Chemical Engineering, 44*, 2071-2076.

He, R., Dai, Y., Lu, J., & Mou, C. (2018). Developing ladder network for intelligent evaluation system: Case of remaining useful life prediction for centrifugal pumps. *Reliability Engineering & System Safety, 180*, 385-393.

He, S.-G., He, Z., & Wang, G. A. (2013). Online monitoring and fault identification of mean shifts in bivariate processes using decision tree learning techniques. *Journal of Intelligent Manufacturing, 24*(1), 25-34.

He, Y., Zhu, C., He, Z., Gu, C., & Cui, J. (2017). Big data oriented root cause identification approach based on Axiomatic domain mapping and weighted association rule mining for product infant failure. *Computers & Industrial Engineering, 109*, 253-265.

Hegenbarth, Y., Bartsch, T., & Ristow, G. H. (2018). Efficient and fast monitoring and disruption management for a pressure diecast system. *Information Technology, 60*(3), 165-171. doi:10.1515/itit-2017-0039

Heger, J., Hildebrandt, T., & Scholz-Reiter, B. (2015). Dispatching rule selection with Gaussian processes. *Central European Journal of Operations Research, 23*(1), 235-249.

Hermann, M., Pentek, T., & Otto, B. (2015). Design principles for Industrie 4.0 scenarios: a literature review. In. Working Paper No. 01: Technische Universität Dortmund.

Hermann, M., Pentek, T., & Otto, B. (2016). *Design principles for industrie 4.0 scenarios*. Paper presented at the 49th Hawaii International Conference on System Sciences, Koloa, HI.

Hinchi, A. Z., & Tkiouat, M. (2018). Rolling element bearing remaining useful life estimation based on a convolutional long-short-term memory network. *Procedia Computer Science, 127*, 123-132.

Hirsch, V., Reimann, P., Kirn, O., & Mitschang, B. (2018). Analytical Approach to Support Fault Diagnosis and Quality Control in End-Of-Line Testing. *Procedia CIRP, 72*, 1333-1338.

Holsapple, C., Lee-Post, A., & Pakath, R. (2014). A unified foundation for business analytics. *Decision Support Systems, 64*, 130-141.

Hranisavljevic, N., Niggemann, O., & Maier, A. (2016). *A novel anomaly detection algorithm for hybrid production systems based on deep learning and timed automata.* Paper presented at the International Workshop on the Principles of Diagnosis

Hseush, W., Huang, Y.-C., Hsu, S.-C., & Pu, C. (2013). *Real-Time Collaborative Planning with Big Data.* Paper presented at the IEEE International Conference on Collaborative Computing.

Hsu, C.-Y. (2014). Integrated data envelopment analysis and neural network model for forecasting performance of wafer fabrication operations. *Journal of Intelligent Manufacturing, 25*(5), 945-960. doi:10.1007/s10845-013-0808-0

Hsu, C.-Y., Kang, L.-W., & Weng, M.-F. (2016). *Big data analytics: Prediction of surface defects on steel slabs based on one class support vector machine.* Paper presented at the ASME 2016 Conference on Information Storage and Processing Systems.

Hu, H., Jia, X., He, Q., Fu, S., & Liu, K. (2020). Deep reinforcement learning based AGVs real-time scheduling with mixed rule for flexible shop floor in industry 4.0. *Computers & Industrial Engineering, 149*, 106749.

Hu, J., Lewis, F. L., Gan, O. P., Phua, G. H., & Aw, L. L. (2014). Discrete-event shop-floor monitoring system in RFID-enabled manufacturing. *IEEE Transactions on Industrial Electronics, 61*(12), 7083-7091.

Hu, L., Miao, Y., Wu, G., Hassan, M. M., & Humar, I. (2019). iRobot-Factory: An intelligent robot factory based on cognitive manufacturing and edge computing. *Future Generation Computer Systems, 90*, 569-577.
Hu, S., Zhao, L., Yao, Y., & Dou, R. (2016). A variance change point estimation method based on intelligent ensemble model for quality fluctuation analysis. *International Journal of Production Research, 54*(19), 5783-5797.
Huang, D., Lin, C., Chen, C., & Sze, J. (2018, 25-27 May 2018). *The Internet technology for defect detection system with deep learning method in smart factory.* Paper presented at the 2018 4th International Conference on Information Management.
Huang, H., Ding, S., Zhao, L., Huang, H., Chen, L., Gao, H., & Ahmed, S. H. (2019). Real-time fault detection for iiot facilities using gbrbm-based dnn. *IEEE Internet of Things Journal, 7*(7), 5713-5722.
Hummel, D., Schacht, S., & Maedche, A. (2016). *Determinants of multi-channel behavior: exploring avenues for future research in the services industry*. Paper presented at the Proceedings of the 37th International Conference on Information Systems.
Hur, M., Lee, S.-k., Kim, B., Cho, S., Lee, D., & Lee, D. (2015). A study on the man-hour prediction system for shipbuilding. *Journal of Intelligent Manufacturing, 26*(6), 1267-1279.
Hussain, M., Mansoor, A., & Nisar, S. (2018). Bearing Degradation Prognosis Using Structural Break Classifier. *Mechanika, 24*(4), 456-462.
Huynh, K. T., Grall, A., & Bérenguer, C. (2018). A Parametric Predictive Maintenance Decision-Making Framework Considering Improved System Health Prognosis Precision. *IEEE Transactions on Reliability, 68*(1), 375-396.
Iannino, V., Mocci, C., Vannocci, M., Colla, V., Caputo, A., & Ferraris, F. (2020). An Event-Driven Agent-Based Simulation Model for Industrial Processes. *Applied Sciences, 10*(12), 4343.
Ireland, R., & Liu, A. (2018). Application of data analytics for product design: Sentiment analysis of online product reviews. *CIRP Journal of Manufacturing Science and Technology, 23*, 128-144.
Ishii, S., Yoshida, W., & Yoshimoto, J. (2002). Control of exploitation–exploration meta-parameter in reinforcement learning. *Neural networks, 15*(4-6), 665-687.
Ismail, A., Idris, M. Y. I., Ayub, M. N., & Yee, L. (2019). Investigation of fusion features for apple classification in smart manufacturing. *Symmetry, 11*(10), 1194.
Ivanov, D., Dolgui, A., Sokolov, B., Werner, F., & Ivanova, M. (2016). A dynamic model and an algorithm for short-term supply chain scheduling in the smart factory industry 4.0. *International Journal of Production Research, 54*(2), 386-402.
Jahirabadkar, S., & Kulkarni, P. (2013). SCAF An effective approach to Classify Subspace Clustering algorithms. *arXiv preprint arXiv:1304.3603*.
Jain, A. K., & Lad, B. K. (2017). A novel integrated tool condition monitoring system. *Journal of Intelligent Manufacturing, 30*(3), 1423-1436.
Jain, R., Singh, A., & Mishra, P. (2013). Prioritization of supplier selection criteria: A fuzzy-AHP approach. *MIT International Journal of Mechanical Engineering, 3*(1), 34-42.
Jain, R., Singh, A., Yadav, H., & Mishra, P. (2014). Using data mining synergies for evaluating criteria at pre-qualification stage of supplier selection. *Journal of Intelligent Manufacturing, 25*(1), 165-175. doi:10.1007/s10845-012-0684-z
Jain, S., Lechevalier, D., & Narayanan, A. (2017). *Towards smart manufacturing with virtual factory and data analytics*. Paper presented at the Proceedings of the 2017 Winter Simulation Conference.
Jain, S., Shao, G., & Shin, S.-J. (2017). Manufacturing data analytics using a virtual factory representation. *International Journal of Production Research, 55*(18), 5450-5464.
Jain, V., Kundu, A., Chan, F., & Patel, M. (2015). A Chaotic Bee Colony approach for supplier selection-order allocation with different discounting policies in a coopetitive multi-echelon supply chain. *Journal of Intelligent Manufacturing, 26*(6), 1131-1144. doi:10.1007/s10845-013-0845-8
Janssens, O., Slavkovikj, V., Vervisch, B., Stockman, K., Loccufier, M., Verstockt, S., . . . Van Hoecke, S. (2016). Convolutional neural network based fault detection for rotating machinery. *Journal of Sound and Vibration, 377*, 331-345.
Jaramillo, V. H., Ottewill, J. R., Dudek, R., Lepiarczyk, D., & Pawlik, P. (2017). Condition monitoring of distributed systems using two-stage Bayesian inference data fusion. *Mechanical Systems and Signal Processing, 87*, 91-110.
Javed, K., Gouriveau, R., Li, X., & Zerhouni, N. (2018). Tool wear monitoring and prognostics challenges: a comparison of connectionist methods toward an adaptive ensemble model. *Journal of Intelligent Manufacturing, 29*(8), 1873-1890.

Jayaram, A. (2017, 5-6 May 2017). *An IIoT quality global enterprise inventory management model for automation and demand forecasting based on cloud.* Paper presented at the 2017 International Conference on Computing, Communication and Automation (ICCCA).

Jha, S. B., Babiceanu, R. F., & Seker, R. (2019). Formal modeling of cyber-physical resource scheduling in IIoT cloud environments. *Journal of Intelligent Manufacturing*, 1-16.

Ji-Hyeong, H., & Su-Young, C. (2016, 5-8 July 2016). *Consideration of manufacturing data to apply machine learning methods for predictive manufacturing.* Paper presented at the 2016 Eighth International Conference on Ubiquitous and Future Networks.

Ji, W., & Wang, L. (2017). Big data analytics based fault prediction for shop floor scheduling. *Journal of Manufacturing Systems, 43*, 187-194.

Ji, W., Yin, S., & Wang, L. (2018). A big data analytics based machining optimisation approach. *Journal of Intelligent Manufacturing, 30*(3), 1483–1495. doi:https://doi.org/10.1007/s10845-018-1440-9

Ji, Y., Yu, C., Xu, X., Yu, S., & Zhang, W. (2018). Data mining based multi-level aggregate service planning for cloud manufacturing. *Journal of Intelligent Manufacturing, 29*(6), 1351-1361. doi:10.1007/s10845-015-1184-8

Jia, F., Lei, Y., Lin, J., Zhou, X., & Lu, N. (2016). Deep neural networks: A promising tool for fault characteristic mining and intelligent diagnosis of rotating machinery with massive data. *Mechanical Systems and Signal Processing, 72*, 303-315.

Jia, X., Hu, B., Marchant, B. P., Zhou, L., Shi, Z., & Zhu, Y. (2019). A methodological framework for identifying potential sources of soil heavy metal pollution based on machine learning: A case study in the Yangtze Delta, China. *Environmental Pollution, 250*, 601-609.

Jiang, J., & Kuo, C. (2017, 17-20 Nov. 2017). *Enhancing Convolutional Neural Network Deep Learning for Remaining Useful Life Estimation in Smart Factory Applications.* Paper presented at the 2017 International Conference on Information, Communication and Engineering.

Jiang, W., Zhang, N., Xue, X., Xu, Y., Zhou, J., & Wang, X. (2020). Intelligent Deep Learning Method for Forecasting the Health Evolution Trend of Aero-Engine With Dispersion Entropy-Based Multi-Scale Series Aggregation and LSTM Neural Network. *IEEE Access, 8*, 34350-34361.

Jiao, J., Lin, J., Zhao, M., & Liang, K. (2020). Double-level adversarial domain adaptation network for intelligent fault diagnosis. *Knowledge-Based Systems, 205*, 106236.

Jiao, Y., Yang, Y., Zhong, J., & Zhang, H. (2017). *A Comparative Analysis of Intelligent Classifiers for Mapping Customer Requirements to Product Configurations.* Paper presented at the Proceedings of the 2017 International Conference on Big Data Research.

Jin, L., Zhang, C., Wen, X., & Christopher, G. G. (2020). A Neutrosophic Number-Based Memetic Algorithm for the Integrated Process Planning and Scheduling Problem With Uncertain Processing Times. *IEEE Access, 8*, 96628-96648.

Jin, X., Fan, J., & Chow, T. W. (2018). Fault detection for rolling-element bearings using multivariate statistical process control methods. *IEEE Transactions on Instrumentation and Measurement, 68*(9), 3128-3136.

Jin, X., Que, Z., Sun, Y., Guo, Y., & Qiao, W. (2019). A data-driven approach for bearing fault prognostics. *IEEE Transactions on Industry Applications, 55*(4), 3394-3401.

Jing, S., Ma, L., Hu, K., Zhu, Y., & Chen, H. (2018). A restructured artificial bee colony optimizer combining life-cycle, local search and crossover operations for droplet property prediction in printable electronics fabrication. *Journal of Intelligent Manufacturing, 29*(1), 109-134. doi:10.1007/s10845-015-1092-y

Jöhnk, J., Röglinger, M., Thimmel, M., & Urbach, N. (2017). *How to implement agile IT setups: A taxonomy of design options.* Paper presented at the 25th European Conference on Information Systems (ECIS).

Johns, G. (2006). The essential impact of context on organizational behavior. *Academy of management review, 31*(2), 386-408.

Jomthanachai, S., Rattanamanee, W., Sinthavalai, R., & Wong, W.-P. (2020). The application of genetic algorithm and data analytics for total resource management at the firm level. *Resources, Conservation and Recycling, 161*, 104985.

Jung, S., Tsai, Y., Chiu, W., Hu, J., & Sun, C. (2018). *Defect Detection on Randomly Textured Surfaces by Convolutional Neural Networks.* Paper presented at the 2018 IEEE/ASME International Conference on Advanced Intelligent Mechatronics.

Kabugo, J. C., Jämsä-Jounela, S.-L., Schiemann, R., & Binder, C. (2020). Industry 4.0 based process data analytics platform: A waste-to-energy plant case study. *International Journal of Electrical Power & Energy Systems, 115*, 105508.

Kadar, M., Jardim-Gonçalves, R., Covaciu, C., & Bullon, S. (2017, 27-29 June 2017). *Intelligent defect management system for porcelain industry through cyber-physical systems.* Paper presented at the 2017 International Conference on Engineering, Technology and Innovation.

Kagermann, H., Wahlster, W., & Helbig, J. (2013). *Umsetzungsempfehlung für das Zukunftsprojekt Industrie 4.0 - Deutschlands Zukunft als Produktionsstandort sichern - Abschlussbereicht des Arbeitskreises Industrie 4.0.* Retrieved from Frankfurt a. M.:

Kajmakovic, A., Zupanc, R., Mayer, S., Kajtazovic, N., Hoeffernig, M., & Vogl, H. (2018, 6-8 June 2018). *Predictive Fail-Safe Improving the Safety of Industrial Environments through Model-based Analytics on hidden Data Sources.* Paper presented at the 13th International Symposium on Industrial Embedded Systems.

Kamsu-Foguem, B., Rigal, F., & Mauget, F. (2013). Mining association rules for the quality improvement of the production process. *Expert Systems with Applications, 40*(4), 1034-1045.

Kan, C., Yang, H., & Kumara, S. (2018). Parallel computing and network analytics for fast Industrial Internet-of-Things (IIoT) machine information processing and condition monitoring. *Journal of Manufacturing Systems, 46*, 282-293. doi:https://doi.org/10.1016/j.jmsy.2018.01.010

Kanawaday, A., & Sane, A. (2017, 24-26 Nov. 2017). *Machine learning for predictive maintenance of industrial machines using IoT sensor data.* Paper presented at the 8th International Conference on Software Engineering and Service Science.

Kang, H.-S., Lee, J. Y., Choi, S., Kim, H., Park, J. H., Son, J. Y., . . . Noh, S. D. (2016). Smart Manufacturing: Past Research, Present Findings, and Future Directions. *International Journal of Precision Engineering and Manufacturing - Green Technology, 3*(1), 111-128.

Kannan, K., Arunachalam, N., Chawla, A., & Natarajan, S. (2018). Multi-Sensor Data Analytics for Grinding Wheel Redress Life Estimation- An Approach towards Industry 4.0. *Procedia Manufacturing, 26*, 1230-1241. doi:https://doi.org/10.1016/j.promfg.2018.07.160

Kannan, R., Manohar, S. S., & Kumaran, M. S. (2018). Nominal features-based class specific learning model for fault diagnosis in industrial applications. *Computers & Industrial Engineering, 116*, 163-177.

Kannan, R., Manohar, S. S., & Kumaran, M. S. (2019). *IoT-Based Condition Monitoring and Fault Detection for Induction Motor.* Paper presented at the Proceedings of 2nd International Conference on Communication, Computing and Networking.

Kao, H.-A., Hsieh, Y.-S., Chen, C.-H., & Lee, J. (2017). *Quality prediction modeling for multistage manufacturing based on classification and association rule mining.* Paper presented at the The 2nd International Conference on Precision Machinery and Manufacturing Technology.

Karabadji, N. E. I., Seridi, H., Khelf, I., Azizi, N., & Boulkroune, R. (2014). Improved decision tree construction based on attribute selection and data sampling for fault diagnosis in rotating machines. *Engineering Applications of Artificial Intelligence, 35*, 71-83.

Karandikar, J., McLeay, T., Turner, S., & Schmitz, T. (2015). Tool wear monitoring using naive Bayes classifiers. *The International Journal of Advanced Manufacturing Technology, 77*(9-12), 1613-1626.

Kashkoush, M., & ElMaraghy, H. (2017). An integer programming model for discovering associations between manufacturing system capabilities and product features. *Journal of Intelligent Manufacturing, 28*(4), 1031-1044. doi:10.1007/s10845-015-1044-6

Kaufman, L., & Rousseeuw, P. J. (2009). *Finding groups in data: an introduction to cluster analysis* (Vol. 344): John Wiley & Sons.

Kaur, K., Selway, M., Grossmann, G., Stumptner, M., & Johnston, A. (2018). *Towards an open-standards based framework for achieving condition-based predictive maintenance.* Paper presented at the Proceedings of the 8th International Conference on the Internet of Things.

Kaur, N., & Sood, S. K. (2015). Cognitive decision making in smart industry. *Computers in Industry, 74*, 151-161.

Kazi, M.-K., Eljack, F., & Mahdi, E. (2020). Data-driven modeling to predict the load vs. displacement curves of targeted composite materials for industry 4.0 and smart manufacturing. *Composite Structures*, 113207.

Ke, G., Chen, R.-S., Chen, Y.-C., Wang, S., & Zhang, X. (2020). Using ant colony optimisation for improving the execution of material requirements planning for smart manufacturing. *Enterprise Information Systems*, 1-23.

Kedadouche, M., Thomas, M., & Tahan, A. (2016). A comparative study between Empirical Wavelet Transforms and Empirical Mode Decomposition Methods: Application to bearing defect diagnosis. *Mechanical Systems and Signal Processing, 81*, 88-107.

Keizer, M. C. O., Teunter, R. H., & Veldman, J. (2017). Joint condition-based maintenance and inventory optimization for systems with multiple components. *European Journal of Operational Research, 257*(1), 209-222.

Khakifirooz, M., Chien, C. F., & Chen, Y.-J. (2018). Bayesian inference for mining semiconductor manufacturing big data for yield enhancement and smart production to empower industry 4.0. *Applied Soft Computing, 68*, 990-999.

Khalili, A., & Sami, A. (2015). SysDetect: A systematic approach to critical state determination for Industrial Intrusion Detection Systems using Apriori algorithm. *Journal of Process Control, 32*, 154-160. doi:10.1016/j.jprocont.2015.04.005

Khalilijafarabad, A., Helfert, M., & Ge, M. (2016). *Developing a Data Quality Research Taxonomy-an organizational perspective.* Paper presented at the Proceedings of the 21st International Conference on Information Quality.

Khan, S., & Yairi, T. (2018). A review on the application of deep learning in system health management. *Mechanical Systems and Signal Processing, 107*, 241-265.

Khazaee, M., Banakar, A., Ghobadian, B., Agha Mirsalim, M., Minaei, S., & Jafari, S. M. (2017). Detection of inappropriate working conditions for the timing belt in internal-combustion engines using vibration signals and data mining. *Proceedings of the Institution of Mechanical Engineers, Part D: Journal of Automobile Engineering, 231*(3), 418-432.

Khelif, R., Chebel-Morello, B., Malinowski, S., Laajili, E., Fnaiech, F., & Zerhouni, N. (2017). Direct Remaining Useful Life Estimation Based on Support Vector Regression. *IEEE Trans. Industrial Electronics, 64*(3), 2276-2285.

Khoda, M. E., Imam, T., Kamruzzaman, J., Gondal, I., & Rahman, A. (2019). Robust malware defense in industrial IoT applications using machine learning with selective adversarial samples. *IEEE Transactions on Industry Applications, 56*(4), 4415-4424.

Kiangala, K. S., & Wang, Z. (2018). Initiating predictive maintenance for a conveyor motor in a bottling plant using industry 4.0 concepts. *The International Journal of Advanced Manufacturing Technology, 97*(9-12), 3251-3271.

Kiangala, K. S., & Wang, Z. (2020). An Effective Predictive Maintenance Framework for Conveyor Motors Using Dual Time-Series Imaging and Convolutional Neural Network in an Industry 4.0 Environment. *IEEE Access, 8*, 121033-121049.

Kibira, D., Hatim, Q., Kumara, S., & Shao, G. (2015). *Integrating data analytics and simulation methods to support manufacturing decision making.* Paper presented at the Proceedings of the 2015 Winter Simulation Conference, Huntington Beach, California.

Kibira, D., & Shao, G. (2017). Integrating Data Mining and Simulation Optimization for Decision Making in Manufacturing. In M. M. Mujica & D. L. M. I. Flores (Eds.), *Applied Simulation and Optimization 2* (pp. 81-105). Cham: Springer.

Kim, A., Oh, K., Jung, J.-Y., & Kim, B. (2018). Imbalanced classification of manufacturing quality conditions using cost-sensitive decision tree ensembles. *International Journal of Computer Integrated Manufacturing, 31*(8), 701-717. doi:10.1080/0951192X.2017.1407447

Kim, C., Lee, H., Kim, K., Lee, Y., & Lee, W. B. (2018). Efficient process monitoring via the integrated use of Markov random fields learning and the graphical lasso. *Industrial & Engineering Chemistry Research, 57*(39), 13144-13155.

Kim, D.-H., Kim, T. J., Wang, X., Kim, M., Quan, Y.-J., Oh, J. W., . . . Yang, I. (2018). Smart Machining Process Using Machine Learning: A Review and Perspective on Machining Industry. *International Journal of Precision Engineering and Manufacturing-Green Technology, 5*(4), 555-568.

Kim, D., Han, S. C., Lin, Y., Kang, B. H., & Lee, S. (2018). RDR-based knowledge based system to the failure detection in industrial cyber physical systems. *Knowledge-Based Systems, 150*, 1-13. doi:https://doi.org/10.1016/j.knosys.2018.02.009

Kim, D., Yang, H., Chung, M., Cho, S., Kim, H., Kim, M., . . . Kim, E. (2018, 23-25 March 2018). *Squeezed Convolutional Variational AutoEncoder for unsupervised anomaly detection in edge device industrial Internet of Things.* Paper presented at the 2018 International Conference on Information and Computer Technologies.

Kim, D. B. (2019). An approach for composing predictive models from disparate knowledge sources in smart manufacturing environments. *Journal of Intelligent Manufacturing, 30*(4), 1999-2012.

Kim, J., & Hwangbo, H. (2018). Sensor-Based Real-Time Detection in Vulcanization Control Using Machine Learning and Pattern Clustering. *Sensors, 18*(9), E3123. doi:10.3390/s18093123

Kim, J., & Hwangbo, H. (2019). Real-time early warning system for sustainable and intelligent plastic film manufacturing. *Sustainability, 11*(5), 1490.

Kim, M. S., Choi, Y. J., Park, I. S., Kong, N., Lee, M., & Park, P. (2018, 15-18 May 2018). *Sensitivity analysis on a neural network for analyzing the camber in hot rolling process.* Paper presented at the 2018 IEEE Industrial Cyber-Physical Systems.

Kim, S. W., Lee, Y. G., Tama, B. A., & Lee, S. (2020). Reliability-Enhanced Camera Lens Module Classification Using Semi-Supervised Regression Method. *Applied Sciences, 10*(11), 3832.

Kireev, V. S., Filippov, S. A., Guseva, A. I., Bochkaryov, P. V., Kuznetsov, I. A., Migalin, V., & Filin, S. S. (2018). *Predictive repair and support of engineering systems based on distributed data processing model within an IoT concept.* Paper presented at the 6th International Conference on Future Internet of Things and Cloud Workshops.

Kisskalt, D., Fleischmann, H., Kreitlein, S., Knott, M., & Franke, J. (2018). A novel approach for data-driven process and condition monitoring systems on the example of mill-turn centers. *Production Engineering-Research and Development, 12*(3-4), 525-533. doi:10.1007/s11740-018-0797-0

Klöber-Koch, J., Braunreuther, S., & Reinhart, G. (2017). Predictive production planning considering the operative risk in a manufacturing system. *Procedia CIRP, 63*, 360-365.

Kohlert, M., & Konig, A. (2016). Advanced multi-sensory process data analysis and on-line evaluation by innovative human-machine-based process monitoring and control for yield optimization in polymer film industry. *Technisches Messen, 83*(9), 474-483. doi:10.1515/teme-2015-0120

König, C., & Helmi, A. M. (2020). Sensitivity analysis of sensors in a hydraulic condition monitoring system using cnn models. *Sensors, 20*(11), 3307.

Konstantakopoulos, I. C., Barkan, A. R., He, S., Veeravalli, T., Liu, H., & Spanos, C. (2019). A deep learning and gamification approach to improving human-building interaction and energy efficiency in smart infrastructure. *Applied Energy, 237*, 810-821.

Koulali, M., Koulali, S., Tembine, H., & Kobbane, A. (2018). Industrial Internet of Things-Based Prognostic Health Management: A Mean-Field Stochastic Game Approach. *IEEE Access, 6*, 54388-54395. doi:10.1109/ACCESS.2018.2871859

Kozjek, D., Kralj, D., & Butala, P. (2017). A Data-Driven Holistic Approach to Fault Prognostics in a Cyclic Manufacturing Process. *Procedia CIRP, 63*, 664-669.

Kozjek, D., Rihtaršič, B., & Butala, P. (2018). Big data analytics for operations management in engineer-to-order manufacturing. *Procedia CIRP, 72*, 209-214.

Kozjek, D., Vrabic, R., Kralj, D., & Butala, P. (2017). Interpretative identification of the faulty conditions in a cyclic manufacturing process. *Journal of Manufacturing Systems, 43*, 214-224. doi:10.1016/j.jmsy.2017.03.001

Krishnakumari, A., Elayaperumal, A., Saravanan, M., & Arvindan, C. (2017). Fault diagnostics of spur gear using decision tree and fuzzy classifier. *The International Journal of Advanced Manufacturing Technology, 89*(9-12), 3487-3494.

Krishnamurthy, P., Karri, R., & Khorrami, F. (2019). Anomaly detection in real-time multi-threaded processes using hardware performance counters. *IEEE Transactions on Information Forensics and Security, 15*, 666-680.

Krumeich, J., Werth, D., & Loos, P. (2016). Prescriptive Control of Business Processes - New Potentials Through Predictive Analytics of Big Data in the Process Manufacturing Industry. *Business & Information Systems Engineering, 58*(4), 261-280.

Krumeich, J., Werth, D., Loos, P., Schimmelpfennig, J., & Jacobi, S. (2014). *Advanced planning and control of manufacturing processes in steel industry through big data analytics: Case study and architecture proposal.* Paper presented at the Big Data (Big Data), 2014 IEEE International Conference on.

Küfner, T., Uhlemann, T. H. J., & Ziegler, B. (2018). Lean Data in Manufacturing Systems: Using Artificial Intelligence for Decentralized Data Reduction and Information Extraction. *Procedia CIRP, 72*, 219-224. doi:https://doi.org/10.1016/j.procir.2018.03.125

Kumar, A., Chinnam, R. B., & Tseng, F. (2018). An HMM and polynomial regression based approach for remaining useful life and health state estimation of cutting tools. *Computers & Industrial Engineering, 128*, 1008-1014. doi:https://doi.org/10.1016/j.cie.2018.05.017

Kumar, A., Dimitrakopoulos, R., & Maulen, M. (2020). Adaptive self-learning mechanisms for updating short-term production decisions in an industrial mining complex. *Journal of Intelligent Manufacturing, 31*(7), 1795-1811.

Kumar, A., Shankar, R., Choudhary, A., & Thakur, L. S. (2016). A big data MapReduce framework for fault diagnosis in cloud-based manufacturing. *International Journal of Production Research, 54*(23), 7060-7073.

Kumar, A., Shankar, R., & Thakur, L. S. (2018). A big data driven sustainable manufacturing framework for condition-based maintenance prediction. *Journal of Computational Science, 27*, 428-439.

Kumar, R., Singh, S. P., & Lamba, K. (2018). Sustainable robust layout using Big Data approach: A key towards industry 4.0. *Journal of Cleaner Production, 204*, 643-659. doi:https://doi.org/10.1016/j.jclepro.2018.08.327

Kumar, S. L. (2017). State of The Art-Intense Review on Artificial Intelligence Systems Application in Process Planning and Manufacturing. *Engineering Applications of Artificial Intelligence, 65*, 294-329.

Kumaraguru, S., & Morris, K. (2014). *Integrating real-time analytics and continuous performance management in smart manufacturing systems.* Paper presented at the IFIP International Conference on Advances in Production Management Systems.

Kumru, M., & Kumru, P. Y. (2014a). Using artificial neural networks to forecast operation times in metal industry. *International Journal of Computer Integrated Manufacturing, 27*(1), 48-59.

Kumru, M., & Kumru, P. Y. (2014b). Using artificial neural networks to forecast operation times in metal industry. *International Journal of Computer Integrated Manufacturing, 27*(1), 48-59.

Kuo, C.-J., Ting, K.-C., Chen, Y.-C., Yang, D.-L., & Chen, H.-M. (2017). Automatic machine status prediction in the era of Industry 4.0: Case study of machines in a spring factory. *Journal of Systems Architecture, 81*, 44-53.

Kuo, C.-M., Chen, W.-Y., Tseng, C.-Y., & Kao, C. T. (2020). Developing a smart system with Industry 4.0 for customer dissatisfaction. *Industrial Management & Data Systems*.

Kuo, Y.-H., & Kusiak, A. (2018a). From data to big data in production research: the past and future trends. *International Journal of Production Research*, https://doi.org/10.1080/00207543.00202018.01443230.

Kuo, Y.-H., & Kusiak, A. (2018b). From data to big data in production research: the past and future trends. *International Journal of Production Research, 57*(15-16), 4828-4853.

Kwon, O., Lee, N., & Shin, B. (2014). Data quality management, data usage experience and acquisition intention of big data analytics. *International Journal of Information Management, 34*(3), 387-394.

Lachenmaier, J. F., Lasi, H., & Kemper, H.-G. (2015). *Entwicklung und Evaluation eines Informationsversorgungskonzepts für die Prozess-und Produktionsplanung im Kontext von Industrie 4.0.* Paper presented at the Wirtschaftsinformatik Proceedings.

Lade, P., Ghosh, R., & Srinivasan, S. (2017). Manufacturing Analytics and Industrial Internet of Things. *IEEE Intelligent Systems, 32*(3), 74-79. doi:10.1109/MIS.2017.49

Lai, C.-F., Chien, W.-C., Yang, L. T., & Qiang, W. (2019). LSTM and edge computing for big data feature recognition of industrial electrical equipment. *IEEE Transactions on Industrial Informatics, 15*(4), 2469-2477.

Lai, P.-J., & Wu, H.-C. (2015). Using heuristic algorithms to solve the scheduling problems with job-dependent and machine-dependent learning effects. *Journal of Intelligent Manufacturing, 26*(4), 691-701. doi:10.1007/s10845-013-0827-x

Lai, X., Zhang, Q., Chen, Q., Huang, Y., Mao, N., & Liu, J. (2018). The analytics of product-design requirements using dynamic internet data: application to Chinese smartphone market. *International Journal of Production Research*, https://doi.org/10.1080/00207543.00202018.01541200.

Langarica, S., Rüffelmacher, C., & Núñez, F. (2018, 15-18 May 2018). *An industrial internet platform for real-time fault detection in industrial motors.* Paper presented at the 2018 IEEE Industrial Cyber-Physical Systems.

Langarica, S., Rüffelmacher, C., & Núñez, F. (2019). An industrial internet application for real-time fault diagnosis in industrial motors. *IEEE Transactions on Automation Science and Engineering, 17*(1), 284-295.

Langone, R., Alzate, C., Bey-Temsamani, A., & Suykens, J. A. (2014). *Alarm prediction in industrial machines using autoregressive LS-SVM models.* Paper presented at the Computational Intelligence and Data Mining (CIDM), 2014 IEEE Symposium on.

Latif, S., Zou, Z., Idrees, Z., & Ahmad, J. (2020). A novel attack detection scheme for the industrial internet of things using a lightweight random neural network. *IEEE Access, 8*, 89337-89350.

Laux, H., Bytyn, A., Ascheid, G., Schmeink, A., Kurt, G. K., & Dartmann, G. (2018). *Learning-based indoor localization for industrial applications*. Paper presented at the Proceedings of the 15th ACM International Conference on Computing Frontiers, Ischia, Italy.

LaValle, S., Lesser, E., Shockley, R., Hopkins, M. S., & Krschwitz, N. (2011). Big Data, Analytics and the Path from Insights to Value. *MIT Sloan Management Review, 52*(2), 20-31.

Lavrova, D., Poltavtseva, M., & Shtyrkina, A. (2018, 15-18 May 2018). *Security analysis of cyber-physical systems network infrastructure.* Paper presented at the 2018 IEEE Industrial Cyber-Physical Systems.

Lee, E. A. (2008). *Cyber physical systems: Design challenges*. Paper presented at the 11th IEEE Symposium on Object Oriented Real-Time Distributed Computing.

Lee, G. Y., Kim, M., Quan, Y. J., Kim, M. S., Kim, T. J. Y., Yoon, H. S., . . . Ahn, S. H. (2018). Machine health management in smart factory: A review. *Journal of Mechanical Science and Technology, 32*(3), 987-1009.

Lee, H. (2017). Framework and development of fault detection classification using IoT device and cloud environment. *Journal of Manufacturing Systems, 43*, 257-270. doi:https://doi.org/10.1016/j.jmsy.2017.02.007

Lee, H. L., & Rosenblatt, M. J. (1987). Simultaneous determination of production cycle and inspection schedules in a production system. *Management Science, 33*(9), 1125-1136.

Lee, J., Jin, C., & Bagheri, B. (2017). Cyber physical systems for predictive production systems. *Production Engineering-Research and Development, 11*(2), 155-165. doi:10.1007/s11740-017-0729-4

Lee, J., Kao, H.-A., & Yang, S. (2014). Service Innovation and Smart Analytics for Industry 4.0 and Big Data Environment. *Procedia CIRP, 16*, 3-8.

Lee, J., Noh, S. D., Kim, H. J., & Kang, Y. S. (2018). Implementation of Cyber-Physical Production Systems for Quality Prediction and Operation Control in Metal Casting. *Sensors, 18*(5). doi:10.3390/s18051428

Lee, J., Wu, F., Zhao, W., Ghaffari, M., Liao, L., & Siegel, D. (2014). Prognostics and health management design for rotary machinery systems—Reviews, methodology and applications. *Mechanical Systems and Signal Processing, 42*(1-2), 314-334.

Lee, J. Y., Yoon, J. S., & Kim, B. H. (2017). A big data analytics platform for smart factories in small and medium-sized manufacturing enterprises: An empirical case study of a die casting factory. *International Journal of Precision Engineering and Manufacturing, 18*(10), 1353-1361. doi:10.1007/s12541-017-0161-x

Lee, K. B., Cheon, S., & Kim, C. O. (2017). A convolutional neural network for fault classification and diagnosis in semiconductor manufacturing processes. *IEEE Transactions on Semiconductor Manufacturing, 30*(2), 135-142.

Lee, W. J., Mendis, G. P., Triebe, M. J., & Sutherland, J. W. (2020). Monitoring of a machining process using kernel principal component analysis and kernel density estimation. *Journal of Intelligent Manufacturing, 31*(5), 1175-1189.

Legat, C., & Vogel-Heuser, B. (2017). A configurable partial-order planning approach for field level operation strategies of PLC-based industry 4.0 automated manufacturing systems. *Engineering Applications of Artificial Intelligence, 66*, 128-144. doi:https://doi.org/10.1016/j.engappai.2017.06.014

Lei, Y., Jia, F., Lin, J., Xing, S., & Ding, S. X. (2016). An intelligent fault diagnosis method using unsupervised feature learning towards mechanical big data. *IEEE Transactions on Industrial Electronics, 63*(5), 3137-3147.

Lei, Y., Jia, F., Zhou, X., & Lin, J. (2015). A deep learning-based method for machinery health monitoring with big data. *Journal of Mechanical Engineering, 51*(21), 49-56.

Lei, Y., Li, N., Gontarz, S., Lin, J., Radkowski, S., & Dybala, J. (2016). A model-based method for remaining useful life prediction of machinery. *IEEE Transactions on Reliability, 65*(3), 1314-1326.

Lei, Y., Li, N., Guo, L., Li, N., Yan, T., & Lin, J. (2018). Machinery health prognostics: A systematic review from data acquisition to RUL prediction. *Mechanical Systems and Signal Processing, 104*, 799-834.

Lenz, J., MacDonald, E., Harik, R., & Wuest, T. (2020). Optimizing smart manufacturing systems by extending the smart products paradigm to the beginning of life. *Journal of Manufacturing Systems, 57*, 274-286.

Lenz, J., & Westkaemper, E. (2017). Wear Prediction of Woodworking Cutting Tools based on History Data. *Procedia CIRP, 63*, 675-679.

Lesany, S. A., Koochakzadeh, A., & Fatemi Ghomi, S. M. T. (2014). Recognition and classification of single and concurrent unnatural patterns in control charts via neural networks and fitted line of samples. *International Journal of Production Research, 52*(6), 1771-1786.

Li, C., Cerrada, M., Cabrera, D., Sanchez, R. V., Pacheco, F., Ulutagay, G., . . . de Oliveira, J. V. (2018). A comparison of fuzzy clustering algorithms for bearing fault diagnosis. *Journal of Intelligent & Fuzzy Systems, 34*(6), 3565-3580. doi:10.3233/JIFS-169534

Li, C., Sánchez, R.-V., Zurita, G., Cerrada, M., & Cabrera, D. (2016). Fault diagnosis for rotating machinery using vibration measurement deep statistical feature learning. *Sensors, 16*(6), 895.

Li, C., Sanchez, R.-V., Zurita, G., Cerrada, M., Cabrera, D., & Vásquez, R. E. (2016). Gearbox fault diagnosis based on deep random forest fusion of acoustic and vibratory signals. *Mechanical Systems and Signal Processing, 76*, 283-293.

Li, C., Tao, Y., Ao, W., Yang, S., & Bai, Y. (2018). Improving forecasting accuracy of daily enterprise electricity consumption using a random forest based on ensemble empirical mode decomposition. *Energy, 165*, 1220-1227.

Li, J.-Q., Yu, F. R., Deng, G., Luo, C., Ming, Z., & Yan, Q. (2017). Industrial Internet: A Survey on the Enabling Technologies, Applications, and Challenges. *IEEE Communications Surveys & Tutorials, 19*(3), 1504-1526.

Li, J., Xu, X., Gao, L., Wang, Z., & Shao, J. (2020). Cognitive visual anomaly detection with constrained latent representations for industrial inspection robot. *Applied Soft Computing, 95*, 106539.

Li, K. (2015). *Made in China 2025*. Retrieved from Beijing:

Li, L., Ota, K., & Dong, M. (2018). Deep Learning for Smart Industry: Efficient Manufacture Inspection System With Fog Computing. *IEEE Transactions on Industrial Informatics, 14*(10), 4665-4673. doi:10.1109/TII.2018.2842821

Li, L., Wang, Z., Wang, X., & Tang, L. (2019). A Multi-Objective Evolutionary Algorithm for Multi-Energy Allocation Problem Considering Production Changeover in the Integrated Iron and Steel Enterprise. *IEEE Access, 7*, 40428-40444.

Li, N., Gebraeel, N., Lei, Y., Fang, X., Cai, X., & Yan, T. (2020). Remaining useful life prediction based on a multi-sensor data fusion model. *Reliability Engineering & System Safety, 208*, 107249.

Li, P., Cheng, K., Jiang, P., & Katchasuwanmanee, K. (2020). Investigation on industrial dataspace for advanced machining workshops: enabling machining operations control with domain knowledge and application case studies. *Journal of Intelligent Manufacturing*, 1-17.

Li, Q., & Liang, S. (2018a). Degradation Trend Prediction for Rotating Machinery Using Long-Range Dependence and Particle Filter Approach. *Algorithms, 11*(7), 89.

Li, Q., & Liang, S. (2018b). Intelligent Prognostics of Degradation Trajectories for Rotating Machinery Based on Asymmetric Penalty Sparse Decomposition Model. *Symmetry, 10*(6), 214.

Li, Q., & Liang, S. Y. (2018). Degradation Trend Prognostics for Rolling Bearing Using Improved R/S Statistic Model and Fractional Brownian Motion Approach. *IEEE Access, 6*, 21103-21114.

Li, Q., Meng, S., Zhang, S., Wu, M., Zhang, J., Ahvanooey, M. T., & Aslam, M. S. (2019). Safety risk monitoring of cyber-physical power systems based on ensemble learning algorithm. *IEEE Access, 7*, 24788-24805.

Li, S., Yang, W., Zhang, A., Liu, H., Huang, J., Li, C., & Hu, J. (2020). A Novel Method of Bearing Fault Diagnosis in Time-Frequency Graphs Using InceptionResnet and Deformable Convolution Networks. *IEEE Access, 8*, 92743-92753.

Li, T., He, Y., & Zhu, C. (2016, 3-4 Dec. 2016). *Big Data Oriented Macro-Quality Index Based on Customer Satisfaction Index and PLS-SEM for Manufacturing Industry.* Paper presented at the 2016 International Conference on Industrial Informatics - Computing Technology, Intelligent Technology, Industrial Information Integration.

Li, T., Zhang, D., Luo, M., & Wu, B. (2017). *Tool Wear Condition Monitoring Based on Wavelet Packet Analysis and RBF Neural Network.* Paper presented at the International Conference on Intelligent Robotics and Applications.

Li, W., Xie, L., & Wang, Z. (2018). Two-Loop Covert Attacks Against Constant Value Control of Industrial Control Systems. *IEEE Transactions on Industrial Informatics, 15*(2), 663-676.

Li, X., Ding, Q., & Sun, J.-Q. (2018). Remaining useful life estimation in prognostics using deep convolution neural networks. *Reliability Engineering & System Safety, 172*, 1-11.

Li, X., Jiang, H., Xiong, X., & Shao, H. (2019). Rolling bearing health prognosis using a modified health index based hierarchical gated recurrent unit network. *Mechanism and Machine Theory, 133*, 229-249.

Li, X., Xu, M., Vijayakumar, P., Kumar, N., & Liu, X. (2020). Detection of Low-Frequency and Multi-Stage Attacks in Industrial Internet of Things. *IEEE Transactions on Vehicular Technology, 69*(8), 8820-8831.

Li, X., Zhang, W., & Ding, Q. (2018). Deep Learning-Based Remaining Useful Life Estimation of Bearings Using Multi-Scale Feature Extraction. *Reliability Engineering & System Safety, 182*, 208-218.

Li, X., Zhang, W., Ding, Q., & Sun, J.-Q. (2018). Intelligent rotating machinery fault diagnosis based on deep learning using data augmentation. *Journal of Intelligent Manufacturing*, https://doi.org/10.1007/s10845-10018-11456-10841.

Li, Y., Carabelli, S., Fadda, E., Manerba, D., Tadei, R., & Terzo, O. (2020). Machine learning and optimization for production rescheduling in industry 4.0. *The International Journal of Advanced Manufacturing Technology, 110*(9), 2445-2463.

Li, Y., & Liu, J. (2018). *Preventive maintenance strategy for detection and buffer optimization in series production system.* Paper presented at the Journal of Physics: Conference Series.

Li, Z., Liu, R., & Wu, D. (2019). Data-driven smart manufacturing: tool wear monitoring with audio signals and machine learning. *Journal of Manufacturing Processes, 48*, 66-76.

Li, Z., Wang, Y., & Wang, K. (2019). A deep learning driven method for fault classification and degradation assessment in mechanical equipment. *Computers in Industry, 104*, 1-10. doi:https://doi.org/10.1016/j.compind.2018.07.002

Li, Z., Wang, Y., & Wang, K. (2020). A data-driven method based on deep belief networks for backlash error prediction in machining centers. *Journal of Intelligent Manufacturing, 31*, 1693–1705.

Li, Z., Wang, Y., & Wang, K. S. (2017). Intelligent predictive maintenance for fault diagnosis and prognosis in machine centers: Industry 4.0 scenario. *Advances in Manufacturing, 5*(4), 377-387. doi:10.1007/s40436-017-0203-8

Liang, Y. C., Lu, X., Li, W. D., & Wang, S. (2018). Cyber Physical System and Big Data enabled energy efficient machining optimisation. *Journal of Cleaner Production, 187*, 46-62.

Liao, H., Zhou, Z., Zhao, X., Zhang, L., Mumtaz, S., Jolfaei, A., . . . Bashir, A. K. (2019). Learning-based context-aware resource allocation for edge-computing-empowered industrial IoT. *IEEE Internet of Things Journal, 7*(5), 4260-4277.

Librantz, A., Araújo, S., Alves, W., Belan, P., Mesquita, R., & Selvatici, A. (2017). Artificial intelligence based system to improve the inspection of plastic mould surfaces. *Journal of Intelligent Manufacturing, 28*(1), 181-190.

Lim, J., Chae, M.-J., Yang, Y., Park, I.-B., Lee, J., & Park, J. (2016). Fast scheduling of semiconductor manufacturing facilities using case-based reasoning. *IEEE Transactions on Semiconductor Manufacturing, 29*(1), 22-32.

Lin, C.-C., Deng, D.-J., Chih, Y.-L., & Chiu, H.-T. (2019). Smart manufacturing scheduling with edge computing using multiclass deep Q network. *IEEE Transactions on Industrial Informatics, 15*(7), 4276-4284.

Lin, C.-C., Deng, D.-J., Kuo, C.-H., & Chen, L. (2019). Concept drift detection and adaption in big imbalance industrial IoT data using an ensemble learning method of offline classifiers. *IEEE Access, 7*, 56198-56207.

Lin, C., Shu, L., Deng, D., Yeh, T., Chen, Y., & Hsieh, H. (2017). A MapReduce-Based Ensemble Learning Method with Multiple Classifier Types and Diversity for Condition-Based Maintenance with Concept Drifts. *IEEE Cloud Computing, 4*(6), 38-48. doi:10.1109/MCC.2018.1081065

Lin, S., He, Z., & Sun, L. (2019). Defect enhancement generative adversarial network for enlarging data set of microcrack defect. *IEEE Access, 7*, 148413-148423.

Lin, Y.-C., Yeh, C.-C., Chen, W.-H., Liu, W.-C., & Wang, J.-J. (2020). The Use of Big Data for Sustainable Development in Motor Production Line Issues. *Sustainability, 12*(13), 5323.

Lingitz, L., Gallina, V., Ansari, F., Gyulai, D., Pfeiffer, A., & Monostori, L. (2018). Lead time prediction using machine learning algorithms: A case study by a semiconductor manufacturer. *Procedia CIRP, 72*, 1051-1056.

Lis, D., & Otto, B. (2021). *Towards a Taxonomy of Ecosystem Data Governance*. Paper presented at the Proceedings of the 54th Hawaii International Conference on System Sciences.

Lithoxoidou, E., Ziogou, C., Vafeiadis, T., Krinidis, S., Ioannidis, D., Voutetakis, S., & Tzovaras, D. (2020). Towards the behavior analysis of chemical reactors utilizing data-driven trend analysis and machine learning techniques. *Applied Soft Computing, 94*, 106464.

Liu, C., Li, H., Tang, Y., Lin, D., & Liu, J. (2019). Next generation integrated smart manufacturing based on big data analytics, reinforced learning, and optimal routes planning methods. *International Journal of Computer Integrated Manufacturing, 32*(9), 820-831.

Liu, C., Li, Y., & Li, Z. (2018). A machining feature definition approach by using two-times unsupervised clustering based on historical data for process knowledge reuse. *Journal of Manufacturing Systems, 49*, 16-24.

Liu, C., Li, Y., Zhou, G., & Shen, W. (2018). A sensor fusion and support vector machine based approach for recognition of complex machining conditions. *Journal of Intelligent Manufacturing, 29*(8), 1739-1752.

Liu, C., Tang, L., & Liu, J. (2019). A stacked autoencoder with sparse Bayesian regression for end-point prediction problems in steelmaking process. *IEEE Transactions on Automation Science and Engineering, 17*(2), 550-561.

Liu, C., Tang, L., Liu, J., & Tang, Z. (2018). A dynamic analytics method based on multistage modeling for a BOF steelmaking process. *IEEE Transactions on Automation Science and Engineering, 16*(3), 1097-1109.

Liu, C., Wang, G., & Li, Z. (2015). Incremental learning for online tool condition monitoring using Ellipsoid ARTMAP network model. *Applied Soft Computing, 35*, 186-198.

Liu, H., Men, X., Li, F., Zhang, J., Wang, X., & Liu, C. (2018). *A new methodology for condition monitoring based on perceptual hashing.* Paper presented at the 2018 13th IEEE Conference on Industrial Electronics and Applications (ICIEA).

Liu, J., An, Y., Dou, R., Ji, H., & Liu, Y. (2018). Helical fault diagnosis model based on data-driven incremental mergence. *Computers & Industrial Engineering, 125*, 517-532. doi:https://doi.org/10.1016/j.cie.2018.02.002

Liu, J., Zhang, W., Ma, T., Tang, Z., Xie, Y., Gui, W., & Niyoyita, J. P. (2020). Toward security monitoring of industrial cyber-physical systems via hierarchically distributed intrusion detection. *Expert Systems with Applications, 158*, 113578.

Liu, J. P., Beyca, O. F., Rao, P. K., Kong, Z. J., & Bukkapatnam, S. T. S. (2017). Dirichlet Process Gaussian Mixture Models for Real-Time Monitoring and Their Application to Chemical Mechanical Planarization. *IEEE Transactions on Automation Science and Engineering, 14*(1), 208-221. doi:10.1109/TASE.2016.2599436

Liu, P., Zhang, Y., Wu, H., & Fu, T. (2020). Optimization of Edge-PLC-Based Fault Diagnosis With Random Forest in Industrial Internet of Things. *IEEE Internet of Things Journal, 7*(10), 9664-9674.

Liu, T.-I., & Jolley, B. (2015). Tool condition monitoring (TCM) using neural networks. *The International Journal of Advanced Manufacturing Technology, 78*(9-12), 1999-2007.

Liu, W., Kong, C., Niu, Q., Jiang, J., & Zhou, X. (2020). A method of NC machine tools intelligent monitoring system in smart factories. *Robotics and Computer-Integrated Manufacturing, 61*, 101842.

Liu, Y., & Jin, S. (2013). Application of Bayesian networks for diagnostics in the assembly process by considering small measurement data sets. *The International Journal of Advanced Manufacturing Technology, 65*(9-12), 1229-1237.

Liu, Y., & Xu, X. (2017). Industry 4.0 and cloud manufacturing: A comparative analysis. *Journal of Manufacturing Science and Engineering, 139*(3), 034701.

Liu, Z., Chen, W., Zhang, C., Yang, C., & Chu, H. (2019). Data super-network fault prediction model and maintenance strategy for mechanical product based on digital twin. *IEEE Access, 7*, 177284-177296.

Liu, Z., & Pu, J. (2019). Analysis and research on intelligent manufacturing medical product design and intelligent hospital system dynamics based on machine learning under big data. *Enterprise Information Systems*, 1-15.

Lolli, F., Balugani, E., Ishizaka, A., Gamberini, R., Rimini, B., & Regattieri, A. (2019). Machine learning for multi-criteria inventory classification applied to intermittent demand. *Production Planning & Control, 30*(1), 76-89.

Longo, F., Nicoletti, L., & Padovano, A. (2019). Emergency preparedness in industrial plants: A forward-looking solution based on industry 4.0 enabling technologies. *Computers in Industry, 105*, 99-122.

Lou, S., Feng, Y., Zheng, H., Gao, Y., & Tan, J. (2018a). Data-driven customer requirements discernment in the product lifecycle management via intuitionistic fuzzy sets and electroencephalogram. *Journal of Intelligent Manufacturing*, https://doi.org/10.1007/s10845-10018-11395-x.

Lou, S., Feng, Y., Zheng, H., Gao, Y., & Tan, J. (2018b). Data-driven customer requirements discernment in the product lifecycle management via intuitionistic fuzzy sets and electroencephalogram. *Journal of Intelligent Manufacturing, 31*, 1721–1736.

Lu, C., Wang, Z., & Zhou, B. (2017). Intelligent fault diagnosis of rolling bearing using hierarchical convolutional network based health state classification. *Advanced Engineering Informatics, 32*, 139-151.

Lu, Z.-J., Xiang, Q., Wu, Y.-m., & Gu, J. (2015). *Application of support vector machine and genetic algorithm optimization for quality prediction within complex industrial process.* Paper presented at the 2015 IEEE 13th International Conference on Industrial Informatics

Luangpaiboon, P. (2015). Evolutionary elements on composite ascent algorithm for multiple response surface optimisation. *Journal of Intelligent Manufacturing, 26*(3), 539-552.

Luo, B., Wang, H., Liu, H., Li, B., & Peng, F. (2018). Early Fault Detection of Machine Tools Based on Deep Learning and Dynamic Identification. *IEEE Transactions on Industrial Electronics, 66* (1), 509-518.

Luo, J., Chen, Q., Yu, F. R., & Tang, L. (2020). Blockchain-enabled software-defined industrial internet of things with deep reinforcement learning. *IEEE Internet of Things Journal, 7*(6), 5466-5480.

Luo, J., Xu, H., Su, Z., Xiao, H., Zheng, K., Zhang, Y., . . . de Oliveira, J. V. (2018). Fault diagnosis based on orthogonal semi-supervised LLTSA for feature extraction and Transductive SVM for fault identification. *Journal of Intelligent & Fuzzy Systems, 34*(6), 3499-3511. doi:10.3233/JIFS-169529

Luo, W., Hu, T., Ye, Y., Zhang, C., & Wei, Y. (2020). A hybrid predictive maintenance approach for CNC machine tool driven by Digital Twin. *Robotics and Computer-Integrated Manufacturing, 65*, 101974.

Lustig, I., Dietrich, B., Johnson, C., & Dziekan, C. (2010). The analytics journey. *Analytics Magazine, 3*(6), 11-13.

Lv, J., Tang, R., Cao, Y., Tang, W., Jia, S., & Liu, Y. (2018). An investigation into methods for predicting material removal energy consumption in turning. *Journal of Cleaner Production, 193*, 128-139. doi:10.1016/j.jclepro.2018.05.035

Lv, Y., & Lin, D. (2017). Design an intelligent real-time operation planning system in distributed manufacturing network. *Industrial Management & Data Systems, 117*(4), 742-753.

Ma, H., Chu, X., Xue, D., & Chen, D. (2017). A systematic decision making approach for product conceptual design based on fuzzy morphological matrix. *Expert Systems with Applications, 81*, 444-456.

Ma, J., Kwak, M., & Kim, H. M. (2014). Demand trend mining for predictive life cycle design. *Journal of Cleaner Production, 68*, 189-199.

Ma, S., Zhang, Y., Liu, Y., Yang, H., Lv, J., & Ren, S. (2020). Data-driven sustainable intelligent manufacturing based on demand response for energy-intensive industries. *Journal of Cleaner Production, 274*, 123155.

Ma, Y., Zhu, W., Benton, M. G., & Romagnoli, J. (2019). Continuous control of a polymerization system with deep reinforcement learning. *Journal of Process Control, 75*, 40-47.

Maggipinto, M., Beghi, A., McLoone, S., & Susto, G. A. (2019). DeepVM: A Deep Learning-based approach with automatic feature extraction for 2D input data Virtual Metrology. *Journal of Process Control, 84*, 24-34.

Maggipinto, M., Terzi, M., Masiero, C., Beghi, A., & Susto, G. A. (2018). A Computer Vision-Inspired Deep Learning Architecture for Virtual Metrology Modeling With 2-Dimensional Data. *IEEE Transactions on Semiconductor Manufacturing, 31*(3), 376-384. doi:10.1109/TSM.2018.2849206

Mahdavi, I., Shirazi, B., Ghorbani, N., & Sahebjamnia, N. (2013). IMAQCS: Design and implementation of an intelligent multi-agent system for monitoring and controlling quality of cement production processes. *Computers in Industry, 64*(3), 290-298.

Malaca, P., Rocha, L. F., Gomes, D., Silva, J., & Veiga, G. (2019). Online inspection system based on machine learning techniques: real case study of fabric textures classification for the automotive industry. *Journal of Intelligent Manufacturing, 30*(1), 351-361.

Manco, G., Ritacco, E., Rullo, P., Gallucci, L., Astill, W., Kimber, D., & Antonelli, M. (2017). Fault detection and explanation through big data analysis on sensor streams. *Expert Systems with Applications, 87*, 141-156.

Marsland, S. (2015). *Machine Learning: An Algorithmic Perspective*. Boca Raton, FL: Chapman & Hall/CRC.

Martinek, P., & Krammer, O. (2019). Analysing machine learning techniques for predicting the hole-filling in pin-in-paste technology. *Computers & Industrial Engineering, 136*, 187-194.

Martínez-Arellano, G., Terrazas, G., & Ratchev, S. (2019). Tool wear classification using time series imaging and deep learning. *The International Journal of Advanced Manufacturing Technology, 104*(9), 3647-3662.

Martinez, P., Al-Hussein, M., & Ahmad, R. (2020). Intelligent vision-based online inspection system of screw-fastening operations in light-gauge steel frame manufacturing. *The International Journal of Advanced Manufacturing Technology, 109*(3), 645-657.

Mashhadi, A. R., & Behdad, S. (2017). Optimal sorting policies in remanufacturing systems: Application of product life-cycle data in quality grading and end-of-use recovery. *Journal of Manufacturing Systems, 43*(Part 1), 15-24.

Mashhadi, A. R., Cade, W., & Behdad, S. (2018). Moving towards Real-time Data-driven Quality Monitoring: A Case Study of Hard Disk Drives. *Procedia Manufacturing, 26*, 1107-1115. doi:https://doi.org/10.1016/j.promfg.2018.07.147

Mayer, C., Mayer, R., & Abdo, M. (2017). *StreamLearner: Distributed Incremental Machine Learning on Event Streams: Grand Challenge*. Paper presented at the Proceedings of the 11th ACM International Conference on Distributed and Event-based Systems, Barcelona, Spain.

Mbuli, J., Trentesaux, D., Clarhaut, J., & Branger, G. (2017). *Decision support in condition-based maintenance of a fleet of cyber-physical systems: a fuzzy logic approach.* Paper presented at the Intelligent Systems Conference.

Mehta, P., Butkewitsch-Choze, S., & Seaman, C. (2018). Smart manufacturing analytics application for semi-continuous manufacturing process – a use case. *Procedia Manufacturing, 26*, 1041-1052. doi:https://doi.org/10.1016/j.promfg.2018.07.138

Mehta, P., Werner, A., & Mears, L. (2015). Condition based maintenance-systems integration and intelligence using Bayesian classification and sensor fusion. *Journal of Intelligent Manufacturing, 26*(2), 331-346. doi:10.1007/s10845-013-0787-1

Mell, P., & Grance, T. (2011). *The NIST Definition of Cloud Computing*. Retrieved from Special Publication 800-145:

Mi, S., Feng, Y., Zheng, H., Li, Z., Gao, Y., & Tan, J. (2020). Integrated intelligent green scheduling of predictive maintenance for complex equipment based on information services. *IEEE Access, 8*, 45797-45812.

Miao, G., Hsieh, S.-J., Segura, J., & Wang, J.-C. (2019). Cyber-physical system for thermal stress prevention in 3D printing process. *The International Journal of Advanced Manufacturing Technology, 100*(1-4), 553-567.

Mileva Boshkoska, B., Bohanec, M., Boškoski, P., & Juričić, Đ. (2015). Copula-based decision support system for quality ranking in the manufacturing of electronically commutated motors. *Journal of Intelligent Manufacturing, 26*(2), 281-293. doi:10.1007/s10845-013-0781-7

Milo, M. W., Roan, M., & Harris, B. (2015). A new statistical approach to automated quality control in manufacturing processes. *Journal of Manufacturing Systems, 36*, 159-167.

Min, Q., Lu, Y., Liu, Z., Su, C., & Wang, B. (2019). Machine learning based digital twin framework for production optimization in petrochemical industry. *International Journal of Information Management, 49*, 502-519.

Mishra, D., Gupta, A., Raj, P., Kumar, A., Anwer, S., Pal, S. K., . . . Pal, A. (2020). Real time monitoring and control of friction stir welding process using multiple sensors. *CIRP Journal of Manufacturing Science and Technology, 30*, 1-11.

Moens, P., Bracke, V., Soete, C., Vanden Hautte, S., Nieves Avendano, D., Ooijevaar, T., . . . Van Hoecke, S. (2020). Scalable Fleet Monitoring and Visualization for Smart Machine Maintenance and Industrial IoT Applications. *Sensors, 20*(15), 4308.

Moharana, U., & Sarmah, S. (2016). Determination of optimal order-up to level quantities for dependent spare parts using data mining. *Computers & Industrial Engineering, 95*, 27-40.

Molka-Danielsen, J., Engelseth, P., & Wang, H. (2018). Large scale integration of wireless sensor network technologies for air quality monitoring at a logistics shipping base. *Journal of Industrial Information Integration, 10*, 20-28.

Moosavian, A., Ahmadi, H., Tabatabaeefar, A., & Khazaee, M. (2013). Comparison of two classifiers; K-nearest neighbor and artificial neural network, for fault diagnosis on a main engine journal-bearing. *Shock and Vibration, 20*(2), 263-272.

Morariu, C., & Borangiu, T. (2018). *Time series forecasting for dynamic scheduling of manufacturing processes.* Paper presented at the International Conference on Automation, Quality and Testing, Robotics.

Moreira, L. C., Li, W., Lu, X., & Fitzpatrick, M. E. (2019). Supervision controller for real-time surface quality assurance in CNC machining using artificial intelligence. *Computers & Industrial Engineering, 127*, 158-168.

Morgan, J., & O'Donnell, G. (2017). Multi-sensor process analysis and performance characterisation in CNC turning-a cyber physical system approach. *International Journal of Advanced Manufacturing Technology, 92*(1-4), 855-868. doi:10.1007/s00170-017-0113-8

Morgan, J., & O'Donnell, G. E. (2018). Cyber physical process monitoring systems. *Journal of Intelligent Manufacturing, 29*(6), 1317-1328. doi:10.1007/s10845-015-1180-z

Mortada, M.-A., Yacout, S., & Lakis, A. (2014). Fault diagnosis in power transformers using multi-class logical analysis of data. *Journal of Intelligent Manufacturing, 25*(6), 1429-1439. doi:10.1007/s10845-013-0750-1

Mortenson, M. J., Doherty, N. F., & Robinson, S. (2015). Operational research from Taylorism to Terabytes: A research agenda for the analytics age. *European Journal of Operational Research, 241*, 583-595.

Mörth, O., Emmanouilidis, C., Hafner, N., & Schadler, M. (2020). Cyber-physical systems for performance monitoring in production intralogistics. *Computers & Industrial Engineering, 142*, 106333.

Mosallam, A., Medjaher, K., & Zerhouni, N. (2016). Data-driven prognostic method based on Bayesian approaches for direct remaining useful life prediction. *Journal of Intelligent Manufacturing, 27*(5), 1037-1048. doi:10.1007/s10845-014-0933-4

Mourtzis, D., Milas, N., & Vlachou, A. (2018). An Internet of Things-Based Monitoring System for Shop-Floor Control. *Journal of Computing and Information Science in Engineering, 18*(2), 021005.

Mourtzis, D., Vlachou, E., Milas, N., & Dimitrakopoulos, G. (2016). Energy consumption estimation for machining processes based on real-time shop floor monitoring via wireless sensor networks. *Procedia CIRP, 57*, 637-642.

Mrugalska, B. (2018). A bounded-error approach to actuator fault diagnosis and remaining useful life prognosis of Takagi-Sugeno fuzzy systems. *ISA transactions, 80*, 257-266.

Muhammad, K., Hussain, T., Del Ser, J., Palade, V., & De Albuquerque, V. H. C. (2019). DeepReS: A deep learning-based video summarization strategy for resource-constrained industrial surveillance scenarios. *IEEE Transactions on Industrial Informatics, 16*(9), 5938-5947.

Müller, O., Junglas, I., Brocke, J. v., & Debortoli, S. (2016). Utilizing big data analytics for information systems research: challenges, promises and guidelines. *European Journal of Information Systems, 25*(4), 289-302.

Mulrennan, K., Donovan, J., Creedon, L., Rogers, I., Lyons, J. G., & McAfee, M. (2018). A soft sensor for prediction of mechanical properties of extruded PLA sheet using an instrumented slit die and machine learning algorithms. *Polymer Testing, 69*, 462-469. doi:https://doi.org/10.1016/j.polymertesting.2018.06.002

Mumtaz, J., Guan, Z., Yue, L., Wang, Z., Ullah, S., & Rauf, M. (2019). Multi-level planning and scheduling for parallel PCB assembly lines using hybrid spider monkey optimization approach. *IEEE Access, 7*, 18685-18700.

Muralidhar, N., Wang, C., Self, N., Momtazpour, M., Nakayama, K., Sharma, R., & Ramakrishnan, N. (2018). illiad: InteLLigent Invariant and Anomaly Detection in Cyber-Physical Systems. *ACM Trans. Intell. Syst. Technol., 9*(3), 1-20. doi:10.1145/3066167

Muralidharan, V., & Sugumaran, V. (2013). Rough set based rule learning and fuzzy classification of wavelet features for fault diagnosis of monoblock centrifugal pump. *Measurement, 46*(9), 3057-3063.

Murphy, W. H., & Gorchels, L. (1996). How to improve product management effectiveness. *Industrial Marketing Management, 25*(1), 47-58.

Nath, A. G., Udmale, S. S., & Singh, S. K. (2020). Role of artificial intelligence in rotor fault diagnosis: A comprehensive review. *Artificial Intelligence Review*, 1-60.

National Research Foundation. (2016). *Research - Innovation - Enterprise (RIE) 2020.* Retrieved from Singapore:

Neef, B., Bartels, J., & Thiede, S. (2018). *Tool Wear and Surface Quality Monitoring Using High Frequency CNC Machine Tool Current Signature.* Paper presented at the 16th International Conference on Industrial Informatics.

Neely, A., Gregory, M., & Platts, K. (1995). Performance measurement system design: a literature review and research agenda. *International Journal of Operations & Production Management, 15*(4), 80-116.

Neto, F. C., Gerônimo, T. M., Cruz, C. E. D., Aguiar, P. R., & Bianchi, E. E. C. (2013). Neural Models for Predicting Hole Diameters in Drilling Processes. *Procedia CIRP, 12*, 49-54. doi:https://doi.org/10.1016/j.procir.2013.09.010

Neupane, D., & Seok, J. (2020). Bearing Fault Detection and Diagnosis Using Case Western Reserve University Dataset With Deep Learning Approaches: A Review. *IEEE Access, 8*, 93155-93178.

Nguyen, H. N., Kim, C.-H., & Kim, J.-M. (2018). Effective Prediction of Bearing Fault Degradation under Different Crack Sizes Using a Deep Neural Network. *Applied Sciences, 8*(11), 2332.

Nguyen, M. T., Truong, L. H., Tran, T. T., & Chien, C.-F. (2020). Artificial intelligence based data processing algorithm for video surveillance to empower industry 3.5. *Computers & Industrial Engineering, 148*, 106671.

Nickerson, R. C., Varshney, U., & Muntermann, J. (2013). A method for taxonomy development and its application in information systems. *European Journal of Information Systems, 22*(3), 336-359.

Nickerson, R. C., Varshney, U., & Muntermann, J. (2017). *Of taxonomies and taxonomic theories.* Paper presented at the AMCIS 2017 PROCEEDINGS.

Nie, Y., & Wan, J. (2015). *Estimation of remaining useful life of bearings using sparse representation method.* Paper presented at the Prognostics and System Health Management Conference.

Niesen, T., Houy, C., Fettke, P., & Loos, P. (2016, 5-8 Jan. 2016). *Towards an Integrative Big Data Analysis Framework for Data-Driven Risk Management in Industry 4.0.* Paper presented at the 2016 49th Hawaii International Conference on System Sciences.

Ning, D., Yu, J., & Huang, J. (2018, 6-7 Sept. 2018). *An Intelligent Device Fault Diagnosis Method in Industrial Internet of Things.* Paper presented at the 2018 International Symposium in Sensing and Instrumentation in IoT Era.

Niu, G., & Jiang, J. (2017). Prognostic control-enhanced maintenance optimization for multi-component systems. *Reliability Engineering & System Safety, 168*, 218-226.

Niu, G., & Li, H. (2017). IETM centered intelligent maintenance system integrating fuzzy semantic inference and data fusion. *Microelectronics Reliability, 75*, 197-204.

Nouiri, M., Bekrar, A., Jemai, A., Niar, S., & Ammari, A. C. (2018). An effective and distributed particle swarm optimization algorithm for flexible job-shop scheduling problem. *Journal of Intelligent Manufacturing, 29*(3), 603-615.

Noyel, M., Thomas, P., Thomas, A., & Charpentier, P. (2016). Reconfiguration process for neuronal classification models: Application to a quality monitoring problem. *Computers in Industry, 83*, 78-91.

Nyanteh, Y. D., Srivastava, S. K., Edrington, C. S., & Cartes, D. A. (2013). Application of artificial intelligence to stator winding fault diagnosis in Permanent Magnet Synchronous Machines. *Electric Power Systems Research, 103*, 201-213.

O'Donovan, P., Leahy, K., Bruton, K., & O'Sullivan, D. T. J. (2015a). Big data in manufacturing: a systematic mapping study. *Journal of Big Data, 2*(1), Article No. 2.

O'Donovan, P., Leahy, K., Bruton, K., & O'Sullivan, D. T. J. (2015b). An industrial big data pipeline for data-driven analytics maintenance applications in large-scale smart manufacturing facilities. *Journal of Big Data, 2*(25).

Oh, Y., Park, H., Yoo, A., Kim, N., Kim, Y., Kim, D., . . . Yang, H. (2013). A product quality prediction model using real-time process monitoring in manufacturing supply chain. *Journal of Korean Institute of Industrial Engineers, 39*(4), 271-277.

Oh, Y., Ransikarbum, K., Busogi, M., Kwon, D., & Kim, N. (2018). Adaptive SVM-based real-time quality assessment for primer-sealer dispensing process of sunroof assembly line. *Reliability Engineering & System Safety, 184*, 202-212. doi:https://doi.org/10.1016/j.ress.2018.03.020

Oliff, H., & Liu, Y. (2017). Towards Industry 4.0 Utilizing Data-Mining Techniques: A Case Study on Quality Improvement. *Procedia CIRP, 63*, 167-172. doi:https://doi.org/10.1016/j.procir.2017.03.311

Olivotti, D., Passlick, J., Dreyer, S., Lebek, B., & Breitner, M. H. (2018). Maintenance Planning Using Condition Monitoring Data. In *Operations Research Proceedings 2017* (pp. 543-548): Springer.

Onel, M., Kieslich, C. A., Guzman, Y. A., Floudas, C. A., & Pistikopoulos, E. N. (2018). Big data approach to batch process monitoring: Simultaneous fault detection and diagnosis using nonlinear support vector machine-based feature selection. *Computers & Chemical Engineering, 115*, 46-63.

Onyeiwu, C., Yang, E., Rodden, T., Yan, X.-T., Zante, R. C., & Ion, W. (2017). *In-process monitoring and quality control of hot forging processes towards Industry 4.0.* Paper presented at the Industrial Systems in the Digital Age Conference 2017.

Orman, M., Rzeszucinski, P., Tkaczyk, A., Krishnamoorthi, K., Pinto, C. T., & Sulowicz, M. (2015). *Bearing fault detection with the use of acoustic signals recorded by a hand-held mobile phone.* Paper presented at the 2015 International Conference on Condition Assessment Techniques in Electrical Systems.

Ortego, P., Diez-Olivan, A., Del Ser, J., Veiga, F., Penalva, M., & Sierra, B. (2020). Evolutionary LSTM-FCN networks for pattern classification in industrial processes. *Swarm and Evolutionary Computation, 54*, 100650.

Oses, N., Legarretaetxebarria, A., Quartulli, M., Garcia, I., & Serrano, M. (2016). Uncertainty reduction in measuring and verification of energy savings by statistical learning in manufacturing environments. *International Journal of Interactive Design and Manufacturing, 10*(3), 291-299.

Ouyang, Z. Y., Sun, X. K., Chen, J. G., Yue, D., & Zhang, T. F. (2018). Multi-View Stacking Ensemble for Power Consumption Anomaly Detection in the Context of Industrial Internet of Things. *IEEE Access, 6*, 9623-9631.

Oyekanlu, E. (2017, 11-14 Dec. 2017). *Predictive edge computing for time series of industrial IoT and large scale critical infrastructure based on open-source software analytic of big data.* Paper presented at the International Conference on Big Data.

Ozturk, G., Bahadir, O., & Teymourifar, A. (2018). Extracting priority rules for dynamic multi-objective flexible job shop scheduling problems using gene expression programming. *International Journal of Production Research*, https://doi.org/10.1080/00207543.00202018.01543964.

Paelke, V. (2014). *Augmented reality in the smart factory: Supporting workers in an industry 4.0. environment*. Paper presented at the Proceedings of the 2014 IEEE Emerging Technology and Factory Automation (ETFA).

Palacios, J., González-Rodríguez, I., Vela, C., & Puente, J. (2015). Swarm lexicographic goal programming for fuzzy open shop scheduling. *Journal of Intelligent Manufacturing, 26*(6), 1201-1215. doi:10.1007/s10845-013-0850-y

Pandiyan, V., Caesarendra, W., Tjahjowidodo, T., & Tan, H. H. (2018). In-process tool condition monitoring in compliant abrasive belt grinding process using support vector machine and genetic algorithm. *Journal of Manufacturing Processes, 31*, 199-213.

Papananias, M., McLeay, T. E., Mahfouf, M., & Kadirkamanathan, V. (2019). A Bayesian framework to estimate part quality and associated uncertainties in multistage manufacturing. *Computers in Industry, 105*, 35-47.

Papananias, M., McLeay, T. E., Obajemu, O., Mahfouf, M., & Kadirkamanathan, V. (2020). Inspection by exception: A new machine learning-based approach for multistage manufacturing. *Applied Soft Computing, 97*, 106787.

Para, J., Del Ser, J., Aguirre, A., & Nebro, A. J. (2018). *Decision Making in Industry 4.0 Scenarios Supported by Imbalanced Data Classification.* Paper presented at the International Symposium on Intelligent and Distributed Computing.

Para, J., Del Ser, J., Nebro, A. J., Zurutuza, U., & Herrera, F. (2019). Analyze, sense, preprocess, predict, implement, and deploy (ASPPID): An incremental methodology based on data analytics for cost-efficiently monitoring the industry 4.0. *Engineering Applications of Artificial Intelligence, 82*, 30-43.

Park, C. Y., Kim, J. W., Kim, B., & Lee, J. (2020). Prediction for manufacturing factors in a steel plate rolling smart factory using data clustering-based machine learning. *IEEE Access, 8*, 60890-60905.

Park, C. Y., Laskey, K. B., Salim, S., & Lee, J. Y. (2017, 10-13 July 2017). *Predictive situation awareness model for smart manufacturing.* Paper presented at the 20th International Conference on Information Fusion

Park, J.-K., Kwon, B.-K., Park, J.-H., & Kang, D.-J. (2016). Machine learning-based imaging system for surface defect inspection. *International Journal of Precision Engineering and Manufacturing-Green Technology, 3*(3), 303-310.

Park, K. T., Kang, Y. T., Yang, S. G., Zhao, W. B., Kang, Y.-S., Im, S. J., . . . Do Noh, S. (2020). Cyber physical energy system for saving energy of the dyeing process with industrial Internet of Things and manufacturing big data. *International Journal of Precision Engineering and Manufacturing-Green Technology, 7*(1), 219-238.

Park, S.-T., Li, G., & Hong, J.-C. (2020). A study on smart factory-based ambient intelligence context-aware intrusion detection system using machine learning. *Journal of Ambient Intelligence and Humanized Computing, 11*(4), 1405-1412.

Patel, J., & Choi, S.-K. (2014). An enhanced classification approach for reliability estimation of structural systems. *Journal of Intelligent Manufacturing, 25*(3), 505-519. doi:10.1007/s10845-012-0702-1

Patel, J., & Upadhyay, S. (2016). Comparison between artificial neural network and support vector method for a fault diagnostics in rolling element bearings. *Procedia engineering, 144*, 390-397.

Pawellek, G. (2016). *Integrierte Instandhaltung und Ersatzteillogistik: Vorgehensweisen, Methoden, Tools*: Springer-Verlag.

Peng, H., Wang, J., PéRez-JiméNez, M. J., Wang, H., Shao, J., & Wang, T. (2013). Fuzzy reasoning spiking neural P system for fault diagnosis. *Information Sciences, 235*, 106-116.

Peng, P., Zhang, Y., Liu, F., Wang, H., & Zhang, H. (2019). A robust and sparse process fault detection method based on RSPCA. *IEEE Access, 7*, 133799-133811.

Peng, W., Ye, Z.-S., & Chen, N. (2018). Joint Online RUL Prediction for Multi-Deteriorating Systems. *IEEE Transactions on Industrial Informatics, Early Access*, DOI: 10.1109/TII.2018.2869429.

Penumuru, D. P., Muthuswamy, S., & Karumbu, P. (2019). Identification and classification of materials using machine vision and machine learning in the context of industry 4.0. *Journal of Intelligent Manufacturing*, 1-13.

Peres, R. S., Barata, J., Leitao, P., & Garcia, G. (2019). Multistage quality control using machine learning in the automotive industry. *IEEE Access, 7*, 79908-79916.

Peres, R. S., Dionisio Rocha, A., Leitao, P., & Barata, J. (2018). IDARTS – Towards intelligent data analysis and real-time supervision for industry 4.0. *Computers in Industry, 101*, 138-146.

Petrović, M., Miljković, Z., & Jokić, A. (2019). A novel methodology for optimal single mobile robot scheduling using whale optimization algorithm. *Applied Soft Computing, 81*, 105520.

Pierezan, J., Maidl, G., Yamao, E. M., dos Santos Coelho, L., & Mariani, V. C. (2019). Cultural coyote optimization algorithm applied to a heavy duty gas turbine operation. *Energy Conversion and Management, 199*, 111932.

Pillai, S., Punnoose, N. J., Vadakkepat, P., Loh, A., & Lee, K. J. (2018, 4-7 Sept. 2018). *An Ensemble of fuzzy Class-Biased Networks for Product Quality Estimation.* Paper presented at the 2018 IEEE 23rd International Conference on Emerging Technologies and Factory Automation (ETFA).

Pimenov, D. Y., Bustillo, A., & Mikolajczyk, T. (2018). Artificial intelligence for automatic prediction of required surface roughness by monitoring wear on face mill teeth. *Journal of Intelligent Manufacturing, 29*(5), 1045-1061. doi:10.1007/s10845-017-1381-8

Pittino, F., Puggl, M., Moldaschl, T., & Hirschl, C. (2020). Automatic anomaly detection on in-production manufacturing machines using statistical learning methods. *Sensors, 20*(8), 2344.

Plehiers, P. P., Symoens, S. H., Amghizar, I., Marin, G. B., Stevens, C. V., & Van Geem, K. M. (2019). Artificial intelligence in steam cracking modeling: a deep learning algorithm for detailed effluent prediction. *Engineering, 5*(6), 1027-1040.

Posada, J., Toro, C., Barandiaran, I., Oyarzun, D., Stricker, D., de Amicis, R., . . . Vallarino. (2015). Visual Computing as Key Enabling Technology for Industrie 4.0 & Industrial Internet. *IEEE Computer Graphics and Applications, 25*(2), 26-40.

Precup, R.-E., Angelov, P., Costa, B. S. J., & Sayed-Mouchaweh, M. (2015). An overview on fault diagnosis and nature-inspired optimal control of industrial process applications. *Computers in Industry, 74*, 75-94.

Priore, P., Gómez, A., Pino, R., & Rosillo, R. (2014). Dynamic scheduling of manufacturing systems using machine learning: An updated review. *Artificial Intelligence for Engineering Design, Analysis and Manufacturing, 28*(1), 83-97.

Prosvirin, A., Islam, M., Kim, J., & Kim, J.-M. (2018). Rub-Impact Fault Diagnosis Using an Effective IMF Selection Technique in Ensemble Empirical Mode Decomposition and Hybrid Feature Models. *Sensors, 18*(7), 2040.

Proto, S., Di Corso, E., Apiletti, D., Cagliero, L., Cerquitelli, T., Malnati, G., & Mazzucchi, D. (2020). REDTag: a predictive maintenance framework for parcel delivery services. *IEEE Access, 8*, 14953-14964.

Psarommatis, F., & Kiritsis, D. (2018). *A Scheduling Tool for Achieving Zero Defect Manufacturing (ZDM): A Conceptual Framework.* Paper presented at the IFIP International Conference on Advances in Production Management Systems.

Purarjomandlangrudi, A., Ghapanchi, A. H., & Esmalifalak, M. (2014). A data mining approach for fault diagnosis: An application of anomaly detection algorithm. *Measurement, 55*, 343-352.

Püschel, L., Schlott, H., & Röglinger, M. (2016a). *What's in a smart thing? Development of a multi-layer taxonomy.* Paper presented at the Thirty Seventh International Conference on Information Systems.

Püschel, L., Schlott, H., & Röglinger, M. (2016b). *What's in a smart thing? Development of a multi-layer taxonomy.* Paper presented at the Proceedings of the 37th International Conference on Information Systems, Dublin, IR.

Püschel, L. C., Röglinger, M., & Brandt, R. (2020). Unblackboxing Smart Things—A Multilayer Taxonomy and Clusters of Nontechnical Smart Thing Characteristics. *IEEE Transactions on Engineering Management*.

Qiao, H., Wang, T., Wang, P., Qiao, S., & Zhang, L. (2018). A Time-Distributed Spatiotemporal Feature Learning Method for Machine Health Monitoring with Multi-Sensor Time Series. *Sensors, 18*(9), E2932.

Qu, S., Chu, T., Wang, J., Leckie, J., & Jian, W. (2015, 8-11 Sept. 2015). *A centralized reinforcement learning approach for proactive scheduling in manufacturing.* Paper presented at the 20th Conference on Emerging Technologies & Factory Automation.

Qu, S., Wang, J., Govil, S., & Leckie, J. O. (2016). Optimized Adaptive Scheduling of a Manufacturing Process System with Multi-skill Workforce and Multiple Machine Types: An Ontology-based, Multi-agent Reinforcement Learning Approach. *Procedia CIRP, 57*, 55-60.

Qu, S., Wang, J., & Jasperneite, J. (2018). *Dynamic scheduling in large-scale stochastic processing networks for demand-driven manufacturing using distributed reinforcement learning.* Paper presented at the 23rd International Conference on Emerging Technologies and Factory Automation.

Ragab, A., El-koujok, M., Amazouz, M., & Yacout, S. (2017). *Fault detection and diagnosis in the Tennessee Eastman Process using interpretable knowledge discovery.* Paper presented at the Annual Reliability and Maintainability Symposium.

Ragab, A., El-Koujok, M., Poulin, B., Amazouz, M., & Yacout, S. (2018). Fault diagnosis in industrial chemical processes using interpretable patterns based on Logical Analysis of Data. *Expert Systems with Applications, 95*, 368-383.

Ragab, A., Ouali, M.-S., Yacout, S., & Osman, H. (2014). *Condition-based maintenance prognostics using logical analysis of data.* Paper presented at the IIE Annual Conference. Proceedings.

Ragab, A., Ouali, M.-S., Yacout, S., & Osman, H. (2016). Remaining useful life prediction using prognostic methodology based on logical analysis of data and Kaplan–Meier estimation. *Journal of Intelligent Manufacturing, 27*(5), 943-958.

Ragab, A., Yacout, S., Ouali, M.-S., & Osman, H. (2016). Prognostics of multiple failure modes in rotating machinery using a pattern-based classifier and cumulative incidence functions. *Journal of Intelligent Manufacturing, 30*(1), 255-274.

Ragab, A., Yacout, S., Ouali, M. S., & Osman, H. (2017). Pattern-based prognostic methodology for condition-based maintenance using selected and weighted survival curves. *Quality and Reliability Engineering International, 33*(8), 1753-1772.

Rahman, H. F., Janardhanan, M. N., & Nielsen, I. E. (2019). Real-time order acceptance and scheduling problems in a flow shop environment using hybrid GA-PSO algorithm. *IEEE Access, 7*, 112742-112755.

Ranjit, M., Gazula, H., Hsiang, S. M., Yu, Y., Borhani, M., Spahr, S., . . . Elliott, B. (2015). Fault Detection Using Human–Machine Co-Construct Intelligence in Semiconductor Manufacturing Processes. *IEEE Transactions on Semiconductor Manufacturing, 28*(3), 297-305.

Rao, P. K., Liu, J. P., Roberson, D., Kong, Z. J., & Williams, C. (2015). Online real-time quality monitoring in additive manufacturing processes using heterogeneous sensors. *Journal of Manufacturing Science and Engineering, 137*(6), 061007.

Rashid, M. M., Amar, M., Gondal, I., & Kamruzzaman, J. (2016). A data mining approach for machine fault diagnosis based on associated frequency patterns. *Applied Intelligence, 45*(3), 638-651.

Rato, T. J., & Reis, M. S. (2020). An integrated multiresolution framework for quality prediction and process monitoring in batch processes. *Journal of Manufacturing Systems, 57*, 198-216.

Rauf, M., Guan, Z., Yue, L., Guo, Z., Mumtaz, J., & Ullah, S. (2020). Integrated planning and scheduling of multiple manufacturing projects under resource constraints using raccoon family optimization algorithm. *IEEE Access, 8*, 151279-151295.

Ray, P., & Mishra, D. P. (2016). Support vector machine based fault classification and location of a long transmission line. *Engineering science and technology, an international journal, 19*(3), 1368-1380.

Reina, A., Cho, S. J., May, G., Coscia, E., Cassina, J., & Kiritsis, D. (2018). Maintenance Planning Support Tool Based on Condition Monitoring with Semantic Modeling of Systems. In M. Zelm, F. W. Jaekel, G. Doumeingts, & M. Wollschlaeger (Eds.), *Enterprise Interoperability: Smart Services and Business Impact of Enterprise Interoperability* (pp. 271-276). Hoboken, NJ: John Wiley & Sons.

Reis, M. S., & Gins, G. (2017). Industrial Process Monitoring in the Big Data/Industry 4.0 Era: From Detection, to Diagnosis, to Prognosis. *Processes, 5*(3), 35.

Reis, M. S., & Rato, T. J. (2018). Multiresolution Analytics for Large Scale Industrial Processes. *IFAC-PapersOnLine, 51*(18), 464-469. doi:https://doi.org/10.1016/j.ifacol.2018.09.381

Ren, L., Cui, J., Sun, Y., & Cheng, X. (2017). Multi-bearing remaining useful life collaborative prediction: A deep learning approach. *Journal of Manufacturing Systems, 43*, 248-256. doi:10.1016/j.jmsy.2017.02.013

Ren, L., Lv, W., & Jiang, S. (2018). Machine prognostics based on sparse representation model. *Journal of Intelligent Manufacturing, 29*(2), 277-285. doi:10.1007/s10845-015-1107-8

Ren, L., Meng, Z., Wang, X., Lu, R., & Yang, L. T. (2020). A Wide-Deep-Sequence Model-Based Quality Prediction Method in Industrial Process Analysis. *IEEE transactions on neural networks and learning systems, 31*(9), 3721-3731.

Ren, L., Sun, Y., Cui, J., & Zhang, L. (2018). Bearing remaining useful life prediction based on deep autoencoder and deep neural networks. *Journal of Manufacturing Systems, 48*, 71-77.

Ren, L., Sun, Y., Wang, H., & Zhang, L. (2018). Prediction of Bearing Remaining Useful Life With Deep Convolution Neural Network. *IEEE Access, 6*, 13041-13049. doi:10.1109/ACCESS.2018.2804930

Ren, R., Hung, T., & Tan, K. C. (2018). A generic deep-learning-based approach for automated surface inspection. *IEEE Transactions on Cybernetics, 48*(3), 929-940.

Rendall, R., Castillo, I., Lu, B., Colegrove, B., Broadway, M., Chiang, L. H., & Reis, M. S. (2018). Image-based manufacturing analytics: Improving the accuracy of an industrial pellet classification system using deep neural networks. *Chemometrics and Intelligent Laboratory Systems, 180*, 26-35. doi:https://doi.org/10.1016/j.chemolab.2018.07.001

Ringsquandl, M., Kharlamov, E., Stepanova, D., Lamparter, S., Lepratti, R., Horrocks, I., & Kröger, P. (2017, 11-14 Dec. 2017). *On event-driven knowledge graph completion in digital factories.* Paper presented at the International Conference on Big Data.

Rivera Torres, P. J., Anido Rifón, L., & Serrano Mercado, E. I. (2018). Probabilistic Boolean network modeling and model checking as an approach for DFMEA for manufacturing systems. *Journal of Intelligent Manufacturing, 29*(6), 1393-1413. doi:10.1007/s10845-015-1183-9

Rivera Torres, P. J., Serrano Mercado, E. I., & Anido Rifón, L. (2018). Probabilistic Boolean network modeling of an industrial machine. *Journal of Intelligent Manufacturing, 29*(4), 875-890. doi:10.1007/s10845-015-1143-4

Rivera Torres, P. J., Serrano Mercado, E. I., Llanes Santiago, O., & Anido Rifón, L. (2018). Modeling preventive maintenance of manufacturing processes with probabilistic Boolean networks with interventions. *Journal of Intelligent Manufacturing, 29*(8), 1941-1952. doi:10.1007/s10845-016-1226-x

Rodríguez, G. G., Gonzalez-Cava, J. M., & Pérez, J. A. M. (2019). An intelligent decision support system for production planning based on machine learning. *Journal of Intelligent Manufacturing*, 1-17.

Rodríguez, I., Nottensteiner, K., Leidner, D., Durner, M., Stulp, F., & Albu-Schäffer, A. (2020). Pattern recognition for knowledge transfer in robotic assembly sequence planning. *IEEE Robotics and Automation Letters, 5*(2), 3666-3673.

Rodríguez, J. J., Quintana, G., Bustillo, A., & Ciurana, J. (2017). A decision-making tool based on decision trees for roughness prediction in face milling. *International Journal of Computer Integrated Manufacturing, 30*(9), 943-957.

Rødseth, H., & Schjølberg, P. (2016). Data-driven predictive maintenance for green manufacturing. *Advanced manufacturing and automation VI, 24*, 36-41.

Rødseth, H., Schjølberg, P., & Marhaug, A. (2017). Deep digital maintenance. *Advances in Manufacturing, 5*(4), 299-310. doi:10.1007/s40436-017-0202-9

Roh, S.-B., & Oh, S.-K. (2016). Identification of plastic wastes by using fuzzy radial basis function neural networks classifier with conditional fuzzy C-means clustering. *Journal of Electrical Engineering & Technology, 11*(6), 1872-1879.

Romeo, L., Loncarski, J., Paolanti, M., Bocchini, G., Mancini, A., & Frontoni, E. (2020). Machine learning-based design support system for the prediction of heterogeneous machine parameters in industry 4.0. *Expert Systems with Applications, 140*, 112869.

Romero-Hdz, J., Saha, B. N., Tstutsumi, S., & Fincato, R. (2020). Incorporating domain knowledge into reinforcement learning to expedite welding sequence optimization. *Engineering Applications of Artificial Intelligence, 91*, 103612.

Rossit, D. A., Tohmé, F., & Frutos, M. (2019). A data-driven scheduling approach to smart manufacturing. *Journal of Industrial Information Integration, 15*, 69-79.

Rousseau, D. M., & Fried, Y. (2001). Location, location, location: Contextualizing organizational research. *Journal of organizational behavior*, 1-13.

Roy, U., Li, Y., & Zhu, B. (2014, 27-30 Oct. 2014). *Building a rigorous foundation for performance assurance assessment techniques for "smart" manufacturing systems.* Paper presented at the nternational Conference on Big Data.

Rude, D. J., Adams, S., & Beling, P. A. (2018). Task recognition from joint tracking data in an operational manufacturing cell. *Journal of Intelligent Manufacturing, 29*(6), 1203-1217. doi:10.1007/s10845-015-1168-8

Ruiz-Sarmiento, J.-R., Monroy, J., Moreno, F.-A., Galindo, C., Bonelo, J.-M., & Gonzalez-Jimenez, J. (2020). A predictive model for the maintenance of industrial machinery in the context of industry 4.0. *Engineering Applications of Artificial Intelligence, 87*, 103289.

Russo, B. (2016). The need for data analysis patterns (in software engineering). In T. Menzies, L. Williams, & T. Zimmermann (Eds.), *Perspectives on Data Science for Software Engineering* (pp. 19-23): Elsevier.

Sacco, C., Radwan, A. B., Anderson, A., Harik, R., & Gregory, E. (2020). Machine learning in composites manufacturing: A case study of Automated Fiber Placement inspection. *Composite Structures, 250*, 112514.
Sądel, B., & Śnieżyński, B. (2017). Online Supervised Learning Approach for Machine Scheduling. *Schedae Informaticae, 25*, 165-176.
Saez, M., Maturana, F. P., Barton, K., & Tilbury, D. M. (2018). Real-Time Manufacturing Machine and System Performance Monitoring Using Internet of Things. *IEEE Transactions on Automation Science & Engineering, 15*(4), 1735-1748. doi:10.1109/TASE.2017.2784826
Safizadeh, M., & Latifi, S. (2014). Using multi-sensor data fusion for vibration fault diagnosis of rolling element bearings by accelerometer and load cell. *Information Fusion, 18*, 1-8.
Saha, C., Aqlan, F., Lam, S. S., & Boldrin, W. (2016). A decision support system for real-time order management in a heterogeneous production environment. *Expert Systems with Applications, 60*, 16-26.
Sahay, B., & Ranjan, J. (2008). Real time business intelligence in supply chain analytics. *Information Management & Computer Security, 16*(1), 28-48.
Said, M., ben Abdellafou, K., & Taouali, O. (2019). Machine learning technique for data-driven fault detection of nonlinear processes. *Journal of Intelligent Manufacturing*, 1-20.
Salary, R. R., Lombardi, J. P., Weerawarne, D. L., Tootooni, M. S., Rao, P. K., & Poliks, M. D. (2020). A Sparse Representation Classification Approach for Near Real-Time, Physics-Based Functional Monitoring of Aerosol Jet-Fabricated Electronics. *Journal of Manufacturing Science and Engineering, 142*(8).
Saldivar, A. A. F., Goh, C., Chen, W., & Li, Y. (2016, 24-29 July 2016). *Self-organizing tool for smart design with predictive customer needs and wants to realize Industry 4.0.* Paper presented at the 2016 IEEE Congress on Evolutionary Computation, Vancouver, Canada.
Saldivar, A. A. F., Goh, C., Li, Y., Chen, Y., & Yu, H. (2016, 7-8 Sept. 2016). *Identifying smart design attributes for Industry 4.0 customization using a clustering Genetic Algorithm.* Paper presented at the 22nd International Conference on Automation and Computing.
Saldivar, A. A. F., Goh, C., Li, Y., Yu, H., & Chen, Y. (2016, 15-17 Dec. 2016). *Attribute identification and predictive customisation using fuzzy clustering and genetic search for Industry 4.0 environments.* Paper presented at the 10th International Conference on Software, Knowledge, Information Management & Applications.
Samuel, A. L. (1959). Some studies in machine learning using the game of checkers. *IBM Journal of Research and Development, 3*(3), 210–229.
Samui, P. (2014). Determination of Surface and Hole Quality in Drilling of AISI D2 Cold Work Tool Steel Using MPMR, MARS and LSSVM. *Journal of Advanced Manufacturing Systems, 13*(4), 237-246. doi:10.1142/S0219686714500140
Sanchez, J. A., Conde, A., Arriandiaga, A., Wang, J., & Plaza, S. (2018). Unexpected Event Prediction in Wire Electrical Discharge Machining Using Deep Learning Techniques. *Materials, 11*(7).
Santhana Babu, A. V., Giridharan, P. K., Ramesh Narayanan, P., & Narayana Murty, S. V. S. (2016). Prediction of Bead Geometry for Flux Bounded TIG Welding of AA 2219-T87 Aluminum Alloy. *Journal of Advanced Manufacturing Systems, 15*(2), 69-84. doi:10.1142/S0219686716500074
Saucedo-Dorantes, J. J., Delgado-Prieto, M., Osornio-Rios, R. A., & de Jesus Romero-Troncoso, R. (2020). Industrial data-driven monitoring based on incremental learning applied to the detection of novel faults. *IEEE Transactions on Industrial Informatics, 16*(9), 5985-5995.
Saucedo-Espinosa, M., Escalante, H., & Berrones, A. (2017). Detection of defective embedded bearings by sound analysis: a machine learning approach. *Journal of Intelligent Manufacturing, 28*(2), 489-500.
Scalabrini Sampaio, G., Vallim Filho, A. R. d. A., Santos da Silva, L., & Augusto da Silva, L. (2019). Prediction of motor failure time using an artificial neural network. *Sensors, 19*(19), 4342.
Schabus, S., & Scholz, J. (2015, 21-23 July 2015). *Geographic Information Science and technology as key approach to unveil the potential of Industry 4.0: How location and time can support smart manufacturing.* Paper presented at the 12th International Conference on Informatics in Control, Automation and Robotics.
Schlegel, P., Briele, K., & Schmitt, R. H. (2018). *Autonomous Data-Driven Quality Control in Self-learning Production Systems.* Paper presented at the Congress of the German Academic Association for Production Technology.
Schoormann, T., Behrens, D., & Knackstedt, R. (2017). *Sustainability in Business Process Models: A Taxonomy-Driven Approach to Synthesize Knowledge and Structure the Field.* Paper presented at the Proceedings of the 38th International Conference on Information Systems, Seoul, South Korea.
Schuh, G., Prote, J.-P., Luckert, M., & Hünnekes, P. (2017). Knowledge Discovery Approach for Automated Process Planning. *Procedia CIRP, 63*, 539-544. doi:https://doi.org/10.1016/j.procir.2017.03.092
Schutze, A., & Helwig, N. (2017). Sensorik und Messtechnik für die Industrie 4.0. *Technisches Messen, 84*(5), 310-319. doi:10.1515/teme-2016-0047

Seera, M., Lim, C. P., & Loo, C. K. (2016). Motor fault detection and diagnosis using a hybrid FMM-CART model with online learning. *Journal of Intelligent Manufacturing, 27*(6), 1273-1285.

Sezer, E., Romero, D., Guedea, F., Macchi, M., & Emmanouilidis, C. (2018, 17-20 June 2018). *An Industry 4.0-Enabled Low Cost Predictive Maintenance Approach for SMEs.* Paper presented at the 2018 IEEE International Conference on Engineering, Technology and Innovation (ICE/ITMC).

Shaban, Y., Yacout, S., Balazinski, M., & Jemielniak, K. (2017). Cutting tool wear detection using multiclass logical analysis of data. *Machining Science and Technology, 21*(4), 526-541.

Shaban, Y., Yacout, S., Balazinski, M., Meshreki, M., & Attia, H. (2015). *Diagnosis of machining outcomes based on machine learning with Logical Analysis of Data.* Paper presented at the International Conference on Industrial Engineering and Operations Management.

Shaban, Y., Yacout, S., Meshreki, M., Attia, H., & Balazinski, M. (2017). Process control based on pattern recognition for routing carbon fiber reinforced polymer. *Journal of Intelligent Manufacturing, 28*(1), 165-179. doi:10.1007/s10845-014-0968-6

Shao, G., Brodsky, A., Shin, S.-J., & Kim, D. (2017). Decision guidance methodology for sustainable manufacturing using process analytics formalism. *Journal of Intelligent Manufacturing, 28*(2), 455-472.

Shao, H., Jiang, H., Zhang, H., Duan, W., Liang, T., & Wu, S. (2018). Rolling bearing fault feature learning using improved convolutional deep belief network with compressed sensing. *Mechanical Systems and Signal Processing, 100*, 743-765.

Sharp, M., Ak, R., & Hedberg, T. (2018). A survey of the advancing use and development of machine learning in smart manufacturing. *Journal of Manufacturing Systems, 48*, 170-179.

Shatnawi, Y., & Al-Khassaweneh, M. (2014). Fault diagnosis in internal combustion engines using extension neural network. *IEEE Transactions on Industrial Electronics, 61*(3), 1434-1443.

Shaw, M. J., Park, S., & Raman, N. (1992). Intelligent scheduling with machine learning capabilities: the induction of scheduling knowledge. *IIE transactions, 24*(2), 156-168.

Shen, Y., Yang, F., Habibullah, M. S., Ahmed, J., Das, A. K., Zhou, Y., & Ho, C. L. (2020). Predicting tool wear size across multi-cutting conditions using advanced machine learning techniques. *Journal of Intelligent Manufacturing*, 1-14.

Shi, H., & Zeng, J. (2016). Real-time prediction of remaining useful life and preventive opportunistic maintenance strategy for multi-component systems considering stochastic dependence. *Computers & Industrial Engineering, 93*, 192-204.

Shimada, J., & Sakajo, S. (2016). *A statistical approach to reduce failure facilities based on predictive maintenance.* Paper presented at the 2016 International Joint Conference on Neural Networks.

Shin, H.-J., Cho, K.-W., & Oh, C.-H. (2018). SVM-Based Dynamic Reconfiguration CPS for Manufacturing System in Industry 4.0. *Wireless Communications and Mobile Computing, 2018*, Article ID 5795037.

Shin, J.-H., Kiritsis, D., & Xirouchakis, P. (2015). Design modification supporting method based on product usage data in closed-loop PLM. *International Journal of Computer Integrated Manufacturing, 28*(6), 551-568.

Shin, S.-J., Kim, D., Shao, G., Brodsky, A., & Lechevalier, D. (2017). Developing a decision support system for improving sustainability performance of manufacturing processes. *Journal of Intelligent Manufacturing, 28*(6), 1421-1440.

Shin, S.-J., Woo, J., & Rachuri, S. (2014). Predictive Analytics Model for Power Consumption in Manufacturing. *Procedia CIRP, 15*, 153-158. doi:https://doi.org/10.1016/j.procir.2014.06.036

Shiue, Y.-R., Lee, K.-C., & Su, C.-T. (2018). Real-time scheduling for a smart factory using a reinforcement learning approach. *Computers & Industrial Engineering, 125*, 604-614.

Shmueli, G. (2010). To explain or to predict? *Statistical science, 25*(3), 289-310.

Shuhui, Q., Jie, W., & Shivani, G. (2016, 6-9 Sept. 2016). *Learning adaptive dispatching rules for a manufacturing process system by using reinforcement learning approach.* Paper presented at the 21st International Conference on Emerging Technologies and Factory Automation.

Si, X.-S., Wang, W., Hu, C.-H., Chen, M.-Y., & Zhou, D.-H. (2013). A Wiener-process-based degradation model with a recursive filter algorithm for remaining useful life estimation. *Mechanical Systems and Signal Processing, 35*(1-2), 219-237.

Smith, J. S. (2003). Survey on the use of simulation for manufacturing system design and operation. *Journal of Manufacturing Systems, 22*(2), 157-171.

Song, D., & Yang, Y. (2018, 25-27 July 2018). *Temperature prediction of molten iron transporting process based on extended Kalman filter.* Paper presented at the 37th Chinese Control Conference.

Song, L., Lin, W., Yang, Y.-G., Zhu, X., Guo, Q., & Xi, J. (2019). Weak micro-scratch detection based on deep convolutional neural network. *IEEE Access, 7*, 27547-27554.

Song, Y., Li, Y., Jia, L., & Qiu, M. (2019). Retraining strategy-based domain adaption network for intelligent fault diagnosis. *IEEE Transactions on Industrial Informatics, 16*(9), 6163-6171.

Sonntag, D., Zillner, S., van der Smagt, P., & Lörincz, A. (2017). Overview of the CPS for smart factories project: deep learning, knowledge acquisition, anomaly detection and intelligent user interfaces. In *Industrial Internet of Things* (pp. 487-504): Springer.

Soualhi, A., Medjaher, K., & Zerhouni, N. (2015). Bearing health monitoring based on Hilbert–Huang transform, support vector machine, and regression. *IEEE Transactions on Instrumentation and Measurement, 64*(1), 52-62.

Soualhi, A., Razik, H., Clerc, G., & Doan, D. D. (2014). Prognosis of bearing failures using hidden Markov models and the adaptive neuro-fuzzy inference system. *IEEE Transactions on Industrial Electronics, 61*(6), 2864-2874.

Sowa, J. F., & Zachman, J. A. (1992). Extending and formalizing the framework for information systems architecture. *IBM systems journal, 31*(3), 590-616.

Spendla, L., Kebisek, M., Tanuska, P., & Hrcka, L. (2017, 26-28 Jan. 2017). *Concept of predictive maintenance of production systems in accordance with industry 4.0.* Paper presented at the 15th International Symposium on Applied Machine Intelligence and Informatics.

Spezzano, G., & Vinci, A. (2015). Pattern Detection in Cyber-Physical Systems. *Procedia Computer Science, 52*, 1016-1021. doi:https://doi.org/10.1016/j.procs.2015.05.096

Sreenuch, T., Tsourdos, A., & Jennions, I. K. (2013). Distributed embedded condition monitoring systems based on OSA-CBM standard. *Computer Standards & Interfaces, 35*(2), 238-246.

Srivastava, N., & Salakhutdinov, R. (2014). Multimodal Learning with Deep Boltzmann Machines. *Journal of Machine Learning Research, 15*, 2949-2980.

Ståhl, N., Mathiason, G., Falkman, G., & Karlsson, A. (2019). Using recurrent neural networks with attention for detecting problematic slab shapes in steel rolling. *Applied Mathematical Modelling, 70*, 365-377.

Stefanovic, N. (2015). Collaborative predictive business intelligence model for spare parts inventory replenishment. *Computer Science and Information Systems, 12*(3), 911-930.

Stein, N., & Flath, C. M. (2017). *Applying Data Science for the Shop-Floor Performance Prediction*. Paper presented at the Twenty-Fifth European Conference on Information Systems.

Stein, N., Meller, J., & Flath, C. M. (2018). Big data on the shop-floor: sensor-based decision-support for manual processes. *Journal of Business Economics, 88*(5), 593-616.

Stojanovic, L., Dinic, M., Stojanovic, N., & Stojadinovic, A. (2016). *Big-data-driven anomaly detection in industry (4.0): An approach and a case study.* Paper presented at the International Conference on Big Data.

Stojanovic, L., & Stojanovic, N. (2017). *PREMIuM: Big data platform for enabling self-healing manufacturing.* Paper presented at the 2017 International Conference on Engineering, Technology and Innovation.

Stoyanov, S., Ahsan, M., Bailey, C., Wotherspoon, T., & Hunt, C. (2019). Predictive analytics methodology for smart qualification testing of electronic components. *Journal of Intelligent Manufacturing, 30*(3), 1497-1514.

Subakti, H., & Jiang, J. (2018, 23-27 July 2018). *Indoor Augmented Reality Using Deep Learning for Industry 4.0 Smart Factories.* Paper presented at the 42nd Annual Computer Software and Applications Conference.

Subramaniyan, M., Skoogh, A., Gopalakrishnan, M., & Hanna, A. (2016). Real-time data-driven average active period method for bottleneck detection. *International Journal of Design & Nature and Ecodynamics, 11*(3), 428-437.

Subramaniyan, M., Skoogh, A., Gopalakrishnan, M., Salomonsson, H., Hanna, A., & Lämkull, D. (2016). An algorithm for data-driven shifting bottleneck detection. *Cogent Engineering, 3*(1), 1239516.

Subramaniyan, M., Skoogh, A., Salomonsson, H., Bangalore, P., & Bokrantz, J. (2018). A data-driven algorithm to predict throughput bottlenecks in a production system based on active periods of the machines. *Computers & Industrial Engineering, 125*, 533-544.

Subramaniyan, M., Skoogh, A., Salomonsson, H., Bangalore, P., Gopalakrishnan, M., & Sheikh Muhammad, A. (2018). Data-driven algorithm for throughput bottleneck analysis of production systems. *Production & Manufacturing Research, 6*(1), 225-246.

Sun, D., Huang, R., Chen, Y., Wang, Y., Zeng, J., Yuan, M., . . . Qu, H. (2019). PlanningVis: A visual analytics approach to production planning in smart factories. *IEEE Transactions on Visualization and Computer Graphics, 26*(1), 579-589.

Sun, D., Lee, V. C., & Lu, Y. (2016). *An intelligent data fusion framework for structural health monitoring.* Paper presented at the 11th Conference on Industrial Electronics and Applications.

Sun, K. H., Huh, H., Tama, B. A., Lee, S. Y., Jung, J. H., & Lee, S. (2020). Vision-Based Fault Diagnostics Using Explainable Deep Learning With Class Activation Maps. *IEEE Access, 8*, 129169-129179.

Sun, T.-H., Tien, F.-C., Tien, F.-C., & Kuo, R.-J. (2016). Automated thermal fuse inspection using machine vision and artificial neural networks. *Journal of Intelligent Manufacturing, 27*(3), 639-651. doi:10.1007/s10845-014-0902-y

Susto, G. A., & Beghi, A. (2016, 6-9 Sept. 2016). *Dealing with time-series data in Predictive Maintenance problems.* Paper presented at the 21st International Conference on Emerging Technologies and Factory Automation (ETFA).

Susto, G. A., Beghi, A., & McLoone, S. (2017, 15-18 May 2017). *Anomaly detection through on-line isolation Forest: An application to plasma etching.* Paper presented at the 28th Annual SEMI Advanced Semiconductor Manufacturing Conference

Susto, G. A., Maggipinto, M., Zocco, F., & McLoone, S. (2019). Induced Start Dynamic Sampling for Wafer Metrology Optimization. *IEEE Transactions on Automation Science and Engineering, 17*(1), 418-432.

Susto, G. A., Schirru, A., Pampuri, S., McLoone, S., & Beghi, A. (2015). Machine learning for predictive maintenance: A multiple classifier approach. *IEEE Transactions on Industrial Informatics, 11*(3), 812-820.

Susto, G. A., Terzi, M., & Beghi, A. (2017). Anomaly Detection Approaches for Semiconductor Manufacturing. *Procedia Manufacturing, 11*, 2018-2024.

Susto, G. A., Wan, J., Pampuri, S., Zanon, M., Johnston, A. B., O'Hara, P. G., & McLoone, S. (2014). *An adaptive machine learning decision system for flexible predictive maintenance.* Paper presented at the Automation Science and Engineering (CASE), 2014 IEEE International Conference on.

Sutharssan, T., Stoyanov, S., Bailey, C., & Yin, C. (2015). Prognostic and health management for engineering systems: a review of the data-driven approach and algorithms. *The Journal of Engineering, 1*(1), 215 – 222.

Syafrudin, M., Alfian, G., Fitriyani, N., & Rhee, J. (2018). Performance Analysis of IoT-Based Sensor, Big Data Processing, and Machine Learning Model for Real-Time Monitoring System in Automotive Manufacturing. *Sensors, 18*(9), 2946.

Syafrudin, M., Fitriyani, N. L., Alfian, G., & Rhee, J. (2019). An affordable fast early warning system for edge computing in assembly line. *Applied Sciences, 9*(1), 84.

Syafrudin, M., Fitriyani, N. L., Li, D., Alfian, G., Rhee, J., & Kang, Y. S. (2017). An Open Source-Based Real-Time Data Processing Architecture Framework for Manufacturing Sustainability. *Sustainability, 9*(11). doi:10.3390/su9112139

Szopinski, D., Schoormann, T., & Kundisch, D. (2019). *Because your taxonomy is worth it: Towards a framework for taxonomy evaluation.* Paper presented at the European Conference on Information Systems (ECIS).

Tabernik, D., Šela, S., Skvarč, J., & Skočaj, D. (2020). Segmentation-based deep-learning approach for surface-defect detection. *Journal of Intelligent Manufacturing, 31*(3), 759-776.

Tamilselvan, P., & Wang, P. (2013). Failure diagnosis using deep belief learning based health state classification. *Reliability Engineering & System Safety, 115*, 124-135.

Tamura, Y., Iizuka, H., Yamamoto, M., & Furukawa, M. (2015). Application of local clustering organization to reactive job-shop scheduling. *Soft Computing, 19*(4), 891-899.

Tan, C. J., Neoh, S. C., Lim, C. P., Hanoun, S., Wong, W. P., Loo, C. K., . . . Nahavandi, S. (2019). Application of an evolutionary algorithm-based ensemble model to job-shop scheduling. *Journal of Intelligent Manufacturing, 30*(2), 879-890.

Tang, X., Gu, X., Wang, J., He, Q., Zhang, F., & Lu, J. (2020). A bearing fault diagnosis method based on feature selection feedback network and improved DS evidence fusion. *IEEE Access, 8*, 20523-20536.

Tangjitsitcharoen, S., Thesniyom, P., & Ratanakuakangwan, S. (2017). Prediction of surface roughness in ball-end milling process by utilizing dynamic cutting force ratio. *Journal of Intelligent Manufacturing, 28*(1), 13-21. doi:10.1007/s10845-014-0958-8

Tangjitsitcharoen, S., & Wongtangthinthan, C. (2016). Advanced in tool wear prediction in CNC turning by utilizing average variances of dynamic cutting forces. *Full Paper Proceeding ECBA-2016, 380*(11), 1-8.

Tao, J., Wang, K., Li, B., Liu, L., & Cai, Q. (2016). Hierarchical models for the spatial–temporal carbon nanotube height variations. *International Journal of Production Research, 54*(21), 6613-6632.

Tao, Y., Wang, X., Sánchez, R.-V., Yang, S., & Bai, Y. (2019). Spur gear fault diagnosis using a multilayer gated recurrent unit approach with vibration signal. *IEEE Access, 7*, 56880-56889.

Tayal, A., & Singh, S. P. (2018). Integrating big data analytic and hybrid firefly-chaotic simulated annealing approach for facility layout problem. *Annals of Operations Research, 270*(1-2), 489-514.

Terrazas, G., Martínez-Arellano, G., Benardos, P., & Ratchev, S. (2018). Online Tool Wear Classification during Dry Machining Using Real Time Cutting Force Measurements and a CNN Approach. *Journal of Manufacturing and Materials Processing, 2*(4), 72.

Terzidis, O., Oberle, D., Friesen, A., Janiesch, C., & Barros, A. (2012). The internet of services and usdl. In *Handbook of service description* (pp. 1-16): Springer.

Thiesse, F., Wirth, M., Kemper, H. G., Moisa, M., Morar, D., Lasi, H., . . . Minshall, T. (2015). Economic implications of additive manufacturing and the contribution of MIS. *Business & Information Systems Engineering, 57*(2), 139-148.

Tian, J., Azarian, M. H., Pecht, M., Niu, G., & Li, C. (2017, 9-12 July 2017). *An ensemble learning-based fault diagnosis method for rotating machinery.* Paper presented at the 2017 Prognostics and System Health Management Conference.

Tian, J., Morillo, C., Azarian, M. H., & Pecht, M. (2016). Motor bearing fault detection using spectral kurtosis-based feature extraction coupled with K-nearest neighbor distance analysis. *IEEE Transactions on Industrial Electronics, 63*(3), 1793-1803.

Tong, X., Teng, R., Sun, L., & Guan, T. (2018, 8-9 Feb. 2018). *Intelligent design and optimization of assembly process for urban rail vehicle oriented to intelligent manufacturing.* Paper presented at the 2018 IEEE International Conference on Smart Manufacturing, Industrial & Logistics Engineering

Tristo, G., Bissacco, G., Lebar, A., & Valentinčič, J. (2015). Real time power consumption monitoring for energy efficiency analysis in micro EDM milling. *The International Journal of Advanced Manufacturing Technology, 78*(9-12), 1511-1521.

Trkman, P., McCormack, K., de Oliveira, M., & Ladeira, M. (2010). The impact of business analytics on supply chain performance. *Decision Support Systems, 49*(3), 318-327.

Trunzer, E., Weiß, I., Folmer, J., Schrüfer, C., Vogel-Heuser, B., Erben, S., . . . Vermum, C. (2017, 10-13 Dec. 2017). *Failure mode classification for control valves for supporting data-driven fault detection.* Paper presented at the 2017 IEEE International Conference on Industrial Engineering and Engineering Management.

Truong, H. (2018, 21-23 Oct. 2018). *Integrated Analytics for IIoT Predictive Maintenance Using IoT Big Data Cloud Systems.* Paper presented at the 2018 IEEE International Conference on Industrial Internet (ICII).

Tsai, C.-W., Lai, C.-F., Chiang, M.-C., & Yang, L. T. (2013). Data mining for internet of things: A survey. *IEEE Communications Surveys & Tutorials, 16*(1), 77-97.

Tsai, C.-Y., Chen, C.-J., & Lo, Y.-T. (2014). A cost-based module mining method for the assemble-to-order strategy. *Journal of Intelligent Manufacturing, 25*(6), 1377-1392.

Tsai, Y.-T., Lee, C.-H., Liu, T.-Y., Chang, T.-J., Wang, C.-S., Pawar, S., . . . Huang, J.-H. (2020). Utilization of a reinforcement learning algorithm for the accurate alignment of a robotic arm in a complete soft fabric shoe tongues automation process. *Journal of Manufacturing Systems, 56*, 501-513.

Uhlmann, E., Laghmouchi, A., Geisert, C., & Hohwieler, E. (2017). Decentralized Data Analytics for Maintenance in Industrie 4.0. *Procedia Manufacturing, 11*, 1120-1126. doi:https://doi.org/10.1016/j.promfg.2017.07.233

Unal, M., Sahin, Y., Onat, M., Demetgul, M., & Kucuk, H. (2017). Fault diagnosis of rolling bearings using data mining techniques and boosting. *Journal of Dynamic Systems, Measurement, and Control, 139*(2), 021003.

Unnikrishnan, S., Donovan, J., Macpherson, R., & Tormey, D. (2020). An Integrated Histogram-Based Vision and Machine-Learning Classification Model for Industrial Emulsion Processing. *IEEE Transactions on Industrial Informatics, 16*(9), 5948-5955.

Vafeiadis, T., Dimitriou, N., Ioannidis, D., Wotherspoon, T., Tinker, G., & Tzovaras, D. (2018). A framework for inspection of dies attachment on PCB utilizing machine learning techniques. *Journal of Management Analytics, 5*(2), 81-94.

Vaidya, S., Ambad, P., & Bhosle, S. (2018). Industry 4.0 - A Glimpse. *Procedia Manufacturing, 20*, 233-238.

Van Horenbeek, A., & Pintelon, L. (2013). A dynamic predictive maintenance policy for complex multi-component systems. *Reliability Engineering & System Safety, 120*, 39-50.

van Staden, H. E., & Boute, R. N. (2020). The effect of multi-sensor data on condition-based maintenance policies. *European Journal of Operational Research, 290*(2), 585-600.

Vazan, P., Janikova, D., Tanuska, P., Kebisek, M., & Cervenanska, Z. (2017). Using data mining methods for manufacturing process control. *IFAC-PapersOnLine, 50*(1), 6178-6183. doi:https://doi.org/10.1016/j.ifacol.2017.08.986

Veeramani, S., Muthuswamy, S., Sagar, K., & Zoppi, M. (2019). Artificial intelligence planners for multi-head path planning of SwarmItFIX agents. *Journal of Intelligent Manufacturing*, 1-18.

Verma, S., Kawamoto, Y., Fadlullah, Z. M., Nishiyama, H., & Kato, N. (2017). A survey on network methodologies for real-time analytics of massive IoT data and open research issues. *IEEE Communications Surveys & Tutorials, 19*(3), 1457-1477.

Verstraete, D., Ferrada, A., Droguett, E. L., Meruane, V., & Modarres, M. (2017). Deep learning enabled fault diagnosis using time-frequency image analysis of rolling element bearings. *Shock and Vibration, 2017.*

Villalonga, A., Beruvides, G., Castaño, F., Haber, R. E., & Novo, M. (2018). Condition-Based Monitoring Architecture for CNC Machine Tools Based on Global Knowledge. *IFAC-PapersOnLine, 51*(11), 200-204.

Vogel, P. U. (2020). *Trending in der pharmazeutischen Industrie*: Springer.
Voisin, A., Laloix, T., Iung, B., & Romagne, E. (2018). Predictive Maintenance and part quality control from joint product-process-machine requirements: application to a machine tool. *Procedia Manufacturing, 16*, 147-154.
vom Brocke, J., Simons, A., Riemer, K., Niehaves, B., Plattfaut, R., & Cleven, A. (2015). Standing on the shoulders of giants: Challenges and recommendations of literature search in information systems research. *Communications of the association for information systems, 37*(1), 9.
Vrabic, R., Kozjek, D., & Butala, P. (2017). Knowledge elicitation for fault diagnostics in plastic injection moulding: A case for machine-to-machine communication. *Cirp Annals - Manufacturing Technology, 66*(1), 433-436. doi:10.1016/j.cirp.2017.04.001
Vukicevic, A. M., Djapan, M., Todorovic, P., Erić, M., Stefanovic, M., & Macuzic, I. (2019). Decision Support System for Dimensional Inspection of Extruded Rubber Profiles. *IEEE Access, 7*, 112605-112616.
Vununu, C., Moon, K. S., Lee, S. H., & Kwon, K. R. (2018). A Deep Feature Learning Method for Drill Bits Monitoring Using the Spectral Analysis of the Acoustic Signals. *Sensors, 18*(8), E2634. doi:10.3390/s18082634
Wan, J., Tang, S., Li, D., Wang, S., Liu, C., Abbas, H., & Vasilakos, A. V. (2017). A Manufacturing Big Data Solution for Active Preventive Maintenance. *IEEE Transactions on Industrial Informatics, 13*(4), 2039-2047. doi:10.1109/TII.2017.2670505
Wang, C.-Y., Chen, Y.-J., & Chien, C.-F. (2020). Industry 3.5 to empower smart production for poultry farming and an empirical study for broiler live weight prediction. *Computers & Industrial Engineering*, 106931.
Wang, C., Cheng, K., Nelson, N., Sawangsri, W., & Rakowski, R. (2015). Cutting force–based analysis and correlative observations on the tool wear in diamond turning of single-crystal silicon. *Proceedings of the Institution of Mechanical Engineers, Part B: Journal of Engineering Manufacture, 229*(10), 1867-1873.
Wang, C., Gan, M., & Zhu, C. a. (2016). A supervised sparsity-based wavelet feature for bearing fault diagnosis. *Journal of Intelligent Manufacturing, 30*(1), 229-239.
Wang, C., Gan, M., & Zhu, C. a. (2017). Intelligent fault diagnosis of rolling element bearings using sparse wavelet energy based on overcomplete DWT and basis pursuit. *Journal of Intelligent Manufacturing, 28*(6), 1377-1391.
Wang, C., Gan, M., & Zhu, C. a. (2018). Fault feature extraction of rolling element bearings based on wavelet packet transform and sparse representation theory. *Journal of Intelligent Manufacturing, 29*(4), 937-951.
Wang, C., & Jiang, P. (2017). Deep neural networks based order completion time prediction by using real-time job shop RFID data. *Journal of Intelligent Manufacturing, 30*(3), 1303–1318. doi:https://doi.org/10.1007/s10845-017-1325-3
Wang, G., Guo, Z., & Qian, L. (2014). Tool wear prediction considering uncovered data based on partial least square regression. *Journal of Mechanical Science and Technology, 28*(1), 317-322.
Wang, G., Guo, Z., & Yang, Y. (2013). Force sensor based online tool wear monitoring using distributed Gaussian ARTMAP network. *Sensors and Actuators A: Physical, 192*, 111-118.
Wang, G., Liu, C., Cui, Y., & Feng, X. (2014). Tool wear monitoring based on cointegration modelling of multisensory information. *International Journal of Computer Integrated Manufacturing, 27*(5), 479-487.
Wang, G., Nixon, M., & Boudreaux, M. (2019). Toward cloud-assisted industrial IoT platform for large-scale continuous condition monitoring. *Proceedings of the IEEE, 107*(6), 1193-1205.
Wang, G., Yang, Y., Xie, Q., & Zhang, Y. (2014). Force based tool wear monitoring system for milling process based on relevance vector machine. *Advances in Engineering Software, 71*, 46-51.
Wang, G., Zhang, Y., Liu, C., Xie, Q., & Xu, Y. (2016). A new tool wear monitoring method based on multi-scale PCA. *Journal of Intelligent Manufacturing, 30*(1), 113–122.
Wang, G. F., Xie, Q. L., & Zhang, Y. C. (2017). Tool condition monitoring system based on support vector machine and differential evolution optimization. *Proceedings of the Institution of Mechanical Engineers, Part B: Journal of Engineering Manufacture, 231*(5), 805-813.
Wang, H.-K., & Chien, C.-F. (2020). An inverse-distance weighting genetic algorithm for optimizing the wafer exposure pattern for enhancing OWE for smart manufacturing. *Applied Soft Computing, 94*, 106430.
Wang, H., Li, S., Song, L., Cui, L., & Wang, P. (2019). An enhanced intelligent diagnosis method based on multi-sensor image fusion via improved deep learning network. *IEEE Transactions on Instrumentation and Measurement, 69*(6), 2648-2657.
Wang, H., Ren, B., Song, L., & Cui, L. (2019). A novel weighted sparse representation classification strategy based on dictionary learning for rotating machinery. *IEEE Transactions on Instrumentation and Measurement, 69*(3), 712-720.
Wang, H., Wang, W., Sun, H., Cui, Z., Rahnamayan, S., & Zeng, S. (2017). A new cuckoo search algorithm with hybrid strategies for flow shop scheduling problems. *Soft Computing, 21*(15), 4297-4307.

Wang, J., Gao, R. X., Yuan, Z., Fan, Z., & Zhang, L. (2016). A joint particle filter and expectation maximization approach to machine condition prognosis. *Journal of Intelligent Manufacturing, 30*(2), 605-621.
Wang, J., Qu, S., Wang, J., Leckie, J. O., & Xu, R. (2017). *Real-Time Decision Support with Reinforcement Learning for Dynamic Flowshop Scheduling.* Paper presented at the European Conference on Smart Objects, Systems and Technologies.
Wang, J., Sun, Y., Zhang, W., Thomas, I., Duan, S., & Shi, Y. (2016). Large-Scale Online Multitask Learning and Decision Making for Flexible Manufacturing. *IEEE Transactions on Industrial Informatics, 12*(6), 2139-2147. doi:10.1109/TII.2016.2549919
Wang, J., Wang, K., Wang, Y., Huang, Z., & Xue, R. (2018). Deep Boltzmann machine based condition prediction for smart manufacturing. *Journal of Ambient Intelligence and Humanized Computing, 1ß*(3), 851-861.
Wang, J., Yan, J., Li, C., Gao, R. X., & Zhao, R. (2019). Deep heterogeneous GRU model for predictive analytics in smart manufacturing: Application to tool wear prediction. *Computers in Industry, 111*, 1-14.
Wang, J., Ye, L., Gao, R. X., Li, C., & Zhang, L. (2019). Digital Twin for rotating machinery fault diagnosis in smart manufacturing. *International Journal of Production Research, 57*(12), 3920-3934.
Wang, K., Jiang, W., & Li, B. (2016). A spatial variable selection method for monitoring product surface. *International Journal of Production Research, 54*(14), 4161-4181. doi:10.1080/00207543.2015.1109723
Wang, K., Zhou, Y., Liu, Z., Shao, Z., Luo, X., & Yang, Y. (2020). Online task scheduling and resource allocation for intelligent NOMA-based industrial internet of things. *IEEE Journal on Selected Areas in Communications, 38*(5), 803-815.
Wang, P., & Gao, R. X. (2020). Transfer learning for enhanced machine fault diagnosis in manufacturing. *CIRP Annals, 69*(1), 413-416.
Wang, P., Liu, H., Wang, L., & Gao, R. X. (2018). Deep learning-based human motion recognition for predictive context-aware human-robot collaboration. *CIRP Annals, 67*(1), 17-20. doi:https://doi.org/10.1016/j.cirp.2018.04.066
Wang, P., Yan, R., & Gao, R. X. (2017). Virtualization and deep recognition for system fault classification. *Journal of Manufacturing Systems, 44*, 310-316.
Wang, Q., Jiao, W., Wang, P., & Zhang, Y. (2020). A tutorial on deep learning-based data analytics in manufacturing through a welding case study. *Journal of Manufacturing Processes.*
Wang, S., & Liu, M. (2015). Multi-objective optimization of parallel machine scheduling integrated with multi-resources preventive maintenance planning. *Journal of Manufacturing Systems, 37*, 182-192.
Wang, S., Wan, J., Li, D., & Liu, C. (2018). Knowledge Reasoning with Semantic Data for Real-Time Data Processing in Smart Factory. *Sensors, 18*(2), 471.
Wang, S., Yang, L., Chen, X., Tong, C., Ding, B., & Xiang, J. (2017). Nonlinear squeezing time-frequency transform and application in rotor rub-impact fault diagnosis. *Journal of Manufacturing Science and Engineering, 139*(10), 101005.
Wang, T., Qiao, M., Zhang, M., Yang, Y., & Snoussi, H. (2018). Data-driven prognostic method based on self-supervised learning approaches for fault detection. *Journal of Intelligent Manufacturing*, https://doi.org/10.1007/s10845-10018-11431-x.
Wang, X., Li, Y., Rui, T., Zhu, H., & Fei, J. (2015). Bearing fault diagnosis method based on Hilbert envelope spectrum and deep belief network. *Journal of Vibroengineering, 17*(3), 1295-1308.
Wang, X., Wang, H., & Qi, C. (2016). Multi-agent reinforcement learning based maintenance policy for a resource constrained flow line system. *Journal of Intelligent Manufacturing, 27*(2), 325-333.
Wang, Y., Hulstijn, J., & Tan, Y.-h. (2018). *Analyzing Transaction Codes in Manufacturing for Compliance Monitoring.* Paper presented at the Twenty-fourth Americas Confernce on Information Systems (AMCIS).
Wang, Y., Li, K., Gan, S., & Cameron, C. (2019). Analysis of energy saving potentials in intelligent manufacturing: A case study of bakery plants. *Energy, 172*, 477-486.
Wang, Y., Ma, Q., Zhu, Q., Liu, X., & Zhao, L. (2014). An intelligent approach for engine fault diagnosis based on Hilbert–Huang transform and support vector machine. *Applied Acoustics, 75*, 1-9.
Wang, Y., & Tseng, M. (2015). A Naïve Bayes approach to map customer requirements to product variants. *Journal of Intelligent Manufacturing, 26*(3), 501-509. doi:10.1007/s10845-013-0806-2
Wang, Y., & Tseng, M. M. (2014). Identifying Emerging Customer Requirements in an Early Design Stage by Applying Bayes Factor-Based Sequential Analysis. *IEEE Transactions on Engineering Management, 61*(1), 129-137. doi:10.1109/TEM.2013.2248729
Wang, Y., Xu, G., Liang, L., & Jiang, K. (2015). Detection of weak transient signals based on wavelet packet transform and manifold learning for rolling element bearing fault diagnosis. *Mechanical Systems and Signal Processing, 54*, 259-276.

Wang, Y., Yuan, S.-M., Ling, D., Zhao, Y.-B., Gu, X.-G., & Li, B.-Y. (2019). Fault monitoring based on adaptive partition non-negative matrix factorization for non-Gaussian processes. *IEEE Access, 7*, 32783-32795.

Wanner, J., Wissuchek, C., & Janiesch, C. (2019). *Machine Learning und Complex Event Processing: Effiziente Echtzeitauswertung am Beispiel Smart Factory.* Paper presented at the 14. Internationalen Tagung Wirtschaftsinformatik (WI), Siegen, Germany.

Waschneck, B., Reichstaller, A., Belzner, L., Altenmüller, T., Bauernhansl, T., Knapp, A., & Kyek, A. (2018a). *Deep reinforcement learning for semiconductor production scheduling.* Paper presented at the 29th Annual SEMI Advanced Semiconductor Manufacturing Conference.

Waschneck, B., Reichstaller, A., Belzner, L., Altenmüller, T., Bauernhansl, T., Knapp, A., & Kyek, A. (2018b). Optimization of global production scheduling with deep reinforcement learning. *Procedia CIRP, 72*, 1264-1269. doi:https://doi.org/10.1016/j.procir.2018.03.212

Wedel, M., von Hacht, M., Hieber, R., Metternich, J., & Abele, E. (2015). Real-time bottleneck detection and prediction to prioritize fault repair in interlinked production lines. *Procedia CIRP, 37*, 140-145.

Weigelt, M., Mayr, A., Seefried, J., Heisler, P., & Franke, J. (2018). Conceptual design of an intelligent ultrasonic crimping process using machine learning algorithms. *Procedia Manufacturing, 17*, 78-85. doi:https://doi.org/10.1016/j.promfg.2018.10.015

Weimer, D., Scholz-Reiter, B., & Shpitalni, M. (2016). Design of deep convolutional neural network architectures for automated feature extraction in industrial inspection. *CIRP Annals, 65*(1), 417-420.

Weiß, I., & Vogel-Heuser, B. (2018). Assessment of variance & distribution in data for effective use of statistical methods for product quality prediction. *Automatisierungstechnik, 66*(4), 344-355.

Weiss, S., Dhurandhar, A., Baseman, R., White, B., Logan, R., Winslow, J., & Poindexter, D. (2016). Continuous prediction of manufacturing performance throughout the production lifecycle. *Journal of Intelligent Manufacturing, 27*(4), 751-763. doi:10.1007/s10845-014-0911-x

Wells, L. J., Camelio, J. A., Williams, C. B., & White, J. (2014). Cyber-physical security challenges in manufacturing systems. *Manufacturing Letters, 2*(2), 74-77.

Wen, J., Gao, H., & Zhang, J. (2018). Bearing Remaining Useful Life Prediction Based on a Nonlinear Wiener Process Model. *Shock and Vibration, 2018*, Article ID 4068431.

Wen, L., Gao, L., & Li, X. (2017). A new deep transfer learning based on sparse auto-encoder for fault diagnosis. *IEEE Transactions on Systems, Man, and Cybernetics: Systems, 49*(1), 136-144.

Wen, L., Li, X., Gao, L., & Zhang, Y. (2018). A New Convolutional Neural Network-Based Data-Driven Fault Diagnosis Method. *IEEE Transactions on Industrial Electronics, 65*(7), 5990-5998. doi:10.1109/TIE.2017.2774777

Wen, X., & Gong, Y. (2017). Modeling and prediction research on wear of electroplated diamond micro-grinding tool in soda lime glass grinding. *The International Journal of Advanced Manufacturing Technology, 91*(9-12), 3467-3479.

Westbrink, F., Chadha, G. S., & Schwung, A. (2018, 15-18 May 2018). *Integrated IPC for data-driven fault detection.* Paper presented at the Industrial Cyber-Physical Systems.

Windmann, S., Jungbluth, F., & Niggemann, O. (2015, 8-11 Sept. 2015). *A HMM-based fault detection method for piecewise stationary industrial processes.* Paper presented at the 2015 IEEE 20th Conference on Emerging Technologies & Factory Automation (ETFA).

Windmann, S., & Niggemann, O. (2015). *Efficient fault detection for industrial automation processes with observable process variables.* Paper presented at the 13th International Conference on Industrial Informatics.

Witte, A.-K., & Zarnekow, R. (2018). *Is Open Always Better? - A Taxonomy-based Analysis of Platform Ecosystems for Fitness Trackers*. Paper presented at the Multikonferenz Wirtschaftsinformatik.

Wu, C., Chen, T., & Jiang, R. (2017). Bearing fault diagnosis via kernel matrix construction based support vector machine. *Journal of Vibroengineering, 19*(5).

Wu, C., Chen, T., Jiang, R., Ning, L., & Jiang, Z. (2017). A novel approach to wavelet selection and tree kernel construction for diagnosis of rolling element bearing fault. *Journal of Intelligent Manufacturing, 28*(8), 1847-1858. doi:10.1007/s10845-015-1070-4

Wu, D., Jennings, C., Terpenny, J., & Kumara, S. (2016, 5-8 Dec. 2016). *Cloud-based machine learning for predictive analytics: Tool wear prediction in milling.* Paper presented at the International Conference on Big Data

Wu, D., Jiang, Z., Xie, X., Wei, X., Yu, W., & Li, R. (2019). LSTM learning with Bayesian and Gaussian processing for anomaly detection in industrial IoT. *IEEE Transactions on Industrial Informatics, 16*(8), 5244-5253.

Wu, D., Liu, S., Zhang, L., Terpenny, J., Gao, R. X., Kurfess, T., & Guzzo, J. A. (2017). A fog computing-based framework for process monitoring and prognosis in cyber-manufacturing. *Journal of Manufacturing Systems, 43*, 25-34. doi:https://doi.org/10.1016/j.jmsy.2017.02.011

Wu, D. Z., Jennings, C., Terpenny, J., Gao, R. X., & Kumara, S. (2017). A Comparative Study on Machine Learning Algorithms for Smart Manufacturing: Tool Wear Prediction Using Random Forests. *Journal of Manufacturing Science and Engineering, 139*(7), Paper No: MANU-16-1567. doi:10.1115/1.4036350

Wu, D. Z., Jennings, C., Terpenny, J., Kumara, S., & Gao, R. X. (2018). Cloud-Based Parallel Machine Learning for Tool Wear Prediction. *Journal of Manufacturing Science and Engineering, 140*(4). doi:10.1115/1.4038002

Wu, J., Su, Y., Cheng, Y., Shao, X., Deng, C., & Liu, C. (2018). Multi-sensor information fusion for remaining useful life prediction of machining tools by adaptive network based fuzzy inference system. *Applied Soft Computing, 68*, 13-23.

Wu, M., & Moon, Y. B. (2019). Intrusion detection system for cyber-manufacturing system. *Journal of Manufacturing Science and Engineering, 141*(3).

Wu, M., Song, Z., & Moon, Y. B. (2019). Detecting cyber-physical attacks in CyberManufacturing systems with machine learning methods. *Journal of Intelligent Manufacturing, 30*(3), 1111-1123.

Wu, Q., Ding, K., & Huang, B. (2018). Approach for fault prognosis using recurrent neural network. *Journal of Intelligent Manufacturing*, 1-13.

Wu, W., Zheng, Y., Chen, K., Wang, X., & Cao, N. (2018, 10-13 April 2018). *A Visual Analytics Approach for Equipment Condition Monitoring in Smart Factories of Process Industry.* Paper presented at the 2018 IEEE Pacific Visualization Symposium.

Wu, X., Tian, S., & Zhang, L. (2019). The Internet of Things enabled shop floor scheduling and process control method based on Petri nets. *IEEE Access, 7*, 27432-27442.

Wu, Y., Wang, S., Chen, L., & Yu, C. (2017, 11-14 Dec. 2017). *Streaming analytics processing in manufacturing performance monitoring and prediction.* Paper presented at the International Conference on Big Data.

Wu, Y., Yuan, M., Dong, S., Lin, L., & Liu, Y. (2018). Remaining useful life estimation of engineered systems using vanilla LSTM neural networks. *Neurocomputing, 275*, 167-179.

Wuest, T., Irgens, C., & Thoben, K.-D. (2014). An approach to monitoring quality in manufacturing using supervised machine learning on product state data. *Journal of Intelligent Manufacturing, 25*(5), 1167-1180. doi:10.1007/s10845-013-0761-y

Wuest, T., Weimer, D., Irgens, C., & Thoben, K. D. (2016). Machine learning in manufacturing: advantages, challenges, and applications. *Production and Manufacturing Research, 4*(1), 23-45.

Xanthopoulos, A. S., Kiatipis, A., Koulouriotis, D. E., & Stieger, S. (2018). Reinforcement Learning-Based and Parametric Production-Maintenance Control Policies for a Deteriorating Manufacturing System. *IEEE Access, 6*, 576-588. doi:10.1109/ACCESS.2017.2771827

Xia, T., Xi, L., Zhou, X., & Lee, J. (2013). Condition-based maintenance for intelligent monitored series system with independent machine failure modes. *International Journal of Production Research, 51*(15), 4585-4596.

Xiao, L., Chen, X., Zhang, X., & Liu, M. (2017). A novel approach for bearing remaining useful life estimation under neither failure nor suspension histories condition. *Journal of Intelligent Manufacturing, 28*(8), 1893-1914.

Xie, H., Tong, X., Meng, W., Liang, D., Wang, Z., & Shi, W. (2015). A multilevel stratified spatial sampling approach for the quality assessment of remote-sensing-derived products. *IEEE Journal of Selected Topics in Applied Earth Observations and Remote Sensing, 8*(10), 4699-4713.

Xiong, J., Zhang, Q., Sun, G., Zhu, X., Liu, M., & Li, Z. (2016). An information fusion fault diagnosis method based on dimensionless indicators with static discounting factor and KNN. *IEEE Sensors Journal, 16*(7), 2060-2069.

Xu, C., & Zhu, G. (2020). Intelligent manufacturing lie group machine learning: Real-time and efficient inspection system based on fog computing. *Journal of Intelligent Manufacturing*, 1-13.

Xu, H., Liu, X., Yu, W., Griffith, D., & Golmie, N. (2020). Reinforcement learning-based control and networking co-design for industrial internet of things. *IEEE Journal on Selected Areas in Communications, 38*(5), 885-898.

Xu, P., Mei, H., Ren, L., & Chen, W. (2017). ViDX: Visual Diagnostics of Assembly Line Performance in Smart Factories. *IEEE Transactions on Visualization and Computer Graphics, 23*(1), 291-300. doi:10.1109/TVCG.2016.2598664

Xu, R., & Wunsch, D. (2005). Survey of clustering algorithms. *IEEE Transactions on neural networks, 16*(3), 645-678.

Xu, S. S.-D., Huang, H.-C., Kung, Y.-C., & Lin, S.-K. (2019). Collision-free fuzzy formation control of swarm robotic cyber-physical systems using a robust orthogonal firefly algorithm. *IEEE Access, 7*, 9205-9214.

Xu, X., Tao, Z., Ming, W., An, Q., & Chen, M. (2020). Intelligent monitoring and diagnostics using a novel integrated model based on deep learning and multi-sensor feature fusion. *Measurement, 165*, 108086.

Xu, X., Zhong, M., Wan, J., Yi, M., & Gao, T. (2016). Health monitoring and management for manufacturing workers in adverse working conditions. *Journal of medical systems, 40*(10), 222.

Xu, X. Y., & Hua, Q. S. (2017). Industrial Big Data Analysis in Smart Factory: Current Status and Research Strategies. *IEEE Access, 5*, 17543-17551.

Xu, Y., Sun, Y., Liu, X., & Zheng, Y. (2019). A digital-twin-assisted fault diagnosis using deep transfer learning. *IEEE Access, 7*, 19990-19999.

Xu, Y., Sun, Y., Wan, J., Liu, X., & Song, Z. (2017). Industrial big data for fault diagnosis: Taxonomy, review, and applications. *IEEE Access, 5*, 17368-17380.

Xun, P., Zhu, P. D., Zhang, Z. Y., Cui, P. S., & Xiong, Y. Q. (2018). Detectors on Edge Nodes against False Data Injection on Transmission Lines of Smart Grid. *Electronics, 7*(6), 89.

Yacob, F., Semere, D., & Nordgren, E. (2019). Anomaly detection in Skin Model Shapes using machine learning classifiers. *The International Journal of Advanced Manufacturing Technology, 105*(9), 3677-3689.

Yan, H. H., Wan, J. F., Zhang, C. H., Tang, S. L., Hua, Q. S., & Wang, Z. R. (2018). Industrial Big Data Analytics for Prediction of Remaining Useful Life Based on Deep Learning. *IEEE Access, 6*, 17190-17197. doi:10.1109/access.2018.2809681

Yan, J., Meng, Y., Lu, L., & Guo, C. (2017, 9-12 July 2017). *Big-data-driven based intelligent prognostics scheme in industry 4.0 environment.* Paper presented at the 2017 Prognostics and System Health Management Conference (PHM-Harbin).

Yan, J., Meng, Y., Lu, L., & Li, L. (2017). Industrial Big Data in an Industry 4.0 Environment: Challenges, Schemes, and Applications for Predictive Maintenance. *IEEE Access, 5*, 23484-23491. doi:10.1109/ACCESS.2017.2765544

Yan, T., Lei, Y., & Li, N. (2018). *Remaining Useful Life Prediction of Machinery Subjected to Two-Phase Degradation Process.* Paper presented at the International Conference on Prognostics and Health Management.

Yan, X., Xu, Y., Xing, X., Cui, B., Guo, Z., & Guo, T. (2020). Trustworthy network anomaly detection based on an adaptive learning rate and momentum in iiot. *IEEE Transactions on Industrial Informatics, 16*(9), 6182-6192.

Yang, B., Cao, X., Li, X., Zhang, Q., & Qian, L. (2019). Mobile-Edge-Computing-Based hierarchical machine learning tasks distribution for IIoT. *IEEE Internet of Things Journal, 7*(3), 2169-2180.

Yang, H., Alphones, A., Zhong, W.-D., Chen, C., & Xie, X. (2019). Learning-based energy-efficient resource management by heterogeneous RF/VLC for ultra-reliable low-latency industrial IoT networks. *IEEE Transactions on Industrial Informatics, 16*(8), 5565-5576.

Yang, J., Li, S., Wang, Z., & Yang, G. (2019). Real-time tiny part defect detection system in manufacturing using deep learning. *IEEE Access, 7*, 89278-89291.

Yang, J., Zhou, C., Yang, S., Xu, H., & Hu, B. (2018). Anomaly detection based on zone partition for security protection of industrial cyber-physical systems. *IEEE Transactions on Industrial Electronics, 65*(5), 4257-4267.

Yang, W.-A., Zhou, W., Liao, W., & Guo, Y. (2016). Prediction of drill flank wear using ensemble of co-evolutionary particle swarm optimization based-selective neural network ensembles. *Journal of Intelligent Manufacturing, 27*(2), 343-361.

Yang, Z., Eddy, D., Krishnamurty, S., Grosse, I., Denno, P., Witherell, P. W., & Lopez, F. (2018). Dynamic Metamodeling for Predictive Analytics in Advanced Manufacturing. *Smart and Sustainable Manufacturing Systems, 2*(1), 18-39. doi:10.1520/ssms20170013

Yao, F., Keller, A., Ahmad, M., Ahmad, B., Harrison, R., & Colombo, A. W. (2018, 18-20 July 2018). *Optimizing the Scheduling of Autonomous Guided Vehicle in a Manufacturing Process.* Paper presented at the 6th International Conference on Industrial Informatics.

Yao, H., Gao, P., Zhang, P., Wang, J., Jiang, C., & Lu, L. (2019). Hybrid intrusion detection system for edge-based IIoT relying on machine-learning-aided detection. *IEEE Network, 33*(5), 75-81.

Yao, Y., Wang, J., Long, P., Xie, M., & Wang, J. (2020). Small-batch-size convolutional neural network based fault diagnosis system for nuclear energy production safety with big-data environment. *International Journal of Energy Research, 44*(7), 5841-5855.

Ye, R., Pan, C.-S., Chang, M., & Yu, Q. (2018). Intelligent defect classification system based on deep learning. *Advances in Mechanical Engineering, 10*(3), 1-7.

Yin, G., Zhang, Y.-T., Li, Z.-N., Ren, G.-Q., & Fan, H.-B. (2014). Online fault diagnosis method based on incremental support vector data description and extreme learning machine with incremental output structure. *Neurocomputing, 128*, 224-231.

Yin, X., He, Z., Niu, Z., & Li, Z. (2018). A hybrid intelligent optimization approach to improving quality for serial multistage and multi-response coal preparation production systems. *Journal of Manufacturing Systems, 47*, 199-216. doi:https://doi.org/10.1016/j.jmsy.2018.05.006

Yoo, Y., & Baek, J.-G. (2018). A Novel Image Feature for the Remaining Useful Lifetime Prediction of Bearings Based on Continuous Wavelet Transform and Convolutional Neural Network. *Applied Sciences, 8*(7), 1102.

Yu, H., Khan, F., & Garaniya, V. (2015). Nonlinear Gaussian Belief Network based fault diagnosis for industrial processes. *Journal of Process Control, 35*, 178-200.

Yu, H., Li, H.-r., Tian, Z.-k., & Wang, Y.-K. (2018). Rolling Bearing Fault Trend Prediction Based on Composite Weighted KELM. *International Journal of Acoustics and Vibration, 23*(2), 217-226.

Yu, J.-B., Yu, Y., Wang, L.-N., Yuan, Z., & Ji, X. (2016). The knowledge modeling system of ready-mixed concrete enterprise and artificial intelligence with ANN-GA for manufacturing production. *Journal of Intelligent Manufacturing, 27*(4), 905-914. doi:10.1007/s10845-014-0923-6

Yu, T., Huang, J., & Chang, Q. (2020). Mastering the working sequence in human-robot collaborative assembly based on reinforcement learning. *IEEE Access, 8*, 163868-163877.

Yu, W., Dillon, T., Mostafa, F., Rahayu, W., & Liu, Y. (2019). A global manufacturing big data ecosystem for fault detection in predictive maintenance. *IEEE Transactions on Industrial Informatics, 16*(1), 183-192.

Yuan, Z., Zhang, L., & Duan, L. (2018). A novel fusion diagnosis method for rotor system fault based on deep learning and multi-sourced heterogeneous monitoring data. *Measurement Science and Technology, 29*(11), 115005.

Yun, J. P., Shin, W. C., Koo, G., Kim, M. S., Lee, C., & Lee, S. J. (2020). Automated defect inspection system for metal surfaces based on deep learning and data augmentation. *Journal of Manufacturing Systems, 55*, 317-324.

Yunusa-Kaltungo, A., & Sinha, J. K. (2017). Effective vibration-based condition monitoring (eVCM) of rotating machines. *Journal of Quality in Maintenance Engineering, 23*(3), 279-296.

Yuwono, M., Qin, Y., Zhou, J., Guo, Y., Celler, B. G., & Su, S. W. (2016). Automatic bearing fault diagnosis using particle swarm clustering and Hidden Markov Model. *Engineering Applications of Artificial Intelligence, 47*, 88-100.

Zarandi, M. H. F., Asl, A. A. S., Sotudian, S., & Castillo, O. (2018a). A state of the art review of intelligent scheduling. *Artificial Intelligence Review, 53*(1), 501-593.

Zarandi, M. H. F., Asl, A. A. S., Sotudian, S., & Castillo, O. (2018b). A state of the art review of intelligent scheduling. *Artificial Intelligence Review*, https://doi.org/10.1007/s10462-10018-19667-10466.

Zarei, J., Tajeddini, M. A., & Karimi, H. R. (2014). Vibration analysis for bearing fault detection and classification using an intelligent filter. *Mechatronics, 24*(2), 151-157.

Zenisek, J., Holzinger, F., & Affenzeller, M. (2019). Machine learning based concept drift detection for predictive maintenance. *Computers & Industrial Engineering, 137*, 106031.

Zhang, A., Li, S., Cui, Y., Yang, W., Dong, R., & Hu, J. (2019). Limited data rolling bearing fault diagnosis with few-shot learning. *IEEE Access, 7*, 110895-110904.

Zhang, A., Wang, H., Li, S., Cui, Y., Liu, Z., Yang, G., & Hu, J. (2018). Transfer Learning with Deep Recurrent Neural Networks for Remaining Useful Life Estimation. *Applied Sciences, 8*(12), 2416.

Zhang, B., Katinas, C., & Shin, Y. C. (2018). Robust Tool Wear Monitoring Using Systematic Feature Selection in Turning Processes With Consideration of Uncertainties. *Journal of Manufacturing Science and Engineering, 140*(8), 081010.

Zhang, B., & Shin, Y. C. (2018). A multimodal intelligent monitoring system for turning processes. *Journal of Manufacturing Processes, 35*, 547-558.

Zhang, C., Lim, P., Qin, A., & Tan, K. C. (2017). Multiobjective deep belief networks ensemble for remaining useful life estimation in prognostics. *IEEE transactions on neural networks and learning systems, 28*(10), 2306-2318.

Zhang, C., Sun, J. H., & Tan, K. C. (2015). *Deep belief networks ensemble with multi-objective optimization for failure diagnosis.* Paper presented at the IEEE International Conference on Systems, Man, and Cybernetics

Zhang, C., Yan, H., Lee, S., & Shi, J. (2018). Multiple profiles sensor-based monitoring and anomaly detection. *Journal of Quality Technology, 50*(4), 344-362. doi:10.1080/00224065.2018.1508275

Zhang, C., & Zhang, H. (2016). Modelling and prediction of tool wear using LS-SVM in milling operation. *International Journal of Computer Integrated Manufacturing, 29*(1), 76-91.

Zhang, J., Ahmad, B., Vera, D., & Harrison, R. (2018, 18-20 July 2018). *Automatic Data Representation Analysis for Reconfigurable Systems Integration.* Paper presented at the 16th International Conference on Industrial Informatics.

Zhang, J., Wang, P., Yan, R., & Gao, R. X. (2018). Deep Learning for Improved System Remaining Life Prediction. *Procedia CIRP, 72*, 1033-1038.

Zhang, J., Wang, P., Yan, R., & Gao, R. X. (2018). Long short-term memory for machine remaining life prediction. *Journal of Manufacturing Systems, 48*, 78-86. doi:https://doi.org/10.1016/j.jmsy.2018.05.011

Zhang, L., Gao, H., Dong, D., Fu, G., & Liu, Q. (2018). Wear Calculation-Based Degradation Analysis and Modeling for Remaining Useful Life Prediction of Ball Screw. *Mathematical Problems in Engineering, 2018*, Article ID 2969854.

Zhang, M., Chen, J., He, S., Yang, L., Gong, X., & Zhang, J. (2019). Privacy-preserving database assisted spectrum access for industrial Internet of Things: A distributed learning approach. *IEEE Transactions on Industrial Electronics, 67*(8), 7094-7103.

Zhang, S., Liu, C., Su, S., Han, Y., & Li, X. (2018). A feature extraction method for predictive maintenance with time-lagged correlation–based curve-registration model. *International Journal of Network Management, 28*(5), e2025.

Zhang, S., Zhang, Y., & Zhu, J. (2018). Residual life prediction based on dynamic weighted Markov model and particle filtering. *Journal of Intelligent Manufacturing, 29*(4), 753-761.

Zhang, W., Yang, D., & Wang, H. (2019). Data-driven methods for predictive maintenance of industrial equipment: a survey. *IEEE Systems Journal, 13*(3), 2213-2227.

Zhang, X., Chen, X., Liu, J. K., & Xiang, Y. (2019). DeepPAR and DeepDPA: privacy preserving and asynchronous deep learning for industrial IoT. *IEEE Transactions on Industrial Informatics, 16*(3), 2081-2090.

Zhang, X., Jiang, D., Han, T., Wang, N., Yang, W., & Yang, Y. (2017). Rotating Machinery Fault Diagnosis for Imbalanced Data Based on Fast Clustering Algorithm and Support Vector Machine. *Journal of Sensors, 2017*.

Zhang, X., Kano, M., & Li, Y. (2018). Principal Polynomial Analysis for Fault Detection and Diagnosis of Industrial Processes. *IEEE Access, 6*, 52298-52307.

Zhang, X., Wang, B., & Chen, X. (2015). Intelligent fault diagnosis of roller bearings with multivariable ensemble-based incremental support vector machine. *Knowledge-Based Systems, 89*, 56-85.

Zhang, Y., Beudaert, X., Argandoña, J., Ratchev, S., & Munoa, J. (2020). A CPPS based on GBDT for predicting failure events in milling. *The International Journal of Advanced Manufacturing Technology, 111*(1), 341-357.

Zhang, Y., Ma, S., Yang, H., Lv, J., & Liu, Y. (2018). A big data driven analytical framework for energy-intensive manufacturing industries. *Journal of Cleaner Production, 197*(1), 56-72.

Zhang, Y., Ren, S., Liu, Y., Sakao, T., & Huisingh, D. (2017). A framework for Big Data driven product lifecycle management. *Journal of Cleaner Production, 159*, 229-240.

Zhang, Y., Ren, S., Liu, Y., & Si, S. (2017). A big data analytics architecture for cleaner manufacturing and maintenance processes of complex products. *Journal of Cleaner Production, 142*, 626-641.

Zhang, Y., Soon, H. G., Ye, D., Fuh, J. Y. H., & Zhu, K. (2019). Powder-bed fusion process monitoring by machine vision with hybrid convolutional neural networks. *IEEE Transactions on Industrial Informatics, 16*(9), 5769-5779.

Zhang, Y. F., Wang, W. B., Du, W., Qian, C., & Yang, H. D. (2018). Coloured Petri net-based active sensing system of real-time and multi-source manufacturing information for smart factory. *International Journal of Advanced Manufacturing Technology, 94*(9-12), 3427-3439. doi:10.1007/s00170-017-0800-5

Zhang, Z., Qin, Y., Jia, L., & Chen, X. a. (2018). Visibility Graph Feature Model of Vibration Signals: A Novel Bearing Fault Diagnosis Approach. *Materials, 11*(11), 2262.

Zhang, Z., Yang, Z., Ren, W., & Wen, G. (2019). Random forest-based real-time defect detection of Al alloy in robotic arc welding using optical spectrum. *Journal of Manufacturing Processes, 42*, 51-59.

Zhang, Z. J., & Zhang, P. Z. (2015). Seeing around the corner: an analytic approach for predictive maintenance using sensor data. *Journal of Management Analytics, 2*(4), 333-350. doi:10.1080/23270012.2015.1086704

Zhao, G., Liu, X., Zhang, B., Zhang, G., Niu, G., & Hu, C. (2017). *Bearing Health Condition Prediction Using Deep Belief Network.* Paper presented at the Proceedings of the Annual Conference of Prognostics and Health Management Society, .

Zhao, G., Zhang, G., Ge, Q., & Liu, X. (2016). *Research Advances in Fault Diagnosis and Prognostic based on Deep Learning.* Paper presented at the Proceedings of 2016 Prognostics and System Health Management Conference, Chengdu, Sichuan, China.

Zhao, L., & Wang, X. (2018). A Deep Feature Optimization Fusion Method for Extracting Bearing Degradation Features. *IEEE Access, 6*, 19640-19653.

Zhao, L., Yan, F., Wang, L., & Yao, Y. (2018, 15-18 July 2018). *Research On Intelligent Evaluation Method For Machining State Oriented To Process Quality Control.* Paper presented at the 2018 International Conference on Machine Learning and Cybernetics.

Zhao, P., Kurihara, M., Tanaka, J., Noda, T., Chikuma, S., & Suzuki, T. (2017). *Advanced correlation-based anomaly detection method for predictive maintenance.* Paper presented at the International Conference on Prognostics and Health Management.

Zhao, R., Wang, D., Yan, R., Mao, K., Shen, F., & Wang, J. (2018). Machine health monitoring using local feature-based gated recurrent unit networks. *IEEE Transactions on Industrial Electronics, 65*(2), 1539-1548.

Zhao, R., Yan, R., Wang, J., & Mao, K. (2017). Learning to monitor machine health with convolutional bi-directional lstm networks. *Sensors, 17*(2), 273.

Zhao, Y., Wang, L., Li, S., Zhou, F., Lin, X., Lu, Q., & Ren, L. (2019). A visual analysis approach for understanding durability test data of automotive products. *ACM Transactions on Intelligent Systems and Technology (TIST), 10*(6), 1-23.

Zhao, Y., Yang, L. T., & Sun, J. (2018). Privacy-preserving tensor-based multiple clusterings on cloud for industrial IoT. *IEEE Transactions on Industrial Informatics, 15*(4), 2372-2381.

Zheng, C., Dai, M., Zhang, Z., Hu, Y., & Guo, Y. (2017). *Real-time remote data acquisition and process monitoring system for automatic filling line*. Paper presented at the 24th International Conference on Mechatronics and Machine Vision in Practice

Zheng, H., Feng, Y. X., Gao, Y. C., & Tan, J. R. (2018). A Robust Predicted Performance Analysis Approach for Data-Driven Product Development in the Industrial Internet of Things. *Sensors, 18*(9), 2871. doi:10.3390/s18092871

Zheng, L., Zeng, C., Li, L., Jiang, Y., Xue, W., Li, J., . . . Wang, P. (2014). *Applying data mining techniques to address critical process optimization needs in advanced manufacturing*. Paper presented at the Proceedings of the 20th ACM SIGKDD international conference on Knowledge discovery and data mining, New York, New York, USA.

Zheng, M., & Wu, K. (2017). Smart spare parts management systems in semiconductor manufacturing. *Industrial Management & Data Systems, 117*(4), 754-763. doi:10.1108/IMDS-06-2016-0242

Zheng, X. C., Wang, M. Q., & Ordieres-Mere, J. (2018). Comparison of Data Preprocessing Approaches for Applying Deep Learning to Human Activity Recognition in the Context of Industry 4.0. *Sensors, 18*(7), E2146.

Zhixiang, L., & Jie, L. (2015, 15-17 Aug. 2015). *Reliability evaluation of intelligent manufacturing equipment.* Paper presented at the 11th International Conference on Natural Computation.

Zhong, R., Huang, G., Dai, Q., & Zhang, T. (2014). Mining SOTs and dispatching rules from RFID-enabled real-time shopfloor production data. *Journal of Intelligent Manufacturing, 25*(4), 825-843. doi:10.1007/s10845-012-0721-y

Zhong, R. Y., Dai, Q., Qu, T., Hu, G., & Huang, G. Q. (2013). RFID-enabled real-time manufacturing execution system for mass-customization production. *Robotics and Computer-Integrated Manufacturing, 29*(2), 283-292.

Zhong, R. Y., Huang, G. Q., Lan, S., Dai, Q. Y., Chen, X., & Zhang, T. (2015). A big data approach for logistics trajectory discovery from RFID-enabled production data. *International Journal of Production Economics, 165*, 260-272. doi:https://doi.org/10.1016/j.ijpe.2015.02.014

Zhong, R. Y., Lan, S. L., Xu, C., Dai, Q. Y., & Huang, G. Q. (2016). Visualization of RFID-enabled shopfloor logistics Big Data in Cloud Manufacturing. *International Journal of Advanced Manufacturing Technology, 84*(1-4), 5-16. doi:10.1007/s00170-015-7702-1

Zhong, R. Y., Li, Z., Pang, L., Pan, Y., Qu, T., & Huang, G. Q. (2013). RFID-enabled real-time advanced planning and scheduling shell for production decision making. *International Journal of Computer Integrated Manufacturing, 26*(7), 649-662.

Zhong, R. Y., Wang, L., & Xu, X. (2017). An IoT-enabled Real-time Machine Status Monitoring Approach for Cloud Manufacturing. *Procedia CIRP, 63*, 709-714.

Zhong, R. Y., Xu, X., Klotz, E., & Newman, S. T. (2017). Intelligent Manufacturing in the Context of Industry 4.0: A Review. *Engineering, 3*(5), 616-630.

Zhou, B., & Cheng, Y. (2016). Fault diagnosis for rolling bearing under variable conditions based on image recognition. *Shock and Vibration, 2016*, Article ID 1948029.

Zhou, K., Liu, T., & Liang, L. (2016). From cyber-physical systems to Industry 4.0: make future manufacturing become possible. *International Journal of Manufacturing Research, 11*(2), 167-188.

Zhou, Q., Yan, P., Liu, H., Xin, Y., & Chen, Y. (2018). Research on a configurable method for fault diagnosis knowledge of machine tools and its application. *The International Journal of Advanced Manufacturing Technology, 95*(1-4), 937-960.

Zhou, X., Huang, K., Xi, L., & Lee, J. (2015). Preventive maintenance modeling for multi-component systems with considering stochastic failures and disassembly sequence. *Reliability Engineering & System Safety, 142*, 231-237.

Zhou, X., Zhang, Y., Mao, T., & Zhou, H. (2017). Monitoring and dynamic control of quality stability for injection molding process. *Journal of Materials Processing Technology, 249*, 358-366.

Zhou, Y., & Xue, W. (2018). Review of tool condition monitoring methods in milling processes. *The International Journal of Advanced Manufacturing Technology, 96*, 2509–2523.

Zhou, Z.-H., Chawla, N. V., & Jin, Y. (2014). Big Data Opportunities and Challenges: Discussions from Data Analytics Perspectives. *IEEE Computational intelligence Magazine, 9*(4), 62-72.

Zhu, J., Chen, N., & Peng, W. (2018). Estimation of Bearing Remaining Useful Life based on Multiscale Convolutional Neural Network. *IEEE Transactions on Industrial Electronics, 66*(4), 3208-3216.

Zhu, K., Li, G., & Zhang, Y. (2019). Big data oriented smart tool condition monitoring system. *IEEE Transactions on Industrial Informatics, 16*(6), 4007-4016.

Zhu, K., & Lin, X. (2019). Tool condition monitoring with multiscale discriminant sparse decomposition. *IEEE Transactions on Industrial Informatics, 15*(5), 2819-2827.

Zhu, K., & Liu, T. (2018). Online Tool Wear Monitoring Via Hidden Semi-Markov Model With Dependent Durations. *IEEE Transactions on Industrial Informatics, 14*(1), 69-78.

Zhu, W., Ma, Y., Benton, M., Romagnoli, J., & Zhan, Y. (2019). Deep learning for pyrolysis reactor monitoring: From thermal imaging toward smart monitoring system. *AIChE Journal, 65*(2), 582-591.

Zhu, X., Xiong, J., & Liang, Q. (2018). Fault Diagnosis of Rotation Machinery Based on Support Vector Machine Optimized by Quantum Genetic Algorithm. *IEEE Access, 6*, 33583-33588.

Zhu, X. C., Qiao, F., & Cao, Q. S. (2017). Industrial big data-based scheduling modeling framework for complex manufacturing system. *Advances in Mechanical Engineering, 9*(8), 1–12.

Zhuang, C. B., Liu, J. H., & Xiong, H. (2018). Digital twin-based smart production management and control framework for the complex product assembly shop-floor. *International Journal of Advanced Manufacturing Technology, 96*(1-4), 1149-1163. doi:10.1007/s00170-018-1617-6

Ziani, R., Felkaoui, A., & Zegadi, R. (2017). Bearing fault diagnosis using multiclass support vector machines with binary particle swarm optimization and regularized Fisher's criterion. *Journal of Intelligent Manufacturing, 28*(2), 405-417.

Židek, K., Hosovsky, A., Piteľ, J., & Bednár, S. (2019). Recognition of Assembly Parts by Convolutional Neural Networks. In S. Hloch, D. Klichová, G. M. Krolczyk, S. Chattopadhyaya, & L. Ruppenthalová (Eds.), *Advances in Manufacturing Engineering and Materials* (pp. 281-289). Basel: Springer International Publishing.

Zolanvari, M., Teixeira, M. A., Gupta, L., Khan, K. M., & Jain, R. (2019). Machine learning-based network vulnerability analysis of industrial Internet of Things. *IEEE Internet of Things Journal, 6*(4), 6822-6834.

Zonta, T., da Costa, C. A., da Rosa Righi, R., de Lima, M. J., da Trindade, E. S., & Li, G. P. (2020). Predictive maintenance in the Industry 4.0: A systematic literature review. *Computers & Industrial Engineering*, 106889.

Zou, W., Xia, Y., & Li, H. (2018). Fault Diagnosis of Tennessee-Eastman Process Using Orthogonal Incremental Extreme Learning Machine Based on Driving Amount. *IEEE Transactions on Cybernetics, 48*(12), 3403 - 3410.

Zschech, P. (2018). *A Taxonomy of Recurring Data Analysis Problems in Maintenance Analytics.* Paper presented at the Proceedings of the 26th European Conference on Information Systems (ECIS), Portsmouth, UK.

Zuo, Y., Tao, F., & Nee, A. Y. (2018). An Internet of things and cloud-based approach for energy consumption evaluation and analysis for a product. *International Journal of Computer Integrated Manufacturing, 31*(4-5), 337-348.

Zurita, D., Delgado, M., Carino, J. A., Ortega, J. A., & Clerc, G. (2016). Industrial Time Series Modelling by Means of the Neo-Fuzzy Neuron. *IEEE Access, 4*, 6151-6160.

## About the Authors

**Jonas Wanner** is a PhD student in Business Informatics at the University of Würzburg, Germany. He has a master's degree in Business Management from the same university and previously worked in the IT environment at Allianz Deutschland AG as well as the Daimler AG. His current research focus is on explainable artificial intelligence (XAI) in smart manufacturing. Further work has been published in various conference proceedings such as the International Conference on Information System (ICIS) or the European Conference on Information Systems (ECIS) as well as in journals such as Business Research or the International Journal of Conceptual Modeling (EMISAJ).

**Christopher Wissuchek** is an external PhD student in Business Information Systems at Friedrich-Alexander-Universität Erlangen-Nürnberg (FAU). He has a master's degree in Business Information Systems from the University of Würzburg. He currently works as a Customer Success Account Manager at Microsoft Germany supporting strategic customers in the financial services industry. Christopher's research focuses on business analytics, machine learning, artificial intelligence, and prescriptive analytics systems. Further research has been published in the International Journal of Conceptual Modeling (EMISAJ) and the International Conference on Business Informatics.

**Giacomo Welsch** is software engineer at a medium-sized German company and PhD student in Information Systems at the University of Würzburg, Germany. He holds a master's degree in Business Information Systems from the same university. His dissertation in the area of prediction-oriented machine learning and explainable AI is currently under review. His research has been published in the proceedings of the European Conference on Information Systems and the International Conference on Wirtschaftsinformatik.

**Christian Janiesch** is Professor for Enterprise Computing at the TU Dortmund University. His research focuses on intelligent systems at the intersection of business process management and artificial intelligence with frequent applications in the Industrial Internet of Things. He is on the BPM Department Editorial Board for BISE journal, and he has authored over 150 scholarly publications. His has appeared in journals such as the Journal of the Association for Information Systems, Communications of the Association for Information Systems, Information & Management, Business & Information Systems Engineering, Information Systems, Decision Support Systems, Future Generation Computer Systems as well as in various major international conferences including ICIS, ECIS, BPM, and HICSS and has been registered as U.S. patents.

## Appendix A – Bibliography (n=904)

**Table A.1: Overview of Survey Publications (*n*=39)**

| References | Iteration |
|---|---|
| (Bang et al., 2019; Baum et al., 2018; Bordeleau et al., 2018; Bousdekis, Magoutas, Mentzas, & (2018), 2015; Çaliş & Bulkan, 2015; Cardin et al., 2017; Cerrada et al., 2018; Y. Cheng, K. Chen, et al., 2018; Diez-Olivan et al., 2019; Fay & Kazantsev, 2018; Gölzer et al., 2015; Gölzer & Fritzsche, 2017; Khan & Yairi, 2018; D.-H. Kim et al., 2018; S. L. Kumar, 2017; Y.-H. Kuo & Kusiak, 2018a; G. Y. Lee et al., 2018; J. Lee, Wu, et al., 2014; Yaguo Lei et al., 2018; O'Donovan et al., 2015a; Precup et al., 2015; Priore et al., 2014; M. S. Reis & Gins, 2017; Sharp et al., 2018; Sutharssan et al., 2015; T. Wuest et al., 2016; X. Y. Xu & Hua, 2017; Y. Xu et al., 2017; Zarandi, Asl, Sotudian, & Castillo, 2018b; G. Zhao et al., 2016; Y. Zhou & Xue, 2018; Zschech, 2018) | **Initial Literature Survey (*n*=32)** |
| (Cadavid et al., 2020; Dalzochio et al., 2020; H. Ding et al., 2020; Dowdeswell et al., 2020; Nath et al., 2020; W. Zhang et al., 2019; Zonta et al., 2020) | **Second Literature Survey (*n*=7)** |

**Table A.2 Coded Publications from Initial Literature Survey with Clusters (*n*=633)[3]**

| No. | Reference | Function | Maturity | Objective | Data Source | Integration | Frequency | Method | Cluster |
|---|---|---|---|---|---|---|---|---|---|
| 1 | (Abbasi, Lim, Rosli, Ismail, & Ibrahim, 2018) | Condition Analysis | Predictive | Time; Cost | Machine/ Tool | No Integration | Historical/ Batch | Regression | 6 |
| 2 | (Adly et al., 2014) | Quality Control | Descriptive | Cost; Conformance; Customer Satisfaction | Product | No Integration | Historical/ Batch | Classification; Clustering | 1 |
| 3 | (Afshari & Peng, 2015) | Design Analysis | Predictive | Customer Satisfaction | Customer | End-to-End | Historical/ Batch | Regression | 6 |
| 4 | (Agarwal & Shivpuri, 2015) | Quality Control | Predictive | Conformance; Customer Satisfaction | Process; Product; Reference | Horizontal | Real-time | Custom Development | 1 |
| 5 | (Amarnath, Sugumaran, & Kumar, 2013) | Defect Analysis | Diagnostic | Time; Cost | Machine/ Tool | No Integration | Historical/ Batch | Classification | 5 |

[3] 11 references could not be coded for function and were excluded from clustering as outlined in the paper.

| No. | Reference | Function | Maturity | Objective | Data Source | Integration | Frequency | Method | Cluster |
|---|---|---|---|---|---|---|---|---|---|
| 6 | (Aqlan, Saha, & Ramakrishnan, 2015) | Quality Control | Predictive | Conformance; Customer Satisfaction | Product | No Integration | Historical/ Batch | Classification; Clustering | 1 |
| 7 | (Arabzad, Ghorbani, Razmi, & Shirouyehzad, 2015) | - | Prescriptive | Time; Cost | Process; ERP | No Integration | Historical/ Batch | Fuzzy Logic | - |
| 8 | (Arık & Toksarı, 2018) | Production Planning | Prescriptive | Time; Cost | Process; ERP | No Integration | Historical/ Batch | Mathematical Optimization | 2 |
| 9 | (Aydın et al., 2015) | Condition Analysis | Diagnostic | Time; Cost | Machine/ Tool | No Integration | Real-time | Classification; Fuzzy Logic | 5 |
| 10 | (O. Aydin & Guldamlasioglu, 2017) | Condition Analysis | Predictive | Time; Cost | Machine/ Tool | No Integration | Real-time | Deep Learning | 4 |
| 11 | (Azadeh, Seif, Sheikhalishahi, & Yazdani, 2016) | Condition Analysis | Predictive | Time; Cost | Machine/ Tool | No Integration | Historical/ Batch | Regression | 6 |
| 12 | (Bagheri, Yang, Kao, & Lee, 2015) | Condition Analysis | Predictive | Time; Cost | Machine/ Tool | Vertical | Historical/ Batch | Custom Development | 6 |
| 13 | (Balogun & Mativenga, 2013) | Energy Cons. Analysis | Predictive | Cost; Sustainability | Machine/ Tool; Process | No Integration | Historical/ Batch | Custom Development | 6 |
| 14 | (Bastani, Barazandeh, & Kong, 2018) | Defect Analysis | Diagnostic | Time; Cost | Machine/ Tool | No Integration | Historical/ Batch | Regression | 5 |
| 15 | (Bastania, Rao, & Zhenyu, 2016) | Monitoring | Descriptive | Time; Cost | Machine/ Tool; Process | Horizontal | Real-time | Classification | 3 |
| 16 | (Bauza et al., 2018) | Quality Control | Descriptive | Conformance; Customer Satisfaction | Product | No Integration | Historical/ Batch | Custom Development | 1 |
| 17 | (Benmoussa & Djeziri, 2017) | Condition Analysis | Predictive | Time; Cost | Machine/ Tool | No Integration | Real-time | Clustering | 4 |
| 18 | (Besenhard, Scheibelhofer, François, Joksch, & Kavsek, 2018) | Monitoring | Diagnostic | Time; Cost | Process | Horizontal | Real-time | Custom Development | 3 |
| 19 | (Bevilacqua, Ciarapica, Diamantini, & Potena, 2017) | Energy Cons. Analysis | Descriptive | Cost; Sustainability | Machine/ Tool | Vertical | Real-time | Custom Development | 3 |
| 20 | (Bink & Zschech, 2018) | Condition Analysis | Predictive | Time; Cost | Machine/ Tool | No Integration | Historical/ Batch | Classification; Clustering | 6 |
| 21 | (Borgi, Hidri, Neef, & Naceur, 2017) | Condition Analysis | Predictive | Time; Cost | Machine/ Tool | No Integration | Real-time | Regression | 4 |
| 22 | (Bouazza, Sallez, & Beldjilali, 2017) | Production Planning | Prescriptive | Time; Cost | Machine/ Tool; Process; ERP | Vertical | Historical/ Batch | Reinforcement Learning | 2 |
| 23 | (Bulnes, Usamentiaga, Garcia, & Molleda, 2016) | Quality Control | Descriptive | Conformance; Customer Satisfaction | Product | No Integration | Real-time | Clustering | 1 |
| 24 | (Bumblauskas, Gemmill, Igou, & Anzengruber, 2017) | Maintenance Planning | Prescriptive | Time; Cost | Machine/ Tool | Vertical | Real-time | Custom Development | 2 |
| 25 | (Bustillo, Urbikain, Perez, Pereira, & Lopez de Lacalle, 2018) | Quality Opt. | Prescriptive | Conformance; Customer Satisfaction | Machine/ Tool | No Integration | Historical/ Batch | Classification; Regression | 1 |
| 26 | (Cachada et al., 2018) | Condition Analysis | Predictive | Time; Cost | Machine/ Tool | Vertical | Real-time | Deep Learning | 3 |
| 27 | (B. Cai, Liu, & Xie, 2016) | Monitoring | Diagnostic | Time; Cost | Machine/ Tool; Process | No Integration | Real-time | Probabilistic Methods | 5 |
| 28 | (Canito et al., 2017) | Condition Analysis | Predictive | Time; Cost | Machine/ Tool | No Integration | Real-time | Custom Development | 4 |
| 29 | (Carbery, Woods, & Marshall, 2018) | Quality Control | Predictive | Conformance; Customer Satisfaction | Process; Product | No Integration | Historical/ Batch | Probabilistic Methods | 1 |
| 30 | (Caricato & Grieco, 2017) | Production Planning | Prescriptive | Time; Cost | Machine/ Tool; Process; ERP | Vertical; Horizontal | Historical/ Batch | Mathematical Optimization | 2 |
| 31 | (Chaki, Bathe, Ghosal, & Padmanabham, 2018) | Quality Opt. | Prescriptive | Conformance; Customer Satisfaction | Machine/ Tool; Human | Vertical; Horizontal | Real-time | Mathematical Optimization | 1 |
| 32 | (Chakravorti et al., 2018) | Defect Analysis | Predictive | Time; Cost | Machine/ Tool; Human | Vertical; Horizontal | Real-time | Custom Development | 3 |
| 33 | (Chamnanlor, Sethanan, Gen, & Chien, 2017) | Production Planning | Prescriptive | Time; Cost | Process; ERP | No Integration | Historical/ Batch | Evolutional Algorithm; Swarm Intelligence | 2 |
| 34 | (B. Chen & Chang, 2017) | Defect Analysis | Predictive | Time; Cost | Machine/ Tool | No Integration | Real-time | Classification | 4 |
| 35 | (Chong Chen, Liu, Kumar, & Qin, 2018) | Energy Cons. Analysis | Predictive | Cost; Sustainability | Machine/ Tool | No Integration | Historical/ Batch | Deep Learning | 6 |
| 36 | (Y. Chen, Jin, & Jiri, 2018) | Condition Analysis | Predictive | Time; Cost | Machine/ Tool | No Integration | Real-time | Deep Learning | 4 |
| 37 | (Yiwei Cheng, Zhu, Wu, & Shao, 2018) | Condition Analysis | Predictive | Time; Cost | Machine/ Tool | No Integration | Historical/ Batch | Clustering; Deep Learning | 6 |
| 38 | (Chiang et al., 2015) | Monitoring | Diagnostic | Time; Cost | Process | No Integration | Real-time | Custom Development | 3 |

| No. | Reference | Function | Maturity | Objective | Data Source | Integration | Frequency | Method | Cluster |
|---|---|---|---|---|---|---|---|---|---|
| 39 | (C.-F. Chien, Chang, et al., 2014) | Performance Analysis | Diagnostic | Time; Cost | Machine/ Tool; Process | No Integration | Historical/ Batch | Regression | 6 |
| 40 | (C.-F. Chien, Diaz, et al., 2014) | Monitoring | Descriptive | Time; Cost | Machine/ Tool; Process | No Integration | Real-time | Classification | 3 |
| 41 | (C.-F. Chien et al., 2013a) | Monitoring | Descriptive | Time; Cost | Process | Vertical | Real-time | Classification; Clustering | 3 |
| 42 | (C. Chien et al., 2018) | - | Prescriptive | Cost; Sustainability | Machine/ Tool | No Integration | Historical/ Batch | Regression | - |
| 43 | (S. Cho et al., 2018) | Condition Analysis | Predictive | Time; Cost | Machine/ Tool | Vertical | Real-time | Custom Development | 3 |
| 44 | (J.-H. Choi, Lee, Jung, & Cho, 2017) | Condition Analysis | Descriptive | Time; Cost | Machine/ Tool | No Integration | Real-time | Probabilistic Methods; Clustering | 4 |
| 45 | (Chou & Su, 2017) | Quality Control | Descriptive | Time; Cost; Conformance | Product | No Integration | Historical/ Batch | Classification | 1 |
| 46 | (Codjo, Jaafar, Makich, Knittel, & Nouari, 2018) | Quality Control | Predictive | Conformance; Customer Satisfaction | Machine/ Tool; Product | No Integration | Real-time | Classification | 1 |
| 47 | (Conde et al., 2018) | Quality Opt. | Prescriptive | Conformance; Customer Satisfaction | Machine/ Tool | No Integration | Historical/ Batch | Classification; Mathematical Optimization | 1 |
| 48 | (D'Addona, Ullah, & Matarazzo, 2017) | Condition Analysis | Predictive | Time; Cost | Machine/ Tool | No Integration | Historical/ Batch | Classification | 6 |
| 49 | (da Silva, Gabbar, Junior, & da Costa Junior, 2018) | Defect Analysis | Diagnostic | Time; Cost | Machine/ Tool | No Integration | Real-time | Classification | 4 |
| 50 | (Demetgul, Yildiz, Taskin, Tansel, & Yazicioglu, 2014) | Defect Analysis | Diagnostic | Time; Cost | Machine/ Tool | No Integration | Historical/ Batch | Classification; Dimension Reduction | 5 |
| 51 | (Deutsch, He, & He, 2017) | Condition Analysis | Predictive | Time; Cost | Machine/ Tool | No Integration | Historical/ Batch | Deep Learning | 6 |
| 52 | (K. Ding & Jiang, 2018) | Monitoring | Descriptive | Time; Cost | Machine/ Tool; Process | Vertical | Real-time | Custom Development | 3 |
| 53 | (Domova & Dagnino, 2017) | Security/ Risk Analysis | Diagnostic | Security | Machine/ Tool; Process | Vertical | Real-time | Custom Development | 3 |
| 54 | (Dou & Zhou, 2016) | Defect Analysis | Diagnostic | Time; Cost | Machine/ Tool | No Integration | Real-time | Classification | 4 |
| 55 | (Emec, Krüger, & Seliger, 2016) | Condition Analysis | Descriptive | Time; Cost | Machine/ Tool | No Integration | Real-time | Custom Development | 4 |
| 56 | (Fan, Zhu, Kuo, Lu, & Wu, 2017) | Monitoring | Descriptive | Time; Cost | Machine/ Tool; Process | No Integration | Real-time | Custom Development | 3 |
| 57 | (Fink, Zio, & Weidmann, 2014) | Condition Analysis | Predictive | Time; Cost | Machine/ Tool | No Integration | Historical/ Batch | Classification | 6 |
| 58 | (Fumagalli, Macchi, Colace, Rondi, & Alfieri, 2016) | Condition Analysis | Descriptive | Time; Cost | Machine/ Tool | No Integration | Real-time | Custom Development | 4 |
| 59 | (Gan & Wang, 2016) | Defect Analysis | Diagnostic | Time; Cost | Machine/ Tool | No Integration | Historical/ Batch | Deep Learning | 5 |
| 60 | (Giannetti & Ransing, 2016) | Performance Analysis | Predictive | Conformance; Customer Satisfaction | Process | Vertical | Historical/ Batch | Regression | 3 |
| 61 | (Goryachev et al., 2013) | Production Planning | Prescriptive | Time; Cost | Machine/ Tool; Process; ERP; Human | Vertical | Real-time | Multi-Agent System | 2 |
| 62 | (Granados, Lacroix, & Medjaher, 2018) | Quality Control | Predictive | Conformance | Process | No Integration | Real-time | Custom Development | 1 |
| 63 | (Gröger, Kassner, et al., 2016) | Performance Opt. | Predictive | Time; Cost |  | Vertical; Horizontal | Real-time | Custom Development | 3 |
| 64 | (Gugulothu et al., 2017) | Condition Analysis | Predictive | Time; Cost | Machine/ Tool | No Integration | Historical/ Batch | Regression | 6 |
| 65 | (L. Guo, Lei, Xing, Yan, & Li, 2018) | Defect Analysis | Diagnostic | Time; Cost | Machine/ Tool | No Integration | Historical/ Batch | Deep Learning | 5 |
| 66 | (L. Guo, Li, Jia, Lei, & Lin, 2017) | Defect Analysis | Predictive | Time; Cost | Machine/ Tool | No Integration | Historical/ Batch | Deep Learning | 6 |
| 67 | (Z. Guo, Ngai, Yang, & Liang, 2015) | Production Planning | Prescriptive | Time; Cost | Machine/ Tool; Process; ERP; Human | Vertical; Horizontal | Real-time | Mathematical Optimization | 2 |
| 68 | (D. Han, Zhao, & Shi, 2017) | Defect Analysis | Diagnostic | Time; Cost | Machine/ Tool | No Integration | Historical/ Batch | Classification; Deep Learning | 5 |
| 69 | (Y. He et al., 2017) | Condition Analysis | Predictive | Time; Cost | Machine/ Tool; Process; Product | No Integration | Real-time | Custom Development | 4 |
| 70 | (Y. He et al., 2017) | Quality Control | Diagnostic | Conformance; Customer Satisfaction | Product; Customer | End-to-End | Historical/ Batch | Custom Development | 1 |
| 71 | (Heger, Hildebrandt, & Scholz-Reiter, 2015) | Production Planning | Prescriptive | Time; Cost | Machine/ Tool; Process; ERP | No Integration | Historical/ Batch | Regression | 2 |
| 72 | (Hseush, Huang, Hsu, & Pu, 2013) | Production Planning | Descriptive | Time; Cost | ERP | Vertical; Horizontal | Real-time | Custom Development | 3 |
| 73 | (Chao-Yung Hsu et al., 2016) | Quality Control | Predictive | Conformance; Customer Satisfaction | Product | No Integration | Real-time | Classification | 1 |
| 74 | (J. Hu, Lewis, Gan, Phua, & Aw, 2014) | Monitoring | Descriptive | Time; Cost | Process | No Integration | Real-time | Probabilistic Methods | 3 |
| 75 | (S. Hu et al., 2016) | Monitoring | Diagnostic | Time; Cost | Process | Vertical | Real-time | Classification; Swarm Intelligence | 3 |

| No. | Reference | Function | Maturity | Objective | Data Source | Integration | Frequency | Method | Cluster |
|---|---|---|---|---|---|---|---|---|---|
| 76 | (D. Huang, Lin, Chen, & Sze, 2018) | Quality Control | Descriptive | Conformance; Customer Satisfaction | Machine/ Tool | Vertical | Historical/ Batch | Deep Learning | 1 |
| 77 | (Hur et al., 2015) | Performance Analysis | Predictive | Time; Cost | Process; Human | No Integration | Historical/ Batch | Regression | 6 |
| 78 | (Ivanov, Dolgui, Sokolov, Werner, & Ivanova, 2016) | Production Planning | Prescriptive | Time; Cost | Process; ERP | No Integration | Historical/ Batch | Mathematical Optimization | 2 |
| 79 | (A. K. Jain & Lad, 2017) | Condition Analysis | Descriptive | Time; Cost | Machine/ Tool; Product | No Integration | Real-time | Classification | 4 |
| 80 | (S. Jain, Lechevalier, & Narayanan, 2017) | Condition Analysis | Predictive | Time; Cost | Machine/ Tool | No Integration | Historical/ Batch | Regression | 6 |
| 81 | (V. Jain, Kundu, Chan, & Patel, 2015) | - | Prescriptive | Time; Cost | ERP | Horizontal | Historical/ Batch | Swarm Intelligence | - |
| 82 | (Javed, Gouriveau, Li, & Zerhouni, 2018) | Performance Analysis | Predictive | Time; Cost | Process; Product | No Integration | Historical/ Batch | Regression | 6 |
| 83 | (Ji-Hyeong & Su-Young, 2016) | Condition Analysis | Predictive | Time; Cost | Machine/ Tool | No Integration | Historical/ Batch | Regression | 6 |
| 84 | (W. Ji & Wang, 2017) | Monitoring | Predictive | Time; Cost | Machine/ Tool; Process | Vertical | Real-time | Classification | 3 |
| 85 | (F. Jia, Lei, Lin, Zhou, & Lu, 2016) | Condition Analysis | Predictive | Time; Cost | Machine/ Tool | No Integration | Historical/ Batch | Deep Learning | 6 |
| 86 | (J. Jiang & Kuo, 2017) | Design Analysis | Diagnostic | Customer Satisfaction | Product; Customer | End-to-End | Historical/ Batch | Classification | 6 |
| 87 | (Y. Jiao, Yang, Zhong, & Zhang, 2017) | Condition Analysis | Predictive | Time; Cost | Machine/ Tool | No Integration | Real-time | Deep Learning | 4 |
| 88 | (Jing, Ma, Hu, Zhu, & Chen, 2018) | Quality Control | Predictive | Conformance; Customer Satisfaction | Product | No Integration | Historical/ Batch | Swarm Intelligence | 1 |
| 89 | (Jung, Tsai, Chiu, Hu, & Sun, 2018) | Quality Control | Diagnostic | Conformance; Customer Satisfaction | Product | No Integration | Historical/ Batch | Deep Learning | 1 |
| 90 | (Kadar, Jardim-Gonçalves, Covaciu, & Bullon, 2017) | Quality Control | Descriptive | Conformance; Customer Satisfaction | Product | No Integration | Real-time | Classification | 1 |
| 91 | (Karabadji, Seridi, Khelf, Azizi, & Boulkroune, 2014) | Defect Analysis | Diagnostic | Time; Cost | Machine/ Tool | No Integration | Historical/ Batch | Classification | 5 |
| 92 | (Kashkoush & ElMaraghy, 2017) | Production Planning | Prescriptive | Time; Cost | Process; Product | Vertical; Horizontal | Historical/ Batch | Mathematical Optimization | 2 |
| 93 | (K. Kaur, Selway, Grossmann, Stumptner, & Johnston, 2018) | Condition Analysis | Predictive | Time; Cost | Machine/ Tool | Vertical | Real-time | Custom Development | 3 |
| 94 | (Khazaee et al., 2017) | Defect Analysis | Diagnostic | Time; Cost | Machine/ Tool | No Integration | Historical/ Batch | Classification | 5 |
| 95 | (Kiangala & Wang, 2018) | Condition Analysis | Predictive | Time; Cost | Machine/ Tool | Vertical | Real-time | Custom Development | 3 |
| 96 | (Kibira & Shao, 2017) | Energy Cons. Opt. | Prescriptive | Cost; Sustainability | Machine/ Tool | No Integration | Real-time | Mathematical Optimization | 6 |
| 97 | (A. Kim, Oh, Jung, & Kim, 2018) | Quality Control | Descriptive | Conformance | Machine/ Tool; Process | No Integration | Real-time | Classification | 1 |
| 98 | (M. S. Kim et al., 2018) | Quality Control | Diagnostic | Conformance; Customer Satisfaction | Product | No Integration | Real-time | Regression | 1 |
| 99 | (Kohlert & Konig, 2016) | Performance Opt. | Descriptive | Time; Cost | Machine/ Tool; Human | Vertical | Real-time | Classification | 4 |
| 100 | (Koulali, Koulali, Tembine, & Kobbane, 2018) | Maintenance Planning | Prescriptive | Time; Cost | Machine/ Tool | Vertical | Historical/ Batch | Custom Development | 2 |
| 101 | (Dominik Kozjek et al., 2017) | Monitoring | Diagnostic | Time; Cost | Machine/ Tool; Process | Vertical | Real-time | Classification | 3 |
| 102 | (Dominik Kozjek, Rihtaršič, & Butala, 2018) | Production Planning | Predictive | Time; Cost | Process; Product | Vertical; Horizontal | Historical/ Batch | Classification | 2 |
| 103 | (D. Kozjek, Vrabic, Kralj, & Butala, 2017) | Monitoring | Diagnostic | Time; Cost | Machine/ Tool; Process | Vertical | Real-time | Classification | 3 |
| 104 | (Krumeich, Werth, & Loos, 2016) | Performance Opt. | Prescriptive | Time; Cost | Process | No Integration | Real-time | Custom Development | 2 |
| 105 | (Ajay Kumar, Shankar, Choudhary, & Thakur, 2016) | Monitoring | Diagnostic | Time; Cost | Machine/ Tool; Process | No Integration | Historical/ Batch | Classification | 5 |
| 106 | (Ajay Kumar, Shankar, & Thakur, 2018) | Condition Analysis | Predictive | Time; Cost | Machine/ Tool | No Integration | Historical/ Batch | Classification | 6 |
| 107 | (Kumaraguru & Morris, 2014) | Performance Analysis | Predictive | Time; Cost | Machine/ Tool; Process | Vertical | Real-time | Custom Development | 3 |
| 108 | (Kumru & Kumru, 2014b) | Performance Analysis | Predictive | Time; Cost |  | No Integration | Historical/ Batch | Classification | 6 |
| 109 | (Lachenmaier, Lasi, & Kemper, 2015) | Production Planning | Descriptive | Time; Cost | ERP | Vertical | Historical/ Batch | Custom Development | 3 |
| 110 | (Lade, Ghosh, & Srinivasan, 2017) | Quality Control | Diagnostic | Conformance; Customer Satisfaction | Product | No Integration | Historical/ Batch | Custom Development | 1 |

| No. | Reference | Function | Maturity | Objective | Data Source | Integration | Frequency | Method | Cluster |
|---|---|---|---|---|---|---|---|---|---|
| 111 | (P.-J. Lai & Wu, 2015) | Production Planning | Prescriptive | Time; Cost | Machine/ Tool | No Integration | Historical/ Batch | Evolutional Algorithm; Swarm Intelligence | 2 |
| 112 | (S. Langarica, Rüffelmacher, & Núñez, 2018) | Condition Analysis | Descriptive | Time; Cost | Machine/ Tool | No Integration | Real-time | Dimension Reduction | 4 |
| 113 | (J. Lee, Jin, & Bagheri, 2017) | Condition Analysis | Predictive | Time; Cost | Machine/ Tool | No Integration | Real-time | Deep Learning | 4 |
| 114 | (J. Y. Lee, Yoon, & Kim, 2017) | Quality Control | Descriptive | Conformance | Process; Product | Vertical | Real-time | Custom Development | 3 |
| 115 | (Legat & Vogel-Heuser, 2017) | Production Planning | Prescriptive | Time; Cost; Flexibility | Machine/ Tool; Process | Vertical | Historical/ Batch | Custom Development | 2 |
| 116 | (Yaguo Lei, Jia, Lin, Xing, & Ding, 2016) | Defect Analysis | Diagnostic | Time; Cost | Machine/ Tool | No Integration | Historical/ Batch | Custom Development | 5 |
| 117 | (Y Lei, Jia, Zhou, & Lin, 2015) | Defect Analysis | Diagnostic | Time; Cost | Machine/ Tool | No Integration | Real-time | Deep Learning | 4 |
| 118 | (Yaguo Lei, Li, et al., 2016) | Condition Analysis | Predictive | Time; Cost | Machine/ Tool | No Integration | Real-time | Custom Development | 4 |
| 119 | (Lesany, Koochakzadeh, & Fatemi Ghomi, 2014) | Monitoring | Descriptive | Conformance | Process | Horizontal | Real-time | Classification | 3 |
| 120 | (C. Li, R.-V. Sanchez, et al., 2016) | Defect Analysis | Diagnostic | Time; Cost | Machine/ Tool | No Integration | Historical/ Batch | Deep Learning | 5 |
| 121 | (C. Li, Y. Tao, et al., 2018) | Energy Cons. Analysis | Predictive | Cost; Sustainability | Machine/ Tool | Vertical | Historical/ Batch | Regression | 6 |
| 122 | (Q. Li & S. Liang, 2018b) | Condition Analysis | Predictive | Time; Cost | Machine/ Tool | No Integration | Real-time | Custom Development | 4 |
| 123 | (T. Li, He, & Zhu, 2016) | Quality Control | Descriptive | Conformance; Customer Satisfaction | Process; Customer | End-to-End | Historical/ Batch | Custom Development | 1 |
| 124 | (Xiang Li, Zhang, & Ding, 2018) | Condition Analysis | Predictive | Time; Cost | Machine/ Tool | No Integration | Historical/ Batch | Deep Learning | 6 |
| 125 | (Liang et al., 2018) | - | Prescriptive | Cost; Sustainability | Machine/ Tool | No Integration | Historical/ Batch | Swarm Intelligence | - |
| 126 | (Lim et al., 2016) | Production Planning | Prescriptive | Time; Cost | Process; ERP | No Integration | Historical/ Batch | Custom Development | 2 |
| 127 | (C. Lin et al., 2017) | Condition Analysis | Predictive | Time; Cost | Machine/ Tool | No Integration | Real-time | Classification | 4 |
| 128 | (Changqing Liu, Li, Zhou, & Shen, 2018) | Condition Analysis | Diagnostic | Time; Cost | Machine/ Tool | No Integration | Real-time | Classification | 5 |
| 129 | (C Liu, Wang, & Li, 2015) | Condition Analysis | Descriptive | Time; Cost | Machine/ Tool | No Integration | Real-time | Classification | 4 |
| 130 | (Y. Zhang, S. Ren, Y. Liu, & S. Si, 2017) | Condition Analysis | Predictive | Time; Cost | Machine/ Tool | No Integration | Historical/ Batch | Classification | 6 |
| 131 | (T.-I. Liu & Jolley, 2015) | Condition Analysis | Descriptive | Time; Cost | Machine/ Tool | No Integration | Real-time | Classification | 4 |
| 132 | Liu, 2013 #3319} | Monitoring | Diagnostic | Time; Cost | Process | Horizontal | Historical/ Batch | Probabilistic Methods | 3 |
| 133 | (Z.-J. Lu, Xiang, Wu, & Gu, 2015) | Monitoring | Descriptive | Time; Cost | Machine/ Tool; Human | Vertical | Real-time | Custom Development | 3 |
| 134 | (Z.-J. Lu et al., 2015) | Quality Control | Predictive | Conformance; Customer Satisfaction | Process; Product | No Integration | Historical/ Batch | Classification; Evolutional Algorithm | 1 |
| 135 | (B. Luo, Wang, Liu, Li, & Peng, 2018) | Defect Analysis | Predictive | Time; Cost | Machine/ Tool | No Integration | Real-time | Deep Learning | 4 |
| 136 | (Y. Lv & Lin, 2017) | Production Planning | Prescriptive | Time; Cost | Machine/ Tool; Process; Customer; ERP | Vertical; Horizontal | Real-time | Fuzzy Logic | 2 |
| 137 | (H. Ma et al., 2017) | Design Analysis | Prescriptive | Customer Satisfaction | Product; Customer | End-to-End | Historical/ Batch | Mathematical Optimization; Fuzzy Logic | 6 |
| 138 | (J. Ma, Kwak, & Kim, 2014) | Design Analysis | Predictive | Customer Satisfaction | Product; Customer | End-to-End | Historical/ Batch | Classification | 6 |
| 139 | (Maggipinto, Terzi, Masiero, Beghi, & Susto, 2018) | Quality Control | Descriptive | Conformance; Customer Satisfaction | Product | No Integration | Historical/ Batch | Deep Learning | 1 |
| 140 | (Manco et al., 2017) | Defect Analysis | Predictive | Time; Cost | Machine/ Tool | No Integration | Historical/ Batch | Classification | 6 |
| 141 | (Mileva Boshkoska, Bohanec, Boškoski, & Juričić, 2015) | Quality Control | Descriptive | Conformance; Customer Satisfaction | Product | No Integration | Historical/ Batch | Regression | 1 |
| 142 | (Milo, Roan, & Harris, 2015) | Monitoring | Descriptive | Time; Cost | Process | No Integration | Real-time | Classification | 3 |
| 143 | (Molka-Danielsen et al., 2018) | Security/ Risk Analysis | Descriptive | Security | Environment | No Integration | Real-time | Custom Development | 3 |
| 144 | (Mourtzis, Vlachou, Milas, & Dimitrakopoulos, 2016) | Energy Cons. Analysis | Descriptive | Cost; Sustainability | Machine/ Tool | Vertical | Real-time | Custom Development | 3 |
| 145 | (Mulrennan et al., 2018) | Quality Control | Predictive | Conformance; Customer Satisfaction | Product | No Integration | Real-time | Custom Development | 1 |

| No. | Reference | Function | Maturity | Objective | Data Source | Integration | Frequency | Method | Cluster |
|---|---|---|---|---|---|---|---|---|---|
| 146 | (Muralidhar et al., 2018) | Monitoring | Descriptive | Time; Cost | Machine/ Tool; Process | No Integration | Real-time | Custom Development | 3 |
| 147 | (Muralidharan & Sugumaran, 2013) | Defect Analysis | Diagnostic | Time; Cost | Machine/ Tool | No Integration | Historical/ Batch | Classification | 5 |
| 148 | (H. N. Nguyen, Kim, & Kim, 2018) | Condition Analysis | Predictive | Time; Cost | Machine/ Tool | No Integration | Real-time | Classification | 4 |
| 149 | (Nie & Wan, 2015) | Condition Analysis | Predictive | Time; Cost | Machine/ Tool | No Integration | Historical/ Batch | Regression; Clustering | 6 |
| 150 | (Ning, Yu, & Huang, 2018) | Defect Analysis | Predictive | Time; Cost | Machine/ Tool | No Integration | Real-time | Custom Development | 3 |
| 151 | (Niu & Li, 2017) | Condition Analysis | Diagnostic | Time; Cost | Machine/ Tool | No Integration | Historical/ Batch | Fuzzy Logic | 5 |
| 152 | (Noyel, Thomas, Thomas, & Charpentier, 2016) | Quality Control | Predictive | Time; Cost; Conformance | Process | No Integration | Historical/ Batch | Classification | 1 |
| 153 | (Shiyong Wang, Wan, Li, & Liu, 2018) | Condition Analysis | Predictive | Time; Cost | Machine/ Tool; Process; Human | Vertical | Real-time | Custom Development | 3 |
| 154 | (Nyanteh, Srivastava, Edrington, & Cartes, 2013) | Defect Analysis | Diagnostic | Time; Cost | Machine/ Tool | No Integration | Real-time | Classification; Swarm Intelligence | 4 |
| 155 | (Oh, Ransikarbum, Busogi, Kwon, & Kim, 2018) | Quality Control | Descriptive | Conformance; Customer Satisfaction | Product | No Integration | Real-time | Classification | 1 |
| 156 | (Onel, Kieslich, Guzman, Floudas, & Pistikopoulos, 2018) | Monitoring | Descriptive | Time; Cost | Process | No Integration | Real-time | Classification | 3 |
| 157 | (Oses et al., 2016) | Energy Cons. Analysis | Predictive | Cost; Sustainability | Machine/ Tool | No Integration | Historical/ Batch | Regression | 6 |
| 158 | (Pandiyan et al., 2018) | Condition Analysis | Predictive | Time; Cost | Machine/ Tool | No Integration | Real-time | Classification; Evolutional Algorithm | 4 |
| 159 | (Jiten Patel & Choi, 2014) | Condition Analysis | Descriptive | Time; Cost | Machine/ Tool | No Integration | Historical/ Batch | Classification | 5 |
| 160 | (W. Peng, Ye, & Chen, 2018) | Condition Analysis | Predictive | Time; Cost | Machine/ Tool | No Integration | Real-time | Custom Development | 4 |
| 161 | (Pimenov, Bustillo, & Mikolajczyk, 2018) | Quality Control | Predictive | Conformance; Customer Satisfaction | Machine/ Tool | No Integration | Real-time | Classification; Regression | 1 |
| 162 | (Shuhui Qu, Wang, & Jasperneite, 2018) | Production Planning | Prescriptive | Time; Cost | Machine/ Tool; Process | Vertical | Real-time | Reinforceme nt Learning | 2 |
| 163 | (Ragab, Ouali, et al., 2016) | Condition Analysis | Predictive | Time; Cost | Machine/ Tool | No Integration | Historical/ Batch | Classification | 6 |
| 164 | (Ragab, Yacout, et al., 2017) | Condition Analysis | Predictive | Time; Cost | Machine/ Tool | No Integration | Historical/ Batch | Classification | 6 |
| 165 | (Ranjit et al., 2015) | Monitoring | Descriptive | Time; Cost | Process; Human | Horizontal | Real-time | Classification | 3 |
| 166 | (Lei Ren, Cui, Sun, & Cheng, 2017) | Condition Analysis | Predictive | Time; Cost | Machine/ Tool | No Integration | Historical/ Batch | Deep Learning | 6 |
| 167 | (Lei Ren et al., 2018) | Condition Analysis | Predictive | Time; Cost | Machine/ Tool | No Integration | Historical/ Batch | Deep Learning | 6 |
| 168 | (Rivera Torres, Anido Rifón, & Serrano Mercado, 2018) | Production Planning | Predictive | Time; Cost | Process | No Integration | Real-time | Probabilistic Methods | 6 |
| 169 | (Rivera Torres, Serrano Mercado, & Anido Rifón, 2018) | Production Planning | Predictive | Time; Cost | Machine/ Tool | Vertical | Historical/ Batch | Probabilistic Methods | 6 |
| 170 | (J. J. Rodríguez, Quintana, Bustillo, & Ciurana, 2017) | Quality Control | Predictive | Conformance | Machine/ Tool | Vertical | Historical/ Batch | Classification | 1 |
| 171 | (Rødseth & Schjølberg, 2016) | Condition Analysis | Predictive | Time; Cost | Machine/ Tool | No Integration | Historical/ Batch | Custom Development | 6 |
| 172 | (Rødseth, Schjølberg, & Marhaug, 2017) | Condition Analysis | Predictive | Time; Cost |  | Vertical | Real-time | Deep Learning | 3 |
| 173 | (Rude, Adams, & Beling, 2018) | Performanc e Analysis | Descriptive | Time; Cost | Machine/ Tool; Process | Vertical | Real-time | Custom Development | 3 |
| 174 | (Saez, Maturana, Barton, & Tilbury, 2018) | Performanc e Analysis | Diagnostic | Time; Cost | Machine/ Tool; Process | Vertical | Real-time | Custom Development | 3 |
| 175 | (Safizadeh & Latifi, 2014) | Defect Analysis | Diagnostic | Time; Cost | Machine/ Tool | No Integration | Historical/ Batch | Classification | 5 |
| 176 | (Saha, Aqlan, Lam, & Boldrin, 2016) | - | Prescriptive | Time; Cost | Customer; ERP | Horizontal | Real-time | Fuzzy Logic | - |
| 177 | (Saldivar, Goh, Chen, et al., 2016) | Design Analysis | Prescriptive | Customer Satisfaction | Product; Customer | End-to-End | Real-time | Dimension Reduction; Mathematical Optimization | 6 |
| 178 | (Santhana Babu, Giridharan, Ramesh Narayanan, & Narayana Murty, 2016) | Quality Control | Predictive | Conformance; Customer Satisfaction | Machine/ Tool | No Integration | Historical/ Batch | Custom Development | 1 |
| 179 | (Saucedo-Espinosa et al., 2017) | Defect Analysis | Descriptive | Time; Cost | Machine/ Tool | No Integration | Historical/ Batch | Classification | 5 |

| No. | Reference | Function | Maturity | Objective | Data Source | Integration | Frequency | Method | Cluster |
|---|---|---|---|---|---|---|---|---|---|
| 180 | (Schuh, Prote, Luckert, & Hünnekes, 2017) | Production Planning | Prescriptive | Time; Cost | Process | No Integration | Historical/ Batch | Classification | 2 |
| 181 | (Shaban, Yacout, Balazinski, et al., 2017) | Condition Analysis | Descriptive | Time; Cost | Machine/ Tool | No Integration | Historical/ Batch | Classification | 5 |
| 182 | (H. Shao et al., 2018) | Defect Analysis | Diagnostic | Time; Cost | Machine/ Tool | No Integration | Historical/ Batch | Deep Learning | 5 |
| 183 | (J.-H. Shin, Kiritsis, & Xirouchakis, 2015) | Design Analysis | Diagnostic | Customer Satisfaction | Product | End-to-End | Real-time | Clustering | 6 |
| 184 | (Shiue et al., 2018) | Production Planning | Prescriptive | Time; Cost | Machine/ Tool; Process | No Integration | Real-time | Reinforceme nt Learning | 2 |
| 185 | (Soualhi, Razik, Clerc, & Doan, 2014) | Defect Analysis | Predictive | Time; Cost | Machine/ Tool | No Integration | Historical/ Batch | Probabilistic Methods | 6 |
| 186 | (Spezzano & Vinci, 2015) | Condition Analysis | Descriptive | Time; Cost | Machine/ Tool | No Integration | Real-time | Clustering | 4 |
| 187 | (Sreenuch, Tsourdos, & Jennions, 2013) | Condition Analysis | Descriptive | Time; Cost | Machine/ Tool | Vertical | Real-time | Classification | 5 |
| 188 | (Stefanovic, 2015) | - | Predictive | Time; Cost; Flexibility | ERP | Vertical | Historical/ Batch | Regression; Clustering | - |
| 189 | (Subramaniyan, Skoogh, Gopalakrishnan, Salomonsson, et al., 2016) | Monitoring | Descriptive | Time; Cost | Process | No Integration | Real-time | Classification | 3 |
| 190 | (Susto, Beghi, & McLoone, 2017) | Monitoring | Diagnostic | Time; Cost | Process | No Integration | Real-time | Classification | 3 |
| 191 | (Muhammad Syafrudin, Alfian, Fitriyani, & Rhee, 2018) | Monitoring | Predictive | Time; Cost | Machine/ Tool | Vertical | Real-time | Classification | 3 |
| 192 | (Somkiat Tangjitsitcharoen & Wongtangthinthan, 2016) | Condition Analysis | Predictive | Time; Cost | Machine/ Tool | No Integration | Real-time | Custom Development | 4 |
| 193 | (Tong, Teng, Sun, & Guan, 2018) | Performanc e Opt. | Prescriptive | Time; Cost | Process | Horizontal | Historical/ Batch | Evolutional Algorithm | 3 |
| 194 | (Tristo et al., 2015) | Energy Cons. Analysis | Descriptive | Cost; Sustainability | Machine/ Tool | No Integration | Real-time | Custom Development | 3 |
| 195 | (Trunzer et al., 2017) | Defect Analysis | Descriptive | Time; Cost | Machine/ Tool | No Integration | Historical/ Batch | Classification | 5 |
| 196 | (Vafeiadis et al., 2018) | Quality Control | Descriptive | Conformance; Customer Satisfaction | Product | No Integration | Historical/ Batch | Classification | 1 |
| 197 | (Villalonga et al., 2018) | Condition Analysis | Descriptive | Time; Cost | Machine/ Tool; Process | Vertical | Real-time | Custom Development | 3 |
| 198 | (Vrabic, Kozjek, & Butala, 2017) | Monitoring | Predictive | Time; Cost | Machine/ Tool; Process | Vertical | Real-time | Classification | 3 |
| 199 | (Vununu, Moon, Lee, & Kwon, 2018) | Defect Analysis | Descriptive | Time; Cost | Machine/ Tool | No Integration | Historical/ Batch | Deep Learning | 5 |
| 200 | (Wan et al., 2017) | Condition Analysis | Predictive | Time; Cost | Machine/ Tool | | Real-time | Classification | 4 |
| 201 | (Guofeng Wang, Guo, & Qian, 2014) | Condition Analysis | Predictive | Time; Cost | Machine/ Tool | No Integration | Historical/ Batch | Regression | 6 |
| 202 | (Guofeng Wang, Guo, & Yang, 2013) | Condition Analysis | Descriptive | Time; Cost | Machine/ Tool | No Integration | Real-time | Classification | 4 |
| 203 | (Guofeng Wang, Liu, Cui, & Feng, 2014) | Condition Analysis | Predictive | Time; Cost | Machine/ Tool | No Integration | Historical/ Batch | Regression | 6 |
| 204 | (Guofeng Wang, Yang, Xie, & Zhang, 2014) | Condition Analysis | Descriptive | Time; Cost | Machine/ Tool | No Integration | Real-time | Classification | 4 |
| 205 | (J. Wang et al., 2016) | Performanc e Opt. | Prescriptive | Time; Cost; Flexibility | Machine/ Tool; Process | Vertical | Real-time | Reinforceme nt Learning | 2 |
| 206 | (Shijin Wang & Liu, 2015) | Production Planning | Prescriptive | Time; Cost | Machine/ Tool; Process; ERP | No Integration | Historical/ Batch | Mathematical Optimization | 2 |
| 207 | (X. Wang et al., 2016) | Maintenanc e Planning | Prescriptive | Time; Cost | Machine/ Tool | Vertical | Historical/ Batch | Reinforceme nt Learning | 2 |
| 208 | (Yue Wang & Tseng, 2014) | Design Analysis | Diagnostic | Customer Satisfaction | Customer | End-to-End | Historical/ Batch | Custom Development | 6 |
| 209 | (Yi Wang, Xu, Liang, & Jiang, 2015) | Defect Analysis | Diagnostic | Time; Cost | Machine/ Tool | No Integration | Historical/ Batch | Dimension Reduction | 5 |
| 210 | (Waschneck et al., 2018b) | Production Planning | Prescriptive | Time; Cost | Machine/ Tool; Process; Customer; ERP | Vertical | Historical/ Batch | Deep Learning; Reinforceme nt Learning | 2 |
| 211 | (Wedel et al., 2015) | Performanc e Analysis | Predictive | Time; Cost | Machine/ Tool | Vertical | Real-time | Custom Development | 3 |
| 212 | (Weimer, Scholz-Reiter, & Shpitalni, 2016) | Quality Control | Descriptive | Conformance | Product | No Integration | Historical/ Batch | Deep Learning | 1 |
| 213 | (J. Wen, Gao, & Zhang, 2018) | Condition Analysis | Predictive | Time; Cost | Machine/ Tool | No Integration | Real-time | Custom Development | 4 |
| 214 | (D. Wu et al., 2017) | Monitoring | Predictive | Time; Cost | Machine/ Tool | No Integration | Real-time | Custom Development | 3 |

| No. | Reference | Function | Maturity | Objective | Data Source | Integration | Frequency | Method | Cluster |
|---|---|---|---|---|---|---|---|---|---|
| 215 | (D. Z. Wu, Jennings, Terpenny, Gao, & Kumara, 2017) | Condition Analysis | Predictive | Time; Cost | Machine/ Tool | No Integration | Real-time | Classification | 4 |
| 216 | (J. Wu et al., 2018) | Condition Analysis | Predictive | Time; Cost | Machine/ Tool | No Integration | Real-time | Fuzzy Logic | 4 |
| 217 | (W. Wu, Zheng, Chen, Wang, & Cao, 2018) | Condition Analysis | Descriptive | Time; Cost | Machine/ Tool | No Integration | Real-time | Probabilistic Methods | 4 |
| 218 | (Yuting Wu, Yuan, Dong, Lin, & Liu, 2018) | Condition Analysis | Predictive | Time; Cost | Machine/ Tool | No Integration | Historical/ Batch | Deep Learning | 6 |
| 219 | (Thorsten Wuest, Irgens, & Thoben, 2014) | Quality Control | Descriptive | Conformance; Customer Satisfaction | Machine/ Tool; Process | No Integration | Real-time | Classification; Clustering | 1 |
| 220 | (Xanthopoulos, Kiatipis, Koulouriotis, & Stieger, 2018) | Maintenanc e Planning | Prescriptive | Time; Cost | Machine/ Tool | Vertical | Historical/ Batch | Reinforceme nt Learning | 2 |
| 221 | (Xiaoya Xu et al., 2016) | Security/ Risk Analysis | Descriptive | Security | Human | No Integration | Real-time | Custom Development | 3 |
| 222 | (Xun et al., 2018) | Security/ Risk Analysis | Descriptive | Security | Machine/ Tool | No Integration | Real-time | Classification | 3 |
| 223 | (H. H. Yan et al., 2018) | Condition Analysis | Predictive | Time; Cost | Machine/ Tool | No Integration | Real-time | Deep Learning | 4 |
| 224 | (J. Yan, Meng, Lu, & Li, 2017) | Condition Analysis | Predictive | Time; Cost | Machine/ Tool | No Integration | Historical/ Batch | Regression | 6 |
| 225 | (J. Yang et al., 2018) | Security/ Risk Analysis | Descriptive | Security | Machine/ Tool; Process | No Integration | Real-time | Custom Development | 3 |
| 226 | (W.-A. Yang, Zhou, Liao, & Guo, 2016) | Condition Analysis | Predictive | Time; Cost | Machine/ Tool | No Integration | Real-time | Classification; Swarm Intelligence | 4 |
| 227 | (Z. Yang et al., 2018) | Monitoring | Predictive | Time; Cost | Process | No Integration | Historical/ Batch | Regression | 6 |
| 228 | (F. Yao et al., 2018) | Production Planning | Prescriptive | Time; Cost | Machine/ Tool; Process | Vertical; Horizontal | Real-time | Custom Development | 2 |
| 229 | (G. Yin, Zhang, Li, Ren, & Fan, 2014) | Defect Analysis | Diagnostic | Time; Cost | Machine/ Tool | No Integration | Real-time | Classification | 4 |
| 230 | (X. Yin, He, Niu, & Li, 2018) | Quality Opt. | Prescriptive | Conformance; Customer Satisfaction | Machine/ Tool; Process | No Integration | Real-time | Regression; Evolutional Algorithm | 1 |
| 231 | (Yuan, Zhang, & Duan, 2018) | Defect Analysis | Diagnostic | Time; Cost | Machine/ Tool | No Integration | Real-time | Deep Learning | 4 |
| 232 | (Yunusa-Kaltungo & Sinha, 2017) | Condition Analysis | Descriptive | Time; Cost | Machine/ Tool | No Integration | Real-time | Classification | 4 |
| 233 | (Yuwono et al., 2016) | Defect Analysis | Diagnostic | Time; Cost | Machine/ Tool | No Integration | Historical/ Batch | Classification; Clustering | 5 |
| 234 | (Zarei, Tajeddini, & Karimi, 2014) | Defect Analysis | Descriptive | Time; Cost | Machine/ Tool | No Integration | Historical/ Batch | Classification | 5 |
| 235 | (C. Zhang, Lim, Qin, & Tan, 2017) | Condition Analysis | Predictive | Time; Cost | Machine/ Tool | No Integration | Historical/ Batch | Deep Learning; Evolutional Algorithm | 6 |
| 236 | (C. Zhang, Yan, Lee, & Shi, 2018) | Monitoring | Descriptive | Time; Cost | Machine/ Tool; Process | Horizontal | Real-time | Dimension Reduction | 3 |
| 237 | (C. Zhang & Zhang, 2016) | Condition Analysis | Predictive | Time; Cost | Machine/ Tool | No Integration | Historical/ Batch | Classification | 6 |
| 238 | (J. Zhang, Ahmad, Vera, & Harrison, 2018) | Design Analysis | Predictive | Customer Satisfaction | Process; Product | End-to-End | Historical/ Batch | Custom Development | 6 |
| 239 | (L. Zhang, Gao, Dong, Fu, & Liu, 2018) | Condition Analysis | Descriptive | Time; Cost | Machine/ Tool | No Integration | Real-time | Custom Development | 4 |
| 240 | (Xinmin Zhang, Kano, & Li, 2018) | Monitoring | Diagnostic | Time; Cost | Process | Horizontal | Real-time | Dimension Reduction | 3 |
| 241 | (Yingfeng Zhang, Ma, Yang, Lv, & Liu, 2018) | - | Prescriptive | Sustainability | Machine/ Tool; Process | Vertical; Horizontal | Real-time | Custom Development | - |
| 242 | (Z. J. Zhang & Zhang, 2015) | Condition Analysis | Predictive | Time; Cost | Machine/ Tool | No Integration | Real-time | Regression | 4 |
| 243 | (L. Zhao et al., 2018) | Monitoring | Descriptive | Time; Cost | Process | Horizontal | Real-time | Classification | 3 |
| 244 | (R. Zhao et al., 2018) | Condition Analysis | Diagnostic | Time; Cost | Machine/ Tool | No Integration | Real-time | Deep Learning | 5 |
| 245 | (C. Zheng, Dai, Zhang, Hu, & Guo, 2017) | Monitoring | Descriptive | Time; Cost | Process | No Integration | Real-time | Classification | 3 |
| 246 | (H. Zheng, Feng, Gao, & Tan, 2018) | Performanc e Analysis | Predictive | Time; Cost; Conformance; Customer Satisfaction | Machine/ Tool; Process | Vertical | Real-time | Regression | 3 |
| 247 | (X. C. Zheng et al., 2018) | Monitoring | Descriptive | Time; Cost; Security | Human | No Integration | Real-time | Deep Learning | 6 |
| 248 | (R. Zhong, Huang, Dai, & Zhang, 2014) | Production Planning | Diagnostic | Time; Cost; Flexibility | Process | Vertical | Real-time | Classification | 2 |
| 249 | (Ray Y Zhong, Wang, & Xu, 2017) | Condition Analysis | Descriptive | Time; Cost | Machine/ Tool | Vertical | Real-time | Custom Development | 3 |

| No. | Reference | Function | Maturity | Objective | Data Source | Integration | Frequency | Method | Cluster |
|---|---|---|---|---|---|---|---|---|---|
| 250 | (J. Zhu, Chen, & Peng, 2018) | Condition Analysis | Predictive | Time; Cost | Machine/ Tool | No Integration | Historical/ Batch | Dimension Reduction | 6 |
| 251 | (K. Zhu & Liu, 2018) | Condition Analysis | Predictive | Time; Cost | Machine/ Tool | No Integration | Real-time | Probabilistic Methods | 4 |
| 252 | (X. Zhu et al., 2018) | Defect Analysis | Diagnostic | Time; Cost | Machine/ Tool | No Integration | Historical/ Batch | Classification | 5 |
| 253 | (Ziani, Felkaoui, & Zegadi, 2017) | Defect Analysis | Diagnostic | Time; Cost | Machine/ Tool | No Integration | Historical/ Batch | Classification | 5 |
| 254 | (Zou, Xia, & Li, 2018) | Monitoring | Diagnostic | Time; Cost | Process | Horizontal | Historical/ Batch | Regression | 3 |
| 255 | (Zurita et al., 2016) | Production Planning | Predictive | Time; Cost | Process | Horizontal | Historical/ Batch | Fuzzy Logic | 2 |
| 256 | (Mrugalska, 2018) | Condition Analysis | Predictive | Time; Cost | Machine/ Tool | No Integration | Historical/ Batch | Fuzzy Logic | 6 |
| 257 | (Kedadouche, Thomas, & Tahan, 2016) | Defect Analysis | Descriptive | Time; Cost | Machine/ Tool | No Integration | Historical/ Batch | Custom Development | 5 |
| 258 | (C. Li, M. Cerrada, et al., 2018) | Defect Analysis | Diagnostic | Time; Cost | Machine/ Tool | No Integration | Historical/ Batch | Clustering; Fuzzy Logic | 5 |
| 259 | (Di, Song, Liu, & Wang, 2017) | Defect Analysis | Diagnostic | Time; Cost | Machine/ Tool | No Integration | Historical/ Batch | Classification | 5 |
| 260 | (Shouli Zhang, Liu, Su, Han, & Li, 2018) | Condition Analysis | Predictive | Time; Cost | Machine/ Tool | No Integration | Historical/ Batch | Classification | 6 |
| 261 | (Glawar et al., 2016) | Maintenanc e Planning | Predictive | Time; Cost | Machine/ Tool; Process; Product; Reference | Vertical | Historical/ Batch | Custom Development | 6 |
| 262 | (Jinjiang Wang, Gao, Yuan, Fan, & Zhang, 2016) | Condition Analysis | Predictive | Time; Cost | Machine/ Tool | No Integration | Real-time | Probabilistic Methods | 4 |
| 263 | (Chouhal, Mouss, Benaggoune, & Mahdaoui, 2016) | Defect Analysis | Diagnostic | Time; Cost | Machine/ Tool | No Integration | Historical/ Batch | Multi-Agent System | 5 |
| 264 | (Harris, Triantafyllopoulos, Stillman, & McLeay, 2016) | Condition Analysis | Descriptive | Time; Cost | Machine/ Tool | No Integration | Real-time | Custom Development | 4 |
| 265 | (Xiao, Chen, Zhang, & Liu, 2017) | Condition Analysis | Predictive | Time; Cost | Machine/ Tool | No Integration | Historical/ Batch | Regression | 6 |
| 266 | (C. Wu, Chen, Jiang, Ning, & Jiang, 2017) | Defect Analysis | Diagnostic | Time; Cost | Machine/ Tool | No Integration | Historical/ Batch | Classification | 5 |
| 267 | (Yoo & Baek, 2018) | Condition Analysis | Predictive | Time; Cost | Machine/ Tool | No Integration | Historical/ Batch | Deep Learning | 6 |
| 268 | (Huynh, Grall, & Bérenguer, 2018) | Condition Analysis | Predictive | Time; Cost | Machine/ Tool | No Integration | Historical/ Batch | Custom Development | 6 |
| 269 | (Bousdekis, Magoutas, Apostolou, & Mentzas, 2015) | Condition Analysis | Predictive | Time; Cost | Machine/ Tool | No Integration | Real-time | Custom Development | 4 |
| 270 | (Bousdekis et al., 2017) | Maintenanc e Planning | Prescriptive | Time; Cost | Machine/ Tool | No Integration | Real-time | Custom Development | 2 |
| 271 | (Fleischmann, Kohl, & Franke, 2016) | Condition Analysis | Descriptive | Time; Cost | Machine/ Tool | No Integration | Real-time | Custom Development | 4 |
| 272 | (Cong Wang, Gan, & Zhu, 2016) | Defect Analysis | Diagnostic | Time; Cost | Machine/ Tool | No Integration | Historical/ Batch | Classification | 5 |
| 273 | (Qiao, Wang, Wang, Qiao, & Zhang, 2018) | Condition Analysis | Descriptive | Time; Cost | Machine/ Tool | No Integration | Real-time | Deep Learning; Mathematical Optimization | 4 |
| 274 | (Si, Wang, Hu, Chen, & Zhou, 2013) | Condition Analysis | Predictive | Time; Cost | Machine/ Tool | No Integration | Historical/ Batch | Custom Development | 6 |
| 275 | (Benjamin Y Choo, Adams, Weiss, Marvel, & Beling, 2016) | Condition Analysis | Prescriptive | Time; Cost | Machine/ Tool; Process | Vertical | Historical/ Batch | Mathematical Optimization | 2 |
| 276 | (P. Zhao et al., 2017) | Condition Analysis | Predictive | Time; Cost | Machine/ Tool | No Integration | Historical/ Batch | Probabilistic Methods; Clustering | 6 |
| 277 | (Langone, Alzate, Bey-Temsamani, & Suykens, 2014) | Defect Analysis | Predictive | Time; Cost | Machine/ Tool | No Integration | Historical/ Batch | Classification | 6 |
| 278 | (I. Aydin, Karakose, & Akin, 2014) | Condition Analysis | Descriptive | Time; Cost | Machine/ Tool | No Integration | Historical/ Batch | Classification; Fuzzy Logic | 5 |
| 279 | (J. Tian et al., 2017) | Defect Analysis | Diagnostic | Time; Cost | Machine/ Tool | No Integration | Historical/ Batch | Classification | 5 |
| 280 | (Akhilesh Kumar, Chinnam, & Tseng, 2018) | Condition Analysis | Predictive | Time; Cost | Machine/ Tool | No Integration | Historical/ Batch | Regression | 6 |
| 281 | (YS Wang, Ma, Zhu, Liu, & Zhao, 2014) | Defect Analysis | Diagnostic | Time; Cost | Machine/ Tool | No Integration | Historical/ Batch | Classification | 5 |
| 282 | (D. Sun et al., 2016) | Condition Analysis | Descriptive | Time; Cost | Machine/ Tool | No Integration | Real-time | Fuzzy Logic | 4 |
| 283 | (C.-J. Kuo, Ting, Chen, Yang, & Chen, 2017) | Condition Analysis | Predictive | Time; Cost | Machine/ Tool | No Integration | Historical/ Batch | Classification; Dimension Reduction | 6 |

| No. | Reference | Function | Maturity | Objective | Data Source | Integration | Frequency | Method | Cluster |
|---|---|---|---|---|---|---|---|---|---|
| 284 | (Xingqing Wang, Li, Rui, Zhu, & Fei, 2015) | Defect Analysis | Diagnostic | Time; Cost | Machine/ Tool | No Integration | Historical/ Batch | Deep Learning | 5 |
| 285 | (C. Wu, Chen, & Jiang, 2017) | Defect Analysis | Diagnostic | Time; Cost | Machine/ Tool | No Integration | Historical/ Batch | Classification | 5 |
| 286 | (Soualhi et al., 2015) | Condition Analysis | Descriptive | Time; Cost | Machine/ Tool | No Integration | Real-time | Classification; Regression | 4 |
| 287 | (D. Z. Wu, Jennings, Terpenny, Kumara, & Gao, 2018) | Condition Analysis | Predictive | Time; Cost | Machine/ Tool | No Integration | Historical/ Batch | Classification; Regression | 6 |
| 288 | (JP Patel & Upadhyay, 2016) | Defect Analysis | Diagnostic | Time; Cost | Machine/ Tool | No Integration | Historical/ Batch | Classification | 5 |
| 289 | (Mehta, Werner, & Mears, 2015) | Condition Analysis | Descriptive | Time; Cost | Machine/ Tool | Vertical | Historical/ Batch | Classification | 5 |
| 290 | (Carstensen et al., 2016) | Condition Analysis | Descriptive | Time; Cost | Machine/ Tool | No Integration | Real-time | Custom Development | 4 |
| 291 | (L. Fu, Wei, Fang, Zhou, & Lou, 2017) | Condition Analysis | Descriptive | Time; Cost | Machine/ Tool | No Integration | Real-time | Custom Development | 4 |
| 292 | (Xia, Xi, Zhou, & Lee, 2013) | Maintenance Planning | Prescriptive | Time; Cost | Machine/ Tool | Horizontal | Historical/ Batch | Mathematical Optimization | 2 |
| 293 | (Engeler, Treyer, Zogg, Wegener, & Kunz, 2016) | Condition Analysis | Predictive | Time; Cost | Machine/ Tool | No Integration | Real-time | Custom Development | 4 |
| 294 | (Janssens et al., 2016) | Defect Analysis | Descriptive | Time; Cost | Machine/ Tool | No Integration | Historical/ Batch | Classification | 5 |
| 295 | (Chao Wang, Cheng, Nelson, Sawangsri, & Rakowski, 2015) | Condition Analysis | Diagnostic | Time; Cost | Machine/ Tool | No Integration | Historical/ Batch | Custom Development | 5 |
| 296 | (Mosallam, Medjaher, & Zerhouni, 2016) | Condition Analysis | Predictive | Time; Cost | Machine/ Tool | No Integration | Real-time | Classification | 4 |
| 297 | (T. Wang, Qiao, Zhang, Yang, & Snoussi, 2018) | Defect Analysis | Predictive | Time; Cost | Machine/ Tool | No Integration | Historical/ Batch | Dimension Reduction | 6 |
| 298 | (Jinjiang Wang, Wang, Wang, Huang, & Xue, 2018) | Condition Analysis | Predictive | Time; Cost | Machine/ Tool | No Integration | Historical/ Batch | Deep Learning | 6 |
| 299 | (Jianjing Zhang, Peng Wang, Ruqiang Yan, & Robert X Gao, 2018) | Condition Analysis | Predictive | Time; Cost | Machine/ Tool | No Integration | Historical/ Batch | Deep Learning | 6 |
| 300 | (Q. Li & S. Y. Liang, 2018) | Condition Analysis | Predictive | Time; Cost | Machine/ Tool | No Integration | Historical/ Batch | Custom Development | 6 |
| 301 | (Khelif et al., 2017) | Condition Analysis | Predictive | Time; Cost | Machine/ Tool | No Integration | Historical/ Batch | Regression | 6 |
| 302 | (Fleischmann, Spreng, Kohl, Kißkalt, & Franke, 2016) | Condition Analysis | Descriptive | Time; Cost | Machine/ Tool | No Integration | Real-time | Classification | 4 |
| 303 | (J. A. Carino et al., 2016) | Condition Analysis | Descriptive | Time; Cost | Machine/ Tool | No Integration | Real-time | Classification; Dimension Reduction | 4 |
| 304 | (J. Luo et al., 2018) | Defect Analysis | Descriptive | Time; Cost | Machine/ Tool | No Integration | Historical/ Batch | Classification | 5 |
| 305 | (B. Zhou & Cheng, 2016) | Condition Analysis | Diagnostic | Time; Cost | Machine/ Tool | No Integration | Historical/ Batch | Classification | 5 |
| 306 | (Unal, Sahin, Onat, Demetgul, & Kucuk, 2017) | Defect Analysis | Diagnostic | Time; Cost | Machine/ Tool | No Integration | Historical/ Batch | Classification | 5 |
| 307 | (H. Peng et al., 2013) | Defect Analysis | Diagnostic | Time; Cost | Machine/ Tool | No Integration | Historical/ Batch | Fuzzy Logic | 5 |
| 308 | (Benkedjouh, Medjaher, Zerhouni, & Rechak, 2015) | Condition Analysis | Predictive | Time; Cost | Machine/ Tool | No Integration | Historical/ Batch | Regression | 6 |
| 309 | (Benkedjouh et al., 2015) | Condition Analysis | Diagnostic | Time; Cost | Machine/ Tool | No Integration | Real-time | Custom Development | 5 |
| 310 | (Baraldi, Cannarile, Di Maio, & Zio, 2016) | Defect Analysis | Diagnostic | Time; Cost | Machine/ Tool | No Integration | Historical/ Batch | Classification; Mathematical Optimization | 5 |
| 311 | (Truong, 2018) | Condition Analysis | Predictive | Time; Cost | Machine/ Tool | No Integration | Real-time | Custom Development | 4 |
| 312 | (C. Lu et al., 2017) | Defect Analysis | Diagnostic | Time; Cost | Machine/ Tool | No Integration | Historical/ Batch | Classification | 5 |
| 313 | (Xiang Li, Zhang, Ding, & Sun, 2018) | Defect Analysis | Diagnostic | Time; Cost | Machine/ Tool | No Integration | Historical/ Batch | Deep Learning | 5 |
| 314 | (R. Kannan, Manohar, & Kumaran, 2019) | Condition Analysis | Descriptive | Time; Cost | Machine/ Tool | No Integration | Real-time | Classification | 4 |
| 315 | (Küfner, Uhlemann, & Ziegler, 2018) | Condition Analysis | Descriptive | Time; Cost | Machine/ Tool | No Integration | Real-time | Classification | 4 |
| 316 | (R. Zhao, Yan, Wang, & Mao, 2017) | Condition Analysis | Descriptive | Time; Cost | Machine/ Tool | No Integration | Real-time | Deep Learning | 4 |
| 317 | (Kanawaday & Sane, 2017) | Condition Analysis | Predictive | Time; Cost | Machine/ Tool | No Integration | Real-time | Classification; Regression | 4 |
| 318 | (Diaz-Rozo, Bielza, & Larrañaga, 2017) | Condition Analysis | Diagnostic | Time; Cost | Machine/ Tool | No Integration | Historical/ Batch | Clustering | 5 |

| No. | Reference | Function | Maturity | Objective | Data Source | Integration | Frequency | Method | Cluster |
|---|---|---|---|---|---|---|---|---|---|
| 319 | (Likun Ren, Lv, & Jiang, 2018) | Condition Analysis | Predictive | Time; Cost | Machine/ Tool | No Integration | Historical/ Batch | Custom Development | 6 |
| 320 | (Olivotti, Passlick, Dreyer, Lebek, & Breitner, 2018) | Maintenance Planning | Prescriptive | Time; Cost | Machine/ Tool | No Integration | Real-time | Mathematical Optimization | 2 |
| 321 | (Balsamo et al., 2016) | Defect Analysis | Descriptive | Time; Cost | Machine/ Tool | No Integration | Real-time | Classification | 4 |
| 322 | (K. Kannan, Arunachalam, Chawla, & Natarajan, 2018) | Condition Analysis | Predictive | Time; Cost | Machine/ Tool | No Integration | Historical/ Batch | Regression | 6 |
| 323 | (Q. Han, Li, Dong, Luo, & Xia, 2017) | Defect Analysis | Predictive | Time; Cost | Machine/ Tool | No Integration | Real-time | Classification | 4 |
| 324 | (Terrazas, Martínez-Arellano, Benardos, & Ratchev, 2018) | Condition Analysis | Descriptive | Time; Cost | Machine/ Tool | No Integration | Real-time | Classification | 4 |
| 325 | (Z. Cheng & Cai, 2018) | Condition Analysis | Predictive | Time; Cost | Machine/ Tool | No Integration | Historical/ Batch | Regression | 6 |
| 326 | (L. Ren, Sun, Wang, & Zhang, 2018) | Condition Analysis | Predictive | Time; Cost | Machine/ Tool | No Integration | Historical/ Batch | Deep Learning | 6 |
| 327 | (Bekar et al., 2018) | Condition Analysis | Descriptive | Time; Cost | Machine/ Tool | No Integration | Real-time | Custom Development | 4 |
| 328 | (Xiaojun Zhou, Huang, Xi, & Lee, 2015) | Maintenance Planning | Prescriptive | Time; Cost | Machine/ Tool | No Integration | Historical/ Batch | Mathematical Optimization | 2 |
| 329 | (Y. Li & Liu, 2018) | Maintenance Planning | Prescriptive | Time; Cost | Machine/ Tool | No Integration | Historical/ Batch | Mathematical Optimization | 2 |
| 330 | (Niu & Jiang, 2017) | Maintenance Planning | Prescriptive | Time; Cost | Machine/ Tool | No Integration | Real-time | Mathematical Optimization | 2 |
| 331 | (Shi & Zeng, 2016) | Maintenance Planning | Prescriptive | Time; Cost | Machine/ Tool | No Integration | Real-time | Swarm Intelligence | 2 |
| 332 | (Zhixiang & Jie, 2015) | Condition Analysis | Descriptive | Time; Cost | Machine/ Tool | No Integration | Historical/ Batch | Regression | 5 |
| 333 | (T. Yan, Lei, & Li, 2018) | Condition Analysis | Predictive | Time; Cost | Machine/ Tool | No Integration | Real-time | Mathematical Optimization | 4 |
| 334 | (Ellefsen, Bjørlykhaug, Æsøy, Ushakov, & Zhang, 2018) | Condition Analysis | Predictive | Time; Cost | Machine/ Tool | No Integration | Historical/ Batch | Deep Learning; Evolutional Algorithm | 6 |
| 335 | (L. Hao, Bian, Gebraeel, & Shi, 2017) | Condition Analysis | Predictive | Time; Cost | Machine/ Tool | No Integration | Real-time | Regression | 4 |
| 336 | (B. Zhang et al., 2018) | Condition Analysis | Descriptive | Time; Cost | Machine/ Tool | No Integration | Real-time | Fuzzy Logic | 4 |
| 337 | (He Yu, Li, Tian, & Wang, 2018) | Defect Analysis | Predictive | Time; Cost | Machine/ Tool | No Integration | Historical/ Batch | Regression; Swarm Intelligence | 6 |
| 338 | (Xingqiu Li, Jiang, Xiong, & Shao, 2019) | Condition Analysis | Predictive | Time; Cost | Machine/ Tool | No Integration | Historical/ Batch | Regression | 6 |
| 339 | (X. Zhang et al., 2017) | Defect Analysis | Diagnostic | Time; Cost | Machine/ Tool | No Integration | Historical/ Batch | Classification; Clustering | 5 |
| 340 | (Schutze & Helwig, 2017) | Condition Analysis | Descriptive | Time; Cost | Machine/ Tool | No Integration | Real-time | Classification | 4 |
| 341 | (J. Lee, Kao, et al., 2014) | Maintenance Planning | Prescriptive | Time; Cost; Sustainability | Machine/ Tool; Product; ERP | Horizontal | Real-time | Classification; Clustering | 2 |
| 342 | (Neef, Bartels, & Thiede, 2018) | Condition Analysis | Descriptive | Time; Cost | Machine/ Tool | No Integration | Real-time | Classification | 4 |
| 343 | (A. Zhang et al., 2018) | Condition Analysis | Predictive | Time; Cost | Machine/ Tool | No Integration | Historical/ Batch | Deep Learning | 6 |
| 344 | (P. Wang, Yan, & Gao, 2017) | Defect Analysis | Descriptive | Time; Cost | Machine/ Tool | No Integration | Historical/ Batch | Deep Learning | 5 |
| 345 | (Zhe Zhang, Qin, Jia, & Chen, 2018) | Defect Analysis | Diagnostic | Time; Cost | Machine/ Tool | No Integration | Historical/ Batch | Classification | 5 |
| 346 | (Lenz & Westkaemper, 2017) | Condition Analysis | Predictive | Time; Cost | Machine/ Tool; ERP | No Integration | Real-time | Regression | 4 |
| 347 | (Hang, 2016) | Energy Cons. Opt. | Prescriptive | Sustainability | Machine/ Tool | No Integration | Historical/ Batch | Classification; Reinforcement Learning | 6 |
| 348 | (Zuo, Tao, & Nee, 2018) | Energy Cons. Analysis | Descriptive | Sustainability | Machine/ Tool; Process | End-to-End | Real-time | Custom Development | 3 |
| 349 | (Yuxin Wang, Hulstijn, & Tan, 2018) | Security/ Risk Analysis | Descriptive | Time; Cost | Process; ERP | No Integration | Historical/ Batch | Custom Development | 3 |
| 350 | (Ak & Bhinge, 2015) | Energy Cons. Analysis | Predictive | Sustainability | Machine/ Tool | No Integration | Historical/ Batch | Regression; Evolutional Algorithm | 6 |
| 351 | (G. Shao et al., 2017) | Energy Cons. Opt. | Prescriptive | Sustainability | Machine/ Tool; Process; Product | No Integration | Historical/ Batch | Custom Development | 6 |
| 352 | (S.-J. Shin et al., 2017) | Energy Cons. Opt. | Prescriptive | Cost; Sustainability | Machine/ Tool; Process | No Integration | Historical/ Batch | Custom Development | 6 |
| 353 | (Filonenko & Jo, 2018) | Security/ Risk Analysis | Descriptive | Security | Environment | No Integration | Real-time | Classification | 3 |

| No. | Reference | Function | Maturity | Objective | Data Source | Integration | Frequency | Method | Cluster |
|---|---|---|---|---|---|---|---|---|---|
| 354 | (Ouyang et al., 2018) | Energy Cons. Analysis | Descriptive | Cost; Sustainability | Machine/ Tool | Vertical; Horizontal | Historical/ Batch | Classification | 6 |
| 355 | (Khalili & Sami, 2015) | Security/ Risk Analysis | Descriptive | Security | Process | No Integration | Real-time | Custom Development | 3 |
| 356 | (Niesen, Houy, Fettke, & Loos, 2016) | Security/ Risk Analysis | Diagnostic | Time; Cost; Conformance | Process; ERP | Vertical | Real-time | Custom Development | 3 |
| 357 | (Y. Zhang, S. Ren, Y. Liu, & S. Si, 2017) | Product Lifecycle Opt. | Prescriptive | Cost; Sustainability | Machine/ Tool; Product; Customer; Human | End-to-End | Real-time | Custom Development | 3 |
| 358 | (C.-Y. Tsai et al., 2014) | Design Analysis | Predictive | Cost | Product | No Integration | Historical/ Batch | Custom Development | 6 |
| 359 | (Ireland & Liu, 2018) | Design Analysis | Diagnostic | Customer Satisfaction | Product; Customer | End-to-End | Historical/ Batch | Classification | 6 |
| 360 | (Lou, Feng, Zheng, Gao, & Tan, 2018a) | Design Analysis | Diagnostic | Customer Satisfaction | Product; Customer | End-to-End | Historical/ Batch | Classification; Fuzzy Logic | 6 |
| 361 | (X. Lai et al., 2018) | Design Analysis | Diagnostic | Customer Satisfaction | Customer | No Integration | Historical/ Batch | Custom Development | 6 |
| 362 | (Alexopoulos, Makris, Xanthakis, Sipsas, & Chryssolouris, 2016) | Monitoring | Descriptive | Time; Cost | Machine/ Tool; ERP | Vertical | Real-time | Custom Development | 3 |
| 363 | (K. B. Lee, Cheon, & Kim, 2017) | Monitoring | Descriptive | Time; Cost | Machine/ Tool; Process | No Integration | Historical/ Batch | Deep Learning | 5 |
| 364 | (H. Cai, Guo, & Lu, 2017) | Monitoring | Predictive | Time; Cost; Conformance | Process; Product | No Integration | Real-time | Custom Development | 1 |
| 365 | (Changqing Liu, Li, & Li, 2018) | Production Planning | Descriptive | Time | Process; Product | No Integration | Historical/ Batch | Clustering | 3 |
| 366 | (H. Wang et al., 2017) | Production Planning | Prescriptive | Time; Cost | ERP | No Integration | Historical/ Batch | Mathematical Optimization | 2 |
| 367 | (Hranisavljevic, Niggemann, & Maier, 2016) | Monitoring | Descriptive | Time; Cost | Machine/ Tool | No Integration | Historical/ Batch | Deep Learning | 5 |
| 368 | (Ci Chen, Xia, Zhou, & Xi, 2015) | Production Planning | Prescriptive | Time; Cost; Flexibility | Process; Product; ERP | No Integration | Real-time | Reinforceme nt Learning | 2 |
| 369 | (Nouiri et al., 2018) | Production Planning | Prescriptive | Time; Cost; Flexibility | Machine/ Tool; Process; Product | No Integration | Real-time | Swarm Intelligence; Multi-Agent System | 2 |
| 370 | (Stojanovic, Dinic, Stojanovic, & Stojadinovic, 2016) | Monitoring | Descriptive | Time; Cost; Conformance | Machine/ Tool | No Integration | Real-time | Clustering | 4 |
| 371 | (Cong & Baranowski, 2018) | Monitoring | Descriptive | Time; Cost | Process | No Integration | Real-time | Classification | 3 |
| 372 | (Caggiano, 2018) | Monitoring | Descriptive | Time; Cost | Machine/ Tool; Process | Vertical | Real-time | Classification | 3 |
| 373 | (Yi-Jyun Chen, Lee, & Chiu, 2018) | Performanc e Opt. | Diagnostic | Time; Cost | Machine/ Tool; Product | Vertical | Historical/ Batch | Regression | 6 |
| 374 | (Jeff Morgan & O'Donnell, 2018) | Monitoring | Descriptive | Time; Cost; Flexibility | Machine/ Tool; Process | Vertical | Real-time | Classification; Fuzzy Logic | 3 |
| 375 | (P. Wang, Liu, Wang, & Gao, 2018) | Monitoring | Predictive | Time; Cost; Flexibility | Machine/ Tool; Product; Human | No Integration | Real-time | Deep Learning | 6 |
| 376 | (Chuang Wang & Jiang, 2017) | Performanc e Analysis | Predictive | Time; Cost; Flexibility | Process | No Integration | Real-time | Deep Learning | 6 |
| 377 | (Eiskop, Snatkin, Kõrgesaar, & Søren, 2014) | Monitoring | Descriptive | Time; Cost | Machine/ Tool; Process; ERP | Vertical | Real-time | Custom Development | 3 |
| 378 | (Chao-Chun et al., 2016) | Performanc e Opt. | Diagnostic | Time; Cost | Machine/ Tool | Vertical | Historical/ Batch | Mathematical Optimization | 6 |
| 379 | (J. P. Liu, Beyca, Rao, Kong, & Bukkapatnam, 2017) | Monitoring | Descriptive | Time; Cost | Process | No Integration | Real-time | Clustering | 3 |
| 380 | (Denno, Dickerson, & Harding, 2018) | Production Planning | Descriptive | Time; Cost | ERP | No Integration | Historical/ Batch | Evolutional Algorithm | 2 |
| 381 | (Hegenbarth, Bartsch, & Ristow, 2018) | Monitoring | Descriptive | Time; Cost | Machine/ Tool; Process | Vertical | Real-time | Custom Development | 3 |
| 382 | (Stefan Windmann & Niggemann, 2015) | Monitoring | Descriptive | Time; Cost | Process | No Integration | Historical/ Batch | Probabilistic Methods | 3 |
| 383 | (Ozturk, Bahadir, & Teymourifar, 2018) | Production Planning | Prescriptive | Time; Cost | ERP | No Integration | Historical/ Batch | Evolutional Algorithm | 2 |
| 384 | (Ragab, El-koujok, Amazouz, & Yacout, 2017) | Monitoring | Descriptive | Time; Cost | Process | No Integration | Historical/ Batch | Classification | 3 |
| 385 | (Andonovski et al., 2018) | Monitoring | Descriptive | Time; Cost | Process | No Integration | Historical/ Batch | Fuzzy Logic | 3 |
| 386 | (Ragab, El-Koujok, Poulin, Amazouz, & Yacout, 2018) | Monitoring | Descriptive | Time; Cost | Process | No Integration | Historical/ Batch | Classification | 3 |
| 387 | (Arık & Toksarı, 2019) | Production Planning | Prescriptive | Time; Cost | Process; ERP | No Integration | Historical/ Batch | Mathematical Optimization; Fuzzy Logic | 2 |
| 388 | (Peres et al., 2018) | Monitoring | Predictive | Time; Cost | Machine/ Tool | Vertical | Real-time | Custom Development | 3 |

| No. | Reference | Function | Maturity | Objective | Data Source | Integration | Frequency | Method | Cluster |
|---|---|---|---|---|---|---|---|---|---|
| 389 | (Chia-Yu Hsu, 2014) | Performance Opt. | Predictive | Time; Cost | Machine/ Tool; Process | No Integration | Historical/ Batch | Regression | 6 |
| 390 | (Westbrink, Chadha, & Schwung, 2018) | Monitoring | Descriptive | Time; Cost | Process | No Integration | Real-time | Dimension Reduction | 3 |
| 391 | (Tayal & Singh, 2018) | Production Planning | Prescriptive | Time; Cost | Machine/ Tool; Process; Product; ERP | No Integration | Historical/ Batch | Swarm Intelligence | 2 |
| 392 | (Kibira, Hatim, Kumara, & Shao, 2015) | Performance Opt. | Prescriptive | Time; Cost; Sustainability | Machine/ Tool; Process; Product | No Integration | Historical/ Batch | Mathematical Optimization | 2 |
| 393 | (S. Wang et al., 2018) | Monitoring | Descriptive | Time; Cost | Machine/ Tool; Product | End-to-End | Real-time | Custom Development | 3 |
| 394 | (Lingitz et al., 2018) | Performance Analysis | Predictive | Time; Cost; Flexibility | Machine/ Tool | No Integration | Historical/ Batch | Regression | 6 |
| 395 | (Shuhui, Jie, & Shivani, 2016) | Production Planning | Prescriptive | Time; Cost; Flexibility | Process; Product; ERP | Vertical | Real-time | Reinforcement Learning | 2 |
| 396 | (S. Jain, Shao, & Shin, 2017) | Performance Analysis | Diagnostic | Time; Cost | | Vertical | Historical/ Batch | Custom Development | 6 |
| 397 | (Marco S. Reis & Rato, 2018) | Monitoring | Predictive | Time; Cost | Process | No Integration | Real-time | Regression | 6 |
| 398 | (Ringsquandl et al., 2017) | Monitoring | Descriptive | Time; Cost | Machine/ Tool | No Integration | Historical/ Batch | Custom Development | 5 |
| 399 | (Rao, Liu, Roberson, Kong, & Williams, 2015) | Monitoring | Descriptive | Time; Cost; Conformance | Machine/ Tool; Process | No Integration | Real-time | Classification | 3 |
| 400 | (Klöber-Koch, Braunreuther, & Reinhart, 2017) | Production Planning | Predictive | Time; Cost; Security | | No Integration | Real-time | Custom Development | 6 |
| 401 | (C. Y. Park et al., 2017) | Monitoring | Predictive | Time; Cost; Conformance | Process; Product | No Integration | Historical/ Batch | Probabilistic Methods | 1 |
| 402 | (Stojanovic & Stojanovic, 2017) | Performance Opt. | Prescriptive | Time; Cost | Machine/ Tool; Process; Human | Vertical | Real-time | Custom Development | 2 |
| 403 | (Hammer, Somers, Karre, & Ramsauer, 2017) | Performance Opt. | Prescriptive | Time | Process | No Integration | Real-time | Mathematical Optimization | 2 |
| 404 | (Subramaniyan, Skoogh, Gopalakrishnan, & Hanna, 2016) | Performance Analysis | Descriptive | Time; Cost; Flexibility | Machine/ Tool | No Integration | Real-time | Custom Development | 3 |
| 405 | (Z. M. Bi et al., 2018) | Monitoring | Descriptive | Time; Cost | Machine/ Tool | No Integration | Real-time | Custom Development | 3 |
| 406 | (Ray Y Zhong, Li, et al., 2013) | Production Planning | Prescriptive | Time; Cost; Flexibility | Process; ERP | Vertical | Real-time | Custom Development | 2 |
| 407 | (J. Kim & Hwangbo, 2018) | Monitoring | Predictive | Time; Cost | Machine/ Tool | No Integration | Real-time | Clustering | 4 |
| 408 | (Berger, Berlak, & Reinhart, 2016) | Production Planning | Prescriptive | Time; Cost; Flexibility | Machine/ Tool; Process | Vertical | Real-time | Custom Development | 2 |
| 409 | (Mehta, Butkewitsch-Choze, & Seaman, 2018) | Monitoring | Predictive | Time; Cost | Process | No Integration | Real-time | Classification | 3 |
| 410 | (D. Kim et al., 2018) | Monitoring | Descriptive | Time; Cost | Machine/ Tool | No Integration | Real-time | Classification | 3 |
| 411 | (Q. P. He & Wang, 2017) | Monitoring | Descriptive | Time; Cost | Process | No Integration | Real-time | Custom Development | 3 |
| 412 | (Y. Wu, Wang, Chen, & Yu, 2017) | Performance Opt. | Predictive | Time; Cost | Machine/ Tool | No Integration | Real-time | Classification | 4 |
| 413 | (R. Kumar, Singh, & Lamba, 2018) | Production Planning | Prescriptive | Time; Cost | | No Integration | Historical/ Batch | Dimension Reduction | 2 |
| 414 | (P. Xu, Mei, Ren, & Chen, 2017) | Performance Analysis | Diagnostic | Time; Cost | Process; Product | Vertical | Real-time | Custom Development | 3 |
| 415 | (Diao, Zhao, & Yao, 2015) | Quality Control | Predictive | Time; Cost; Conformance | Process | No Integration | Historical/ Batch | Classification; Dimension Reduction | 1 |
| 416 | (Psarommatis & Kiritsis, 2018) | Quality Opt. | Prescriptive | Time; Cost; Conformance | Machine/ Tool; Process; ERP | Vertical | Real-time | Custom Development | 2 |
| 417 | (S. Du, Liu, & Xi, 2015) | Quality Control | Descriptive | Time; Cost; Conformance | Product | No Integration | Historical/ Batch | Classification | 1 |
| 418 | (K. Wang, Jiang, & Li, 2016) | Quality Control | Descriptive | Time; Cost; Conformance | Product | No Integration | Real-time | Regression; Clustering | 1 |
| 419 | (M. Syafrudin et al., 2017) | Quality Control | Predictive | Time; Conformance; Sustainability | Machine/ Tool | No Integration | Real-time | Classification | 4 |
| 420 | (Hirsch, Reimann, Kirn, & Mitschang, 2018) | Quality Control | Diagnostic | Time; Cost; Conformance | Process; Product | No Integration | Historical/ Batch | Custom Development | 1 |
| 421 | (Librantz et al., 2017) | Quality Control | Descriptive | Time; Cost; Conformance | Product | No Integration | Historical/ Batch | Classification; Dimension Reduction | 1 |
| 422 | (T.-H. Sun, Tien, Tien, & Kuo, 2016) | Quality Control | Descriptive | Time; Cost; Conformance | Product | No Integration | Historical/ Batch | Classification | 1 |
| 423 | (Nikolai Stein, Meller, & Flath, 2018) | Quality Control | Prescriptive | Time; Cost; Conformance | Process; Product | No Integration | Historical/ Batch | Classification; Regression; Mathematical Optimization | 2 |
| 424 | (Y. Feng & Huang, 2018) | Quality Control | Predictive | Time; Cost; Conformance | Reference | No Integration | Historical/ Batch | Custom Development | 1 |
| 425 | (L. Li, Ota, & Dong, 2018) | Quality Control | Descriptive | Time; Cost; Conformance | Product | No Integration | Historical/ Batch | Deep Learning | 1 |

| No. | Reference | Function | Maturity | Objective | Data Source | Integration | Frequency | Method | Cluster |
|---|---|---|---|---|---|---|---|---|---|
| 426 | (Shatnawi & Al-Khassaweneh, 2014) | Quality Control | Descriptive | Time; Cost; Conformance | Product | No Integration | Historical/ Batch | Classification | 1 |
| 427 | (Rendall et al., 2018) | Quality Control | Predictive | Time; Cost; Conformance | Product | No Integration | Historical/ Batch | Deep Learning | 1 |
| 428 | (J. Lee, Noh, Kim, & Kang, 2018) | Quality Control | Predictive | Time; Cost; Conformance | Process | Vertical | Real-time | Custom Development | 1 |
| 429 | (Onyeiwu et al., 2017) | Quality Control | Descriptive | Time; Cost; Conformance | Process; Product | No Integration | Real-time | Custom Development | 1 |
| 430 | (Aqlan, Ramakrishnan, & Shamsan, 2017) | Quality Control | Descriptive | Time; Cost; Conformance | Product | Horizontal | Historical/ Batch | Classification | 1 |
| 431 | (Agarwal & Shivpuri, 2014) | Quality Control | Diagnostic | Time; Cost; Conformance | Machine/ Tool; Process; Product | No Integration | Historical/ Batch | Regression | 1 |
| 432 | (J.-K. Park, Kwon, Park, & Kang, 2016) | Quality Control | Descriptive | Time; Cost; Conformance | Product | No Integration | Historical/ Batch | Deep Learning | 1 |
| 433 | (Xundao Zhou, Zhang, Mao, & Zhou, 2017) | Quality Control | Predictive | Time; Cost; Conformance | Process; Product | No Integration | Historical/ Batch | Custom Development | 1 |
| 434 | (Neto, Gerônimo, Cruz, Aguiar, & Bianchi, 2013) | Quality Control | Predictive | Time; Cost; Conformance | Machine/ Tool; Product | No Integration | Historical/ Batch | Classification | 1 |
| 435 | (Mashhadi & Behdad, 2017) | Quality Control | Diagnostic | Time; Cost; Conformance | Product | No Integration | Historical/ Batch | Clustering | 1 |
| 436 | (Voisin, Laloix, Iung, & Romagne, 2018) | Quality Control | Predictive | Time; Cost; Conformance | Machine/ Tool; Process; Product | No Integration | Historical/ Batch | Custom Development | 1 |
| 437 | (Shaban, Yacout, Meshreki, Attia, & Balazinski, 2017) | Quality Control | Descriptive | Time; Cost; Conformance | Process; Product | No Integration | Historical/ Batch | Classification | 1 |
| 438 | (Kao, Hsieh, Chen, & Lee, 2017) | Quality Control | Predictive | Time; Cost; Conformance | Machine/ Tool; Process; Product | No Integration | Historical/ Batch | Classification | 1 |
| 439 | (J.-B. Yu, Yu, Wang, Yuan, & Ji, 2016) | Quality Control | Predictive | Time; Cost; Conformance | Machine/ Tool; Process; Product | No Integration | Real-time | Regression; Evolutional Algorithm | 1 |
| 440 | (García, Sánchez, Rodríguez-Picón, Méndez-González, & de Jesús Ochoa-Domínguez, 2018) | Quality Control | Predictive | Time; Cost; Conformance | Process; Product | No Integration | Historical/ Batch | Regression | 1 |
| 441 | (Ghadimi, Toosi, & Heavey, 2018) | - | Prescriptive | Time; Cost; Conformance | ERP | Horizontal | Historical/ Batch | Multi-Agent System | - |
| 442 | (Laux et al., 2018) | - | Diagnostic | Time; Cost; Flexibility | | No Integration | Historical/ Batch | Classification; Dimension Reduction | - |
| 443 | (Badurdeen et al., 2014) | - | Diagnostic | Time; Cost; Security | | Horizontal | Historical/ Batch | Probabilistic Methods | - |
| 444 | (R. Y. Zhong, Lan, Xu, Dai, & Huang, 2016) | - | Descriptive | Time; Cost | Machine/ Tool; Process; Reference | Vertical; Horizontal | Real-time | Custom Development | - |
| 445 | (Susto et al., 2014) | Condition Analysis | Predictive | Time; Cost | Machine/ Tool | Vertical | Historical/ Batch | Regression | 6 |
| 446 | (Krumeich, Werth, Loos, Schimmelpfennig, & Jacobi, 2014) | Production Planning | Predictive | Time; Cost | Product | Horizontal | Historical/ Batch | Custom Development | 6 |
| 447 | (Gian Antonio Susto et al., 2017) | Monitoring | Descriptive | Time; Cost | Machine/ Tool | No Integration | Historical/ Batch | Custom Development | 5 |
| 448 | (Tamura, Iizuka, Yamamoto, & Furukawa, 2015) | Production Planning | Prescriptive | Time; Cost | Process | Horizontal | Historical/ Batch | Custom Development | 2 |
| 449 | (L. Zheng et al., 2014) | Performance Opt. | Prescriptive | Time; Cost | Process; Product | Vertical | Historical/ Batch | Classification; Regression | 2 |
| 450 | (N. Stein & Flath, 2017) | Monitoring | Predictive | Time; Cost | Machine/ Tool; Product; Reference | Horizontal | Historical/ Batch | Classification | 5 |
| 451 | (Q. Wu, Ding, & Huang, 2018) | Condition Analysis | Predictive | Time; Cost; Conformance | Machine/ Tool | Vertical | Historical/ Batch | Custom Development | 6 |
| 452 | (AlThobiani & Ball, 2014) | Defect Analysis | Diagnostic | Time; Cost | Machine/ Tool | No Integration | Historical/ Batch | Deep Learning | 5 |
| 453 | (Weiß & Vogel-Heuser, 2018) | Quality Control | Predictive | Time; Cost | Machine/ Tool | No Integration | Historical/ Batch | Regression | 6 |
| 454 | (Saldivar, Goh, Li, Yu, & Chen, 2016) | Design Analysis | Predictive | Time; Cost; Customer Satisfaction | Product | Vertical | Historical/ Batch | Clustering | 6 |
| 455 | (Schlegel, Briele, & Schmitt, 2018) | Quality Control | Predictive | Time; Cost; Conformance | Reference | Vertical | Historical/ Batch | Custom Development | 1 |
| 456 | (Khakifirooz et al., 2018) | Performance Opt. | Diagnostic | Time; Cost | Machine/ Tool | Vertical | Historical/ Batch | Probabilistic Methods | 6 |
| 457 | (Hussain, Mansoor, & Nisar, 2018) | Condition Analysis | Predictive | Time; Cost | Machine/ Tool | No Integration | Historical/ Batch | Classification | 6 |
| 458 | (Orman et al., 2015) | Defect Analysis | Descriptive | Time; Cost; Flexibility | Machine/ Tool | No Integration | Real-time | Custom Development | 3 |
| 459 | (G. Zhao et al., 2017) | Condition Analysis | Predictive | Time; Cost | Machine/ Tool | No Integration | Historical/ Batch | Deep Learning | 6 |
| 460 | (W. Ji, Yin, & Wang, 2018) | Production Planning | Prescriptive | Time; Cost | Machine/ Tool | End-to-End | Historical/ Batch | Deep Learning; Evolutional Algorithm | 6 |

| No. | Reference | Function | Maturity | Objective | Data Source | Integration | Frequency | Method | Cluster |
|---|---|---|---|---|---|---|---|---|---|
| 461 | (Ray Y. Zhong et al., 2015) | Monitoring | Diagnostic | Time; Cost | Process | Vertical | Historical/ Batch | Custom Development | 6 |
| 462 | (Brandenburger et al., 2016) | Quality Control | Descriptive | Time; Cost; Conformance | Process | Vertical | Historical/ Batch | Custom Development | 3 |
| 463 | (J. Yan, Meng, Lu, & Guo, 2017) | Condition Analysis | Predictive | Time; Cost; Flexibility | Machine/ Tool; Product; Customer | Vertical | Real-time | Deep Learning | 3 |
| 464 | (Roy, Li, & Zhu, 2014) | Performance Opt. | Predictive | Time; Cost; Flexibility | Process; Product | Vertical; Horizontal | Real-time | Custom Development | 3 |
| 465 | (S. Qu, Chu, Wang, Leckie, & Jian, 2015) | Production Planning | Prescriptive | Time; Cost; Flexibility | Process | Horizontal | Real-time | Reinforcement Learning | 2 |
| 466 | (Åkerman et al., 2018) | Condition Analysis | Predictive | Time; Cost | Machine/ Tool | Vertical; Horizontal | Historical/ Batch | Classification | 6 |
| 467 | (Batista, Badri, Sabourin, & Thomas, 2013) | Defect Analysis | Diagnostic | Time; Cost | Machine/ Tool | No Integration | Historical/ Batch | Classification | 5 |
| 468 | (D. Wu, Jennings, Terpenny, & Kumara, 2016) | Condition Analysis | Predictive | Time; Cost | Machine/ Tool | No Integration | Historical/ Batch | Classification | 6 |
| 469 | (Y. F. Zhang, Wang, Du, Qian, & Yang, 2018) | Monitoring | Descriptive | Time; Cost; Flexibility | | Vertical; Horizontal | Real-time | Custom Development | 3 |
| 470 | (Bai et al., 2018) | Quality Control | Predictive | Time; Cost; Conformance; Customer Satisfaction | Reference | No Integration | Historical/ Batch | Classification | 1 |
| 471 | (Moosavian, Ahmadi, Tabatabaeefar, & Khazaee, 2013) | Defect Analysis | Diagnostic | Time; Cost | Machine/ Tool | No Integration | Historical/ Batch | Classification | 5 |
| 472 | (Spendla, Kebisek, Tanuska, & Hrcka, 2017) | Condition Analysis | Predictive | Time; Cost | Machine/ Tool | Vertical | Historical/ Batch | Classification | 6 |
| 473 | (Weigelt, Mayr, Seefried, Heisler, & Franke, 2018) | Quality Control | Predictive | Time; Cost; Conformance; Customer Satisfaction | Process; Reference | Vertical | Real-time | Custom Development | 1 |
| 474 | (Jaramillo, Ottewill, Dudek, Lepiarczyk, & Pawlik, 2017) | Condition Analysis | Descriptive | Time; Cost | Machine/ Tool | Horizontal | Historical/ Batch | Custom Development | 5 |
| 475 | (Ragab, Ouali, Yacout, & Osman, 2014) | Condition Analysis | Predictive | Time; Cost | Machine/ Tool | No Integration | Historical/ Batch | Custom Development | 6 |
| 476 | (Bousdekis & Mentzas, 2017) | Condition Analysis | Predictive | Time; Cost | | Vertical | Real-time | Custom Development | 3 |
| 477 | (Weiss et al., 2016) | Quality Control | Predictive | Time; Cost; Conformance; Customer Satisfaction | Machine/ Tool | Horizontal | Real-time | Custom Development | 1 |
| 478 | (Subramaniyan, Skoogh, Salomonsson, Bangalore, Gopalakrishnan, et al., 2018) | Performance Analysis | Diagnostic | Time; Cost; Flexibility | Machine/ Tool | Vertical | Real-time | Custom Development | 3 |
| 479 | (Subramaniyan, Skoogh, Salomonsson, Bangalore, & Bokrantz, 2018) | Performance Analysis | Predictive | Time; Cost | Machine/ Tool | Horizontal | Historical/ Batch | Custom Development | 6 |
| 480 | (Beghi et al., 2016) | Monitoring | Descriptive | Time; Cost | Machine/ Tool | No Integration | Historical/ Batch | Dimension Reduction | 5 |
| 481 | (Fernandes et al., 2018) | Condition Analysis | Predictive | Time; Cost | Machine/ Tool | No Integration | Historical/ Batch | Classification | 6 |
| 482 | (Purarjomandlangrudi, Ghapanchi, & Esmalifalak, 2014) | Defect Analysis | Diagnostic | Time; Cost | Machine/ Tool | No Integration | Historical/ Batch | Classification | 5 |
| 483 | (Denkena, Schmidt, & Krüger, 2014) | Production Planning | Prescriptive | Time; Cost | Machine/ Tool | Vertical | Historical/ Batch | Custom Development | 2 |
| 484 | (Rashid, Amar, Gondal, & Kamruzzaman, 2016) | Defect Analysis | Diagnostic | Time; Cost | Machine/ Tool | No Integration | Historical/ Batch | Classification | 5 |
| 485 | (Y. Ji, Yu, Xu, Yu, & Zhang, 2018) | Production Planning | Predictive | Time; Cost | Product | Vertical | Historical/ Batch | Custom Development | 6 |
| 486 | (Susto & Beghi, 2016) | Condition Analysis | Predictive | Time; Cost | Machine/ Tool | No Integration | Historical/ Batch | Regression | 6 |
| 487 | (Uhlmann, Laghmouchi, Geisert, & Hohwieler, 2017) | Condition Analysis | Descriptive | Time; Cost | Machine/ Tool | Vertical | Historical/ Batch | Classification | 5 |
| 488 | (Para, Del Ser, Aguirre, & Nebro, 2018) | Quality Control | Predictive | Cost; Conformance; Customer Satisfaction | Product; Reference | No Integration | Historical/ Batch | Classification | 1 |
| 489 | (Mbuli et al., 2017) | Maintenance Planning | Prescriptive | Time; Cost | Product | Vertical | Historical/ Batch | Fuzzy Logic | 2 |
| 490 | (C. Zhang, Sun, & Tan, 2015) | Defect Analysis | Diagnostic | Time; Cost | Machine/ Tool | No Integration | Historical/ Batch | Deep Learning | 5 |
| 491 | (Lin Zhao & Wang, 2018) | Condition Analysis | Predictive | Time; Cost | Machine/ Tool | No Integration | Historical/ Batch | Deep Learning | 6 |

| No. | Reference | Function | Maturity | Objective | Data Source | Integration | Frequency | Method | Cluster |
|---|---|---|---|---|---|---|---|---|---|
| 492 | (Zhe Li, Wang, & Wang, 2019) | Defect Analysis | Diagnostic | Time; Cost | Machine/ Tool | No Integration | Historical/ Batch | Classification | 5 |
| 493 | (Verstraete, Ferrada, Droguett, Meruane, & Modarres, 2017) | Defect Analysis | Diagnostic | Time; Cost | Machine/ Tool | No Integration | Historical/ Batch | Deep Learning | 5 |
| 494 | (Waschneck et al., 2018a) | Production Planning | Prescriptive | Time; Cost | Process | Horizontal | Historical/ Batch | Reinforcement Learning | 2 |
| 495 | (Q. Li & S. Liang, 2018a) | Condition Analysis | Predictive | Time; Cost | Machine/ Tool | No Integration | Historical/ Batch | Custom Development | 6 |
| 496 | (Moharana & Sarmah, 2016) | Maintenance Planning | Predictive | Time; Cost | Product | No Integration | Historical/ Batch | Custom Development | 6 |
| 497 | (Samui, 2014) | Quality Control | Descriptive | Time; Cost; Conformance; Customer Satisfaction | Product; Reference | No Integration | Historical/ Batch | Classification; Regression | 1 |
| 498 | (R. He, Dai, Lu, & Mou, 2018) | Condition Analysis | Predictive | Time; Cost | Machine/ Tool | Vertical | Real-time | Deep Learning | 3 |
| 499 | (Y. Cheng, M. Chen, et al., 2018) | Security/ Risk Analysis | Descriptive | Flexibility; Security | Human | Vertical; Horizontal | Real-time | Classification | 3 |
| 500 | (Shaban, Yacout, Balazinski, Meshreki, & Attia, 2015) | Quality Control | Diagnostic | Cost; Conformance; Customer Satisfaction | Reference | No Integration | Historical/ Batch | Classification | 1 |
| 501 | (Zhuang, Liu, & Xiong, 2018) | Monitoring | Predictive | Time; Cost; Flexibility | Process | Vertical; Horizontal | Real-time | Custom Development | 3 |
| 502 | (Dolata, Mrzygłód, & Reiner, 2017) | Quality Control | Descriptive | Time; Cost; Conformance; Customer Satisfaction | Reference | No Integration | Historical/ Batch | Deep Learning | 1 |
| 503 | (Kang He, Zhao, Jia, & Liu, 2018) | Monitoring | Descriptive | Time; Cost | Machine/ Tool | Horizontal | Historical/ Batch | Custom Development | 5 |
| 504 | (Van Horenbeek & Pintelon, 2013) | Condition Analysis | Predictive | Time; Cost | Machine/ Tool | Horizontal | Historical/ Batch | Custom Development | 6 |
| 505 | (C. Kim, Lee, Kim, Lee, & Lee, 2018) | Monitoring | Descriptive | Time; Cost | Process | No Integration | Historical/ Batch | Custom Development | 3 |
| 506 | (Bousdekis, Papageorgiou, Magoutas, Apostolou, & Mentzas, 2018) | Condition Analysis | Predictive | Time; Cost | Machine/ Tool | Vertical | Real-time | Custom Development | 3 |
| 507 | (H.-W. Cho, 2015) | Quality Control | Predictive | Time; Cost; Conformance; Customer Satisfaction | Process; Reference | No Integration | Historical/ Batch | Evolutional Algorithm | 1 |
| 508 | (Ketai He, Zhang, Zuo, Alhwiti, & Megahed, 2017) | Monitoring | Descriptive | Time; Cost | Machine/ Tool | No Integration | Historical/ Batch | Custom Development | 5 |
| 509 | (Pillai, Punnoose, Vadakkepat, Loh, & Lee, 2018) | Quality Control | Descriptive | Time; Cost; Conformance; Customer Satisfaction | Product; Human | No Integration | Historical/ Batch | Fuzzy Logic | 1 |
| 510 | (Anton et al., 2018) | Security/ Risk Analysis | Descriptive | Security | Machine/ Tool | Vertical | Historical/ Batch | Custom Development | 3 |
| 511 | (Luangpaiboon, 2015) | Quality Opt. | Prescriptive | Time; Cost; Conformance; Customer Satisfaction | Reference | Horizontal | Historical/ Batch | Custom Development | 2 |
| 512 | (Ai-ming, Jian-min, & Kun, 2016) | Quality Control | Descriptive | Cost; Conformance; Customer Satisfaction | Reference | Horizontal | Historical/ Batch | Probabilistic Methods | 1 |
| 513 | (Tamilselvan & Wang, 2013) | Defect Analysis | Diagnostic | Time; Cost | Process | Vertical | Historical/ Batch | Classification | 6 |
| 514 | (Bakdi, Kouadri, & Bensmail, 2017) | Defect Analysis | Diagnostic | Time; Cost | Process | Horizontal | Historical/ Batch | Dimension Reduction | 3 |
| 515 | (Y. Du & Du, 2018) | Monitoring | Descriptive | Time; Cost | Process | Horizontal | Historical/ Batch | Dimension Reduction | 3 |
| 516 | (Jesus A Carino et al., 2018) | Condition Analysis | Descriptive | Time; Cost | Machine/ Tool | No Integration | Historical/ Batch | Classification | 5 |
| 517 | (Haasbroek, Strydom, McCoy, & Auret, 2018) | Monitoring | Descriptive | Time; Cost | Process | Horizontal | Historical/ Batch | Dimension Reduction | 3 |
| 518 | (C. Li, Sánchez, Zurita, Cerrada, & Cabrera, 2016) | Defect Analysis | Diagnostic | Time; Cost | Machine/ Tool | Horizontal | Historical/ Batch | Deep Learning | 5 |
| 519 | (Mortada, Yacout, & Lakis, 2014) | Defect Analysis | Diagnostic | Time; Cost | Machine/ Tool | No Integration | Historical/ Batch | Classification | 5 |
| 520 | (Krishnakumari, Elayaperumal, Saravanan, & Arvindan, 2017) | Defect Analysis | Diagnostic | Time; Cost | Machine/ Tool | No Integration | Historical/ Batch | Classification; Fuzzy Logic | 5 |
| 521 | (Gan, Wang, & Zhu, 2018) | Defect Analysis | Descriptive | Time; Cost | Reference | No Integration | Historical/ Batch | Dimension Reduction | 5 |
| 522 | (Cong Wang, Gan, & Zhu, 2018) | Defect Analysis | Diagnostic | Time; Cost | Machine/ Tool | No Integration | Historical/ Batch | Custom Development | 5 |
| 523 | (H. Lee, 2017) | Monitoring | Descriptive | Time; Cost | Reference | Vertical; Horizontal | Historical/ Batch | Classification | 5 |

| No. | Reference | Function | Maturity | Objective | Data Source | Integration | Frequency | Method | Cluster |
|---|---|---|---|---|---|---|---|---|---|
| 524 | (Yingfeng Zhang, Ren, Liu, Sakao, & Huisingh, 2017) | Product Lifecycle Opt. | Prescriptive | Time; Cost; Customer Satisfaction | Product | End-to-End | Historical/ Batch | Custom Development | 2 |
| 525 | (Frieß, Kolouch, Friedrich, & Zander, 2018) | Condition Analysis | Descriptive | Time; Cost | Machine/ Tool | Vertical | Historical/ Batch | Clustering | 5 |
| 526 | (R. Ren, Hung, & Tan, 2018) | Quality Control | Descriptive | Time; Cost; Conformance; Customer Satisfaction | Reference | No Integration | Historical/ Batch | Deep Learning | 1 |
| 527 | (Schabus & Scholz, 2015) | Performance Opt. | Descriptive | Time; Cost | Process | Vertical | Historical/ Batch | Custom Development | 3 |
| 528 | (B. Y. Choo, Adams, & Beling, 2017) | Maintenance Planning | Prescriptive | Time; Cost | Machine/ Tool | Horizontal | Historical/ Batch | Reinforcement Learning | 2 |
| 529 | (J. Liu, An, Dou, Ji, & Liu, 2018) | Defect Analysis | Diagnostic | Time; Cost | Machine/ Tool | No Integration | Real-time | Classification | 4 |
| 530 | (J. Tao, Wang, Li, Liu, & Cai, 2016) | Quality Control | Descriptive | Time; Cost; Conformance; Customer Satisfaction | Reference | No Integration | Historical/ Batch | Custom Development | 1 |
| 531 | (S. Windmann et al., 2015) | Monitoring | Descriptive | Time; Cost | Process | Horizontal | Historical/ Batch | Probabilistic Methods | 3 |
| 532 | (Roh & Oh, 2016) | Energy Cons. Opt. | Descriptive | Cost; Sustainability | Product | No Integration | Historical/ Batch | Classification; Clustering | 3 |
| 533 | (Saldivar, Goh, Li, Chen, & Yu, 2016) | Design Analysis | Predictive | Time; Cost; Customer Satisfaction | Product; Customer | Vertical | Historical/ Batch | Clustering | 6 |
| 534 | (Jayaram, 2017) | Performance Analysis | Predictive | Time; Cost; Flexibility | | Vertical; End-to-End | Historical/ Batch | Custom Development | 6 |
| 535 | (Mahdavi, Shirazi, Ghorbani, & Sahebjamnia, 2013) | Quality Control | Descriptive | Time; Cost; Conformance; Customer Satisfaction | Process; Reference | Horizontal | Real-time | Multi-Agent System | 1 |
| 536 | (Kai Ding & Jiang, 2016) | Design Analysis | Diagnostic | Time; Cost; Customer Satisfaction | Customer | Vertical | Historical/ Batch | Custom Development | 6 |
| 537 | (Subakti & Jiang, 2018) | Condition Analysis | Descriptive | Time; Cost | Environment | Vertical | Real-time | Deep Learning | 3 |
| 538 | (X. C. Zhu et al., 2017) | Production Planning | Prescriptive | Time; Cost | Machine/ Tool; Process; ERP | Vertical | Real-time | Custom Development | 2 |
| 539 | (Sezer, Romero, Guedea, Macchi, & Emmanouilidis, 2018) | Condition Analysis | Predictive | Time; Cost | Machine/ Tool; Product | Vertical | Historical/ Batch | Regression | 6 |
| 540 | (Xiong et al., 2016) | Defect Analysis | Diagnostic | Time; Cost | Machine/ Tool | No Integration | Historical/ Batch | Classification | 5 |
| 541 | (Gouarir, Martínez-Arellano, Terrazas, Benardos, & Ratchev, 2018) | Condition Analysis | Predictive | Time; Cost | Machine/ Tool | Vertical | Historical/ Batch | Deep Learning | 6 |
| 542 | (Ye, Pan, Chang, & Yu, 2018) | Quality Control | Descriptive | Time; Cost; Conformance; Customer Satisfaction | Machine/ Tool | Vertical | Historical/ Batch | Classification | 1 |
| 543 | (XiaoLi Zhang, Wang, & Chen, 2015) | Defect Analysis | Diagnostic | Time; Cost | Machine/ Tool | Vertical | Historical/ Batch | Classification | 6 |
| 544 | (Cong Wang, Gan, & Zhu, 2017) | Defect Analysis | Diagnostic | Time; Cost | Machine/ Tool | No Integration | Historical/ Batch | Classification | 5 |
| 545 | (Bhinge et al., 2014) | Energy Cons. Analysis | Predictive | Sustainability | Machine/ Tool | Vertical | Historical/ Batch | Regression | 6 |
| 546 | (Z. Li, Wang, & Wang, 2017) | Condition Analysis | Predictive | Time; Cost | Machine/ Tool | Vertical | Historical/ Batch | Custom Development | 6 |
| 547 | (Dutta et al., 2018) | Energy Cons. Opt. | Predictive | Cost; Sustainability | Environment | Vertical | Historical/ Batch | Custom Development | 6 |
| 548 | (Mourtzis, Milas, & Vlachou, 2018) | Monitoring | Descriptive | Time; Cost | Machine/ Tool | Vertical | Real-time | Custom Development | 3 |
| 549 | (J. Lv et al., 2018) | Energy Cons. Analysis | Predictive | Sustainability | Machine/ Tool | No Integration | Historical/ Batch | Custom Development | 6 |
| 550 | (Ayad, Terrissa, & Zerhouni, 2018) | Condition Analysis | Descriptive | Time; Cost | Machine/ Tool | Vertical; Horizontal | Historical/ Batch | Custom Development | 5 |
| 551 | (Keizer, Teunter, & Veldman, 2017) | Maintenance Planning | Prescriptive | Time; Cost | Machine/ Tool | Horizontal | Historical/ Batch | Custom Development | 2 |
| 552 | (Diez-Olivan, Pagan, Khoa, Sanz, & Sierra, 2018) | Condition Analysis | Diagnostic | Time; Cost | Machine/ Tool | No Integration | Historical/ Batch | Classification | 5 |
| 553 | (Jianjing Zhang, Peng Wang, Ruqiang Yan, & Robert X. Gao, 2018) | Condition Analysis | Predictive | Time; Cost | Machine/ Tool | No Integration | Historical/ Batch | Deep Learning | 6 |
| 554 | (Susto, Schirru, Pampuri, McLoone, & Beghi, 2015) | Condition Analysis | Predictive | Time; Cost | Machine/ Tool | No Integration | Historical/ Batch | Classification | 6 |

| No. | Reference | Function | Maturity | Objective | Data Source | Integration | Frequency | Method | Cluster |
|---|---|---|---|---|---|---|---|---|---|
| 555 | (Candanedo, Nieves, González, Martín, & Briones, 2018) | Condition Analysis | Predictive | Time; Cost | Environment | No Integration | Historical/ Batch | Regression | 6 |
| 556 | (Reina et al., 2018) | Condition Analysis | Descriptive | Time; Cost | Machine/ Tool | No Integration | Historical/ Batch | Custom Development | 5 |
| 557 | (Ying-Jen Chen, Fan, & Chang, 2016; Deuse, Lenze, Klenner, & Friedrich, 2016) | Performance Analysis | Descriptive | Time; Cost | Machine/ Tool | No Integration | Historical/ Batch | Custom Development | - |
| 558 | (Ying-Jen Chen et al., 2016) | Quality Control | Descriptive | Time; Cost; Conformance; Customer Satisfaction | Machine/ Tool | Vertical | Historical/ Batch | Classification | 1 |
| 559 | (Kamsu-Foguem, Rigal, & Mauget, 2013) | Monitoring | Descriptive | Time; Cost | Process | Horizontal | Historical/ Batch | Custom Development | 3 |
| 560 | (Gröger, Stach, Mitschang, & Westkämper, 2016) | Performance Opt. | Diagnostic | Time; Cost | Process | Vertical; Horizontal | Historical/ Batch | Custom Development | 3 |
| 561 | (X. Wen & Gong, 2017) | Condition Analysis | Predictive | Time; Cost | | | Historical/ Batch | Custom Development | 6 |
| 562 | (Dong, Zhang, & Geng, 2014) | Monitoring | Diagnostic | Time; Cost | Machine/ Tool | No Integration | Historical/ Batch | Probabilistic Methods | 5 |
| 563 | (Rivera Torres, Serrano Mercado, Llanes Santiago, & Anido Rifón, 2018) | Maintenance Planning | Predictive | Time; Cost | Process | Horizontal | Historical/ Batch | Custom Development | 6 |
| 564 | (Jing Tian, Morillo, Azarian, & Pecht, 2016) | Defect Analysis | Diagnostic | Time; Cost | Machine/ Tool | No Integration | Historical/ Batch | Classification | 5 |
| 565 | (Seera, Lim, & Loo, 2016) | Defect Analysis | Diagnostic | Time; Cost | Machine/ Tool | No Integration | Historical/ Batch | Classification; Regression | 5 |
| 566 | (Mashhadi, Cade, & Behdad, 2018) | Quality Control | Predictive | Time; Cost; Conformance; Customer Satisfaction | Product | Vertical; Horizontal | Real-time | Regression | 1 |
| 567 | (Xie et al., 2015) | Quality Control | Descriptive | Time; Cost; Conformance; Customer Satisfaction | Product; Reference | Horizontal | Historical/ Batch | Custom Development | 1 |
| 568 | (B. Zhang & Shin, 2018) | Monitoring | Descriptive | Time; Cost | Machine/ Tool | Vertical; Horizontal | Historical/ Batch | Classification | 5 |
| 569 | (J. Morgan & O'Donnell, 2017) | Monitoring | Descriptive | Time; Cost | Machine/ Tool | Vertical; Horizontal | Historical/ Batch | Custom Development | 5 |
| 570 | (Yue Wang & Tseng, 2015) | Design Analysis | Diagnostic | Time; Cost; Customer Satisfaction | Customer | No Integration | Historical/ Batch | Classification | 6 |
| 571 | (L. Wen, Li, Gao, & Zhang, 2018) | Defect Analysis | Diagnostic | Time; Cost | Machine/ Tool | No Integration | Historical/ Batch | Deep Learning | 5 |
| 572 | (H. Liu et al., 2018) | Condition Analysis | Descriptive | Time; Cost | Machine/ Tool | No Integration | Historical/ Batch | Custom Development | 5 |
| 573 | (Guofeng Wang, Zhang, Liu, Xie, & Xu, 2016) | Condition Analysis | Descriptive | Time; Cost | Machine/ Tool | Horizontal | Historical/ Batch | Dimension Reduction | 5 |
| 574 | (R. Kannan, Manohar, & Kumaran, 2018) | Monitoring | Diagnostic | Time; Cost; Flexibility | Machine/ Tool | No Integration | Real-time | Classification | 5 |
| 575 | (Hongyang Yu, Khan, & Garaniya, 2015) | Monitoring | Descriptive | Time; Cost; Flexibility | Process | No Integration | Real-time | Custom Development | 3 |
| 576 | (S. Wang et al., 2017) | Defect Analysis | Diagnostic | Time; Cost | Machine/ Tool | No Integration | Historical/ Batch | Custom Development | 5 |
| 577 | (Kisskalt, Fleischmann, Kreitlein, Knott, & Franke, 2018) | Condition Analysis | Descriptive | Time; Cost | Machine/ Tool | Vertical | Historical/ Batch | Probabilistic Methods | 5 |
| 578 | (Gowid, Dixon, & Ghani, 2015) | Condition Analysis | Descriptive | Time; Cost | Machine/ Tool | No Integration | Historical/ Batch | Classification | 5 |
| 579 | (S.-G. He et al., 2013) | Monitoring | Descriptive | Time; Cost | Process | No Integration | Historical/ Batch | Classification | 3 |
| 580 | (Sądel & Śnieżyński, 2017) | Production Planning | Prescriptive | Time; Cost | Process | Horizontal | Historical/ Batch | Reinforcement Learning; Multi-Agent System | 2 |
| 581 | (Duan, Deng, Gong, & Wang, 2018) | Maintenance Planning | Prescriptive | Time; Cost | Machine/ Tool | Horizontal | Historical/ Batch | Evolutional Algorithm | 2 |
| 582 | (Shuhui Qu, Wang, Govil, & Leckie, 2016) | Production Planning | Prescriptive | Time; Cost | Product; Human | Vertical; Horizontal | Real-time | Reinforcement Learning | 2 |
| 583 | (Kan, Yang, & Kumara, 2018) | Condition Analysis | Descriptive | Time; Cost | Machine/ Tool | Vertical; Horizontal | Historical/ Batch | Custom Development | 5 |
| 584 | (S. Tangjitsitcharoen, Thesniyom, & Ratanakuakangwan, 2017) | Quality Control | Predictive | Time; Cost; Conformance; Customer Satisfaction | Machine/ Tool; Product | No Integration | Historical/ Batch | Regression | 1 |
| 585 | (S.-J. Shin, Woo, & Rachuri, 2014) | Energy Cons. Analysis | Predictive | Sustainability | Machine/ Tool | Vertical | Historical/ Batch | Classification | 6 |

| No. | Reference | Function | Maturity | Objective | Data Source | Integration | Frequency | Method | Cluster |
|---|---|---|---|---|---|---|---|---|---|
| 586 | (Oyekanlu, 2017) | Condition Analysis | Descriptive | Time; Cost | Machine/ Tool | Vertical; Horizontal | Real-time | Custom Development | 3 |
| 587 | (Kajmakovic et al., 2018) | Security/ Risk Analysis | Predictive | Security | Machine/ Tool; Environment; Human | Horizontal | Historical/ Batch | Custom Development | 3 |
| 588 | (Akhavei & Bleicher, 2016) | Production Planning | Predictive | Time; Cost | Process | No Integration | Historical/ Batch | Regression | 6 |
| 589 | (Kireev et al., 2018) | Condition Analysis | Predictive | Time; Cost | Machine/ Tool | Vertical | Historical/ Batch | Custom Development | 6 |
| 590 | (R. Jain, Singh, & Mishra, 2013) | Quality Control | Prescriptive |  | Process; Environment; Human | Horizontal | Historical/ Batch | Fuzzy Logic | 3 |
| 591 | (Alexander Brodsky, Shao, & Riddick, 2016) | Monitoring | Descriptive | Time; Cost; Sustainability | Process | No Integration | Historical/ Batch | Custom Development | 3 |
| 592 | (Oh et al., 2013) | Quality Control | Predictive | Time; Cost; Conformance; Customer Satisfaction | Reference | Horizontal | Real-time | Classification | 1 |
| 593 | (Ragab, Yacout, et al., 2016) | Defect Analysis | Predictive | Time; Cost | Machine/ Tool | No Integration | Historical/ Batch | Classification | 6 |
| 594 | (D. Kim, Han, Lin, Kang, & Lee, 2018) | Defect Analysis | Descriptive | Time; Cost | Machine/ Tool | No Integration | Historical/ Batch | Custom Development | 5 |
| 595 | (Jinzhi Wang, Qu, Wang, Leckie, & Xu, 2017) | Production Planning | Prescriptive | Time; Cost; Flexibility | Process | Horizontal | Historical/ Batch | Reinforceme nt Learning | 2 |
| 596 | (Gajjar, Kulahci, & Palazoglu, 2018) | Monitoring | Descriptive | Time; Cost | Process | Horizontal | Real-time | Custom Development | 3 |
| 597 | (M. Canizo, Onieva, Conde, Charramendieta, & Trujillo, 2017) | Condition Analysis | Predictive | Time; Cost | Machine/ Tool | Vertical | Real-time | Classification | 4 |
| 598 | (Židek, Hosovsky, Piteľ, & Bednár, 2019) | Performanc e Opt. | Descriptive | Time; Cost | Process | Horizontal | Historical/ Batch | Deep Learning | 3 |
| 599 | (Duong et al., 2018) | Condition Analysis | Predictive | Time; Cost | Machine/ Tool | No Integration | Historical/ Batch | Regression | 6 |
| 600 | (Ahmad, Khan, Islam, & Kim, 2018) | Condition Analysis | Predictive | Time; Cost | Machine/ Tool | No Integration | Historical/ Batch | Regression | 6 |
| 601 | (X. Chen, Shen, He, Sun, & Liu, 2013) | Condition Analysis | Predictive | Time; Cost | Machine/ Tool | No Integration | Historical/ Batch | Classification | 6 |
| 602 | (Xiang Li, Ding, & Sun, 2018) | Condition Analysis | Predictive | Time; Cost | Machine/ Tool | No Integration | Historical/ Batch | Deep Learning | 6 |
| 603 | (Al Sunny, Liu, & Shahriar, 2018) | Defect Analysis | Diagnostic | Time; Cost | Machine/ Tool | Vertical | Historical/ Batch | Custom Development | 6 |
| 604 | (Q. Zhou, Yan, Liu, Xin, & Chen, 2018) | Defect Analysis | Diagnostic | Time; Cost | Machine/ Tool | Vertical | Historical/ Batch | Custom Development | 6 |
| 605 | (Shuai Zhang et al., 2018) | Condition Analysis | Predictive | Time; Cost | Machine/ Tool | No Integration | Historical/ Batch | Probabilistic Methods | 6 |
| 606 | (Ray Y Zhong, Dai, Qu, Hu, & Huang, 2013) | Production Planning | Prescriptive | Time; Cost; Customer Satisfaction | Process | Vertical; Horizontal | Real-time | Multi-Agent System | 2 |
| 607 | (Hinchi & Tkiouat, 2018) | Condition Analysis | Predictive | Time; Cost | Machine/ Tool | End-to-End | Historical/ Batch | Deep Learning | 6 |
| 608 | (Prosvirin, Islam, Kim, & Kim, 2018) | Defect Analysis | Diagnostic | Time; Cost | Product | No Integration | Historical/ Batch | Classification | 5 |
| 609 | (Gawand et al., 2017) | Security/ Risk Analysis | Descriptive | Security | Machine/ Tool; Environment | Vertical | Historical/ Batch | Custom Development | 3 |
| 610 | (Lavrova et al., 2018) | Security/ Risk Analysis | Descriptive | Security | Process | Vertical | Historical/ Batch | Custom Development | 3 |
| 611 | (M. Zheng & Wu, 2017) | Maintenanc e Planning | Predictive | Time; Cost | Machine/ Tool | Vertical | Real-time | Custom Development | 3 |
| 612 | (Germen, Başaran, & Fidan, 2014) | Defect Analysis | Diagnostic | Time; Cost | Machine/ Tool | No Integration | Historical/ Batch | Dimension Reduction | 5 |
| 613 | (Shimada & Sakajo, 2016) | Condition Analysis | Predictive | Time; Cost | Machine/ Tool | No Integration | Historical/ Batch | Custom Development | 6 |
| 614 | (Q. Peter He & Jin Wang, 2018) | Monitoring | Descriptive | Time; Cost | Process | Vertical | Real-time | Custom Development | 3 |
| 615 | (Q Peter He & Jin Wang, 2018) | Monitoring | Descriptive | Time; Cost | Process | Vertical | Real-time | Custom Development | 3 |
| 616 | (Mayer, Mayer, & Abdo, 2017) | Monitoring | Descriptive | Time; Cost | Machine/ Tool | Vertical | Historical/ Batch | Classification; Clustering | 5 |
| 617 | (Gururajapathy, Mokhlis, Illias, & Awalin, 2017) | Defect Analysis | Descriptive | Time; Cost | Machine/ Tool | No Integration | Historical/ Batch | Classification | 5 |
| 618 | (Ray & Mishra, 2016) | Defect Analysis | Descriptive | Time; Cost | Machine/ Tool | No Integration | Historical/ Batch | Classification | 5 |
| 619 | (H.-J. Shin, Cho, & Oh, 2018) | Defect Analysis | Predictive | Time; Cost | Process | Horizontal | Historical/ Batch | Classification | 2 |
| 620 | (Palacios, González-Rodríguez, Vela, & Puente, 2015) | Production Planning | Prescriptive | Time; Cost | Process | No Integration | Historical/ Batch | Swarm Intelligence | 2 |
| 621 | (C.-F. Chien, S.-C. Hsu, & Y.-J. Chen, 2013b) | Quality Control | Descriptive | Time; Cost; Conformance; Customer Satisfaction | Machine/ Tool | Vertical | Historical/ Batch | Classification | 1 |

| No. | Reference | Function | Maturity | Objective | Data Source | Integration | Frequency | Method | Cluster |
|---|---|---|---|---|---|---|---|---|---|
| 622 | (D. Song & Yang, 2018) | Monitoring | Predictive | Time; Cost | Environment | No Integration | Historical/ Batch | Regression | 6 |
| 623 | (Morariu & Borangiu, 2018) | Production Planning | Prescriptive | Time; Cost; Flexibility | Process | Vertical | Real-time | Deep Learning | 2 |
| 624 | (G. F. Wang, Xie, & Zhang, 2017) | Condition Analysis | Descriptive | Time; Cost | Machine/ Tool | No Integration | Historical/ Batch | Classification; Evolutional Algorithm | 5 |
| 625 | (Tao Li, Zhang, Luo, & Wu, 2017) | Condition Analysis | Descriptive | Time; Cost | Machine/ Tool | No Integration | Historical/ Batch | Classification | 5 |
| 626 | (Karandikar, McLeay, Turner, & Schmitz, 2015) | Condition Analysis | Descriptive | Time; Cost | Machine/ Tool | No Integration | Historical/ Batch | Classification | 5 |
| 627 | (A. Brodsky, Krishnamoorthy, Menascé, Shao, & Rachuri, 2014) | Performance Opt. | Prescriptive | Time; Cost | Process | Vertical; Horizontal | Historical/ Batch | Custom Development | 3 |
| 628 | (Oliff & Liu, 2017) | Quality Control | Descriptive | Time; Cost; Conformance; Customer Satisfaction | Process; Reference | Vertical | Historical/ Batch | Classification | 1 |
| 629 | (Flath & Stein, 2018) | Quality Control | Predictive | Time; Cost; Conformance; Customer Satisfaction | Process | Vertical | Historical/ Batch | Custom Development | 1 |
| 630 | (Sanchez et al., 2018) | Monitoring | Predictive | Time; Cost | Machine/ Tool | Vertical | Historical/ Batch | Deep Learning | 6 |
| 631 | (Vazan, Janikova, Tanuska, Kebisek, & Cervenanska, 2017) | Monitoring | Predictive | Time; Cost | Process | Vertical; Horizontal | Historical/ Batch | Custom Development | 6 |
| 632 | (R. Jain, Singh, Yadav, & Mishra, 2014) | Product Lifecycle Opt. | Prescriptive | Time; Cost | ERP; Human | Horizontal | Historical/ Batch | Classification | 2 |
| 633 | (Baban et al., 2016) | Condition Analysis | Predictive | Time; Cost; Flexibility | Machine/ Tool | No Integration | Historical/ Batch | Fuzzy Logic | 6 |

## Table A.3 Coded Publications from Second Literature Survey with Clusters (*n*=232)

| No. | Reference | Function | Maturity | Objective | Data Source | Integration | Frequency | Method | Cluster |
|---|---|---|---|---|---|---|---|---|---|
| 634 | (A. Zhang et al., 2019) | Defect Analysis | Predictive | Time; Cost | Product | No Integration | Historical/ Batch | Deep Learning | 6 |
| 635 | (Abidi, Alkhalefah, Mohammed, Umer, & Qudeiri, 2020) | Production Planning | Prescriptive | Time; Cost; Flexibility | Process | No Integration | Historical/ Batch | Classification; Regression; Deep Learning; Mathematical Optimization; Fuzzy Logic | 2 |
| 636 | (Alavian, Eun, Meerkov, & Zhang, 2020) | Performance Opt. | Prescriptive | Time; Cost; Flexibility | Machine/ Tool; Process; Product | Vertical | Real-time | Probabilistic Methods; Mathematical Optimization; Custom Development | 2 |
| 637 | (Alexopoulos, Nikolakis, & Chryssolouris, 2020) | Design Analysis | Descriptive | Time; Cost; Conformance | Reference; ERP | End-to-End | Historical/ Batch | Regression; Deep Learning | 6 |
| 638 | (Alfeo, Cimino, Manco, Ritacco, & Vaglini, 2020) | Design Analysis | Diagnostic | Conformance | Machine/ Tool | No Integration | Historical/ Batch | Deep Learning | 1 |
| 639 | (Ansari, Glawar, & Nemeth, 2019) | Maintenance Planning | Prescriptive | Time | Machine/ Tool; ERP | Vertical | Historical/ Batch | Custom Development | 2 |
| 640 | (Arachchige et al., 2020) | Security/ Risk Analysis | Descriptive | Security; Customer Satisfaction | Machine/ Tool; Customer | Vertical; Horizontal | Historical/ Batch | Deep Learning | 3 |
| 641 | (Arellano-Espitia, Delgado-Prieto, Martinez-Viol, Saucedo-Dorantes, & Osornio-Rios, 2020) | Defect Analysis | Diagnostic | Sustainability | Machine/ Tool | No Integration | Real-time | Deep Learning | 4 |
| 642 | (Arpaia et al., 2020) | Security/ Risk Analysis | Predictive | Cost; Security | Human | No Integration | Real-time | Classification | 3 |
| 643 | (Ashish Kumar, Dimitrakopoulos, & Maulen, 2020) | Production Planning | Prescriptive | Time; Cost; Flexibility | ERP; Environment | End-to-End | Real-time | Classification; Probabilistic Methods; Deep Learning; Reinforcement Learning; Mathematical Optimization; Custom Development | 2 |

| No. | Reference | Function | Maturity | Objective | Data Source | Integration | Frequency | Method | Cluster |
|---|---|---|---|---|---|---|---|---|---|
| 644 | (Aydemir & Paynabar, 2019) | Condition Analysis | Predictive | Time; Cost | Machine/ Tool | No Integration | Historical/ Batch | Regression; Deep Learning | 6 |
| 645 | (B. Chen et al., 2019) | Performance Opt. | Descriptive | Time; Cost; Flexibility | Machine/ Tool | No Integration | Real-time | Deep Learning; Reinforcement Learning | 3 |
| 646 | (B. Yang, Cao, Li, Zhang, & Qian, 2019) | Condition Analysis | Prescriptive | Time; Conformance | Process | No Integration | Historical/ Batch | Custom Development | 2 |
| 647 | (B.-A. Han & Yang, 2020) | Production Planning | Prescriptive | Time; Cost; Flexibility | ERP | No Integration | Historical/ Batch | Deep Learning; Reinforcement Learning | 2 |
| 648 | (Barde, Yacout, & Shin, 2019) | Maintenance Planning | Prescriptive | Time; Cost | Machine/ Tool; Process | No Integration | Real-time | Probabilistic Methods; Reinforcement Learning; Mathematical Optimization | 2 |
| 649 | (Bowler, Bakalis, & Watson, 2020) | Monitoring | Predictive | Cost | Reference; Environment | No Integration | Real-time | Classification; Deep Learning | 6 |
| 650 | (C. Xu & Zhu, 2020) | Quality Control | Descriptive | Conformance | Product | Horizontal | Real-time | Classification | 1 |
| 651 | (C.-C. Lin, Deng, Chih, & Chiu, 2019) | Production Planning | Prescriptive | Time; Cost | Machine/ Tool | No Integration | Historical/ Batch | Deep Learning; Reinforcement Learning | 2 |
| 652 | (C.-C. Lin, Deng, Kuo, & Chen, 2019) | Condition Analysis | Predictive | Time; Cost; Sustainability | Machine/ Tool | No Integration | Historical/ Batch | Classification; Regression | 6 |
| 653 | (C.-F. Lai, Chien, Yang, & Qiang, 2019) | Monitoring | Prescriptive | Cost; Conformance | Machine/ Tool | No Integration | Historical/ Batch | Classification | 1 |
| 654 | (C.-M. Kuo, Chen, Tseng, & Kao, 2020) | Design Analysis | Predictive | Customer Satisfaction | Customer; Human | No Integration | Historical/ Batch | Custom Development | 6 |
| 655 | (C.-Y. Wang, Chen, & Chien, 2020) | Performance Opt. | Predictive | Cost | Product | No Integration | Historical/ Batch | Clustering; Mathematical Optimization | 6 |
| 656 | (Cakir, Guvenc, & Mistikoglu, 2020) | Condition Analysis | Predictive | Cost | Machine/ Tool; Environment | No Integration | Real-time | Classification | 4 |
| 657 | (Carvajal Soto, Tavakolizadeh, & Gyulai, 2019) | Quality Control | Predictive | Time; Conformance; Flexibility | Product | No Integration | Real-time | Classification; Deep Learning | 1 |
| 658 | (Chang Liu, Li, Tang, Lin, & Liu, 2019) | Production Planning | Prescriptive | Time; Cost | Process; Customer | Horizontal | Real-time | Reinforcement Learning; Mathematical Optimization | 2 |
| 659 | (Chang Liu, Tang, & Liu, 2019) | Quality Opt. | Prescriptive | Conformance; Security | Process | Horizontal | Historical/ Batch | Regression | 3 |
| 660 | (Chang Liu, Tang, Liu, & Tang, 2018) | Quality Control | Prescriptive | Time; Cost; Conformance | Process | No Integration | Real-time | Regression; Dimension Reduction | 2 |
| 661 | (Che, Liu, Che, & Lang, 2020) | Production Planning | Prescriptive | Time; Cost | Process | No Integration | Historical/ Batch | Mathematical Optimization | 2 |
| 662 | (Cheol Young Park, Kim, Kim, & Lee, 2020) | Quality Opt. | Predictive | Cost; Conformance | Machine/ Tool | No Integration | Real-time | Regression; Clustering | 4 |
| 663 | (Chong Chen, Liu, Kumar, Qin, & Ren, 2019) | Energy Cons. Analysis | Predictive | Cost; Sustainability | Reference | No Integration | Historical/ Batch | Deep Learning | 6 |
| 664 | (Chouliaras & Sotiriadis, 2019) | Monitoring | Predictive | Cost; Conformance | Process | Horizontal | Real-time | Deep Learning | 6 |
| 665 | (Chu, Xie, Wu, Guo, & Yau, 2020) | Quality Opt. | Descriptive | Cost; Conformance; Customer Satisfaction | Reference | No Integration | Historical/ Batch | Regression; Evolutional Algorithm; Multi-Agent System | 1 |
| 666 | (Cui, Ren, Wang, & Zhang, 2019) | Condition Analysis | Predictive | Time; Cost; Security | Machine/ Tool | No Integration | Real-time | Regression; Deep Learning; Custom Development | 4 |
| 667 | (D. B. Kim, 2019) | Performance Opt. | Prescriptive | Conformance; Flexibility | Machine/ Tool; Process; Product | Vertical | Historical/ Batch | Custom Development | 2 |
| 668 | (D. Wu et al., 2019) | Monitoring | Descriptive | Time | Machine/ Tool | No Integration | Real-time | Classification; Probabilistic Methods; Deep Learning | 3 |
| 669 | (Dai, Wang, Huang, Shi, & Zhu, 2020) | Condition Analysis | Descriptive | Time; Cost | Machine/ Tool | No Integration | Historical/ Batch | Dimension Reduction; Deep Learning | 5 |
| 670 | (Dan et al., 2020) | Quality Opt. | Predictive | Time; Cost; Conformance | Machine/ Tool | No Integration | Historical/ Batch | Dimension Reduction; Deep Learning | 1 |
| 671 | (de Farias, de Almeida, Delijaicov, | Condition Analysis | Predictive | Time; Cost | Machine/ Tool | No Integration | Historical/ Batch | Deep Learning | 6 |

| No. | Reference | Function | Maturity | Objective | Data Source | Integration | Frequency | Method | Cluster |
|---|---|---|---|---|---|---|---|---|---|
| | Seriacopi, & Bordinassi, 2020) | | | | | | | | |
| 672 | (de Sa, Carmo, & Machado, 2017) | Security/ Risk Analysis | Diagnostic | Security | ERP | No Integration | Real-time | Mathematical Optimization; Swarm Intelligence | 3 |
| 673 | (Demertzis, Iliadis, Tziritas, & Kikiras, 2020) | Security/ Risk Analysis | Predictive | Security | Process | Vertical | Real-time | Classification; Deep Learning; Custom Development | 3 |
| 674 | (F. Deng et al., 2020) | Monitoring | Prescriptive | Time; Cost | Process | No Integration | Historical/ Batch | Classification; Deep Learning | 2 |
| 675 | (Denkena, Bergmann, & Witt, 2019) | Performance Opt. | Predictive | Conformance; Flexibility | Machine/ Tool; Process | No Integration | Historical/ Batch | Classification; Dimension Reduction; Deep Learning | 1 |
| 676 | (Dimitriou et al., 2019) | Quality Control | Predictive | Time; Cost; Conformance | Machine/ Tool | No Integration | Historical/ Batch | Classification; Deep Learning | 1 |
| 677 | (Doltsinis, Krestenitis, & Doulgeri, 2019) | Quality Control | Predictive | Time; Cost; Conformance | Machine/ Tool | No Integration | Real-time | Classification | 4 |
| 678 | (Dong Sun et al., 2019) | Production Planning | Diagnostic | Time; Cost | ERP | End-to-End | Historical/ Batch | Clustering | 6 |
| 679 | (Elgendi, Hossain, Jamalipour, & Munasinghe, 2019) | Security/ Risk Analysis | Prescriptive | Cost; Security | Process | Vertical | Historical/ Batch | Classification | 3 |
| 680 | (Elsheikh, Yacout, Ouali, & Shaban, 2020) | Condition Analysis | Predictive | Time; Cost; Security | Machine/ Tool | No Integration | Real-time | Classification; Regression; Custom Development | 4 |
| 681 | (Epureanu, Li, Nassehi, & Koren, 2020) | Maintenance Planning | Prescriptive | Time; Cost; Flexibility | Machine/ Tool; Product; ERP | Vertical | Historical/ Batch | Reinforcement Learning | 2 |
| 682 | (Essien & Giannetti, 2020) | Production Planning | Predictive | Time; Cost | Machine/ Tool | End-to-End | Historical/ Batch | Dimension Reduction; Deep Learning | 6 |
| 683 | (Ezeme, Mahmoud, & Azim, 2019) | Monitoring | Diagnostic | Time; Security | Machine/ Tool; Process | No Integration | Historical/ Batch | Deep Learning | 5 |
| 684 | (Faraci, Raciti, Rizzo, & Schembra, 2020) | Energy Cons. Opt. | Prescriptive | Time; Cost; Sustainability | Machine/ Tool | No Integration | Historical/ Batch | Reinforcement Learning | 6 |
| 685 | (Farivar, Haghighi, Jolfaei, & Alazab, 2019) | Security/ Risk Analysis | Predictive | Security | Process | No Integration | Historical/ Batch | Deep Learning | 3 |
| 686 | (Foresti, Rossi, Magnani, Bianco, & Delmonte, 2020) | Maintenance Planning | Predictive | Time; Cost; Conformance; Sustainability | ERP; Human | Vertical | Real-time | Custom Development | 3 |
| 687 | (Frumosu et al., 2020) | Quality Control | Predictive | Time; Cost; Conformance | Process; Reference | No Integration | Historical/ Batch | Classification; Evolutional Algorithm | 1 |
| 688 | (G. G. Rodríguez, Gonzalez-Cava, & Pérez, 2019) | Production Planning | Predictive | Time; Cost | Process; ERP; Environment | End-to-End | Real-time | Fuzzy Logic | 6 |
| 689 | (Gang Wang, Nixon, & Boudreaux, 2019) | Condition Analysis | Predictive | Time | Machine/ Tool | End-to-End | Real-time | Regression | 4 |
| 690 | (Gao, Zhang, Chen, Dong, & Vucetic, 2018) | Energy Cons. Opt. | Predictive | Cost | Machine/ Tool | Horizontal | Real-time | Mathematical Optimization | 6 |
| 691 | (Genge, Haller, & Enăchescu, 2019) | Security/ Risk Analysis | Predictive | Security | Machine/ Tool | No Integration | Historical/ Batch | Custom Development | 3 |
| 692 | (Ghahramani, Qiao, Zhou, Hagan, & Sweeney, 2020) | Performance Opt. | Predictive | Time; Cost | Process | Vertical | Historical/ Batch | Dimension Reduction; Deep Learning | 6 |
| 693 | (Gou et al., 2019) | Condition Analysis | Predictive | Time; Cost; Security | Machine/ Tool | No Integration | Real-time | Mathematical Optimization | 4 |
| 694 | (Grzenda & Bustillo, 2019) | Quality Opt. | Predictive | Time; Cost; Conformance | Machine/ Tool | No Integration | Real-time | Classification; Regression; Custom Development | 4 |
| 695 | (H. Hu et al., 2020) | Production Planning | Prescriptive | Time; Flexibility | Machine/ Tool; Process | Vertical | Real-time | Reinforcement Learning | 2 |
| 696 | (H. Huang et al., 2019) | Defect Analysis | Predictive | Time; Cost | Machine/ Tool | No Integration | Real-time | Classification; Deep Learning | 4 |
| 697 | (H. Wang, Li, Song, Cui, & Wang, 2019) | Defect Analysis | Predictive | Cost | Machine/ Tool; Product | No Integration | Historical/ Batch | Clustering; Deep Learning | 6 |
| 698 | (H. Wang, Ren, Song, & Cui, 2019) | Defect Analysis | Predictive | Time; Cost | Machine/ Tool | No Integration | Historical/ Batch | Classification | 6 |
| 699 | (H. Xu, Liu, Yu, Griffith, & Golmie, 2020) | Performance Opt. | Prescriptive | Flexibility | Reference; Environment | Vertical | Historical/ Batch | Reinforcement Learning | 2 |
| 700 | (H. Yang, Alphones, Zhong, Chen, & Xie, 2019) | Energy Cons. Opt. | Prescriptive | Cost; Sustainability | Machine/ Tool | No Integration | Historical/ Batch | Deep Learning; | 6 |

| No. | Reference | Function | Maturity | Objective | Data Source | Integration | Frequency | Method | Cluster |
|---|---|---|---|---|---|---|---|---|---|
| | | | | | | | | Reinforcement Learning | |
| 701 | (H. Yao et al., 2019) | Security/ Risk Analysis | Descriptive | Security | Machine/ Tool | No Integration | Historical/ Batch | Custom Development | 3 |
| 702 | (H.-K. Wang & Chien, 2020) | Quality Opt. | Prescriptive | Cost; Conformance | Process; Reference | No Integration | Historical/ Batch | Evolutional Algorithm | 1 |
| 703 | (Halawa et al., 2020) | Security/ Risk Analysis | Descriptive | Security | Process | End-to-End | Real-time | Custom Development | 3 |
| 704 | (Hassan, Gumaei, Huda, & Almogren, 2020) | Security/ Risk Analysis | Descriptive | Security; Customer Satisfaction | Reference | End-to-End | Historical/ Batch | Classification; Regression | 3 |
| 705 | (I. Rodríguez et al., 2020) | Production Planning | Prescriptive | Time; Cost; Flexibility | Process | No Integration | Historical/ Batch | Classification | 2 |
| 706 | (Iannino et al., 2020) | Performance Opt. | Predictive | Time; Cost | Process | Horizontal | Real-time | Multi-Agent System | 3 |
| 707 | (Ismail, Idris, Ayub, & Yee, 2019) | Quality Control | Predictive | Time; Conformance | Product; Reference | No Integration | Historical/ Batch | Classification | 1 |
| 708 | (J. Chen, Zhang, & Wu, 2020) | Quality Control | Predictive | Time; Cost; Conformance | Machine/ Tool; Reference | No Integration | Real-time | Classification; Deep Learning | 4 |
| 709 | (J. Feng, Li, Xu, & Zhong, 2018) | Quality Control | Descriptive | Time; Conformance | Process | No Integration | Historical/ Batch | Classification | 1 |
| 710 | (J. Jiao, Lin, Zhao, & Liang, 2020) | Defect Analysis | Predictive | Time; Cost; Security | Machine/ Tool | No Integration | Real-time | Classification; Deep Learning; Reinforcement Learning; Multi-Agent System; Custom Development | 4 |
| 711 | (J. Li, Xu, Gao, Wang, & Shao, 2020) | Quality Control | Predictive | Conformance | Machine/ Tool | No Integration | Real-time | Classification; Dimension Reduction; Deep Learning; Reinforcement Learning; Multi-Agent System; Custom Development | 4 |
| 712 | (J. Liu et al., 2020) | Security/ Risk Analysis | Descriptive | Security | Machine/ Tool | Vertical; Horizontal | Historical/ Batch | Probabilistic Methods | 3 |
| 713 | (J. Luo, Chen, Yu, & Tang, 2020) | Energy Cons. Opt. | Prescriptive | Time; Cost | Process | Vertical | Historical/ Batch | Deep Learning; Reinforcement Learning | 2 |
| 714 | (Jha, Babiceanu, & Seker, 2019) | Production Planning | Prescriptive | Time; Cost; Flexibility | Process; ERP | End-to-End | Real-time | Mathematical Optimization | 2 |
| 715 | (Jing Yang, Li, Wang, & Yang, 2019) | Quality Control | Predictive | Time; Cost; Conformance | Machine/ Tool | No Integration | Real-time | Deep Learning | 4 |
| 716 | (Jinjiang Wang, Yan, Li, Gao, & Zhao, 2019) | Condition Analysis | Predictive | Time; Cost | Machine/ Tool | No Integration | Historical/ Batch | Deep Learning | 6 |
| 717 | (Jinjiang Wang, Ye, Gao, Li, & Zhang, 2019) | Defect Analysis | Predictive | Time; Cost | Machine/ Tool | No Integration | Real-time | Mathematical Optimization; Custom Development | 3 |
| 718 | (Jomthanachai, Rattanamanee, Sinthavalai, & Wong, 2020) | Quality Opt. | Diagnostic | Time; Cost; Conformance | Product | No Integration | Historical/ Batch | Evolutional Algorithm | 1 |
| 719 | (Jonghyuk Kim & Hwangbo, 2019) | Monitoring | Predictive | Time; Sustainability | Process | No Integration | Real-time | Custom Development | 3 |
| 720 | (K. H. Sun et al., 2020) | Defect Analysis | Diagnostic | Time; Cost | Machine/ Tool | No Integration | Historical/ Batch | Deep Learning | 5 |
| 721 | (K. T. Park et al., 2020) | Energy Cons. Opt. | Predictive | Sustainability | Machine/ Tool; Process; ERP; Environment | End-to-End | Real-time | Classification; Regression; Deep Learning | 6 |
| 722 | (K. Wang et al., 2020) | Energy Cons. Opt. | Prescriptive | Cost | Process | Vertical | Real-time | Reinforcement Learning | 2 |
| 723 | (K. Zhu & Lin, 2019) | Condition Analysis | Predictive | Cost | Machine/ Tool | No Integration | Historical/ Batch | Classification; Dimension Reduction | 6 |
| 724 | (K. Zhu, Li, & Zhang, 2019) | Condition Analysis | Predictive | Time; Cost | Machine/ Tool; Process | No Integration | Real-time | Deep Learning | 4 |
| 725 | (Kabugo, Jämsä-Jounela, Schiemann, & Binder, 2020) | Monitoring | Predictive | Time; Cost | Product | End-to-End | Real-time | Deep Learning | 6 |
| 726 | (Kai Ding, Zhang, Chan, Chan, & Wang, 2019) | Production Planning | Prescriptive | Time; Cost | Machine/ Tool | No Integration | Historical/ Batch | Probabilistic Methods | 2 |
| 727 | (Kazi, Eljack, & Mahdi, 2020) | Quality Control | Predictive | Conformance | Product; Reference | No Integration | Historical/ Batch | Deep Learning | 1 |

| No. | Reference | Function | Maturity | Objective | Data Source | Integration | Frequency | Method | Cluster |
|---|---|---|---|---|---|---|---|---|---|
| 728 | (Ke, Chen, Chen, Wang, & Zhang, 2020) | Production Planning | Prescriptive | Time | ERP | No Integration | Historical/ Batch | Swarm Intelligence | 2 |
| 729 | (Khoda, Imam, Kamruzzaman, Gondal, & Rahman, 2019) | Security/ Risk Analysis | Predictive | Security | Reference | No Integration | Historical/ Batch | Classification | 3 |
| 730 | (Kiangala & Wang, 2020) | Condition Analysis | Predictive | Cost; Security | Machine/ Tool; Environment | No Integration | Historical/ Batch | Classification; Dimension Reduction; Deep Learning | 6 |
| 731 | (König & Helmi, 2020) | Condition Analysis | Diagnostic | Time; Cost | Machine/ Tool | No Integration | Real-time | Deep Learning | 5 |
| 732 | (Konstantakopoulos et al., 2019) | Energy Cons. Opt. | Predictive | Sustainability; Customer Satisfaction | Machine/ Tool | No Integration | Historical/ Batch | Deep Learning | 6 |
| 733 | (Krishnamurthy, Karri, & Khorrami, 2019) | Monitoring | Diagnostic | Time; Cost; Security | Machine/ Tool | Vertical | Real-time | Classification | 3 |
| 734 | (L. Hu, Miao, Wu, Hassan, & Humar, 2019) | Performance Opt. | Prescriptive | Time; Flexibility; Customer Satisfaction | Machine/ Tool; Product; Customer | Horizontal | Real-time | Deep Learning; Custom Development | 2 |
| 735 | (L. Jin, Zhang, Wen, & Christopher, 2020) | Production Planning | Prescriptive | Time; Cost | Process | Vertical | Historical/ Batch | Evolutional Algorithm | 2 |
| 736 | (L. Song et al., 2019) | Quality Opt. | Predictive | Cost; Conformance | Machine/ Tool | No Integration | Historical/ Batch | Classification | 1 |
| 737 | (Latif, Zou, Idrees, & Ahmad, 2020) | Security/ Risk Analysis | Predictive | Security | Machine/ Tool | No Integration | Historical/ Batch | Classification; Deep Learning | 3 |
| 738 | (Lei Ren, Meng, Wang, Lu, & Yang, 2020) | Quality Control | - | Cost; Conformance | Process | No Integration | Historical/ Batch | Dimension Reduction | 1 |
| 739 | (Lenz, MacDonald, Harik, & Wuest, 2020) | Product Lifecycle Opt. | Descriptive | Time; Cost; Conformance | Machine/ Tool; Product | Horizontal | Real-time | Custom Development | 3 |
| 740 | (Liao et al., 2019) | Performance Opt. | Prescriptive | Time; Cost | Machine/ Tool | No Integration | Historical/ Batch | Mathematical Optimization | 2 |
| 741 | (Linlin Li, Wang, Wang, & Tang, 2019) | Energy Cons. Opt. | Prescriptive | Cost; Sustainability | Process | Vertical | Historical/ Batch | Mathematical Optimization; Evolutional Algorithm | 2 |
| 742 | (Lithoxoidou et al., 2020) | Condition Analysis | Predictive | Time; Cost; Security | Machine/ Tool | No Integration | Real-time | Classification; Clustering | 4 |
| 743 | (Lolli et al., 2019) | Production Planning | Diagnostic | Time; Cost; Flexibility | Reference | Horizontal | Historical/ Batch | Classification | 2 |
| 744 | (Long Wen, Gao, & Li, 2017) | Quality Control | Predictive | Conformance | Machine/ Tool | No Integration | Historical/ Batch | Deep Learning | 1 |
| 745 | (Longo, Nicoletti, & Padovano, 2019) | Security/ Risk Analysis | Descriptive | Security | Process; Human | No Integration | Historical/ Batch | Custom Development | 3 |
| 746 | (M. Chen, 2019) | Performance Opt. | Predictive | Cost; Customer Satisfaction | Customer | No Integration | Historical/ Batch | Classification | 6 |
| 747 | (M. T. Nguyen, Truong, Tran, & Chien, 2020) | Monitoring | Descriptive | Time; Cost | Process; Environment | Vertical | Real-time | Dimension Reduction | 3 |
| 748 | (M. Wu & Moon, 2019) | Security/ Risk Analysis | Predictive | Cost; Security | Machine/ Tool; Process; ERP | End-to-End | Historical/ Batch | Classification | 3 |
| 749 | (M. Wu, Song, & Moon, 2019) | Security/ Risk Analysis | Descriptive | Security | Reference | No Integration | Real-time | Classification | 3 |
| 750 | (M. Zhang et al., 2019) | Security/ Risk Analysis | Diagnostic | Security | Machine/ Tool | No Integration | Historical/ Batch | Custom Development | 3 |
| 751 | (Maggipinto, Beghi, McLoone, & Susto, 2019) | Design Analysis | Predictive | Time; Cost | Reference; Environment | No Integration | Historical/ Batch | Deep Learning | 6 |
| 752 | (Malaca, Rocha, Gomes, Silva, & Veiga, 2019) | Quality Control | Predictive | Time; Cost; Conformance; Flexibility | Machine/ Tool; Reference; Environment; Human | No Integration | Real-time | Classification; Dimension Reduction; Deep Learning | 4 |
| 753 | (Martinek & Krammer, 2019) | Performance Opt. | Predictive | Time; Cost; Conformance | Machine/ Tool; Process | No Integration | Historical/ Batch | Classification; Deep Learning; Fuzzy Logic | 1 |
| 754 | (Martinez, Al-Hussein, & Ahmad, 2020) | Quality Control | Diagnostic | Cost | Product | No Integration | Real-time | Classification | 1 |
| 755 | (Martínez-Arellano, Terrazas, & Ratchev, 2019) | Condition Analysis | Predictive | Cost; Conformance | Machine/ Tool | No Integration | Historical/ Batch | Classification; Deep Learning | 1 |
| 756 | (Mi et al., 2020) | Maintenance Planning | Prescriptive | Time; Cost; Security; Sustainability | Machine/ Tool | No Integration | Real-time | Mathematical Optimization | 2 |
| 757 | (Miao, Hsieh, Segura, & Wang, 2019) | Performance Opt. | Predictive | Cost | Product; Environment | No Integration | Historical/ Batch | Regression | 6 |

| No. | Reference | Function | Maturity | Objective | Data Source | Integration | Frequency | Method | Cluster |
|---|---|---|---|---|---|---|---|---|---|
| 758 | (Mikel Canizo, Conde, et al., 2019) | Monitoring | Predictive | Cost; Conformance | Machine/ Tool | Horizontal | Historical/ Batch | Deep Learning | 6 |
| 759 | (Mikel Canizo, Triguero, Conde, & Onieva, 2019) | Monitoring | Predictive | Time | Process | No Integration | Real-time | Deep Learning | 3 |
| 760 | (Min, Lu, Liu, Su, & Wang, 2019) | Performanc e Opt. | Prescriptive | Time; Cost | Machine/ Tool | No Integration | Real-time | Classification | 4 |
| 761 | (Mishra et al., 2020) | Monitoring | Predictive | Conformance | Machine/ Tool; Process | Vertical | Real-time | Custom Development | 3 |
| 762 | (Moens et al., 2020) | Condition Analysis | Predictive | Time | Machine/ Tool | Horizontal | Historical/ Batch | Classification; Deep Learning | 6 |
| 763 | (Moreira, Li, Lu, & Fitzpatrick, 2019) | Quality Control | Predictive | Conformance | Machine/ Tool; Product | No Integration | Real-time | Custom Development | 1 |
| 764 | (Mörth, Emmanouilidis, Hafner, & Schadler, 2020) | Performanc e Analysis | Descriptive | Cost | Process | End-to-End | Real-time | Custom Development | 3 |
| 765 | (Muhammad Syafrudin, Fitriyani, Alfian, & Rhee, 2019) | Monitoring | Predictive | Time; Cost; Customer Satisfaction | Process; Environment | No Integration | Real-time | Classification; Custom Development | 3 |
| 766 | (Muhammad, Hussain, Del Ser, Palade, & De Albuquerque, 2019) | Monitoring | Predictive | Cost | Environment | Vertical | Historical/ Batch | Deep Learning | 6 |
| 767 | (Mumtaz et al., 2019) | Production Planning | Prescriptive | Time; Cost | Process | Horizontal | Historical/ Batch | Mathematical Optimization; Swarm Intelligence | 2 |
| 768 | (N. Li et al., 2020) | Condition Analysis | Predictive | Time; Conformance | Machine/ Tool | No Integration | Real-time | Classification; Regression | 4 |
| 769 | (Neupane & Seok, 2020) | Defect Analysis | Descriptive | Time; Cost; Security | Machine/ Tool | No Integration | Historical/ Batch | Deep Learning | 5 |
| 770 | (Ortego et al., 2020) | Performanc e Analysis | Predictive | Time; Cost | Process | No Integration | Historical/ Batch | Classification; Deep Learning; Evolutional Algorithm | 6 |
| 771 | (P. Fang, Yang, Zheng, Zhong, & Jiang, 2020) | Monitoring | Descriptive | Conformance | Process; ERP | Horizontal | Historical/ Batch | Probabilistic Methods | 3 |
| 772 | (P. Li, Cheng, Jiang, & Katchasuwanmane e, 2020) | Monitoring | Descriptive | Cost | Machine/ Tool | No Integration | Real-time | Dimension Reduction; Custom Development | 3 |
| 773 | (P. Liu, Zhang, Wu, & Fu, 2020) | Condition Analysis | Descriptive | Time; Cost | Machine/ Tool | No Integration | Real-time | Dimension Reduction | 4 |
| 774 | (P. Peng, Zhang, Liu, Wang, & Zhang, 2019) | Defect Analysis | Diagnostic | Time; Cost | Machine/ Tool; Process | No Integration | Real-time | Dimension Reduction; Evolutional Algorithm | 4 |
| 775 | (P. Wang & Gao, 2020) | Defect Analysis | Diagnostic | Time; Cost | Machine/ Tool | No Integration | Historical/ Batch | Deep Learning | 5 |
| 776 | (Papananias, McLeay, Mahfouf, & Kadirkamanathan, 2019) | Quality Control | Predictive | Cost; Conformance | Process | Horizontal | Real-time | Regression | 1 |
| 777 | (Papananias, McLeay, Obajemu, Mahfouf, & Kadirkamanathan, 2020) | Quality Control | Predictive | Time; Conformance | Product | No Integration | Historical/ Batch | Regression; Probabilistic Methods | 1 |
| 778 | (Para, Del Ser, Nebro, Zurutuza, & Herrera, 2019) | Monitoring | Descriptive | Cost | Process; Human | Vertical | Historical/ Batch | Classification | 3 |
| 779 | (Penumuru, Muthuswamy, & Karumbu, 2019) | Quality Control | Predictive | Conformance | Reference | No Integration | Historical/ Batch | Classification; Regression | 1 |
| 780 | (Peres, Barata, Leitao, & Garcia, 2019) | Quality Control | Predictive | Conformance | Product | No Integration | Historical/ Batch | Classification | 1 |
| 781 | (Petrović, Miljković, & Jokić, 2019) | Production Planning | Prescriptive | Time; Cost; Flexibility; Sustainability | Process | No Integration | Historical/ Batch | Mathematical Optimization; Swarm Intelligence | 2 |
| 782 | (Pierezan, Maidl, Yamao, dos Santos Coelho, & Mariani, 2019) | Energy Cons. Opt. | Descriptive | Conformance; Sustainability | Machine/ Tool | No Integration | Historical/ Batch | Mathematical Optimization | 1 |
| 783 | (Pittino, Puggl, Moldaschl, & Hirschl, 2020) | Condition Analysis | Diagnostic | Time; Cost | Machine/ Tool | No Integration | Real-time | Deep Learning | 5 |
| 784 | (Plehiers et al., 2019) | Performanc e Opt. | Predictive | Time; Cost; Sustainability | Process; Product; Environment | No Integration | Historical/ Batch | Deep Learning | 6 |
| 785 | (Proto et al., 2020) | Quality Control | Predictive | Conformance; Customer Satisfaction | Process; Product; Customer; Human | Vertical; Horizontal | Real-time | Classification | 1 |
| 786 | (Q. Bi, Wang, Wu, Zhu, & Ding, 2019) | Condition Analysis | Predictive | Time; Conformance | Machine/ Tool | No Integration | Real-time | Regression; Fuzzy Logic | 4 |

| No. | Reference | Function | Maturity | Objective | Data Source | Integration | Frequency | Method | Cluster |
|---|---|---|---|---|---|---|---|---|---|
| 787 | (Q. Li et al., 2019) | Security/ Risk Analysis | Predictive | Cost; Security | Machine/ Tool | No Integration | Real-time | Classification; Regression; Clustering | 3 |
| 788 | (Q. Wang, Jiao, Wang, & Zhang, 2020) | Quality Control | Predictive | Time; Cost; Conformance | Machine/ Tool; Environment | No Integration | Real-time | Classification; Dimension Reduction | 4 |
| 789 | (R. Hao, Lu, Cheng, Li, & Huang, 2020) | Quality Control | Predictive | Conformance | Machine/ Tool | No Integration | Historical/ Batch | Deep Learning | 1 |
| 790 | (Rahman, Janardhanan, & Nielsen, 2019) | Production Planning | Descriptive | Time; Cost; Flexibility | ERP | Horizontal | Real-time | Evolutional Algorithm; Swarm Intelligence | 3 |
| 791 | (Rato & Reis, 2020) | Quality Control | Predictive | Conformance | Process; Product | Vertical | Historical/ Batch | Mathematical Optimization | 1 |
| 792 | (Rauf et al., 2020) | Production Planning | Prescriptive | Time; Cost; Flexibility | Process | Vertical | Historical/ Batch | Mathematical Optimization; Evolutional Algorithm | 2 |
| 793 | (Romeo et al., 2020) | Performanc e Opt. | Predictive | Time; Cost; Conformance; Flexibility | Machine/ Tool; Product | No Integration | Historical/ Batch | Classification; Regression | 1 |
| 794 | (Romero-Hdz, Saha, Tstutsumi, & Fincato, 2020) | Performanc e Opt. | Prescriptive | Time; Cost | Process; Reference | No Integration | Historical/ Batch | Reinforceme nt Learning; Mathematical Optimization | 2 |
| 795 | (Rossit, Tohmé, & Frutos, 2019) | Production Planning | Prescriptive | Flexibility | Machine/ Tool; Process | Vertical; Horizontal | Real-time | Multi-Agent System | 2 |
| 796 | (Ruiz-Sarmiento et al., 2020) | Condition Analysis | Predictive | Time; Cost | Machine/ Tool | No Integration | Historical/ Batch | Probabilistic Methods | 6 |
| 797 | (S. Choi, Youm, & Kang, 2019) | Monitoring | Descriptive | Time | Process | Vertical; Horizontal | Real-time | Multi-Agent System | 3 |
| 798 | (S. Li et al., 2020) | Defect Analysis | Predictive | Time; Cost | Product | No Integration | Historical/ Batch | Classification; Deep Learning | 6 |
| 799 | (S. Lin, He, & Sun, 2019) | Quality Control | Diagnostic | Conformance | Product | No Integration | Real-time | Classification; Dimension Reduction | 1 |
| 800 | (S. Ma et al., 2020) | Energy Cons. Opt. | Predictive | Cost; Sustainability | Machine/ Tool; Environment | End-to-End | Real-time | Swarm Intelligence | 6 |
| 801 | (S. S.-D. Xu, Huang, Kung, & Lin, 2019) | Performanc e Opt. | Prescriptive | Cost; Flexibility | Machine/ Tool; Environment | Horizontal | Real-time | Swarm Intelligence; Fuzzy Logic | 2 |
| 802 | (S. W. Kim, Lee, Tama, & Lee, 2020) | Monitoring | Predictive | Time; Cost | Product | No Integration | Historical/ Batch | Classification; Regression | 6 |
| 803 | (S.-T. Park, Li, & Hong, 2020) | Security/ Risk Analysis | Predictive | Security | Process | Vertical | Real-time | Classification; Regression; Clustering; Deep Learning; Custom Development | 3 |
| 804 | (Sacco, Radwan, Anderson, Harik, & Gregory, 2020) | Quality Opt. | Predictive | Cost; Conformance | Product; Reference | End-to-End | Historical/ Batch | Classification; Dimension Reduction | 1 |
| 805 | (Said, ben Abdellafou, & Taouali, 2019) | Defect Analysis | Predictive | Time; Cost; Security | Machine/ Tool; Environment | No Integration | Real-time | Regression; Dimension Reduction; Mathematical Optimization; Custom Development | 4 |
| 806 | (Salary et al., 2020) | Monitoring | Predictive | Time | Process | Horizontal | Real-time | Classification | 3 |
| 807 | (Saucedo-Dorantes, Delgado-Prieto, Osornio-Rios, & de Jesus Romero-Troncoso, 2020) | Defect Analysis | Diagnostic | Time; Cost | Machine/ Tool | No Integration | Historical/ Batch | Classification; Dimension Reduction | 5 |
| 808 | (Saúl Langarica, Rüffelmacher, & Núñez, 2019) | Defect Analysis | Predictive | Time; Cost; Security | Machine/ Tool | No Integration | Real-time | Probabilistic Methods; Dimension Reduction | 4 |
| 809 | (Scalabrini Sampaio, Vallim Filho, Santos da Silva, & Augusto da Silva, 2019) | Maintenanc e Planning | Predictive | Conformance; Flexibility; Security | Reference; ERP | End-to-End | Historical/ Batch | Custom Development | 6 |
| 810 | (Shen et al., 2020) | Condition Analysis | Predictive | Time; Cost; Conformance | Machine/ Tool | No Integration | Historical/ Batch | Regression | 6 |
| 811 | (Ståhl et al., 2019) | Quality Control | Predictive | Cost; Conformance; Customer Satisfaction | Product | No Integration | Historical/ Batch | Classification; Deep Learning | 1 |
| 812 | (Stoyanov, Ahsan, Bailey, Wotherspoon, & Hunt, 2019) | Quality Opt. | Prescriptive | Time; Cost; Conformance | Machine/ Tool; Process | Vertical | Historical/ Batch | Classification | 2 |

| No. | Reference | Function | Maturity | Objective | Data Source | Integration | Frequency | Method | Cluster |
|---|---|---|---|---|---|---|---|---|---|
| 813 | (Susto, Maggipinto, Zocco, & McLoone, 2019) | Performance Opt. | Prescriptive | Time; Cost | Process | No Integration | Historical/ Batch | Regression; Dimension Reduction | 2 |
| 814 | (T. Yu, Huang, & Chang, 2020) | Performance Opt. | Prescriptive | Time; Cost; Flexibility | Process; Human | No Integration | Historical/ Batch | Deep Learning; Reinforcement Learning | 2 |
| 815 | (Tabernik, Šela, Skvarč, & Skočaj, 2020) | Quality Control | Predictive | Time; Cost; Conformance | Machine/ Tool; Reference | No Integration | Real-time | Classification; Deep Learning | 4 |
| 816 | (Tan et al., 2019) | Production Planning | Prescriptive | Time; Cost | Process; ERP | Vertical | Historical/ Batch | Mathematical Optimization; Evolutional Algorithm | 2 |
| 817 | (Tang et al., 2020) | Defect Analysis | Predictive | Time; Cost | Machine/ Tool | No Integration | Real-time | Classification; Dimension Reduction | 4 |
| 818 | (Unnikrishnan, Donovan, Macpherson, & Tormey, 2020) | Quality Control | Descriptive | Conformance | Environment | No Integration | Real-time | Classification | 1 |
| 819 | (van Staden & Boute, 2020) | Maintenance Planning | Prescriptive | Time; Cost | Machine/ Tool; Process | No Integration | Historical/ Batch | Mathematical Optimization | 2 |
| 820 | (Veeramani, Muthuswamy, Sagar, & Zoppi, 2019) | Performance Opt. | Prescriptive | Time; Cost; Conformance; Flexibility | Machine/ Tool; Product | No Integration | Real-time | Probabilistic Methods; Reinforcement Learning; Custom Development | 2 |
| 821 | (Vukicevic et al., 2019) | Quality Control | Descriptive | Time; Cost; Conformance; Customer Satisfaction | Product | No Integration | Historical/ Batch | Custom Development | 1 |
| 822 | (W. Du, Kang, & Pecht, 2019) | Defect Analysis | Predictive | Time; Cost | Machine/ Tool | No Integration | Real-time | Classification; Deep Learning; Custom Development | 4 |
| 823 | (W. Han, Tian, Shi, Huang, & Li, 2019) | Monitoring | Predictive | Cost | Product; Environment | No Integration | Real-time | Classification; Regression; Deep Learning | 6 |
| 824 | (W. J. Lee, Mendis, Triebe, & Sutherland, 2020) | Monitoring | Predictive | Conformance | Machine/ Tool | No Integration | Real-time | Classification; Probabilistic Methods; Clustering; Dimension Reduction | 4 |
| 825 | (W. Jiang et al., 2020) | Condition Analysis | Predictive | Time; Cost; Security | Machine/ Tool | No Integration | Historical/ Batch | Regression; Deep Learning | 6 |
| 826 | (W. Li, Xie, & Wang, 2018) | Security/ Risk Analysis | Descriptive | Security | Machine/ Tool | No Integration | Historical/ Batch | Classification | 3 |
| 827 | (W. Liu, Kong, Niu, Jiang, & Zhou, 2020) | Monitoring | Descriptive | Conformance | Process | Vertical | Real-time | Custom Development | 3 |
| 828 | (W. Luo, Hu, Ye, Zhang, & Wei, 2020) | Condition Analysis | Predictive | Conformance; Security | Machine/ Tool | No Integration | Real-time | Classification; Regression; Mathematical Optimization; Custom Development | 4 |
| 829 | (W. Yu, Dillon, Mostafa, Rahayu, & Liu, 2019) | Defect Analysis | Predictive | Time; Cost | Process | No Integration | Real-time | Clustering | 6 |
| 830 | (W. Zhu, Ma, Benton, Romagnoli, & Zhan, 2019) | Defect Analysis | Diagnostic | Time; Cost; Security | Machine/ Tool | No Integration | Historical/ Batch | Deep Learning | 5 |
| 831 | (X. Fang et al., 2019) | Energy Cons. Opt. | Diagnostic | Time | Machine/ Tool | No Integration | Historical/ Batch | Custom Development | 5 |
| 832 | (X. Jia et al., 2019) | Security/ Risk Analysis | Prescriptive | Security; Sustainability | Environment | No Integration | Historical/ Batch | Classification; Probabilistic Methods; Deep Learning | 3 |
| 833 | (X. Jin, Fan, & Chow, 2018) | Defect Analysis | Predictive | Time; Cost; Flexibility | Machine/ Tool | No Integration | Historical/ Batch | Custom Development | 6 |
| 834 | (X. Jin, Que, Sun, Guo, & Qiao, 2019) | Defect Analysis | Predictive | Time; Cost | Machine/ Tool | No Integration | Historical/ Batch | Probabilistic Methods; Mathematical Optimization | 6 |
| 835 | (X. Wu, Tian, & Zhang, 2019) | Production Planning | Prescriptive | Time; Cost; Flexibility | Process | Vertical | Real-time | Custom Development | 2 |
| 836 | (X. Yan et al., 2020) | Security/ Risk Analysis | Predictive | Time; Cost; Security | Machine/ Tool | No Integration | Historical/ Batch | Classification; Regression | 3 |
| 837 | (Xiaoyu Zhang, Chen, Liu, & Xiang, 2019) | Security/ Risk Analysis | Prescriptive | Security | Customer | No Integration | Historical/ Batch | Deep Learning | 3 |
| 838 | (Xinghua Li, Xu, Vijayakumar, | Security/ Risk Analysis | Prescriptive | Security | Machine/ Tool | No Integration | Historical/ Batch | Classification; Deep Learning | 3 |

| No. | Reference | Function | Maturity | Objective | Data Source | Integration | Frequency | Method | Cluster |
|---|---|---|---|---|---|---|---|---|---|
| | Kumar, & Liu, 2020) | | | | | | | | |
| 839 | (Xingwei Xu, Tao, Ming, An, & Chen, 2020) | Condition Analysis | Diagnostic | Time; Cost; Conformance; Sustainability | Machine/ Tool | No Integration | Historical/ Batch | Deep Learning | 1 |
| 840 | (Y Zhang, Beudaert, Argandoña, Ratchev, & Munoa, 2020) | Monitoring | Predictive | Time; Cost | Machine/ Tool | Vertical | Real-time | Classification; Dimension Reduction | 3 |
| 841 | (Y. Fu, Zhou, Guo, & Qi, 2019) | Production Planning | Prescriptive | Time; Cost | Process | No Integration | Historical/ Batch | Mathematical Optimization | 2 |
| 842 | (Y. Li et al., 2020) | Production Planning | Predictive | Time; Cost; Flexibility | Process; Environment | No Integration | Historical/ Batch | Classification; Mathematical Optimization | 2 |
| 843 | (Y. Ma, Zhu, Benton, & Romagnoli, 2019) | Monitoring | Descriptive | Time | Process | No Integration | Real-time | Deep Learning; Reinforceme nt Learning | 3 |
| 844 | (Y. Song, Li, Jia, & Qiu, 2019) | Defect Analysis | Predictive | Customer Satisfaction | Machine/ Tool; Reference | No Integration | Historical/ Batch | Classification; Dimension Reduction | 6 |
| 845 | (Y. Tao, Wang, Sánchez, Yang, & Bai, 2019) | Defect Analysis | Descriptive | Time; Cost; Sustainability | Machine/ Tool | No Integration | Historical/ Batch | Classification; Regression | 5 |
| 846 | (Y. Xu, Sun, Liu, & Zheng, 2019) | Defect Analysis | Descriptive | Time; Cost | Machine/ Tool | No Integration | Historical/ Batch | Classification; Deep Learning | 5 |
| 847 | (Y. Yao, Wang, Long, Xie, & Wang, 2020) | Defect Analysis | Diagnostic | Time; Cost | Machine/ Tool; Process | No Integration | Historical/ Batch | Deep Learning | 5 |
| 848 | (Y.-C. Lin, Yeh, Chen, Liu, & Wang, 2020) | Monitoring | Predictive | Time; Cost; Flexibility; Sustainability | Machine/ Tool; Process | Vertical | Historical/ Batch | Custom Development | 6 |
| 849 | (Y.-T. Tsai et al., 2020) | Quality Opt. | Prescriptive | Time; Cost; Conformance; Flexibility | Machine/ Tool; Process; Product; Reference | No Integration | Real-time | Classification; Deep Learning; Reinforceme nt Learning; Mathematical Optimization | 2 |
| 850 | (Yacob, Semere, & Nordgren, 2019) | Quality Opt. | Predictive | Time; Conformance | Product | No Integration | Historical/ Batch | Classification | 1 |
| 851 | (Yaliang Zhao, Yang, & Sun, 2018) | Security/ Risk Analysis | Descriptive | Cost; Security | Process | Horizontal | Real-time | Clustering | 3 |
| 852 | (Yan Wang et al., 2019) | Defect Analysis | Predictive | Time; Cost | Process | No Integration | Historical/ Batch | Mathematical Optimization | 6 |
| 853 | (Yanxia Wang, Li, Gan, & Cameron, 2019) | Energy Cons. Analysis | Diagnostic | Cost; Sustainability | Machine/ Tool | No Integration | Historical/ Batch | Clustering | 6 |
| 854 | (Ying Zhao et al., 2019) | Quality Control | Descriptive | Conformance | Product | No Integration | Historical/ Batch | Clustering | 1 |
| 855 | (Yingjie Zhang, Soon, Ye, Fuh, & Zhu, 2019) | Monitoring | Diagnostic | Time; Cost; Conformance | Machine/ Tool; Process | No Integration | Historical/ Batch | Classification; Deep Learning | 5 |
| 856 | (Yun et al., 2020) | Quality Control | Predictive | Time; Cost; Conformance | Machine/ Tool; Reference | No Integration | Real-time | Classification; Dimension Reduction; Deep Learning | 4 |
| 857 | (Zenisek, Holzinger, & Affenzeller, 2019) | Condition Analysis | Predictive | Time; Cost | Machine/ Tool; Environment | No Integration | Real-time | Regression | 4 |
| 858 | (Zhe Li, Wang, & Wang, 2020) | Defect Analysis | Predictive | Time; Cost | Machine/ Tool | No Integration | Historical/ Batch | Regression; Deep Learning | 6 |
| 859 | (Zhengcai Cao, Zhou, Hu, & Lin, 2019) | Production Planning | Prescriptive | Time; Cost; Flexibility | Process; ERP | Vertical | Real-time | Probabilistic Methods; Mathematical Optimization | 2 |
| 860 | (Zhifen Zhang, Yang, Ren, & Wen, 2019) | Defect Analysis | Diagnostic | Time; Cost | Environment | No Integration | Real-time | Classification | 4 |
| 861 | (Zhifeng Liu, Chen, Zhang, Yang, & Chu, 2019) | Maintenanc e Planning | Predictive | Time; Cost | Reference | Horizontal | Real-time | Custom Development | 3 |
| 862 | (Zhixiong Li, Liu, & Wu, 2019) | Condition Analysis | Predictive | Time; Cost; Conformance | Machine/ Tool; Environment | No Integration | Real-time | Classification; Dimension Reduction | 4 |
| 863 | (Zilong Cao, Zhou, Li, Huang, & Wu, 2020) | Performanc e Opt. | Prescriptive | Time; Conformance | Machine/ Tool | Vertical | Historical/ Batch | Reinforceme nt Learning | 2 |
| 864 | (Zolanvari, Teixeira, Gupta, Khan, & Jain, 2019) | Security/ Risk Analysis | Prescriptive | Security | Machine/ Tool | No Integration | Historical/ Batch | Classification; Regression | 3 |
| 865 | (Zongxin Liu & Pu, 2019) | Product Lifecycle Opt. | Diagnostic | Conformance | Machine/ Tool; Human | No Integration | Real-time | Classification | 1 |

## Appendix B – Taxonomy Building Iterations

Iteration I. For our first iteration, we identified preliminary publications (cf. Section 2.5), which aim at a holistic view, but primarily for specific questions or areas (e.g., business potentials or ML) (Bang et al., 2019; Bordeleau et al., 2018; Y. Cheng, K. Chen, et al., 2018; Diez-Olivan et al., 2019; Fay & Kazantsev, 2018; Gölzer et al., 2015; Gölzer & Fritzsche, 2017; O'Donovan et al., 2015a; Sharp et al., 2018; X. Y. Xu & Hua, 2017). We used this knowledge to employ categorization schemes to derive an initial set of dimensions and characteristics.

Iteration II. In the second iteration, we focused on domain-specific preliminary publications (cf. again Section 2.5). The prerequisite for the analysis is that the respective authors prepared their data in a taxonomic or categorical fashion. We identified 16 relevant contributions (Baum et al., 2018; Bousdekis, Magoutas, Mentzas, et al., 2015; Çaliş & Bulkan, 2015; Cardin et al., 2017; Cerrada et al., 2018; Khan & Yairi, 2018; D.-H. Kim et al., 2018; G. Y. Lee et al., 2018; J. Lee, Wu, et al., 2014; Priore et al., 2014; M. S. Reis & Gins, 2017; Y. Xu et al., 2017; Zarandi et al., 2018a; G. Zhao et al., 2016; Y. Zhou & Xue, 2018; Zschech, 2018). We extracted our categorization from each related work and compared it with the taxonomy from the first iteration. Based on this, we added, divided, or merged dimensions and characteristics.

Iteration III. Due to the extensive structuring of the provisional taxonomy, in this iteration we decided to switch to an empirical-to-conceptual approach. We identified a total of 633 articles, which consider the use of BA within a specific smart manufacturing application case (cf. Section 4.1). We ensured possible modifications of the taxonomy and a possible post-validation by splitting the data. With a random share of 30 % ($n$=189) of the search results for each group, we were confirmed or even supplemented dimensions and characteristics.

Iteration IV. In the fourth iteration, we again selected a random share of 30 % ($n$=189) of the search results from the remaining 446 objects. Again, the objective was the validation or extension of dimensions and characteristics using the empirical-to-conceptual approach. However, the result revealed that no further modification was necessary. All closing conditions were satisfied, and we considered the development of the initial taxonomy complete.

Iteration V. To reconfirm our taxonomy's general structure with new research identified in our second literature survey, we decided to perform another conceptual-to-empirical approach. We identified seven further survey papers that prepared their results in a taxonomic or categorical fashion. None take a holistic approach (see Iteration I), they are all domain-specific (see Iteration II). Taking this into account, the articles did not discuss a domain, which was not yet included in our initial taxonomy. Most authors address industrial maintenance, mostly focusing on surveying analytics techniques (W. Zhang et al., 2019) others additionally take integration, data (Dalzochio et al., 2020) or specific maintenance functions into account (Zonta et al., 2020). H. Ding et al. (2020) employ artificial intelligence algorithms as a starting point and map these to specific use cases in maintenance and quality control. Finally, Cadavid et al. (2020) focus on production planning, but also include functions such as maintenance and product design. As the research is domain-specific, it partly addresses dimensions and characteristics on a lower granularity level. Our current taxonomy addresses these on a higher level with regard to the ending conditions (EC2.1) *concise* and (EC2.3) *comprehensive*, we did not alter the initial taxonomy.

Iteration VI. As we identified more objects on our second survey, ending condition (EC1.1) was not met anymore. Hence, we switched to an empirical-to-conceptual approach, following our procedure in the third and fourth iteration. We analyzed the 232 objects, by mapping them according to our current taxonomy. The results revealed that no further dimensions or characteristics were identifiable and all ending conditions were satisfied, as the research team was able to tag all objects successfully according to the existing dimensions and characteristics. In conclusion, the second literature search confirmed our initial taxonomy.

# Appendix C – Analysis of Temporal Variations and Trends

## C.1 Temporal Variations per Dimension

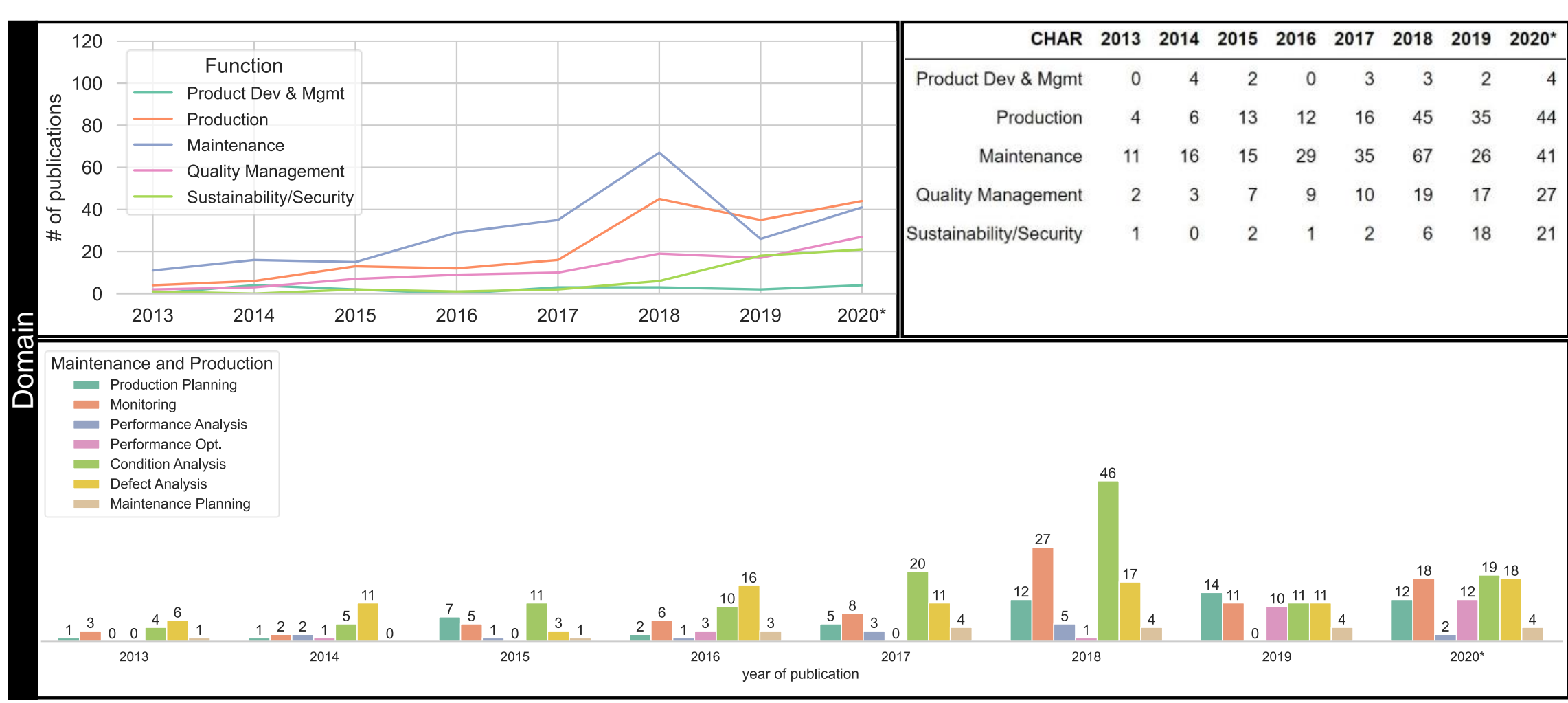

| CHAR | 2013 | 2014 | 2015 | 2016 | 2017 | 2018 | 2019 | 2020* |
|---|---|---|---|---|---|---|---|---|
| Product Dev & Mgmt | 0 | 4 | 2 | 0 | 3 | 3 | 2 | 4 |
| Production | 4 | 6 | 13 | 12 | 16 | 45 | 35 | 44 |
| Maintenance | 11 | 16 | 15 | 29 | 35 | 67 | 26 | 41 |
| Quality Management | 2 | 3 | 7 | 9 | 10 | 19 | 17 | 27 |
| Sustainability/Security | 1 | 0 | 2 | 1 | 2 | 6 | 18 | 21 |

**Figure 2: Temporal Variations of the Dimension *Domain***

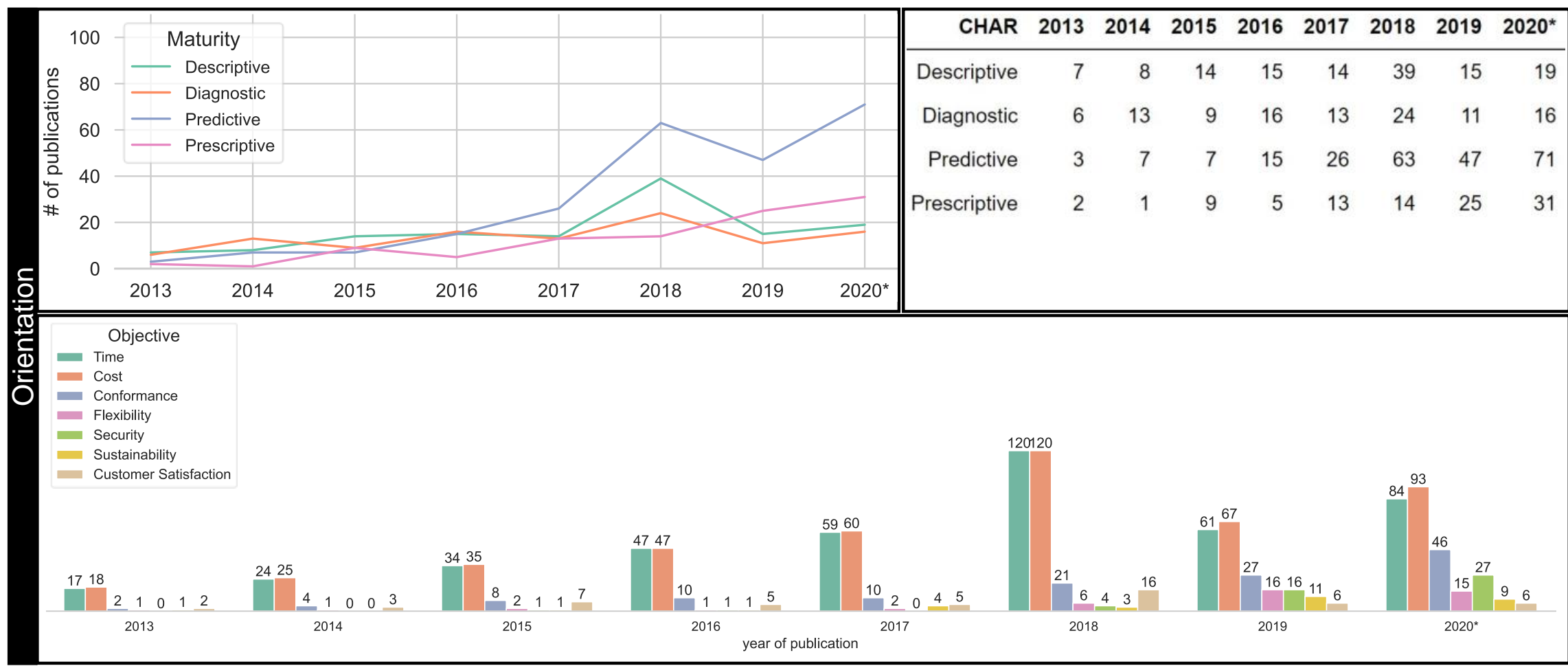

| CHAR | 2013 | 2014 | 2015 | 2016 | 2017 | 2018 | 2019 | 2020* |
|---|---|---|---|---|---|---|---|---|
| Descriptive | 7 | 8 | 14 | 15 | 14 | 39 | 15 | 19 |
| Diagnostic | 6 | 13 | 9 | 16 | 13 | 24 | 11 | 16 |
| Predictive | 3 | 7 | 7 | 15 | 26 | 63 | 47 | 71 |
| Prescriptive | 2 | 1 | 9 | 5 | 13 | 14 | 25 | 31 |

**Figure 3: Temporal Variations of the Dimension *Orientation***

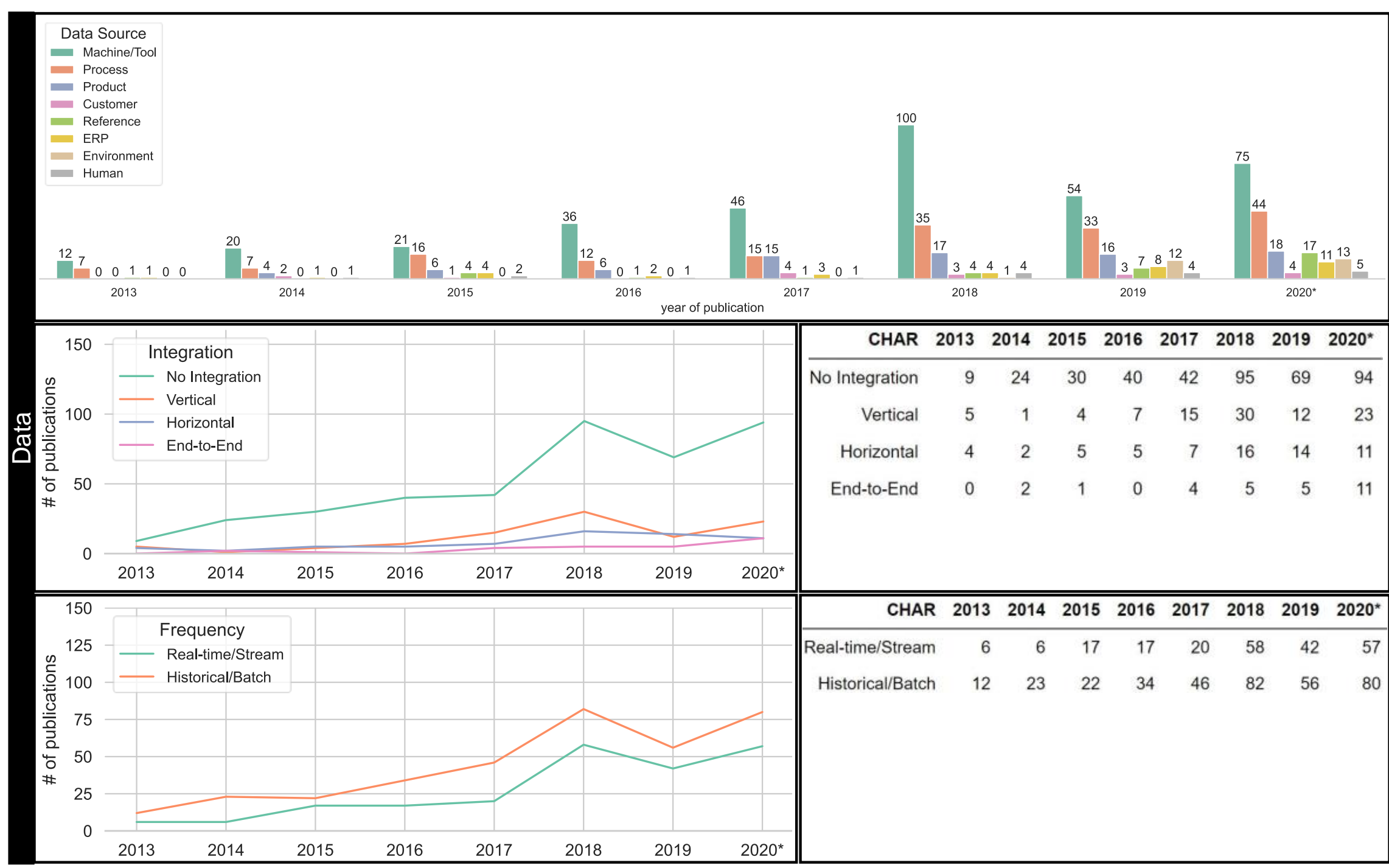

| CHAR | 2013 | 2014 | 2015 | 2016 | 2017 | 2018 | 2019 | 2020* |
|---|---|---|---|---|---|---|---|---|
| No Integration | 9 | 24 | 30 | 40 | 42 | 95 | 69 | 94 |
| Vertical | 5 | 1 | 4 | 7 | 15 | 30 | 12 | 23 |
| Horizontal | 4 | 2 | 5 | 5 | 7 | 16 | 14 | 11 |
| End-to-End | 0 | 2 | 1 | 0 | 4 | 5 | 5 | 11 |

| CHAR | 2013 | 2014 | 2015 | 2016 | 2017 | 2018 | 2019 | 2020* |
|---|---|---|---|---|---|---|---|---|
| Real-time/Stream | 6 | 6 | 17 | 17 | 20 | 58 | 42 | 57 |
| Historical/Batch | 12 | 23 | 22 | 34 | 46 | 82 | 56 | 80 |

**Figure 4: Temporal Variations of the Dimension *Data***

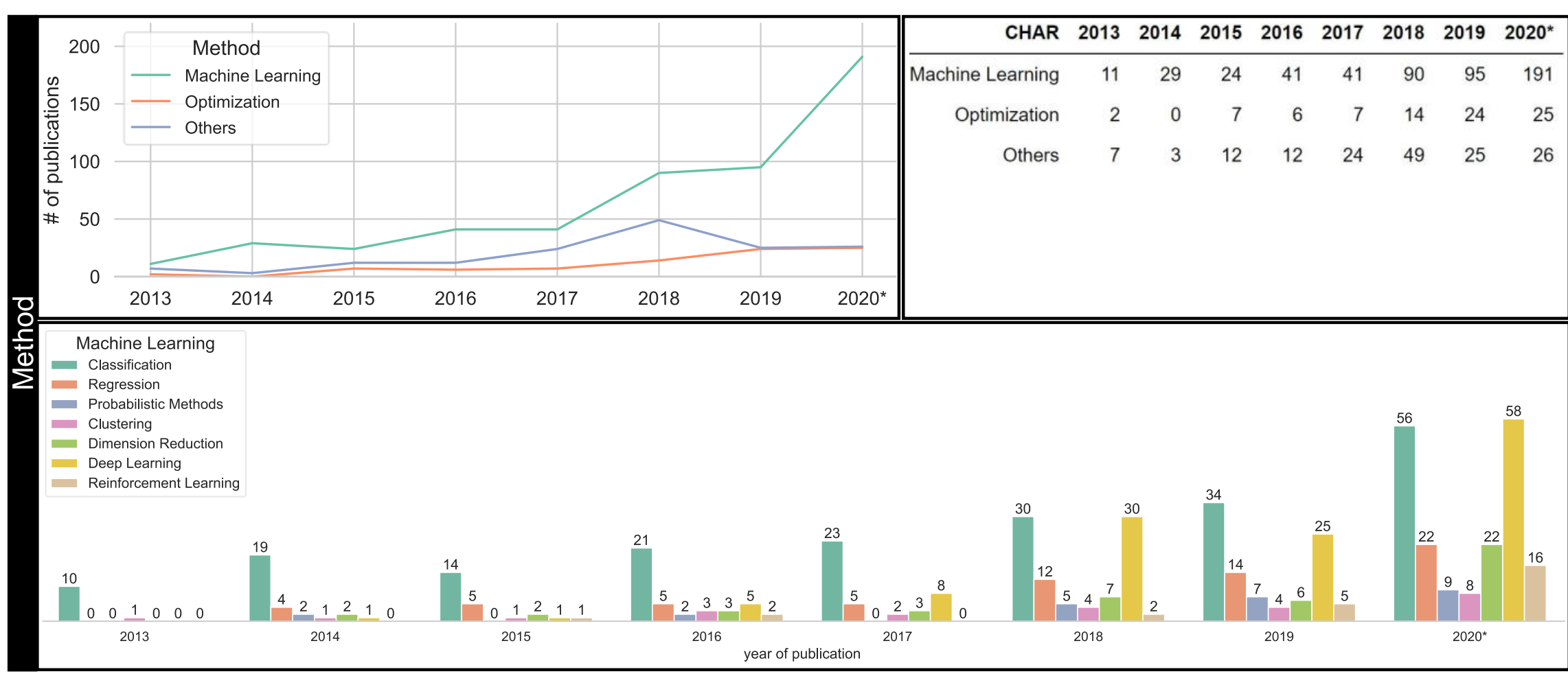

| CHAR | 2013 | 2014 | 2015 | 2016 | 2017 | 2018 | 2019 | 2020* |
|---|---|---|---|---|---|---|---|---|
| Machine Learning | 11 | 29 | 24 | 41 | 41 | 90 | 95 | 191 |
| Optimization | 2 | 0 | 7 | 6 | 7 | 14 | 24 | 25 |
| Others | 7 | 3 | 12 | 12 | 24 | 49 | 25 | 26 |

**Figure 5: Temporal Variations of the Dimension *Method***

## C.2 Temporal Variations per Archetype

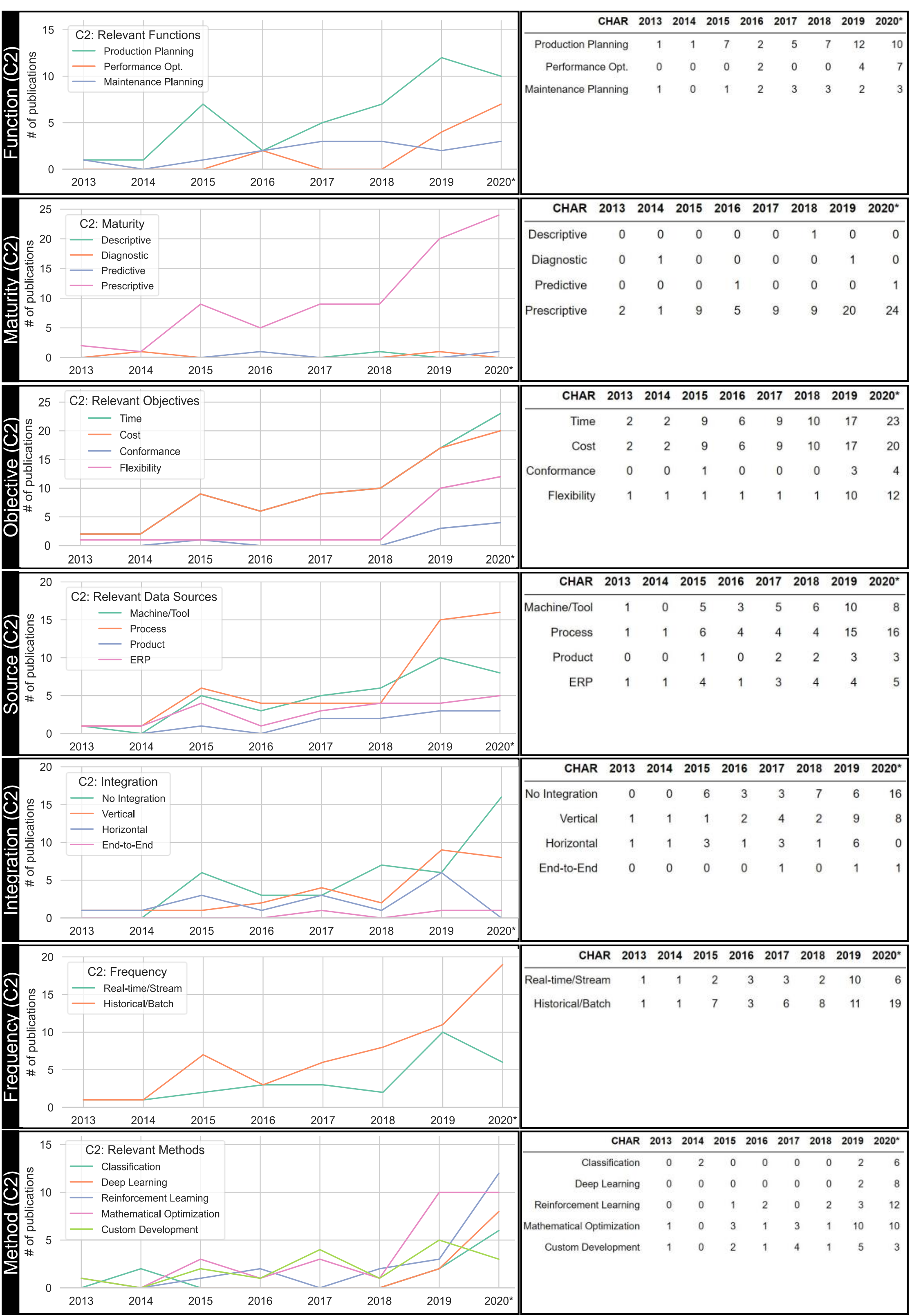

| CHAR | 2013 | 2014 | 2015 | 2016 | 2017 | 2018 | 2019 | 2020* |
|---|---|---|---|---|---|---|---|---|
| Production Planning | 1 | 1 | 7 | 2 | 5 | 7 | 12 | 10 |
| Performance Opt. | 0 | 0 | 0 | 2 | 0 | 0 | 4 | 7 |
| Maintenance Planning | 1 | 0 | 1 | 2 | 3 | 3 | 2 | 3 |

| CHAR | 2013 | 2014 | 2015 | 2016 | 2017 | 2018 | 2019 | 2020* |
|---|---|---|---|---|---|---|---|---|
| Descriptive | 0 | 0 | 0 | 0 | 0 | 1 | 0 | 0 |
| Diagnostic | 0 | 1 | 0 | 0 | 0 | 0 | 1 | 0 |
| Predictive | 0 | 0 | 0 | 1 | 0 | 0 | 0 | 1 |
| Prescriptive | 2 | 1 | 9 | 5 | 9 | 9 | 20 | 24 |

| CHAR | 2013 | 2014 | 2015 | 2016 | 2017 | 2018 | 2019 | 2020* |
|---|---|---|---|---|---|---|---|---|
| Time | 2 | 2 | 9 | 6 | 9 | 10 | 17 | 23 |
| Cost | 2 | 2 | 9 | 6 | 9 | 10 | 17 | 20 |
| Conformance | 0 | 0 | 1 | 0 | 0 | 0 | 3 | 4 |
| Flexibility | 1 | 1 | 1 | 1 | 1 | 1 | 10 | 12 |

| CHAR | 2013 | 2014 | 2015 | 2016 | 2017 | 2018 | 2019 | 2020* |
|---|---|---|---|---|---|---|---|---|
| Machine/Tool | 1 | 0 | 5 | 3 | 5 | 6 | 10 | 8 |
| Process | 1 | 1 | 6 | 4 | 4 | 4 | 15 | 16 |
| Product | 0 | 0 | 1 | 0 | 2 | 2 | 3 | 3 |
| ERP | 1 | 1 | 4 | 1 | 3 | 4 | 4 | 5 |

| CHAR | 2013 | 2014 | 2015 | 2016 | 2017 | 2018 | 2019 | 2020* |
|---|---|---|---|---|---|---|---|---|
| No Integration | 0 | 0 | 6 | 3 | 3 | 7 | 6 | 16 |
| Vertical | 1 | 1 | 1 | 2 | 4 | 2 | 9 | 8 |
| Horizontal | 1 | 1 | 3 | 1 | 3 | 1 | 6 | 0 |
| End-to-End | 0 | 0 | 0 | 0 | 1 | 0 | 1 | 1 |

| CHAR | 2013 | 2014 | 2015 | 2016 | 2017 | 2018 | 2019 | 2020* |
|---|---|---|---|---|---|---|---|---|
| Real-time/Stream | 1 | 1 | 2 | 3 | 3 | 2 | 10 | 6 |
| Historical/Batch | 1 | 1 | 7 | 3 | 6 | 8 | 11 | 19 |

| CHAR | 2013 | 2014 | 2015 | 2016 | 2017 | 2018 | 2019 | 2020* |
|---|---|---|---|---|---|---|---|---|
| Classification | 0 | 2 | 0 | 0 | 0 | 0 | 2 | 6 |
| Deep Learning | 0 | 0 | 0 | 0 | 0 | 0 | 2 | 8 |
| Reinforcement Learning | 0 | 0 | 1 | 2 | 0 | 2 | 3 | 12 |
| Mathematical Optimization | 1 | 0 | 3 | 1 | 3 | 1 | 10 | 10 |
| Custom Development | 1 | 0 | 2 | 1 | 4 | 1 | 5 | 3 |

**Figure 6: Temporal Variations of the Archetype *MRO Planning* (C2)**

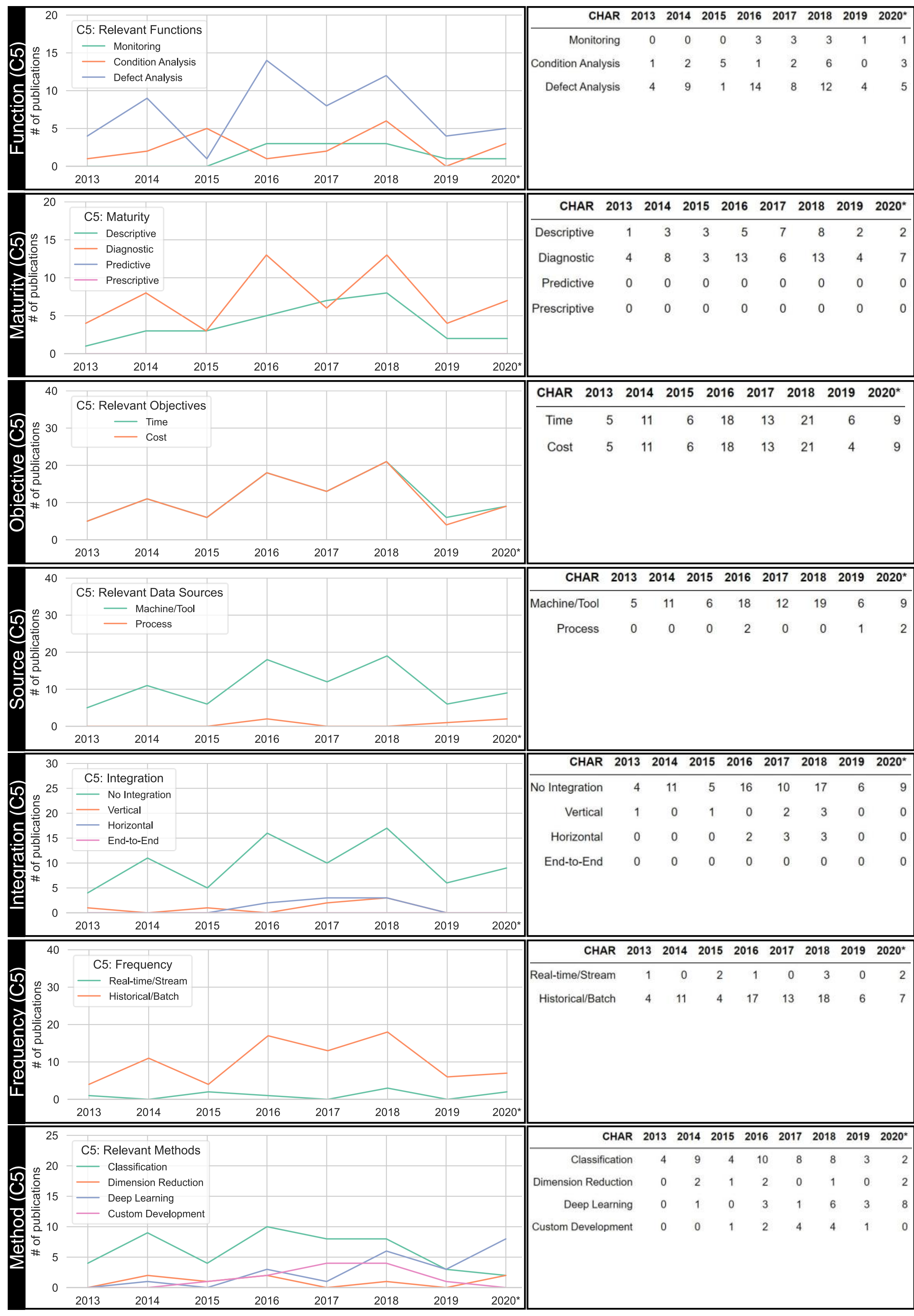

| CHAR | 2013 | 2014 | 2015 | 2016 | 2017 | 2018 | 2019 | 2020* |
|---|---|---|---|---|---|---|---|---|
| Monitoring | 0 | 0 | 0 | 3 | 3 | 3 | 1 | 1 |
| Condition Analysis | 1 | 2 | 5 | 1 | 2 | 6 | 0 | 3 |
| Defect Analysis | 4 | 9 | 1 | 14 | 8 | 12 | 4 | 5 |

| CHAR | 2013 | 2014 | 2015 | 2016 | 2017 | 2018 | 2019 | 2020* |
|---|---|---|---|---|---|---|---|---|
| Descriptive | 1 | 3 | 3 | 5 | 7 | 8 | 2 | 2 |
| Diagnostic | 4 | 8 | 3 | 13 | 6 | 13 | 4 | 7 |
| Predictive | 0 | 0 | 0 | 0 | 0 | 0 | 0 | 0 |
| Prescriptive | 0 | 0 | 0 | 0 | 0 | 0 | 0 | 0 |

| CHAR | 2013 | 2014 | 2015 | 2016 | 2017 | 2018 | 2019 | 2020* |
|---|---|---|---|---|---|---|---|---|
| Time | 5 | 11 | 6 | 18 | 13 | 21 | 6 | 9 |
| Cost | 5 | 11 | 6 | 18 | 13 | 21 | 4 | 9 |

| CHAR | 2013 | 2014 | 2015 | 2016 | 2017 | 2018 | 2019 | 2020* |
|---|---|---|---|---|---|---|---|---|
| Machine/Tool | 5 | 11 | 6 | 18 | 12 | 19 | 6 | 9 |
| Process | 0 | 0 | 0 | 2 | 0 | 0 | 1 | 2 |

| CHAR | 2013 | 2014 | 2015 | 2016 | 2017 | 2018 | 2019 | 2020* |
|---|---|---|---|---|---|---|---|---|
| No Integration | 4 | 11 | 5 | 16 | 10 | 17 | 6 | 9 |
| Vertical | 1 | 0 | 1 | 0 | 2 | 3 | 0 | 0 |
| Horizontal | 0 | 0 | 0 | 2 | 3 | 3 | 0 | 0 |
| End-to-End | 0 | 0 | 0 | 0 | 0 | 0 | 0 | 0 |

| CHAR | 2013 | 2014 | 2015 | 2016 | 2017 | 2018 | 2019 | 2020* |
|---|---|---|---|---|---|---|---|---|
| Real-time/Stream | 1 | 0 | 2 | 1 | 0 | 3 | 0 | 2 |
| Historical/Batch | 4 | 11 | 4 | 17 | 13 | 18 | 6 | 7 |

| CHAR | 2013 | 2014 | 2015 | 2016 | 2017 | 2018 | 2019 | 2020* |
|---|---|---|---|---|---|---|---|---|
| Classification | 4 | 9 | 4 | 10 | 8 | 8 | 3 | 2 |
| Dimension Reduction | 0 | 2 | 1 | 2 | 0 | 1 | 0 | 2 |
| Deep Learning | 0 | 1 | 0 | 3 | 1 | 6 | 3 | 8 |
| Custom Development | 0 | 0 | 1 | 2 | 4 | 4 | 1 | 0 |

**Figure 7: Temporal Variations of the Archetype *Reactive Maintenance* (C5)**

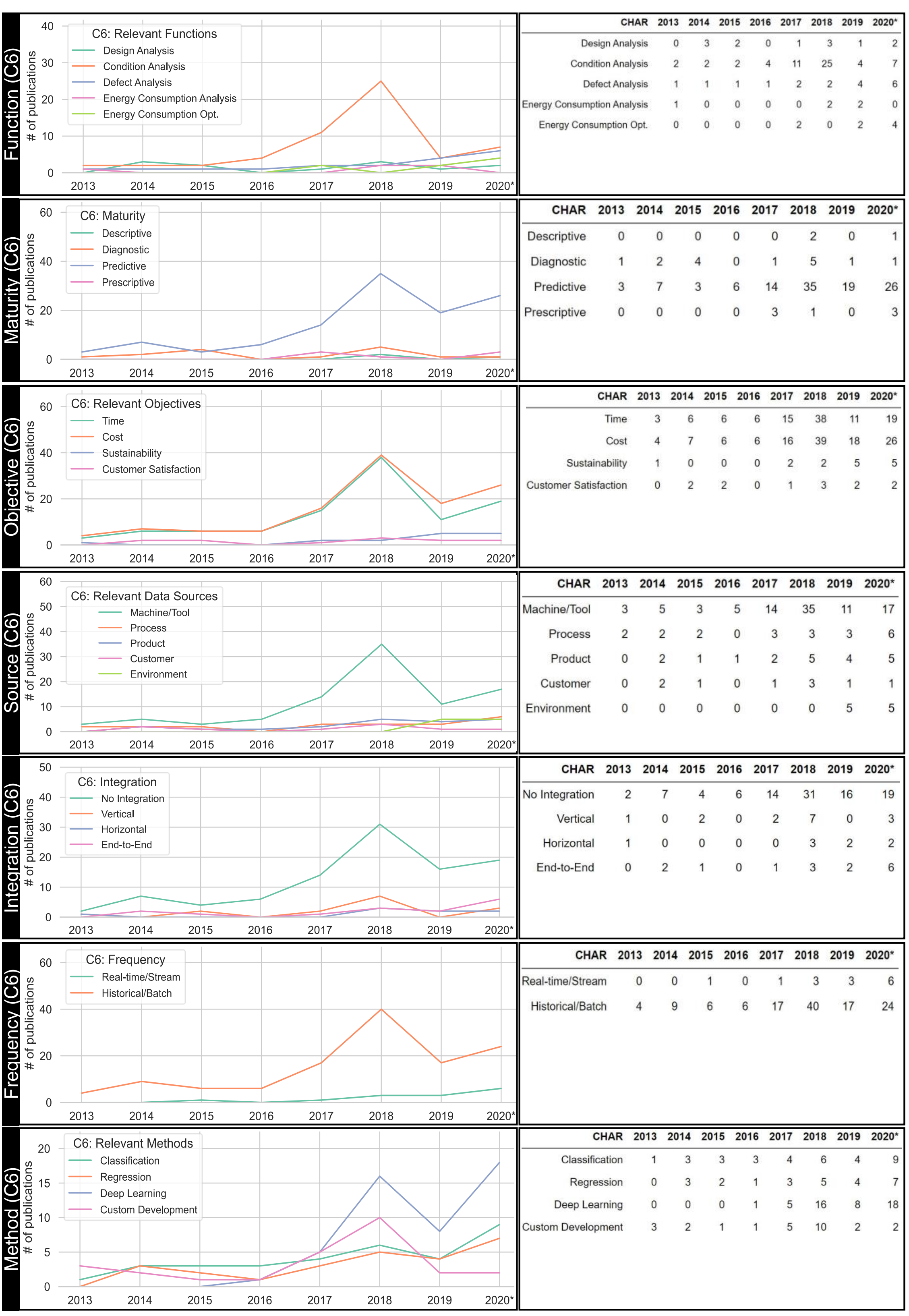

| CHAR | 2013 | 2014 | 2015 | 2016 | 2017 | 2018 | 2019 | 2020* |
|---|---|---|---|---|---|---|---|---|
| Design Analysis | 0 | 3 | 2 | 0 | 1 | 3 | 1 | 2 |
| Condition Analysis | 2 | 2 | 2 | 4 | 11 | 25 | 4 | 7 |
| Defect Analysis | 1 | 1 | 1 | 1 | 2 | 2 | 4 | 6 |
| Energy Consumption Analysis | 1 | 0 | 0 | 0 | 0 | 2 | 2 | 0 |
| Energy Consumption Opt. | 0 | 0 | 0 | 0 | 2 | 0 | 2 | 4 |

| CHAR | 2013 | 2014 | 2015 | 2016 | 2017 | 2018 | 2019 | 2020* |
|---|---|---|---|---|---|---|---|---|
| Descriptive | 0 | 0 | 0 | 0 | 0 | 2 | 0 | 1 |
| Diagnostic | 1 | 2 | 4 | 0 | 1 | 5 | 1 | 1 |
| Predictive | 3 | 7 | 3 | 6 | 14 | 35 | 19 | 26 |
| Prescriptive | 0 | 0 | 0 | 0 | 3 | 1 | 0 | 3 |

| CHAR | 2013 | 2014 | 2015 | 2016 | 2017 | 2018 | 2019 | 2020* |
|---|---|---|---|---|---|---|---|---|
| Time | 3 | 6 | 6 | 6 | 15 | 38 | 11 | 19 |
| Cost | 4 | 7 | 6 | 6 | 16 | 39 | 18 | 26 |
| Sustainability | 1 | 0 | 0 | 0 | 2 | 2 | 5 | 5 |
| Customer Satisfaction | 0 | 2 | 2 | 0 | 1 | 3 | 2 | 2 |

| CHAR | 2013 | 2014 | 2015 | 2016 | 2017 | 2018 | 2019 | 2020* |
|---|---|---|---|---|---|---|---|---|
| Machine/Tool | 3 | 5 | 3 | 5 | 14 | 35 | 11 | 17 |
| Process | 2 | 2 | 2 | 0 | 3 | 3 | 3 | 6 |
| Product | 0 | 2 | 1 | 1 | 2 | 5 | 4 | 5 |
| Customer | 0 | 2 | 1 | 0 | 1 | 3 | 1 | 1 |
| Environment | 0 | 0 | 0 | 0 | 0 | 0 | 5 | 5 |

| CHAR | 2013 | 2014 | 2015 | 2016 | 2017 | 2018 | 2019 | 2020* |
|---|---|---|---|---|---|---|---|---|
| No Integration | 2 | 7 | 4 | 6 | 14 | 31 | 16 | 19 |
| Vertical | 1 | 0 | 2 | 0 | 2 | 7 | 0 | 3 |
| Horizontal | 1 | 0 | 0 | 0 | 0 | 3 | 2 | 2 |
| End-to-End | 0 | 2 | 1 | 0 | 1 | 3 | 2 | 6 |

| CHAR | 2013 | 2014 | 2015 | 2016 | 2017 | 2018 | 2019 | 2020* |
|---|---|---|---|---|---|---|---|---|
| Real-time/Stream | 0 | 0 | 1 | 0 | 1 | 3 | 3 | 6 |
| Historical/Batch | 4 | 9 | 6 | 6 | 17 | 40 | 17 | 24 |

| CHAR | 2013 | 2014 | 2015 | 2016 | 2017 | 2018 | 2019 | 2020* |
|---|---|---|---|---|---|---|---|---|
| Classification | 1 | 3 | 3 | 3 | 4 | 6 | 4 | 9 |
| Regression | 0 | 3 | 2 | 1 | 3 | 5 | 4 | 7 |
| Deep Learning | 0 | 0 | 0 | 1 | 5 | 16 | 8 | 18 |
| Custom Development | 3 | 2 | 1 | 1 | 5 | 10 | 2 | 2 |

**Figure 8: Temporal Variations of the Archetype *Offline Predictive Maintenance* (C6)**

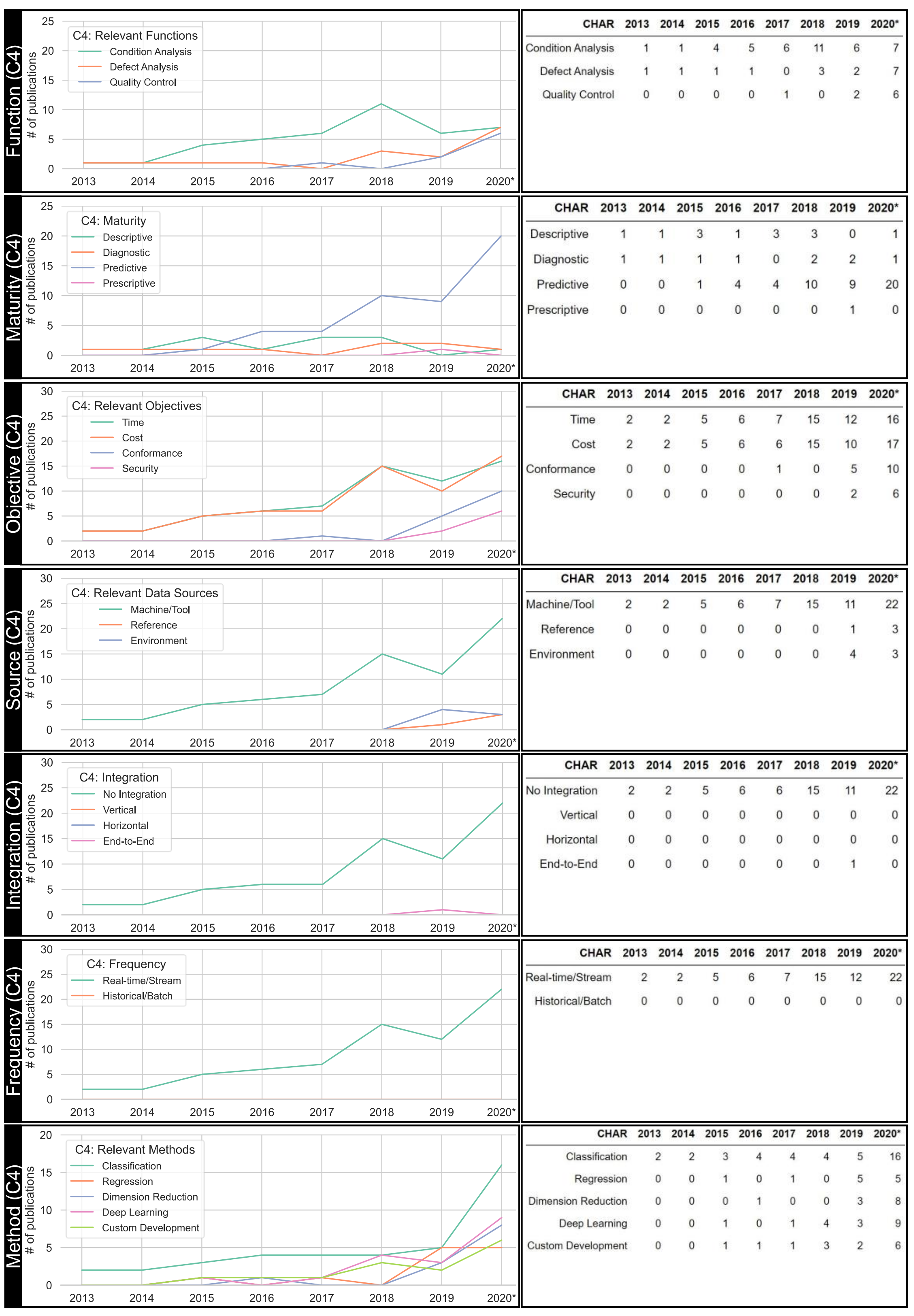

| CHAR | 2013 | 2014 | 2015 | 2016 | 2017 | 2018 | 2019 | 2020* |
|---|---|---|---|---|---|---|---|---|
| Condition Analysis | 1 | 1 | 4 | 5 | 6 | 11 | 6 | 7 |
| Defect Analysis | 1 | 1 | 1 | 1 | 0 | 3 | 2 | 7 |
| Quality Control | 0 | 0 | 0 | 0 | 1 | 0 | 2 | 6 |

| CHAR | 2013 | 2014 | 2015 | 2016 | 2017 | 2018 | 2019 | 2020* |
|---|---|---|---|---|---|---|---|---|
| Descriptive | 1 | 1 | 3 | 1 | 3 | 3 | 0 | 1 |
| Diagnostic | 1 | 1 | 1 | 1 | 0 | 2 | 2 | 1 |
| Predictive | 0 | 0 | 1 | 4 | 4 | 10 | 9 | 20 |
| Prescriptive | 0 | 0 | 0 | 0 | 0 | 0 | 1 | 0 |

| CHAR | 2013 | 2014 | 2015 | 2016 | 2017 | 2018 | 2019 | 2020* |
|---|---|---|---|---|---|---|---|---|
| Time | 2 | 2 | 5 | 6 | 7 | 15 | 12 | 16 |
| Cost | 2 | 2 | 5 | 6 | 6 | 15 | 10 | 17 |
| Conformance | 0 | 0 | 0 | 0 | 1 | 0 | 5 | 10 |
| Security | 0 | 0 | 0 | 0 | 0 | 0 | 2 | 6 |

| CHAR | 2013 | 2014 | 2015 | 2016 | 2017 | 2018 | 2019 | 2020* |
|---|---|---|---|---|---|---|---|---|
| Machine/Tool | 2 | 2 | 5 | 6 | 7 | 15 | 11 | 22 |
| Reference | 0 | 0 | 0 | 0 | 0 | 0 | 1 | 3 |
| Environment | 0 | 0 | 0 | 0 | 0 | 0 | 4 | 3 |

| CHAR | 2013 | 2014 | 2015 | 2016 | 2017 | 2018 | 2019 | 2020* |
|---|---|---|---|---|---|---|---|---|
| No Integration | 2 | 2 | 5 | 6 | 6 | 15 | 11 | 22 |
| Vertical | 0 | 0 | 0 | 0 | 0 | 0 | 0 | 0 |
| Horizontal | 0 | 0 | 0 | 0 | 0 | 0 | 0 | 0 |
| End-to-End | 0 | 0 | 0 | 0 | 0 | 0 | 1 | 0 |

| CHAR | 2013 | 2014 | 2015 | 2016 | 2017 | 2018 | 2019 | 2020* |
|---|---|---|---|---|---|---|---|---|
| Real-time/Stream | 2 | 2 | 5 | 6 | 7 | 15 | 12 | 22 |
| Historical/Batch | 0 | 0 | 0 | 0 | 0 | 0 | 0 | 0 |

| CHAR | 2013 | 2014 | 2015 | 2016 | 2017 | 2018 | 2019 | 2020* |
|---|---|---|---|---|---|---|---|---|
| Classification | 2 | 2 | 3 | 4 | 4 | 4 | 5 | 16 |
| Regression | 0 | 0 | 1 | 0 | 1 | 0 | 5 | 5 |
| Dimension Reduction | 0 | 0 | 0 | 1 | 0 | 0 | 3 | 8 |
| Deep Learning | 0 | 0 | 1 | 0 | 1 | 4 | 3 | 9 |
| Custom Development | 0 | 0 | 1 | 1 | 1 | 3 | 2 | 6 |

**Figure 9: Temporal Variations of the Archetype *Online Predictive Maintenance* (C4)**

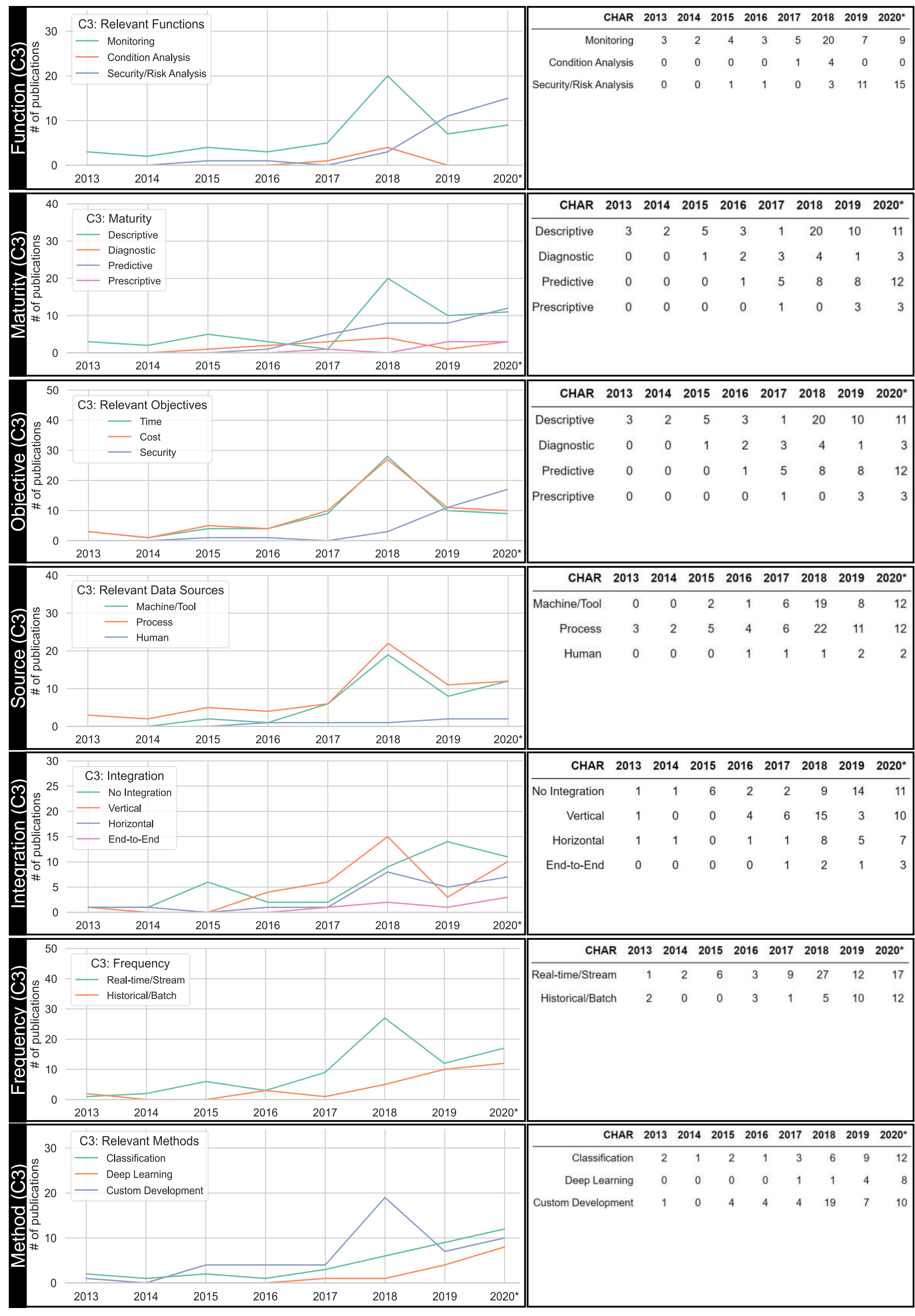

| CHAR | 2013 | 2014 | 2015 | 2016 | 2017 | 2018 | 2019 | 2020* |
|---|---|---|---|---|---|---|---|---|
| Monitoring | 3 | 2 | 4 | 3 | 5 | 20 | 7 | 9 |
| Condition Analysis | 0 | 0 | 0 | 0 | 1 | 4 | 0 | 0 |
| Security/Risk Analysis | 0 | 0 | 1 | 1 | 0 | 3 | 11 | 15 |

| CHAR | 2013 | 2014 | 2015 | 2016 | 2017 | 2018 | 2019 | 2020* |
|---|---|---|---|---|---|---|---|---|
| Descriptive | 3 | 2 | 5 | 3 | 1 | 20 | 10 | 11 |
| Diagnostic | 0 | 0 | 1 | 2 | 3 | 4 | 1 | 3 |
| Predictive | 0 | 0 | 0 | 1 | 5 | 8 | 8 | 12 |
| Prescriptive | 0 | 0 | 0 | 0 | 1 | 0 | 3 | 3 |

| CHAR | 2013 | 2014 | 2015 | 2016 | 2017 | 2018 | 2019 | 2020* |
|---|---|---|---|---|---|---|---|---|
| Descriptive | 3 | 2 | 5 | 3 | 1 | 20 | 10 | 11 |
| Diagnostic | 0 | 0 | 1 | 2 | 3 | 4 | 1 | 3 |
| Predictive | 0 | 0 | 0 | 1 | 5 | 8 | 8 | 12 |
| Prescriptive | 0 | 0 | 0 | 0 | 1 | 0 | 3 | 3 |

| CHAR | 2013 | 2014 | 2015 | 2016 | 2017 | 2018 | 2019 | 2020* |
|---|---|---|---|---|---|---|---|---|
| Machine/Tool | 0 | 0 | 2 | 1 | 6 | 19 | 8 | 12 |
| Process | 3 | 2 | 5 | 4 | 6 | 22 | 11 | 12 |
| Human | 0 | 0 | 0 | 1 | 1 | 1 | 2 | 2 |

| CHAR | 2013 | 2014 | 2015 | 2016 | 2017 | 2018 | 2019 | 2020* |
|---|---|---|---|---|---|---|---|---|
| No Integration | 1 | 1 | 6 | 2 | 2 | 9 | 14 | 11 |
| Vertical | 1 | 0 | 0 | 4 | 6 | 15 | 3 | 10 |
| Horizontal | 1 | 1 | 0 | 1 | 1 | 8 | 5 | 7 |
| End-to-End | 0 | 0 | 0 | 0 | 1 | 2 | 1 | 3 |

| CHAR | 2013 | 2014 | 2015 | 2016 | 2017 | 2018 | 2019 | 2020* |
|---|---|---|---|---|---|---|---|---|
| Real-time/Stream | 1 | 2 | 6 | 3 | 9 | 27 | 12 | 17 |
| Historical/Batch | 2 | 0 | 0 | 3 | 1 | 5 | 10 | 12 |

| CHAR | 2013 | 2014 | 2015 | 2016 | 2017 | 2018 | 2019 | 2020* |
|---|---|---|---|---|---|---|---|---|
| Classification | 2 | 1 | 2 | 1 | 3 | 6 | 9 | 12 |
| Deep Learning | 0 | 0 | 0 | 0 | 1 | 1 | 4 | 8 |
| Custom Development | 1 | 0 | 4 | 4 | 4 | 19 | 7 | 10 |

**Figure 10: Temporal Variations of the Archetype *MRO Monitoring* (C3)**

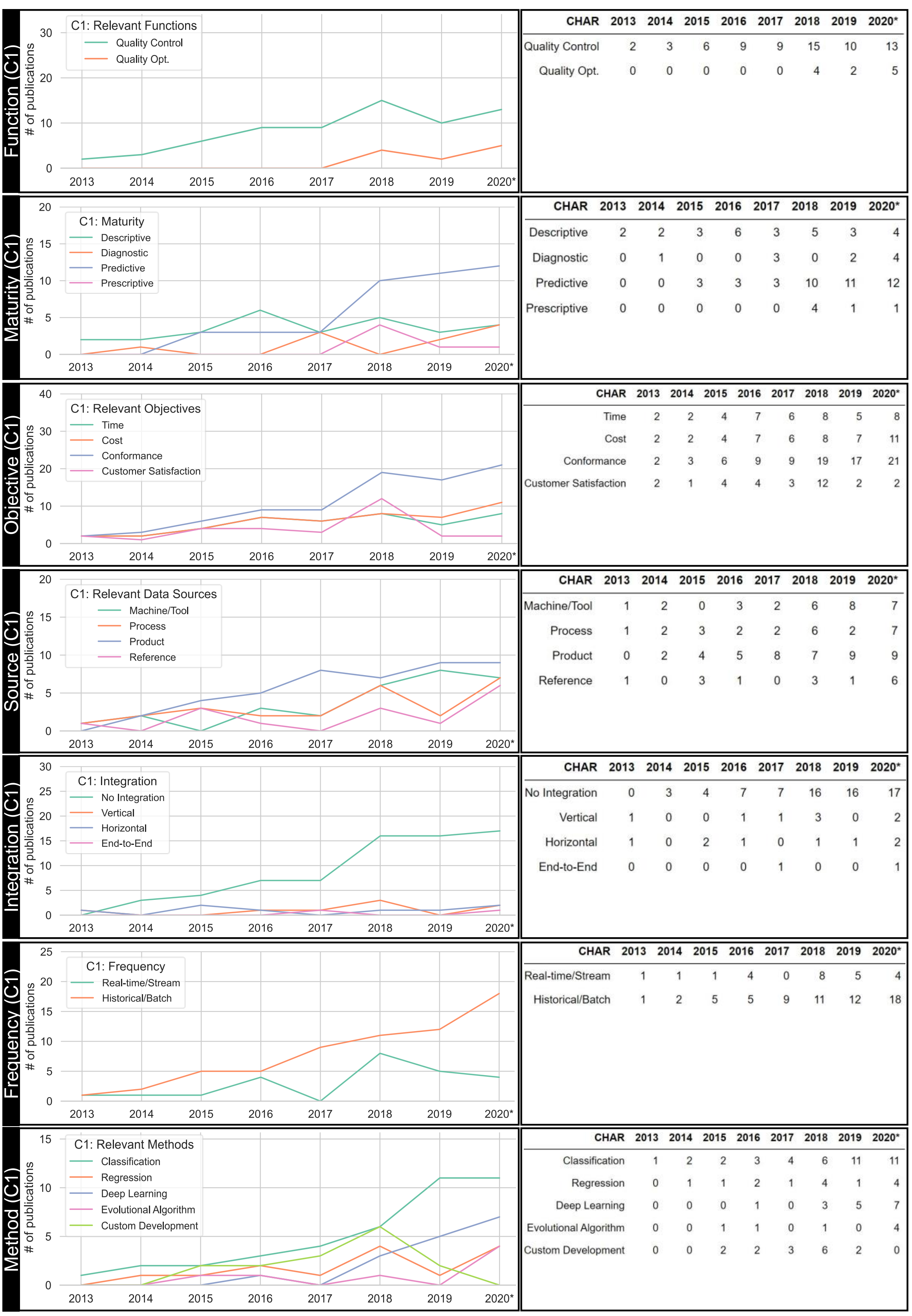

| CHAR | 2013 | 2014 | 2015 | 2016 | 2017 | 2018 | 2019 | 2020* |
|---|---|---|---|---|---|---|---|---|
| Quality Control | 2 | 3 | 6 | 9 | 9 | 15 | 10 | 13 |
| Quality Opt. | 0 | 0 | 0 | 0 | 0 | 4 | 2 | 5 |

| CHAR | 2013 | 2014 | 2015 | 2016 | 2017 | 2018 | 2019 | 2020* |
|---|---|---|---|---|---|---|---|---|
| Descriptive | 2 | 2 | 3 | 6 | 3 | 5 | 3 | 4 |
| Diagnostic | 0 | 1 | 0 | 0 | 3 | 0 | 2 | 4 |
| Predictive | 0 | 0 | 3 | 3 | 3 | 10 | 11 | 12 |
| Prescriptive | 0 | 0 | 0 | 0 | 0 | 4 | 1 | 1 |

| CHAR | 2013 | 2014 | 2015 | 2016 | 2017 | 2018 | 2019 | 2020* |
|---|---|---|---|---|---|---|---|---|
| Time | 2 | 2 | 4 | 7 | 6 | 8 | 5 | 8 |
| Cost | 2 | 2 | 4 | 7 | 6 | 8 | 7 | 11 |
| Conformance | 2 | 3 | 6 | 9 | 9 | 19 | 17 | 21 |
| Customer Satisfaction | 2 | 1 | 4 | 4 | 3 | 12 | 2 | 2 |

| CHAR | 2013 | 2014 | 2015 | 2016 | 2017 | 2018 | 2019 | 2020* |
|---|---|---|---|---|---|---|---|---|
| Machine/Tool | 1 | 2 | 0 | 3 | 2 | 6 | 8 | 7 |
| Process | 1 | 2 | 3 | 2 | 2 | 6 | 2 | 7 |
| Product | 0 | 2 | 4 | 5 | 8 | 7 | 9 | 9 |
| Reference | 1 | 0 | 3 | 1 | 0 | 3 | 1 | 6 |

| CHAR | 2013 | 2014 | 2015 | 2016 | 2017 | 2018 | 2019 | 2020* |
|---|---|---|---|---|---|---|---|---|
| No Integration | 0 | 3 | 4 | 7 | 7 | 16 | 16 | 17 |
| Vertical | 1 | 0 | 0 | 1 | 1 | 3 | 0 | 2 |
| Horizontal | 1 | 0 | 2 | 1 | 0 | 1 | 1 | 2 |
| End-to-End | 0 | 0 | 0 | 0 | 1 | 0 | 0 | 1 |

| CHAR | 2013 | 2014 | 2015 | 2016 | 2017 | 2018 | 2019 | 2020* |
|---|---|---|---|---|---|---|---|---|
| Real-time/Stream | 1 | 1 | 1 | 4 | 0 | 8 | 5 | 4 |
| Historical/Batch | 1 | 2 | 5 | 5 | 9 | 11 | 12 | 18 |

| CHAR | 2013 | 2014 | 2015 | 2016 | 2017 | 2018 | 2019 | 2020* |
|---|---|---|---|---|---|---|---|---|
| Classification | 1 | 2 | 2 | 3 | 4 | 6 | 11 | 11 |
| Regression | 0 | 1 | 1 | 2 | 1 | 4 | 1 | 4 |
| Deep Learning | 0 | 0 | 0 | 1 | 0 | 3 | 5 | 7 |
| Evolutional Algorithm | 0 | 0 | 1 | 1 | 0 | 1 | 0 | 4 |
| Custom Development | 0 | 0 | 2 | 2 | 3 | 6 | 2 | 0 |

**Figure 11: Temporal Variations of the Archetype *Quality Management* (C1)**